\documentclass[letter]{aa} % for the letters 
\usepackage{graphicx}
\usepackage{amsmath}
\usepackage{lscape}

\RequirePackage{color}
\definecolor{darkgreen}{rgb}{0.0, 0.4, 0.0}
\definecolor{darkorange}{rgb}{0.5, 0.3, 0.0}
\definecolor{darkblue}{rgb}{0.17, 0.49, 0.72}

\newcommand{\kms}[1]{{km s$^{-1}$#1}}

\usepackage{multirow}
\usepackage{multicol}
\usepackage{txfonts}
\usepackage{booktabs}

\usepackage{pstricks}
\usepackage{enumitem}
\usepackage{amsmath}
\usepackage{amssymb}
\newcommand{\numcefs}{384}
\newcommand{\nars}{97}
\newcommand{\narswithcef}{81}
\newcommand{\narswithcefper}{83.5\%}
\newcommand{\narswithoutcef}{16}
\newcommand{\narswithoutcefper}{16.5\%}
\newcommand{\clv}{-275\,m\,s$^{-1}$}
\newcommand{\vsdo}{v^{\rm SDO}_{\rm LOS}}
\newcommand{\ms}{m\,s$^{-1}$}
\newcommand{\cefmedian}{6}
\newcommand{\perobsceftotal}{5.9\%}

\newcommand{\totdurobs}{$9.6\pm1.4$\,days}

\newcommand{\ceflifetime}{10.6$_{-6.0}^{+12.4}$\,hr}
\newcommand{\meanceflifetime}{13.5\,hr}

\newcommand{\lbceflifetime}{15.4$\pm10.1$\,hr}
\newcommand{\nLBceflifetime}{11.9$\pm9.3$\,hr}
\newcommand{\difflifetimes}{3.5\,hr}

\linethickness{.8mm}
\usepackage[pdftex]{hyperref}
\hypersetup{%
  colorlinks=true,% hyperlinks will be coloured
  linkcolor=blue,% hyperlink text will be green
  linkbordercolor=blue,% hyperlink border will be red
  citecolor=blue,
  urlcolor=blue,
}
\graphicspath{{figs_cefs/}}

   \title{How rare are counter Evershed flows?}
   \titlerunning{How rare are counter Evershed flows?}
   \authorrunning{{Castellanos~Dur\'an}, Lagg \& Solanki}
   \author{J.~Sebasti\'an {Castellanos~Dur\'an}\inst{1,2}\thanks{\hbox{\email{castellanos@mps.mpg.de}}}, Andreas Lagg\inst{1,4}, and Sami~K. Solanki\inst{1,3}}
   \institute{Max Planck Institute for Solar System Research, Justus-von-Liebig-Weg 3, D-37077. G\"ottingen, Germany.\label{inst1}\and
   Georg-August-Universit\"at G\"ottingen, Friedrich-Hund-Platz 1, D-37077. G\"ottingen, Germany.\label{inst2}\and
   School of Space Research, Kyung Hee University, Yongin, 446-101, Gyeonggi, Republic of Korea.\label{inst3}
   \and
   Department of Computer Science, Aalto University, PO Box 15400, FI-00076 Aalto, Finland \label{inst4}
   }

   \date{Received: 23 April 2021; Accepted 9 June 2021}
\begin{document} 

  % context heading (option
   \abstract
  {One of the main characteristics of the penumbra of sunspots is the radially outward-directed Evershed flow. Only recently have penumbral regions been reported with similar characteristics to normal penumbral filaments, but with an opposite direction of the flow. Such flows directed towards the umbra are known as counter Evershed flows (CEFs). We aim to determine the frequency of occurrence of CEFs in active regions (ARs) and to characterize their lifetime and the prevailing conditions in the ARs. We analysed the continuum images, Dopplergrams, and magnetograms recorded by SDO/HMI of \nars{} ARs that appeared from 2011 to 2017. We followed the ARs for \totdurobs{} on average. We found \numcefs{} CEFs in total, with a median value of \cefmedian{} CEFs per AR. CEFs are a rather common feature, they occur in \narswithcefper{} of all ARs regardless of the magnetic complexity of the AR. However, CEFs were observed on average only during \perobsceftotal{} of the mean total duration of all the observations analyzed here. The lifetime of CEFs follows a log-normal distribution with a median value of \ceflifetime{}. In addition, we report two populations of CEFs depending on whether they are associated with light bridges, or not. We explain that the rarity of reports of CEFs in the literature is a result of highly incomplete coverage of ARs with spectropolarimetric data. By using the continuous observations now routinely available from space, we are able to overcome this limitation.}

   \maketitle

\section{Introduction}\label{sec:intro}

Sunspots are a manifestation of solar magnetism. The two main constituents of sunspots are the dark umbra, harbouring a strong and relatively vertical magnetic field ($B$); and the penumbra, highly filamentary in continuous radiation, with a more horizontal $B$. 

A radially outward-directed flow along the penumbral produces Dopplershifted photospheric spectral lines\footnote{In this paper, the radial direction is taken to be parallel to the solar surface and ascribed from the umbra-penumbra boundary, across the penumbra, towards the quiet-sun.} when the sunspot is observed away from disk center. This Evershed flow \citep[see ][for a review]{Solanki2003} was detected more than a century ago \citep{Evershed1909}. In the past decade, there have been a few reports of {\it peculiar} flows directed towards the umbra \citep{Kleint2013ApJ,Louis2014A&A...CEF,Siu-Tapia2017A&A,Guglielmino2017ApJ,Guglielmino2019ApJ,Louis2020...cefs}. These so-called counter Evershed flows (CEFs) have also been observed in one magneto-hydrodynamic (MHD) simulation \citep{Siu-Tapia2018ApJ}. CEFs studied in the literature have been found in complex active regions (ARs). Recently,  \citet{Louis2020...cefs} reported the appearance of a light bridge associated with a CEF. Common to these works, which all only analyzed one sunspot each, is that  CEFs were described as {\it anomalous}, {\it unusual}, or {\it atypical} flows.

Photospheric CEFs should not be confused with the more commonly reported chromospheric inverse Evershed flows \citep{StJohn1911ApJa....IEF,StJohn1911ApJb....IEF,Maltby1975SoPh...IEF,Choudhary2018ApJ...IEF,Beck2019ApJ...IEF}. Inverse Evershed flows transport material in the chromospheric penumbrae towards the umbra and they are thought to be driven by pressure gradients \citep[siphon flows, e.g.,][]{Thomas1988ApJ..IEF}.

The few existing reports of CEFs might suggest that this type of photospheric flow is a rare phenomenon in sunspots. However, until now, no study systematically looked at a large set of ARs to quantify how often CEFs occur in ARs, and whether their occurrence depends on the magnetic complexity of the host AR. In this Letter, we fill this gap and present the first analysis of CEFs in a large sample of ARs using 6\,years of space-borne data.

\section{Observational data and analysis} \label{sec:obs}

We analyzed \nars{} ARs observed by the Solar Dynamic Observatory \citep[SDO;][]{Pesnell2012} taken by the Helioseismic and Magnetic Imager \cite[HMI;][]{Scherrer2012,Schou2012}. We analyzed the continuum intensity ($I_{\rm c}$), Dopplergrams ($v_{{\rm LOS}}$), and magnetograms ($B_{{\rm LOS}}$) with a spatial resolution of 1\arcsec{}. Data were taken with a cadence of 12 minutes and covered the development of each AR over a time ranging from 153 to 278\,hr. Table~\ref{tab:OtherCEFs} summarizes the observations. Data were processed using standard \texttt{SSWIDL} routines, and temporal series were created by co-aligning the images using cross-correlations between the subsequent frames. 

The \nars{} ARs were observed between 2011 July 30 and 2017 August 24. ARs were tracked continuously while they crossed the solar disk for \totdurobs{} on average. The only criteria applied when selecting the ARs was that it had sunspots and could be followed for at least 6 consecutive days, independently of whether individual sunspots within the AR emerge/decay during this period. This criterion excludes many small ARs with only small, short-lived sunspots, also many ARs that are formed when already well on the disc, or ARs that are already decaying when they appear at the East limb. It might also exclude some ARs that appeared during the SDO's eclipsing seasons. We analyzed $\sim$1.1$\times10^5$ individual time-steps ($\times$3 for the $I_{\rm c}$, $v_{\rm LOS}$, $B_{\rm LOS}$).

Our sample covers all classification types of  ARs, from the most simple to the more complex ones. The magnetic complexity of solar ARs is regularly described by Hale's classification \citep{Hale1919ApJ}. In this classification, an AR is assigned to the category $\alpha$ if it contains one or multiple sunspots of unique polarity, $\beta$ refers to ARs harboring bipolar sunspots or groups, and $\gamma$ describes complex ARs with sunspots and groups of intermixed polarities. An amendment to Hale's classification was done in the 1960s by adding an extra category, $\delta$, to describe those ARs that harbour umbrae of mixed polarities enclosed by a common penumbra within $<\!2^{\circ}$ \citep{Kunzel1960AN...DeltaSpots,Kunzel1965AN....spots}. 
The categories can be appended to describe the complexity of an AR increasing from $\alpha$ to $\beta\gamma\delta$. In our sample, we take into account the change of the magnetic classification for each AR during their lifetimes. Considering that bipolar-$\beta$ ARs are the most common type observed on the Sun \citep[$\sim$65\%;][]{Jaeggli2016ApJL...ClassARs}, our sample is also dominated by this AR type.

The determination of the zero-level of the line-of-sight velocity measurements requires taking into account the following effects: (1) the line-of-sight velocity of the observatory with respect to the Sun ($\vsdo{}$), (2) the large-scale flows (LSF) on the solar surface including solar differential rotation and meridional circulation, (3) the center-to-limb variation of the convective blueshift, and (4) the gravitational redshift.  For SDO, the value for $\vsdo{}$ is estimated by combining the keywords \texttt{OBS\_VR}, \texttt{OBS\_VW}, and \texttt{OBS\_VN} in the header of the HMI data files. Following \citet{Schuck2016ApJ}, $\vsdo{}$ in the helioprojective coordinate system ($\theta_{\rho},\psi)$ is given by
\begin{multline}\label{eq:radial}
    \vsdo{}=\texttt{OBS\_VW}\sin\theta_{\rho}\sin\psi\\-\texttt{OBS\_VN}\sin\theta_{\rho}\cos\psi+\texttt{OBS\_VR}\cos\theta_{\rho},
\end{multline}
where these keywords provide information about the speed of the observatory in the radial direction from the Sun (\texttt{OBS\_VR}), westward in direction of Earth orbit (\texttt{OBS\_VW}), and northward in the direction of the solar north (\texttt{OBS\_VN}). \citet{Thompson2006} provides the conversion between coordinates systems.

\begin{figure}[htbp]
 \begin{center}
 \includegraphics[width=.48\textwidth]{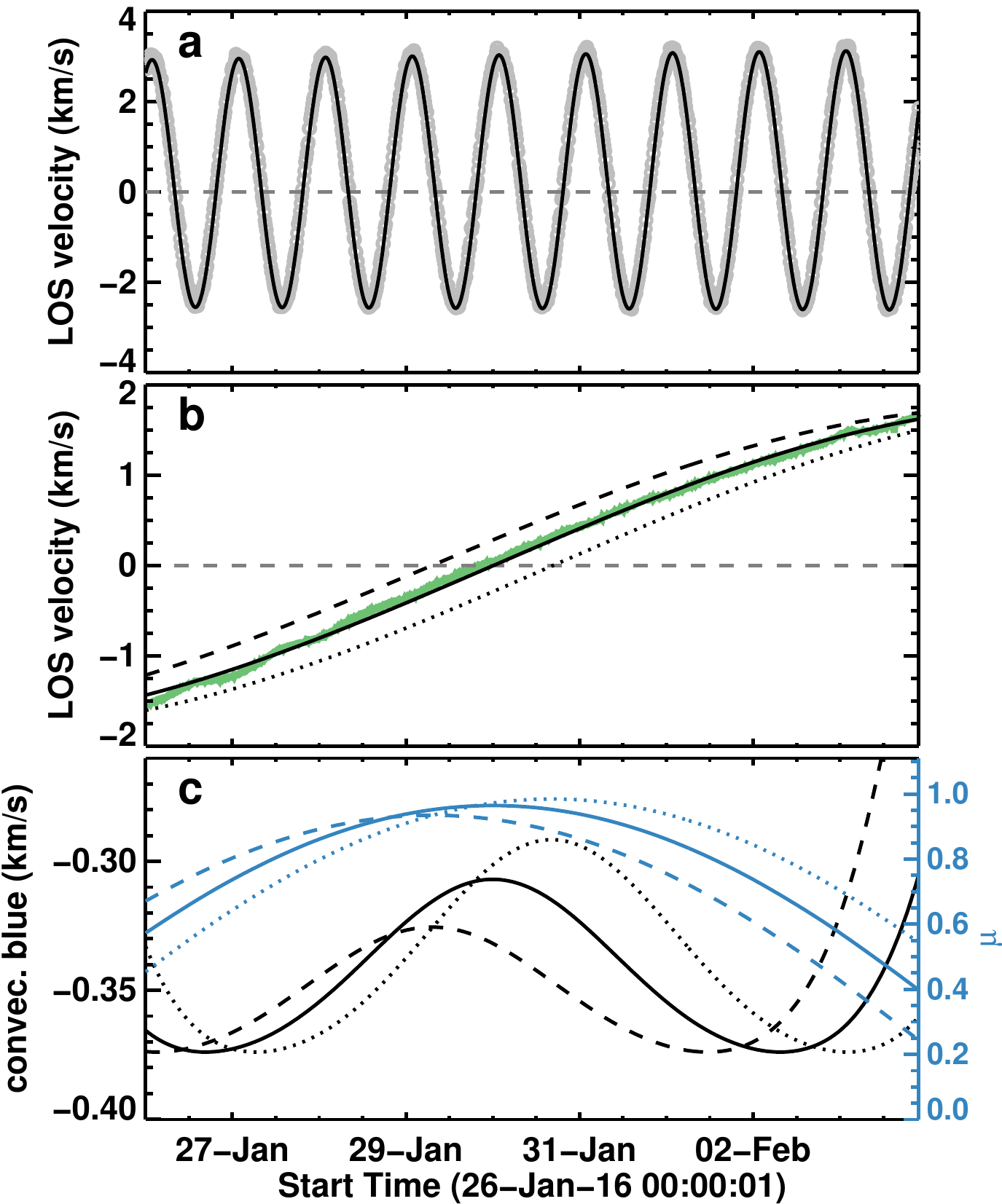}
 \caption{Examples of the calibration of the Dopplergrams measured by SDO/HMI for AR\,12489. (a) Average LOS velocities estimated in the quiet-sun at disk center as a proxy of the observatory velocity $\vsdo{}$. The black line is the $\vsdo{}$ obtained by Eq.~\ref{eq:radial}. (b) Black lines are the solar differential rotation obtained from Eqs.~\ref{eq:solrotLOS}-\ref{eq:meridional} for the lower-left (dotted line), middle (solid line) and top-right (dashed line) pixels inside the FoV with a size of 310\arcsec{}$\times$185\arcsec{}. Green line is the average LOS velocity of the adjacent quiet-sun. (c) Black line shows convective blueshift estimated using Eq.~\ref{eq:convectblue} at the same three pixels inside the FoV with a $\mu$-value given by the blue lines.  \label{fig:loscalibration}}
 \end{center}
 \end{figure}

An alternative method to account for (1) is based on the HMI Doppler- and magnetograms to estimate the average quiet-sun velocity calculated from a region $\pm$15\arcsec{} around disk center. Strong magnetic concentrations are avoided by masking regions with $|B_{\rm LOS}|>500$\,G. The good agreement between the average quiet-sun velocities and the velocity information taken from the header of the data files can be gleaned from Fig.~\ref{fig:loscalibration}a, which shows the diurnal variation of the spacecraft velocity of up to $\pm$3\,\kms{}. Note that the peak-to-peak amplitude and mean of $\vsdo{}$ changes in the course of the year.

The correction for the large-scale flows (2) has two components: differential solar rotation ($v_{\rm rot_{\sun}}$) and the surface meridional flow ($v_{\rm mer_{\sun}}$). We calculate their contribution at every pixel in our maps based on its Stonyhurst heliographic coordinates latitude and longitude $(\Theta,\Phi)$.  The line-of-sight component of these LSFs, for $\theta_{\rho}\approx0$\footnote{See Eqs.~39a-39c in \citet{Schuck2016ApJ} for observations taken closer to the Sun (e.g., Solar Orbiter), where the maximum value of $\theta_{\rho}$, that is obtained at the limb $\sin\theta_{\rho}=R_{\sun}/D_{\sun}$, is not negligible.}, is given by,
\begin{multline}\label{eq:solrotLOS}
\left.v_{\rm LSF_{\sun}}(\Phi,\Theta)\right|_{\rm LOS}=[v_{\rm med_{\sun}}(\Theta)][\sin B_0\cos\Theta-\cos B_0 \cos\Phi\sin\Theta]\\
-[v_{\rm rot_{\sun}}(\Theta)- v_{\rm Carrington}+v_{\rm synodic}]\cos B_{0} \sin \Phi ,
\end{multline}
where $v_{\rm Carrington}=14.184$\,deg\,day$^{-1}$ is the Carrington rotation rate, corresponding to a velocity of $1994.21$\,\ms{} at the equator. $v_{\rm synodic}=0.986$\,deg\,day$^{-1}$ is the mean orbital angular velocity of the observer to account for the synodic rotation between the Sun and Earth. At the equator $v_{\rm synodic}=138.63$\,\ms{}. The differential solar rotation is given by,
\begin{equation}\label{eq:rotation}
v_{\rm rot_{\sun}}(\Theta)=\left(a+b \sin ^{2} \Theta+c \sin ^{4} \Theta\right) \cos \Theta\,[{\rm m\,s}^{-1}],
\end{equation}
and the surface meridional flow is given by,
\begin{equation}\label{eq:meridional}
v_{\rm mer_{\sun}}(\Theta)=\left(d \sin \Theta+e \sin ^{3} \Theta\right) \cos \Theta\,[{\rm m\,s}^{-1}],
\end{equation}
where coefficients in Eqs.~\ref{eq:rotation} and \ref{eq:meridional} are $a=35.6$\,\ms{}, $b=-208.6$\,\ms{}, $c=-420.6$\,\ms{}, $d=29.7$\,\ms{}, and $e=-17.7$\,\ms{} obtained by \citet{Hathaway2011ApJ}. The contribution of the surface meridional flow is tiny and therefore negligible.  Fig.~\ref{fig:loscalibration}b shows examples of the correction for differential rotation for the bottom-left, center, and top-right pixels within the field-of-view (FoV) for the AR\,12489. For comparison, the green line shows the average quiet-sun velocity within the FoV. Notice that this average is influenced by the location of the sunspots within the AR inside the observed FoV, which may change as they evolve and change shape over multiple days.
 
To compensate for the center-to-limb variation of the convective blueshift, we reconstructed the profile by \cite[][their Fig.~2c]{LoehnerBoettcher2013A&A...convectiveblue}, and shifted the profile to match it with the convective blueshift value at disk center measured by the Laser Absolute Reference Spectrograph \citep[LARS;][]{Doerr2015PhDT,LoehnerBoettcher2017A&A...LARS}. LARS performed absolute
wavelength calibrated observations with a resolving power  of $\sim$700\,000 at 6173\,\AA{}, determining the convective blueshift for the \ion{Fe}{I}\,6173.3\,\AA{} line at disk center to be -320\,m\,s$^{-1}$ \citep{Stief2019A&A...blueshift}. At HMI's resolving power of $\sim$81\,000 this value reduces to \clv{} \citep{LoehnerBoettcher2019A&A...624A..57L}. Similar to \citet{Stief2019A&A...blueshift}, we fitted the center-to-limb variation with a fifth degree polynomial, given by
\begin{multline}\label{eq:convectblue}
v_{\mathrm{CLV}}(\mu)=131 -1179 \mu-2029 \mu^{2}\\
+9112 \mu^{3}-10409\mu^{4}+4096\mu^{5}~[\mathrm{m} \mathrm{~s}^{-1}],
\end{multline}
where $\mu=\cos\theta$ and $\theta$ is the heliocentric angle. To correct for this effect, we obtained the $\mu$-values for all pixels inside FoV and subtract the result of Eq.~\ref{eq:convectblue} from the $v_{\rm LOS}$. An example of this correction, taken at the same three pixels inside the FoV, is shown in Fig.~\ref{fig:loscalibration}c.

The gravitational redshift is a relativistic effect and a direct application of the equivalence principle for the light traveling from the Sun to the Earth. The gravitational Doppler shift is given by,
\begin{equation}\label{eq:Grav_redshift}
\Delta\lambda_{\rm G}=\lambda Gm_{\sun}/R_{\sun}c^2
\therefore v_{\rm G}= 636.03\,[{\rm m\,s}^{-1}],
\end{equation} 
where all the constants have their nominal meaning.

Continuum images were corrected for limb darkening following the procedure explained by \citet{CastellanosDuran2020...blow2wl}. To select the penumbra, we use the continuum images and set the commonly used limits $0.5I_{\rm qs}\leq I_{\rm penumbra}\leq0.97I_{\rm qs}$, where $I_{\rm qs}$ is the quiet-sun mean intensity.

Due to the Evershed flow, the limb-side penumbra is observed redshifted, while the center-side is blueshifted. This pattern is reversed if a sector of the penumbra is carrying a CEF. A CEF appears blueshifted if it is located on the limbward side, and redshifted on the center side. When searching for candidates of CEFs, we put the constraint that these events should appear in penumbrae and the penumbral region should have an adjacent umbra. The second condition avoids orphan penumbrae that tend to show strong flows, but where it is non-trivial to establish whether the flow direction corresponds to the normal Evershed flow or not.

\begin{figure}[tbhp]
 \begin{center}
 \includegraphics[width=.48\textwidth]{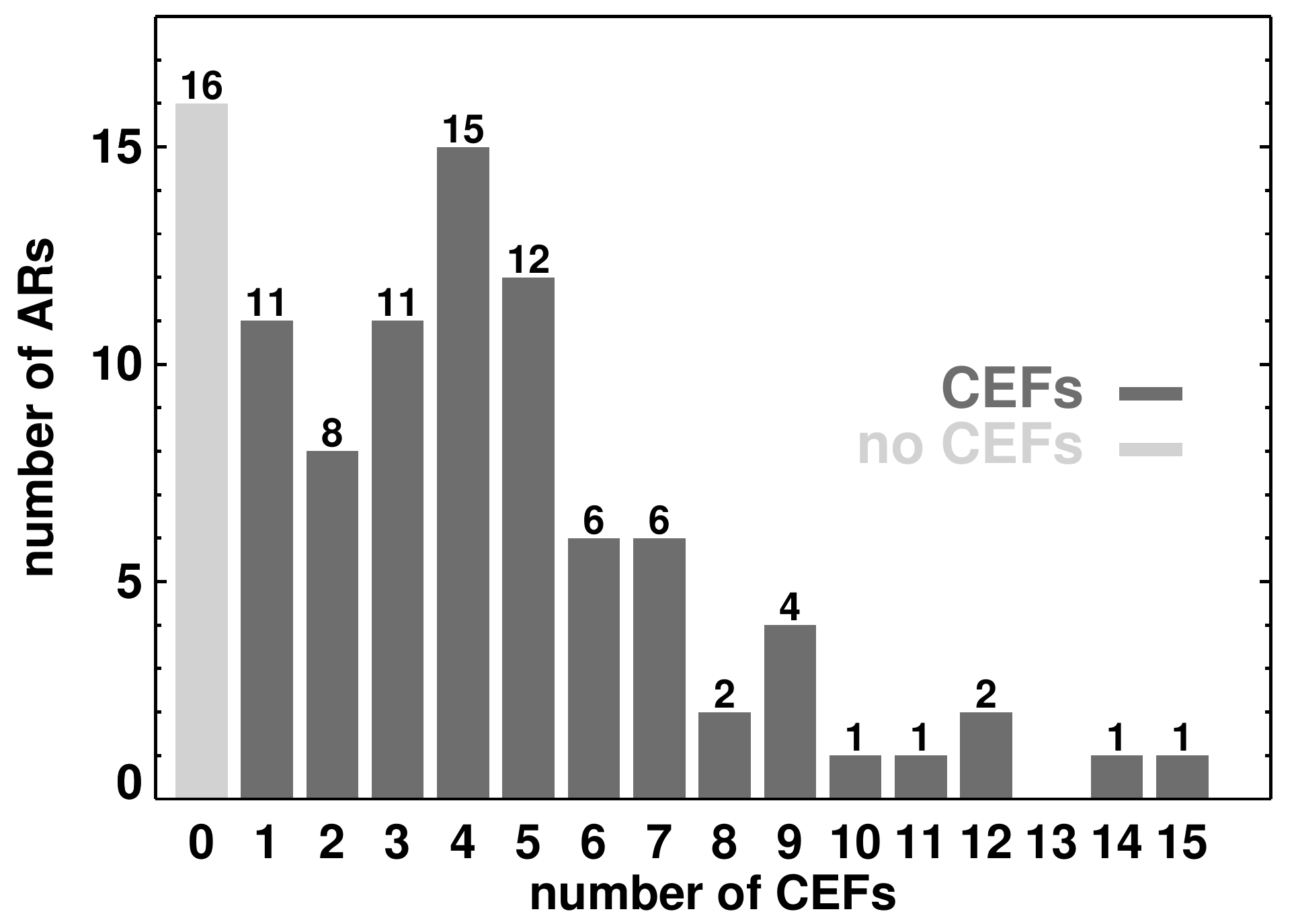}
 \caption{Number of CEFs per AR. The median value of CEFs per AR is \cefmedian{}.   An overview of the \narswithoutcef{} ARs without any CEF is presented in  Appendix~\ref{sec:arnocefs}. 
 \label{fig:numcefperar}}
 \end{center}
 \end{figure} 

Candidates to CEFs were found using the \texttt{SOBEL} edge-enhancement and \texttt{CANNY} edge-detection algorithms, both implemented in the Interactive Data Language \texttt{IDL8.3} (Exelis Visual Information Solutions, Boulder, Colorado)
to detect breaks in the Dopplergrams within the penumbrae. In addition, we calculated the spatial gradient of $v_{\rm LOS}$, where the position of the maximum gradient often outlines the CEF region very well. 
However, this method sometimes failed, especially when the CEF was orientated closely parallel to the direction of the closest limb. It is difficult to detect CEFs that are aligned within $10-15^\circ$ to the direction of the nearest solar limb (these flows are perpendicular to the LOS and hence produce almost no Doppler shift). We, therefore, miss between 5\% and 10\% of all the CEFs in the studied ARs, if we assume that they are isotropically distributed. We additionally checked all sunspots within the ARs by visual inspection of videos showing their temporal evolution for CEFs which were not identified by the method described above.

We used the Solar Region Summary (SRS) provided by the Space Weather Prediction Center (\href{https://www.swpc.noaa.gov/}{SWPC}) to determine parameters such as the NOAA AR number, the heliographic latitude, longitude, and area of the AR, and magnetic classification of the AR. We completed the records for a few ARs that were excluded from the SRS reports because they emerged on the same day, but after the SRS reports were issued at 00:30\,UT.

\begin{figure}[tbhp]
 \begin{center}
 \includegraphics[width=.5\textwidth]{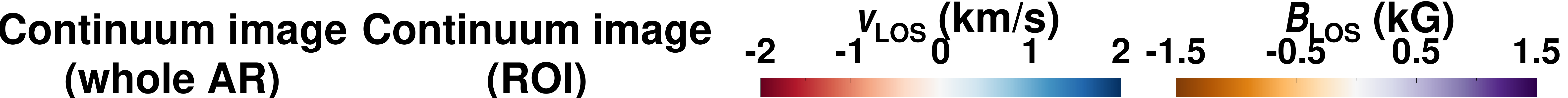}
 \includegraphics[width=.5\textwidth]{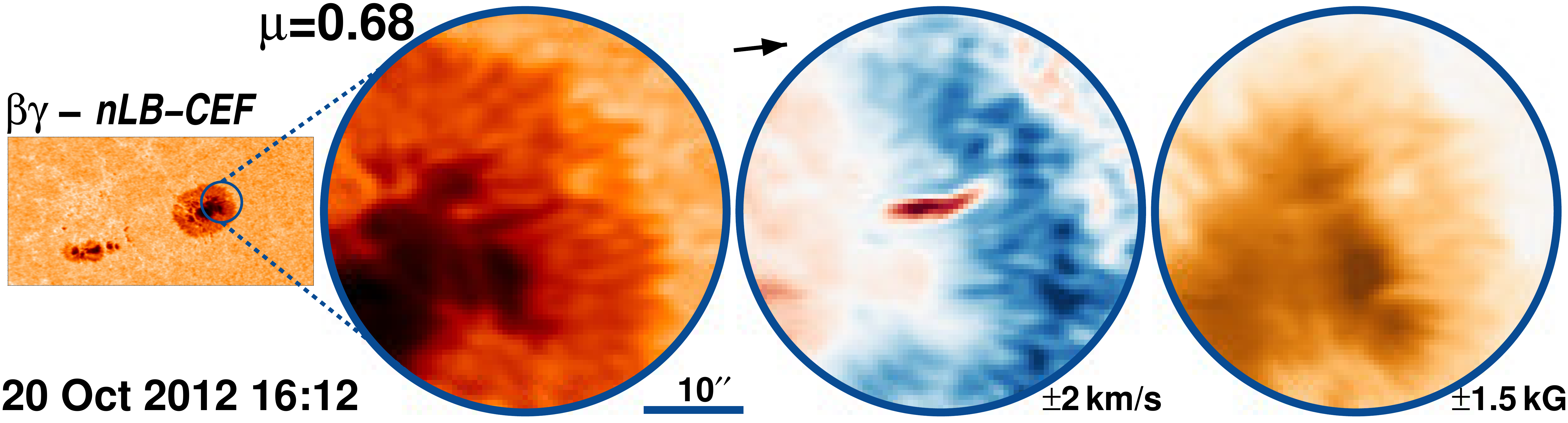}
 \includegraphics[width=.5\textwidth]{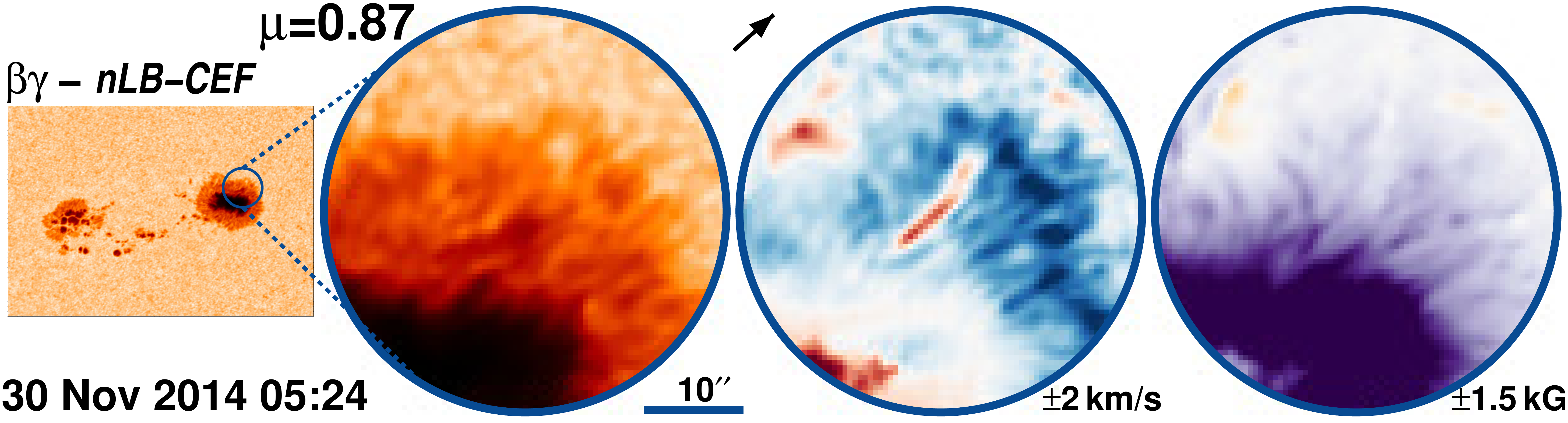}
 \includegraphics[width=.5\textwidth]{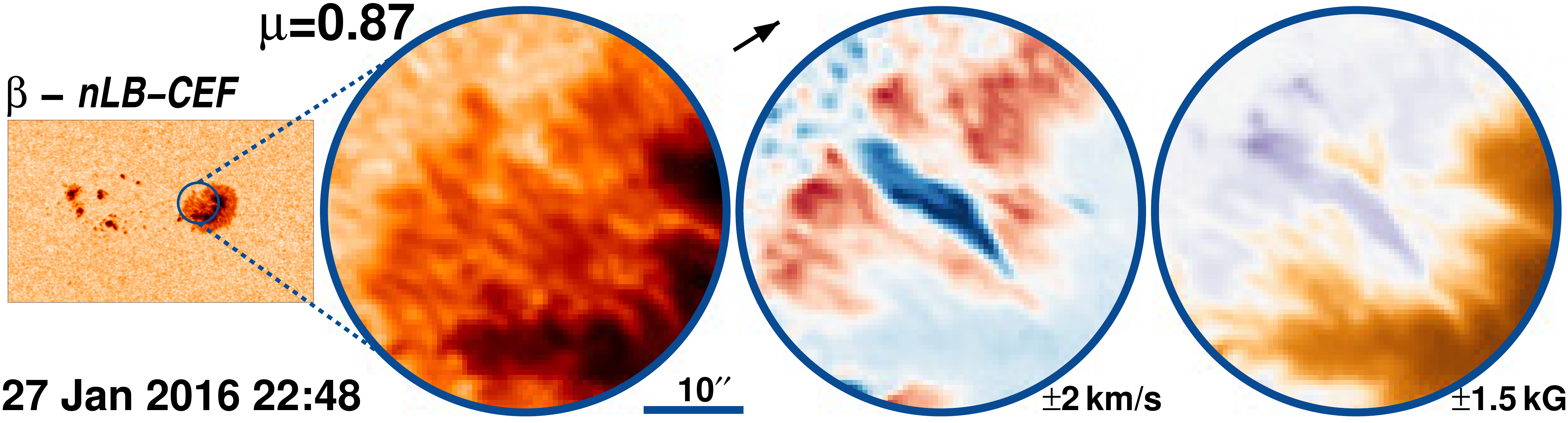}
 \includegraphics[width=.5\textwidth]{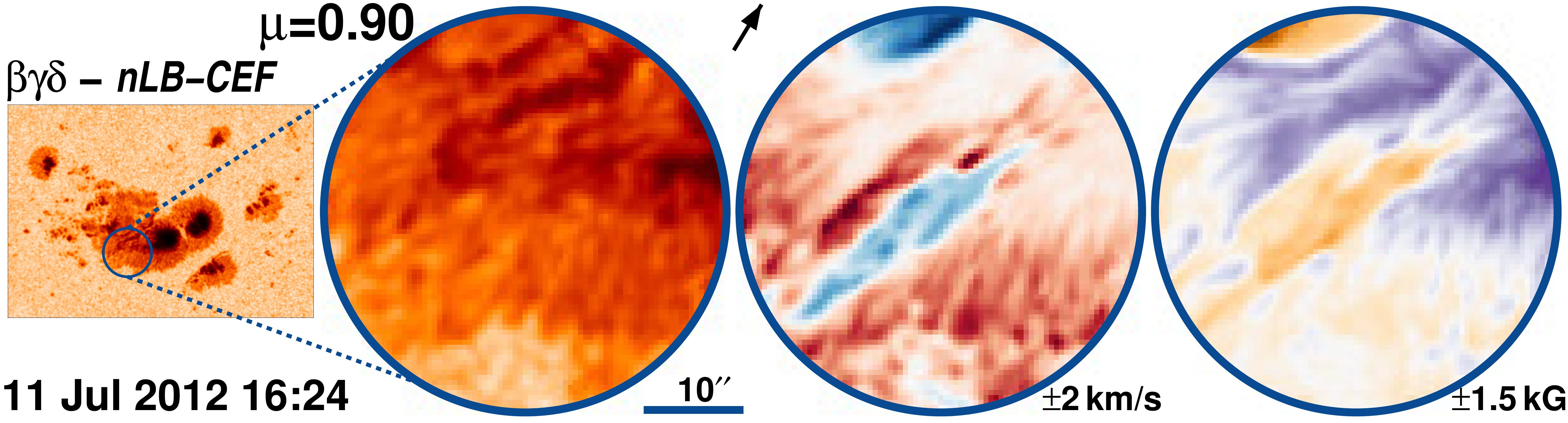}
 \caption{Examples of CEFs not associated with light bridges (nLB-CEFs) found in four ARs. In every row, the left-most map is the continuum image of the whole AR, and the three next maps present a close up of the region surrounding the CEF. From left to right they are: the continuum image, the Dopplergram clipped at $\pm$2\,\kms{},  and the magnetogram clipped at $\pm$1.5\,kG. Arrows point towards disk center. The magnetic classification of the AR and the time of the observation are shown on the top-left and bottom-left of each row.}\label{fig:cefs}
 \end{center}
 \end{figure}

\section{Results} \label{sec:results}

We found CEFs in  \narswithcef{} out of all \nars{} studied ARs between  2011 July 30 to 2017 August 24. Fig.~\ref{fig:numcefperar} shows the distribution of the number of CEFs per AR. The median number of CEFs detected per AR is \cefmedian{}. While 11 ARs presented just one CEF, the peak is found at 4 CEFs, with 15 ARs harbouring that number. There are 6 ARs with 10 or more CEFs, with AR\,11520 harbouring 15 CEFs (see Fig.~\ref{fig:DS97} in the Appendix \ref{sec:allcefs}), the largest number for any AR in our sample. Some CEFs also appear in nests, i.e., multiple CEFs are observed to form in the same part of the penumbra of a given sunspot. In few cases, multiple CEFs were observed at the same time at different locations of the same sunspot.  We observed that the occurrence of CEFs depends on neither the size nor on the number of sunspots belonging to an AR. Table~\ref{tab:OtherCEFs} summarizes parameters of the observations and general properties of the hosting ARs such as the NOAA AR, date and time, the total duration of the observation, size of the FoV, the median latitude of the AR, the start and end longitude of the considered observations of the AR, and magnetic classification of the AR during all the days it was observed.

Four examples of CEFs are shown in Fig.~\ref{fig:cefs}.  From left to right, Fig.~\ref{fig:cefs} displays the continuum image of the whole AR, and zoom into the region of interest showing three quantities, the continuum intensity, $v_{\rm LOS}$, and $B_{\rm LOS}$. CEFs generally appear as elongated structures that, depending on their location are either redshifted (Fig.~\ref{fig:cefs}; two top rows), or blueshifted (Fig.~\ref{fig:cefs}; two bottom rows). All the CEFs found in this study are presented in Figs.~\ref{fig:DS00}-\ref{fig:DS112} in Appendix~\ref{sec:allcefs}. Each Figure groups all CEFs harboured within one AR.  The layout of these figures is the same as of Fig.~\ref{fig:cefs}.

\begin{figure}[tbhp]
 \begin{center}
 \includegraphics[width=.5\textwidth]{colorbars.pdf}
 \includegraphics[width=.5\textwidth]{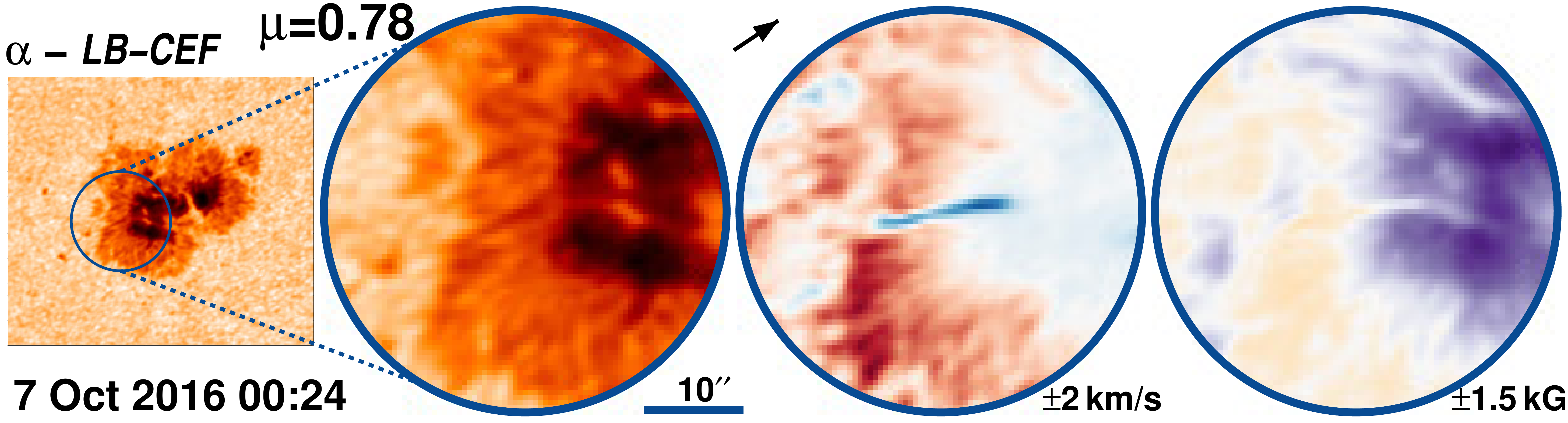}
 \includegraphics[width=.5\textwidth]{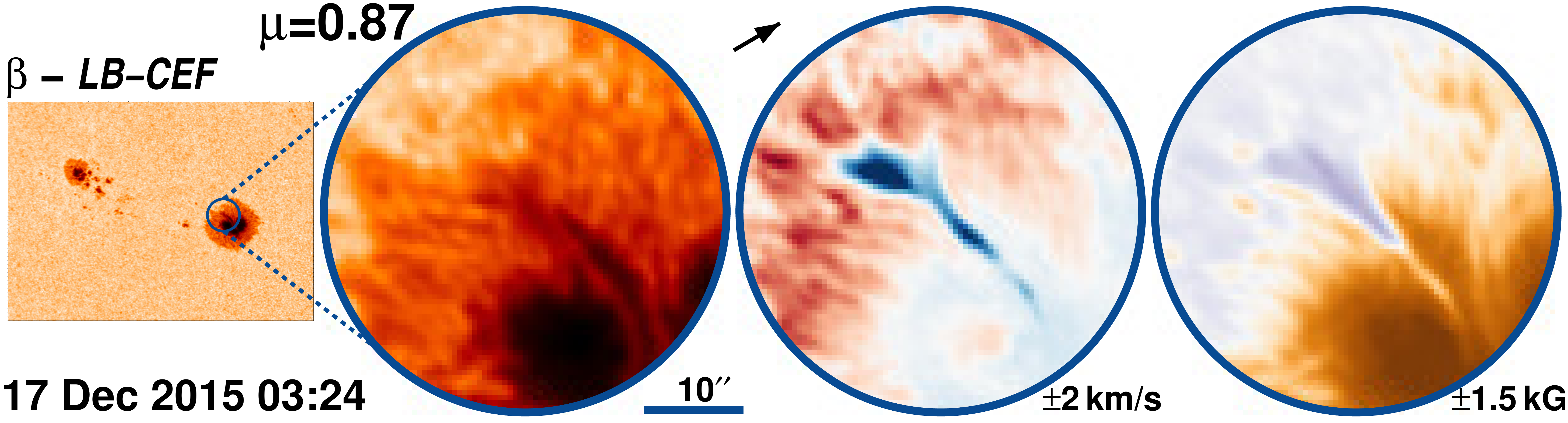}
 \includegraphics[width=.5\textwidth]{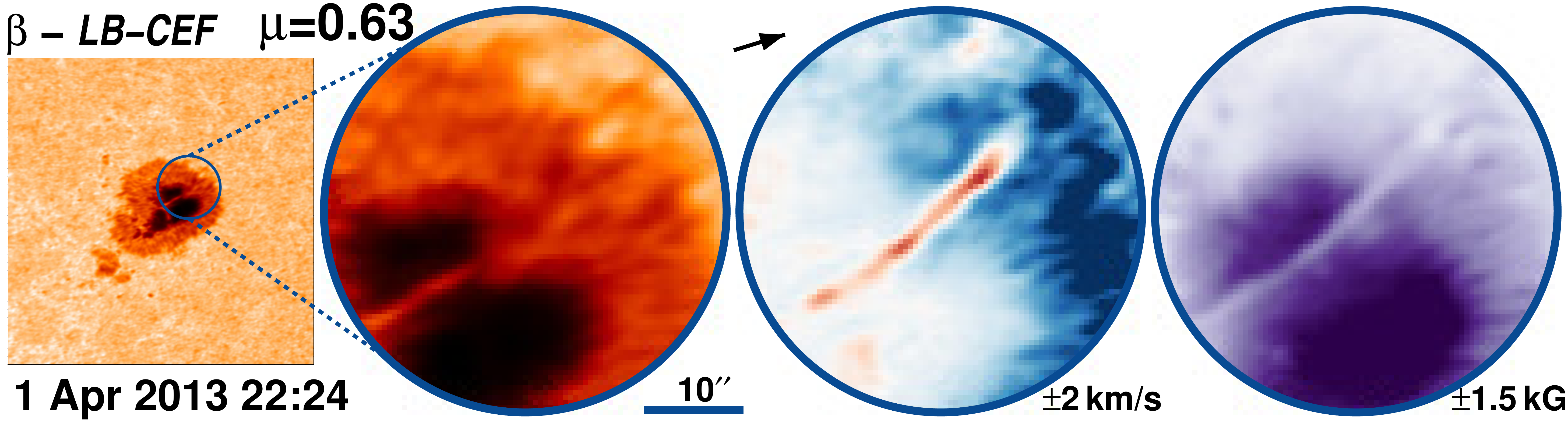}
 \includegraphics[width=.5\textwidth]{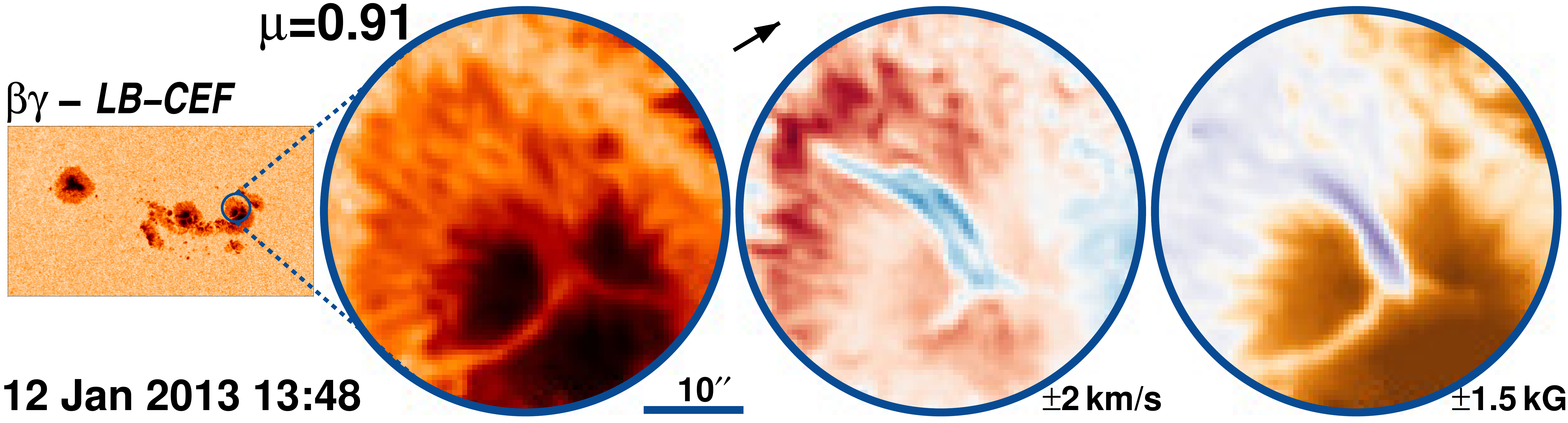}
 \caption{ Examples of four CEFs associated with light bridges (LB-CEFs). Same layout as Fig.~\ref{fig:cefs}.}\label{fig:cefswithlb}
 \end{center}
 \end{figure}

We find that CEFs can be categorized into two different groups: (1) CEFs associated with light bridges: Such CEFs originate in a light bridge or in a penumbral intrusion (i.e. a partial light bridge) and extend into the penumbra with the opposite flow direction compared to the adjacent penumbra. We label these CEFs as LB-CEFs. Four examples of such LB-CEFs are plotted in Fig.~\ref{fig:cefswithlb}. (2) CEFs completely embedded in the penumbra, labelled as nLB-CEFs (non-light bridge CEFs; Fig.~\ref{fig:cefs}). Both types of CEFs are almost equally common in our sample: 198 (51.6\%) nLB-CEFs vs. 186 (48.4\%) LB-CEFs.

CEFs appeared in all magnetic types of ARs: ranging from simple $\alpha$ ARs to the more complex $\delta$ ARs. 
Similarly, the few ARs not harbouring any CEF also show many complexity categories.
The examples of CEFs presented in Figs.~\ref{fig:cefs} and \ref{fig:cefswithlb} indicate the magnetic classification just above the leftmost image of each row, illustrating the large variety of ARs where CEFs appeared. There is no clear dependence on the occurrence of CEFs on the complexity of the AR. Both types of CEFs appear in all magnetic classes.

The \narswithcef{} ARs that hosted CEFs were observed for a total of $\sim$1000 days. The percentage of days on which CEFs are observed does seem to depend on the magnetic complexity of an AR. For the simplest, $\alpha$, regions, CEFs are visible on only 8\% of all days. This increases to roughly 1/3 of all days for $\beta$ and $\beta\gamma$ regions and finally becomes slightly more than 50\% for the $\beta\gamma\delta$ regions.

We did not observe CEFs in \narswithoutcef{} ARs (\narswithoutcefper{}), all except one containing only a single sunspot.  
The magnetic classification of these ARs is dominated by the simple $\alpha$ class (51.8\% of all days these regions were observed). In 39.3\% of those observed days the ARs without CEFs appear in $\beta$ configuration, 7.1\% in $\beta\gamma$, and 1.8\% in $\beta\delta$. 
Fig~\ref{fig:arswithoutcefs} in Appendix~\ref{sec:arnocefs} shows the continuum images of all the ARs without CEFs. The sunspots within these ARs share a similar roundish morphology and are all rather small with a diameter smaller than 40\arcsec{}. The details of these ARs, as well as their evolution through different magnetic classes, are summarized in the top part of Table~\ref{tab:OtherCEFs}.

\begin{figure}[tbhp]
 \begin{center}
 \includegraphics[width=.48\textwidth]{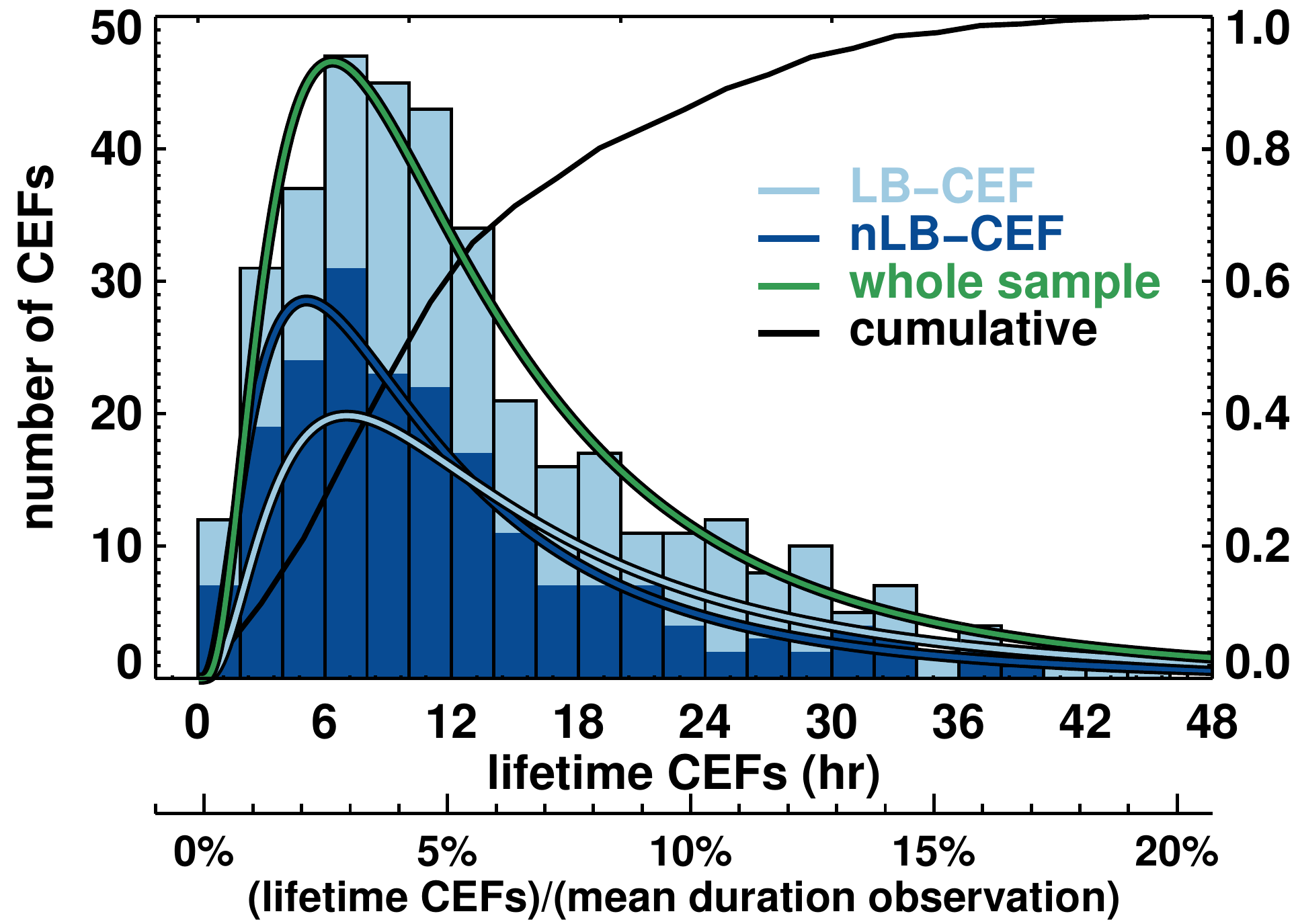}
     \caption{Lifetime of CEFs.  The maximum is $\sim$7\,hr and the median is \ceflifetime{}. Colors mark nLB-CEFs (dark blue) and LB-CEFs (light blue). 
      Thick lines show the best fit of a log-normal function (Eq.~\ref{eq:lognormal}) to the distributions of  nLB-CEFs (dark blue), LB-CEFs (light blue), and the whole sample (green line). The cumulative curve is shown in black.   }\label{fig:durantion}
 \end{center}
 \end{figure}

The lifetimes of CEFs found in the literature range from 1\,hr to approximately 12\,hr \citep[e.g.,][]{Louis2014A&A...CEF,Siu-Tapia2018ApJ,Louis2020...cefs}.  Fig.~\ref{fig:durantion} presents the histogram of the lifetimes of CEFs from our study. The median and average lifetime of CEF are \ceflifetime{} and \meanceflifetime{}. The total lifetime distribution is reasonably well reproduced by a log-normal function given by,
\begin{equation}\label{eq:lognormal}
    f_{\rm log-normal}(t)=\frac{\eta}{\xi\sqrt{2\pi}t}\exp
    \left[-\frac{\left[\ln(t/\zeta\right)]^2}{2\xi^2}\right],
\end{equation}
where $\eta=780.6\pm0.6$, $\zeta=11.6\pm0.4$, and $\xi=0.8\pm0.03$.  85.1\% of CEFs have a lifetime shorter than 1 day and the peak in the lifetime distribution is found at around $\sim$7\,hr. The previously reported lifetimes of CEFs fall close to the median value,  including the MHD-simulated CEF that vanishes after $\sim$10\,hr \citep{Siu-Tapia2018ApJ}. On average we found a CEF in \perobsceftotal{} of the mean total duration of the observations of an AR (see bottom $x-$axis on Fig.
~\ref{fig:durantion}).

In Fig.~\ref{fig:durantion} we differentiate between nLB-CEFs (dark blue) and LB-CEFs (light blue). The mean lifetime of LB-CEFs is \difflifetimes{} longer (\lbceflifetime{} compared to \nLBceflifetime{}). In addition, the lifetime of CEFs does not depend of the magnetic classification on the AR (not shown).

We found a slight positive trend between the lifetime of CEFs and the total area of the AR, but with a large scatter, so that the trend is not statistically significant at the $1\sigma$ confidence level. In addition, ARs are usually formed by a group of many sunspots \citep[e.g.,][]{Sudip2021arXiv...sunspotsarea}, while CEFs often appear in just one of the sunspots within the AR. Therefore, we divided the total area of the AR by the number of visible sunspots belonging to the AR. The scatter remains large. The slight positive trend is enhanced, but still not statistically significant.

Most of the CEFs appear as elongated structures, with the $B_{\rm LOS}$ values being lower than in the surrounding penumbra and sometimes even of opposite magnetic polarity \citep[see for example Fig.~\ref{fig:cefswithlb}, cf.][]{Lim2020ApJ...LB}. 
As we use line-of-sight magnetograms for this study, this polarity statement is only reliable for CEFs located at $\mu$-values larger than $\approx$0.8. From the 181 CEFs fulfilling this criterion, $\sim$42\% are of opposite polarity compared to the surrounding penumbra (38/91 nLB-CEF vs. 38/90 LB-CEF).

\section{Summary and conclusions}\label{sec:discussion}

We presented the first statistical survey of CEFs in a sample of \nars{} ARs. We tracked the ARs for \totdurobs{} days on average using the continuous observations by SDO/HMI. We report the appearance of \numcefs{} CEFs in \narswithcefper{} of the ARs. CEFs were observed in all magnetic types of ARs catalogued, from $\alpha$, $\beta$ to $\beta\gamma\delta$. The number of CEFs produced by an AR ranges from 1 to 15, with a median value of \cefmedian{} CEFs per AR.

We distinguished two populations of CEFs, with one of these populations associated with light bridges the other one not. The two populations are almost equal in size, and $\sim$2 out of 5 of the CEFs within these two populations are of opposite polarity with respect to the surrounding penumbra. Given that not all sunspots have light bridges and these take up only a small part of the sunspot, we conclude that light bridges provide favourable conditions for the formation of CEFs.  LB-CEFs have slightly longer lifetimes, but the separation of these two populations based on their lifetime is not statistically significant. 

Almost all the CEFs in our sample are long and narrow. We could not find another example as broad as (i.e. covering such a large penumbral segment) as the CEF reported by \citet{Siu-Tapia2017A&A}. These results enhance the uniqueness of the AR\,11930 and its CEF \citep{Siutapia2019}.

The few reports in the literature of ARs associated with CEFs suggest that this might be a rather rare phenomenon. In this survey, we showed that almost all ARs of all magnetic types harbour CEFs. The reason for this seeming discrepancy most likely lies in the fact that CEFs are narrow and can easily be missed. Additionally, the average lifetime of CEFs is rather short compared to the total duration of the observation. The ratio between these two quantities highlights the observational difficulty of detecting CEFs (see bottom $x$-axis on Fig.~\ref{fig:durantion}): We needed to track continuously the ARs on average for \totdurobs{} to detect \cefmedian{} CEFs in an ARs (median value). 
From our findings, we expect to see a CEF only on an average of \perobsceftotal{} of the mean total duration of the observation of an AR. Therefore, to detect CEFs, uninterrupted and stable time series of continuum images and Doppler velocity observations are required. These requirements had not been met until observations became available from space, in particular by HMI.

In this survey, we have shown that CEFs are far more common than the few cases reported in the literature suggest. We find that $\sim$85\% of ARs harbour at least one CEF and typically \cefmedian{} in the course of the lifetime of an AR. However, only HMI provides the uninterrupted, stable, and relatively high-resolution measurements of velocity needed to identify a high fraction of CEFs. Based on the probability found in this study, the Japanese solar mission Hinode \citep{Kosugi2007} should have observed more CEFs than those found in the current literature. A quick and simple search in the archive of the Spectro-Polarimeter instrument \citep[SP;][]{Ichimoto2008SoPh} revealed 18 CEFs, indicating that the estimates for the occurrence of CEFs derived in this study are reasonable. The regions harbouring CEFs observed by Hinode/SP were classified from $\alpha$ to $\delta$, and the CEFs belonging to the two categories (LB- and nLB-CEFs), are roughly equal in number, consistent with the results reported in this Letter.

The Polarimetric and Helioseismic Imager \citep[PHI;][]{Solanki2020} onboard Solar Orbiter \citep{Muller2020A&A...SOLARORBITER} will allow us to track individual AR for longer periods of time, thanks to the partial corotation of the spacecraft with the Sun near its perihelion and the possibility of combining measurements from PHI, Hinode/SP, and HMI. Such measurements will allow us to accurately measure the flow velocity, magnetic and thermodynamic conditions inside CEFs.

\begin{acknowledgements}
We thank P.-L. Poulier and H.-P. Doerr for fruitful discussions on the different effects when calibrating HMI's Dopplergrams. J.~S. {Castellanos~Dur\'an} would like to show his gratitude to Y. Katsukawa for all the support and invaluable discussions during the visits to the National Astronomical Observatory of Japan (NAOJ). We thank Zhi-Chao~Liang and R.~Burston for their help with the NetDRMS data management system at MPS. J.~S. {Castellanos~Dur\'an} was funded by the Deutscher Akademischer Austauschdienst (DAAD) and the International Max Planck Research School (IMPRS) for Solar System Science at the University of G\"ottingen. This project has received funding from the European Research Council (ERC) under the European Union’s Horizon 2020 research and innovation program (grant agreement No. 695075).  The data were processed at the German Data Center for SDO (GDC-SDO), funded by the German Aerospace Center (DLR). The HMI data used are courtesy of NASA/SDO and the HMI science team.
\end{acknowledgements}

\bibliographystyle{aa}
\bibliography{references}
%
%%%%% TABLE WITH ALL CEFS   %%%%%
 \longtab{
{\scriptsize
\begin{landscape}
\begin{longtable}{rcccccccclcrrrrrrrrrrrrl}
\caption{Parameters of the observations and of the observed ARs that hosted CEFs.\\ {\tiny \textbf{Notes.} \tablefoottext{a}{NOAA AR number.} \tablefoottext{b}{Starting observation day.} \tablefoottext{c}{Starting time.}  \tablefoottext{d}{Total duration of the observation.} \tablefoottext{e}{Size of the FoV in arcsec.} \tablefoottext{f}{Median latitude of the AR.} \tablefoottext{g,h}{Longitude of the AR when the observations started\tablefoottext{g} and ended.\tablefoottext{h}} \tablefoottext{i}{Median corrected area of the AR in millionths of the solar hemisphere.} \tablefoottext{j}{Number of detected CEFs per AR.}  \tablefoottext{k}{Magnetic classification of the AR of the day. Roman numbers denote the day of the observation starting from the first day, indicated by (I) to (XII). For the days with CEFs, the  magnetic classification is preceded by three numbers: The first number counts how many CEFs appear on that day. The second number inside circular parentheses displays how many CEFs reached their maximum area on that particular day. The last number inside squared parentheses marks how many CEFs disappeared on that day. \tablefoottext{l}{Figure about the AR.}} }  }\label{tab:OtherCEFs}\\
\midrule[1.5pt]
\multirow{2}{*}{ID}&NOAA& \multirow{2}{*}{Date}\tablefootmark{(b)} & \multirow{2}{*}{Time}\tablefootmark{(c)}& $\Delta t_{\rm Obs}$\tablefootmark{(d)} & FoV\tablefootmark{(e)} & {$\phi$}\tablefootmark{(f)} & {$\lambda_{\rm start}$}\tablefootmark{(g)}& {$\lambda_{\rm end}$}\tablefootmark{(h)} & Area\tablefootmark{(i)} & Total &\multicolumn{12}{c}{Magnetic classification of the day \tablefootmark{(k)}} & \multirow{2}{*}{Fig.}\tablefootmark{(l)} \\
&AR\tablefootmark{(a)} & & &(hr)& (\arcsec{})& (deg) & (deg) & (deg) & (MSH)& CEFs\tablefootmark{(j)} & (I)& (II)&(III) &(IV) &(V) &(VI) &(VII) &(VIII) &(IX) &(X) &(XI) &(XII) &\\
\bottomrule[1.5pt]
\endfirsthead
\caption{continued.}\\
\toprule[1.5pt]
\multirow{2}{*}{ID}&NOAA& \multirow{2}{*}{Date}\tablefootmark{(b)} & \multirow{2}{*}{Time}\tablefootmark{(c)}& $\Delta t_{\rm Obs}$\tablefootmark{(d)} & FoV\tablefootmark{(e)} & {$\phi$}\tablefootmark{(f)} & {$\lambda_{\rm start}$}\tablefootmark{(g)}& {$\lambda_{\rm end}$}\tablefootmark{(h)} & Area\tablefootmark{(i)} & Total &\multicolumn{12}{c}{Magnetic classification of the day \tablefootmark{(k)}} & \multirow{2}{*}{Fig.}\tablefootmark{(l)} \\
&AR\tablefootmark{(a)} & & &(hr)& (\arcsec{})& (deg) & (deg) & (deg) & (MSH)& CEFs\tablefootmark{(j)} & (I)& (II)&(III) &(IV) &(V) &(VI) &(VII) &(VIII) &(IX) &(X) &(XI) &(XII) &\\
\midrule[1.5pt]
\endhead
\bottomrule[1.5pt]
\endfoot
  1& 12670& 2017-08-07& 16:30& 217.8& 120$\times$120& S06& E59& W61& 140$^{+30}_{-10}$&  0& $\beta$& $\alpha$& $\alpha$& $\alpha$& $\alpha$& $\alpha$& $\alpha$& $\alpha$& $\alpha$& $\alpha$& $\cdots$& $\cdots$& \ref{fig:arswithoutcefs}a\\
  2& 12664& 2017-06-26& 18:30& 221.8& 120$\times$120& N18& E60& W60&  70$^{+20}_{-20}$&  0& $\alpha$& $\alpha$& $\alpha$& $\beta$& $\beta$& $\beta\gamma$& $\beta$& $\beta$& $\beta$& $\alpha$& $\cdots$& $\cdots$& \ref{fig:arswithoutcefs}b\\
  3& 12600& 2016-10-12& 11:30& 217.8& 120$\times$130& N11& E60& W59&  90$^{+30}_{-10}$&  0& $\alpha$& $\alpha$& $\beta$& $\beta$& $\alpha$& $\alpha$& $\alpha$& $\alpha$& $\alpha$& $\alpha$& $\cdots$& $\cdots$& \ref{fig:arswithoutcefs}c\\
  4& 12574& 2016-08-13& 20:00& 177.2& 110$\times$140& N05& E42& W55& 200$^{+80}_{-40}$&  0& $\beta$& $\beta$& $\beta$& $\beta$& $\beta$& $\beta$& $\beta$& $\alpha$& $\cdots$& $\cdots$& $\cdots$& $\cdots$& \ref{fig:arswithoutcefs}d\\
  5& 12526& 2016-03-30& 10:18& 283.8& 170$\times$140& S05& E72& W84& 190$^{+40}_{-10}$&  0& $\alpha$& $\alpha$& $\alpha$& $\alpha$& $\alpha$& $\beta$& $\alpha$& $\alpha$& $\alpha$& $\alpha$& $\alpha$& $\alpha$& \ref{fig:arswithoutcefs}e\\
  6& 12341& 2015-05-13& 18:41& 252.8& 150$\times$135& S19& E68& W68&  70$^{+10}_{-20}$&  0& $\alpha$& $\alpha$& $\alpha$& $\alpha$& $\alpha$& $\alpha$& $\alpha$& $\alpha$& $\alpha$& $\alpha$& $\alpha$& $\cdots$& \ref{fig:arswithoutcefs}f\\
  7& 12227& 2014-12-08& 15:43& 237.6& 140$\times$130& S04& E69& W62& 120$^{+40}_{-30}$&  0& $\alpha$& $\alpha$& $\alpha$& $\alpha$& $\alpha$& $\alpha$& $\beta$& $\beta$& $\beta$& $\beta$& $\alpha$& $\cdots$& \ref{fig:arswithoutcefs}g\\
  8& 12218& 2014-11-30& 08:40& 218.4& 130$\times$120& N16& E58& W61& 140$^{+10}_{-20}$&  0& $\beta$& $\beta$& $\beta$& $\alpha$& $\alpha$& $\alpha$& $\alpha$& $\alpha$& $\alpha$& $\alpha$& $\cdots$& $\cdots$& \ref{fig:arswithoutcefs}h\\
  9& 12186& 2014-10-13& 16:29& 228.6& 130$\times$120& S20& E60& W63& 170$^{+20}_{-30}$&  0& $\alpha$& $\beta$& $\beta$& $\beta$& $\beta$& $\beta$& $\beta$& $\beta$& $\alpha$& $\beta$& $\alpha$& $\cdots$& \ref{fig:arswithoutcefs}i\\
 10& 12151& 2014-08-29& 09:53& 259.2& 170$\times$130& S07& E73& W70& 110$^{+20}_{-30}$&  0& $\alpha$& $\alpha$& $\beta$& $\alpha$& $\alpha$& $\alpha$& $\alpha$& $\alpha$& $\beta$& $\alpha$& $\alpha$& $\cdots$& \ref{fig:arswithoutcefs}j\\
 11& 12107& 2014-07-05& 23:10& 268.8& 230$\times$150& S20& E69& W76& 250$^{+50}_{-60}$&  0& $\beta$& $\beta\gamma$& $\beta\delta$& $\beta$& $\beta\gamma$& $\beta\gamma$& $\beta\gamma$& $\beta\gamma$& $\beta$& $\beta$& $\alpha$& $\alpha$& \ref{fig:arswithoutcefs}k\\
 12& 12032& 2014-04-13& 21:21& 238.8& 140$\times$150& N13& E63& W68& 150$^{+40}_{-10}$&  0& $\beta$& $\beta\gamma$& $\beta\gamma$& $\beta\gamma$& $\beta\gamma$& $\alpha$& $\alpha$& $\beta$& $\alpha$& $\alpha$& $\cdots$& $\cdots$& \ref{fig:arswithoutcefs}l\\
 13& 11931& 2013-12-24& 17:55&  86.2& 125$\times$120& S15& E65& W72& 190$^{+50}_{-30}$&  0& $\alpha$& $\alpha$& $\alpha$& $\alpha$& $\alpha$& $\alpha$& $\alpha$& $\alpha$& $\beta$& $\beta$& $\alpha$& $\cdots$& \ref{fig:arswithoutcefs}m\\
 14& 11692& 2013-03-15& 19:18& 244.6& 180$\times$195& N09& E62& W72& 220$^{+30}_{-20}$&  0& $\alpha$& $\alpha$& $\alpha$& $\beta$& $\alpha$& $\alpha$& $\alpha$& $\beta$& $\beta$& $\beta$& $\beta$& $\cdots$& \ref{fig:arswithoutcefs}n\\
 15& 11582& 2012-10-02& 07:32& 265.8& 170$\times$170& S12& E73& W73& 270$^{+50}_{-80}$&  0& $\alpha$& $\alpha$& $\alpha$& $\alpha$& $\alpha$& $\alpha$& $\beta$& $\beta$& $\alpha$& $\alpha$& $\alpha$& $\cdots$& \ref{fig:arswithoutcefs}o\\
 16& 11312& 2011-10-11& 00:00& 285.8& 160$\times$125& N23& E76& W78& 210$^{+20}_{-60}$&  0& $\alpha$& $\alpha$& $\alpha$& $\alpha$& $\alpha$& $\alpha$& $\alpha$& $\alpha$& $\beta$& $\alpha$& $\alpha$& $\alpha$& \ref{fig:arswithoutcefs}p\\

%%%%%%%%%%%%%%%%%%%%%%%%%%%%%%%%%%%%%%%%%%%%
\toprule[1.5pt]
%%%%%%%%%%%%%%%%%%%%%%%%%%%%%%%%%%%%%%%%%%%%

 17& 12671& 2017-08-16& 21:24& 205.6& 400$\times$150& N11& E59& W54&  360$^{+140}_{- 50}$&  4& \textbf{2}(\textbf{1})[\textbf{1}]$\beta$& (\textbf{1})[\textbf{1}]$\beta\gamma\delta$& $\beta\gamma$& \textbf{2}(\textbf{2})[\textbf{2}]$\beta\gamma$& $\beta\gamma$& $\beta\gamma$& $\beta\gamma$& $\beta\gamma$& $\beta\gamma$& $\cdots$& $\cdots$& $\cdots$& \ref{fig:DS00}\\
 18& 12665& 2017-07-09& 21:36& 226.8& 390$\times$270& S06& E57& W68&  480$^{+170}_{-210}$&  6& $\beta$& $\beta$& \textbf{1}(\textbf{1})$\beta$& [\textbf{1}]$\beta\gamma$& $\beta\gamma$& $\beta\gamma$& $\beta$& \textbf{4}(\textbf{2})[\textbf{2}]$\beta$& (\textbf{2})$\beta$& \textbf{1}(\textbf{1})[\textbf{3}]$\beta$& $\cdots$& $\cdots$& \ref{fig:DS02}\\
 19& 12599& 2016-10-05& 05:36& 221.8& 125$\times$110& S15& E58& W64&  400$^{+210}_{- 10}$&  8& \textbf{1}(\textbf{1})[\textbf{1}]$\beta$& \textbf{1}$\beta$& (\textbf{1})[\textbf{1}]$\alpha$& $\alpha$& $\beta$& \textbf{1}$\beta$& \textbf{2}(\textbf{3})[\textbf{2}]$\beta$& \textbf{3}(\textbf{3})[\textbf{3}]$\beta$& [\textbf{1}]$\beta$& $\beta$& $\cdots$& $\cdots$& \ref{fig:DS05}\\
 20& 12585& 2016-09-01& 08:48& 225.8& 240$\times$120& N08& E70& W55&  430$^{+ 50}_{-120}$& 10& \textbf{3}(\textbf{2})[\textbf{1}]$\beta$& \textbf{2}(\textbf{2})[\textbf{3}]$\beta$& \textbf{2}(\textbf{2})[\textbf{2}]$\beta\gamma$& \textbf{1}(\textbf{2})[\textbf{2}]$\beta\gamma$& $\beta\gamma$& $\beta$& $\beta\gamma$& \textbf{1}$\beta\gamma$& \textbf{1}(\textbf{1})[\textbf{1}]$\beta\delta$& \textbf{1}(\textbf{1})[\textbf{1}]$\beta\delta$& $\cdots$& $\cdots$& \ref{fig:DS06}\\
 21& 12576& 2016-08-15& 08:00& 153.2&  60$\times$ 50& S12& E55& W29&  130$^{+ 30}_{- 10}$&  1& $\alpha$& $\alpha$& $\alpha$& $\alpha$& $\alpha$& \textbf{1}(\textbf{1})[\textbf{1}]$\alpha$& $\alpha$& $\cdots$& $\cdots$& $\cdots$& $\cdots$& $\cdots$& \ref{fig:DS07}\\
 22& 12565& 2016-07-21& 04:00& 207.8& 145$\times$120& N04& E36& W78&  300$^{+ 80}_{- 20}$&  2& $\alpha$& $\beta$& $\beta$& $\beta$& $\beta$& \textbf{1}$\beta$& $\beta$& \textbf{1}(\textbf{2})[\textbf{2}]$\beta$& $\alpha$& $\alpha$& $\cdots$& $\cdots$& \ref{fig:DS09}\\
 23& 12567& 2016-07-15& 21:00& 253.8& 230$\times$135& N05& E71& W69&  330$^{+140}_{- 60}$&  9& $\beta$& \textbf{2}(\textbf{2})$\beta\gamma$& \textbf{1}(\textbf{1})[\textbf{3}]$\beta\gamma\delta$& $\beta\gamma$& $\beta\gamma$& \textbf{3}(\textbf{3})[\textbf{1}]$\beta\gamma$& \textbf{3}(\textbf{2})[\textbf{4}]$\beta$& (\textbf{1})$\beta\gamma$& [\textbf{1}]$\beta$& $\cdots$& $\cdots$& $\cdots$& \ref{fig:DS10}\\
 24& 12542& 2016-05-06& 17:12& 194.8& 135$\times$140& N11& E57& W49&  150$^{+ 30}_{- 20}$&  2& $\beta$& \textbf{1}(\textbf{1})$\beta$& [\textbf{1}]$\beta$& $\alpha$& $\beta$& \textbf{1}(\textbf{1})[\textbf{1}]$\beta\gamma$& $\beta\gamma$& $\beta$& $\beta\gamma$& $\beta$& $\cdots$& $\cdots$& \ref{fig:DS13}\\
 25& 12529& 2016-04-09& 11:48& 219.8& 170$\times$140& N10& E60& W60&  790$^{+ 70}_{- 30}$& 14& \textbf{1}(\textbf{1})[\textbf{1}]$\beta$& \textbf{3}(\textbf{3})[\textbf{1}]$\beta$& \textbf{2}(\textbf{2})[\textbf{4}]$\beta$& \textbf{3}(\textbf{3})[\textbf{2}]$\beta$& \textbf{1}(\textbf{1})[\textbf{1}]$\beta$& [\textbf{1}]$\beta$& \textbf{1}(\textbf{1})$\beta$& \textbf{2}[\textbf{1}]$\beta\gamma$& \textbf{1}(\textbf{3})[\textbf{3}]$\beta\gamma$& $\beta\gamma$& $\cdots$& $\cdots$& \ref{fig:DS14}\\
 26& 12489& 2016-01-26& 18:36& 213.8& 310$\times$185& N10& E53& W65&  250$^{+150}_{- 50}$&  4& \textbf{1}(\textbf{1})[\textbf{1}]$\beta$& \textbf{1}(\textbf{1})$\beta$& [\textbf{1}]$\beta$& $\beta$& $\beta$& $\beta$& \textbf{1}(\textbf{1})[\textbf{1}]$\beta$& \textbf{1}(\textbf{1})[\textbf{1}]$\beta$& $\alpha$& $\cdots$& $\cdots$& $\cdots$& \ref{fig:DS16}\\
 27& 12483& 2016-01-11& 02:24& 205.2& 240$\times$130& N17& E51& W61&   90$^{+ 80}_{- 50}$&  1& $\beta$& $\beta$& \textbf{1}$\beta$& (\textbf{1})[\textbf{1}]$\beta$& $\beta$& $\beta$& $\beta$& $\beta$& $\alpha$& $\cdots$& $\cdots$& $\cdots$& \ref{fig:DS17}\\
 28& 12480& 2016-01-13& 00:24& 226.2& 295$\times$110& N03& E65& W59&  170$^{+ 40}_{- 10}$&  1& $\beta$& $\beta$& $\beta$& $\beta$& $\beta$& \textbf{1}$\beta$& (\textbf{1})[\textbf{1}]$\beta$& $\beta$& $\alpha$& $\alpha$& $\cdots$& $\cdots$& \ref{fig:DS18}\\
 29& 12470& 2015-12-15& 02:24& 262.2& 390$\times$280& N14& E72& W71&  430$^{+160}_{-180}$&  2& $\alpha$& \textbf{1}$\beta$& (\textbf{1})[\textbf{1}]$\beta$& \textbf{1}$\beta$& (\textbf{1})[\textbf{1}]$\beta$& $\beta$& $\beta$& $\beta$& $\beta$& $\beta$& $\alpha$& $\alpha$& \ref{fig:DS19}\\
 30& 12381& 2015-07-05& 23:48& 257.6& 350$\times$170& N15& E53& W88&  360$^{+260}_{-140}$&  9& \textbf{3}(\textbf{1})$\beta$& \textbf{2}(\textbf{3})[\textbf{2}]$\beta$& \textbf{1}(\textbf{2})[\textbf{3}]$\beta\gamma$& [\textbf{1}]$\beta\gamma$& \textbf{1}$\beta$& (\textbf{1})[\textbf{1}]$\beta$& \textbf{1}(\textbf{1})$\beta$& [\textbf{1}]$\beta$& \textbf{1}(\textbf{1})[\textbf{1}]$\beta$& $\beta$& $\beta$& $\cdots$& \ref{fig:DS23}\\
 31& 12371& 2015-06-17& 15:24& 230.8& 360$\times$200& N12& E64& W62&  950$^{+430}_{-170}$&  9& \textbf{1}(\textbf{1})$\beta$& \textbf{2}(\textbf{1})[\textbf{1}]$\beta\gamma$& \textbf{1}(\textbf{1})[\textbf{2}]$\beta\gamma\delta$& \textbf{1}(\textbf{1})[\textbf{1}]$\beta\gamma\delta$& (\textbf{1})[\textbf{1}]$\beta\gamma\delta$& $\beta\gamma\delta$& \textbf{1}(\textbf{1})$\beta\gamma\delta$& \textbf{2}(\textbf{2})[\textbf{2}]$\beta\gamma\delta$& \textbf{1}(\textbf{1})[\textbf{2}]$\beta\gamma$& $\beta\gamma$& $\cdots$& $\cdots$& \ref{fig:DS24}\\
 32& 12339& 2015-05-08& 17:00& 246.0& 405$\times$170& N13& E70& W65&  670$^{+360}_{-170}$&  4& $\beta\gamma$& $\beta\gamma$& \textbf{2}(\textbf{1})$\beta\gamma$& (\textbf{1})[\textbf{2}]$\beta\gamma$& $\beta\gamma$& \textbf{1}$\beta\gamma$& \textbf{1}(\textbf{2})[\textbf{1}]$\beta\gamma$& [\textbf{1}]$\beta\gamma$& $\beta\gamma$& $\beta$& $\beta\gamma$& $\cdots$& \ref{fig:DS26}\\
 33& 12321& 2015-04-13& 08:12& 218.8& 310$\times$190& N12& E54& W65&  380$^{+300}_{-240}$&  3& \textbf{1}(\textbf{1})[\textbf{1}]$\beta\gamma\delta$& $\beta\gamma\delta$& \textbf{1}(\textbf{1})[\textbf{1}]$\beta\gamma$& $\beta\gamma$& \textbf{1}(\textbf{1})[\textbf{1}]$\beta\gamma$& $\beta\gamma$& $\beta\gamma$& $\beta\gamma$& $\beta$& $\alpha$& $\cdots$& $\cdots$& \ref{fig:DS27}\\
 34& 12324& 2015-04-16& 09:24& 216.6& 175$\times$130& N19& E61& W56&  300$^{+130}_{- 90}$&  3& $\beta$& \textbf{1}(\textbf{1})[\textbf{1}]$\beta$& \textbf{2}(\textbf{2})[\textbf{2}]$\beta\gamma$& $\beta$& $\beta\gamma$& $\beta$& $\beta$& $\beta$& $\beta$& $\cdots$& $\cdots$& $\cdots$& \ref{fig:DS28}\\
 35& 12325& 2015-04-17& 11:36& 224.8& 215$\times$130& N05& E68& W56&  205$^{+ 65}_{- 15}$&  4& $\alpha$& $\beta$& \textbf{1}(\textbf{1})[\textbf{1}]$\beta$& \textbf{1}(\textbf{1})[\textbf{1}]$\beta$& $\beta$& $\beta$& \textbf{1}(\textbf{1})[\textbf{1}]$\beta$& \textbf{1}(\textbf{1})$\beta$& [\textbf{1}]$\beta$& $\alpha$& $\cdots$& $\cdots$& \ref{fig:DS29}\\
 36& 12305& 2015-03-22& 14:00& 238.4& 320$\times$160& S08& E67& W64&  340$^{+120}_{- 70}$&  5& \textbf{1}(\textbf{1})[\textbf{1}]$\beta$& $\beta\gamma$& $\beta\gamma$& $\beta\gamma\delta$& $\beta\gamma\delta$& $\beta\gamma$& \textbf{2}(\textbf{1})$\beta\gamma$& (\textbf{1})[\textbf{2}]$\beta\gamma$& \textbf{2}(\textbf{1})[\textbf{1}]$\beta$& (\textbf{1})$\beta$& [\textbf{1}]$\beta$& $\cdots$& \ref{fig:DS30}\\
 37& 12282& 2015-02-14& 09:24& 252.8& 265$\times$200& N11& E72& W67&  210$^{+120}_{- 30}$&  5& $\alpha$& $\beta$& $\beta$& $\beta\gamma$& $\beta$& $\beta\gamma$& \textbf{3}(\textbf{3})[\textbf{3}]$\beta\gamma$& $\beta\gamma$& $\beta$& \textbf{1}(\textbf{1})[\textbf{1}]$\beta$& \textbf{1}(\textbf{1})[\textbf{1}]$\beta$& $\beta$& \ref{fig:DS32}\\
 38& 12277& 2015-02-01& 04:48& 245.8& 180$\times$140& N08& E71& W63&  400$^{+200}_{- 90}$&  2& $\beta\gamma$& $\beta\gamma$& \textbf{1}(\textbf{1})[\textbf{1}]$\beta\gamma$& $\beta\gamma$& \textbf{1}$\beta$& (\textbf{1})[\textbf{1}]$\beta\gamma$& $\beta\gamma$& $\beta\gamma$& $\beta\gamma$& $\beta\gamma$& $\beta\gamma$& $\cdots$& \ref{fig:DS33}\\
 39& 12268& 2015-01-23& 20:36& 254.8& 410$\times$190& S10& E77& W63&  430$^{+110}_{- 70}$&  4& \textbf{1}(\textbf{1})[\textbf{1}]$\beta$& \textbf{1}$\beta\gamma$& (\textbf{1})[\textbf{1}]$\beta\gamma$& $\beta\gamma$& $\beta\gamma$& $\beta\gamma$& $\beta\gamma$& \textbf{1}(\textbf{1})[\textbf{1}]$\beta\gamma$& \textbf{1}(\textbf{1})[\textbf{1}]$\beta\gamma$& $\beta\gamma$& $\beta\gamma$& $\cdots$& \ref{fig:DS34}\\
 40& 12259& 2015-01-12& 08:24& 277.8& 190$\times$150& S16& E75& W76&  260$^{+170}_{- 90}$&  1& $\beta$& $\beta\gamma$& \textbf{1}$\beta\gamma$& (\textbf{1})[\textbf{1}]$\beta\gamma$& $\beta\gamma$& $\beta\gamma$& $\beta\gamma\delta$& $\beta\delta$& $\beta\gamma$& $\beta$& $\beta$& $\alpha$& \ref{fig:DS36}\\
 41& 12257& 2015-01-08& 18:24& 163.0& 250$\times$150& N07& E17& W73&  380$^{+340}_{- 90}$&  4& $\beta$& $\beta$& \textbf{1}(\textbf{1})[\textbf{1}]$\beta$& \textbf{2}(\textbf{2})[\textbf{2}]$\beta$& \textbf{1}(\textbf{1})$\beta\delta$& [\textbf{1}]$\beta\delta$& $\beta\gamma\delta$& $\beta\gamma\delta$& $\cdots$& $\cdots$& $\cdots$& $\cdots$& \ref{fig:DS37}\\
 42& 12241& 2014-12-15& 05:24& 211.8& 320$\times$210& S10& E57& W59&  450$^{+300}_{-210}$&  7& \textbf{3}(\textbf{3})[\textbf{1}]$\beta\gamma$& \textbf{3}(\textbf{1})[\textbf{2}]$\beta\gamma$& (\textbf{2})[\textbf{3}]$\beta\gamma$& $\beta\gamma\delta$& $\beta\gamma\delta$& $\beta\gamma\delta$& \textbf{1}(\textbf{1})$\beta\gamma\delta$& [\textbf{1}]$\beta\gamma\delta$& $\beta\gamma$& $\cdots$& $\cdots$& $\cdots$& \ref{fig:DS38}\\
 43& 12237& 2014-12-15& 22:00& 196.0& 375$\times$220& S13& E49& W57&   40$^{+ 20}_{- 50}$&  7& $\beta$& $\beta$& \textbf{2}(\textbf{2})$\beta$& \textbf{2}(\textbf{2})[\textbf{3}]$\alpha$& [\textbf{1}]$\alpha$& \textbf{2}(\textbf{2})[\textbf{1}]$\alpha$& \textbf{1}(\textbf{1})[\textbf{2}]$\alpha$& $\alpha$& $\alpha$& $\cdots$& $\cdots$& $\cdots$& \ref{fig:DS39}\\
 44& 12236& 2014-12-12& 20:24& 241.0& 150$\times$100& N30& E71& W56&  120$^{+ 40}_{- 60}$&  2& \textbf{2}(\textbf{1})$\beta$& (\textbf{1})[\textbf{2}]$\beta$& $\beta$& $\beta$& $\beta$& $\beta$& $\beta$& $\alpha$& $\alpha$& $\alpha$& $\alpha$& $\cdots$& \ref{fig:DS40}\\
 45& 12230& 2014-12-11& 02:36& 178.8& 200$\times$145& S15& E40& W58&  150$^{+120}_{- 30}$&  1& $\beta$& $\beta$& $\beta\gamma$& \textbf{1}(\textbf{1})[\textbf{1}]$\beta\gamma$& $\beta\gamma$& $\beta\gamma$& $\beta\gamma$& $\beta$& $\beta$& $\cdots$& $\cdots$& $\cdots$& \ref{fig:DS41}\\
 46& 12222& 2014-11-29& 06:36& 210.8& 320$\times$220& S20& E59& W55&  570$^{+450}_{-110}$& 11& $\alpha$& \textbf{1}$\beta\gamma$& \textbf{1}(\textbf{2})[\textbf{1}]$\beta\gamma$& \textbf{3}(\textbf{3})[\textbf{4}]$\beta\gamma$& $\beta\gamma$& $\beta\gamma$& \textbf{2}(\textbf{2})[\textbf{2}]$\beta\gamma$& \textbf{1}(\textbf{1})[\textbf{1}]$\beta\gamma$& \textbf{2}(\textbf{1})$\beta\gamma$& \textbf{1}(\textbf{2})[\textbf{3}]$\beta\gamma$& $\cdots$& $\cdots$& \ref{fig:DS43}\\
 47& 12216& 2014-11-21& 20:24& 231.2& 240$\times$170& S13& E68& W59&  420$^{+220}_{-220}$&  8& \textbf{2}(\textbf{1})$\beta$& (\textbf{1})[\textbf{1}]$\beta\delta$& [\textbf{1}]$\beta\gamma$& $\beta\gamma\delta$& $\beta\gamma$& $\beta\gamma$& \textbf{4}(\textbf{3})[\textbf{3}]$\beta$& (\textbf{1})[\textbf{1}]$\beta$& \textbf{2}(\textbf{2})[\textbf{2}]$\beta$& $\beta$& $\cdots$& $\cdots$& \ref{fig:DS45}\\
 48& 12209& 2014-11-17& 02:48& 238.0& 320$\times$220& S15& E55& W75&  950$^{+150}_{- 50}$& 12& $\beta\gamma$& \textbf{1}$\beta\gamma\delta$& \textbf{2}(\textbf{3})[\textbf{1}]$\beta\gamma\delta$& \textbf{2}(\textbf{1})[\textbf{2}]$\beta\gamma\delta$& (\textbf{1})[\textbf{1}]$\beta\gamma\delta$& \textbf{1}(\textbf{1})[\textbf{1}]$\beta\gamma\delta$& \textbf{2}[\textbf{1}]$\beta\gamma\delta$& \textbf{2}(\textbf{3})[\textbf{2}]$\beta\gamma\delta$& \textbf{1}(\textbf{1})[\textbf{2}]$\beta\gamma\delta$& \textbf{1}(\textbf{1})[\textbf{1}]$\beta\gamma\delta$& (\textbf{1})[\textbf{1}]$\beta\gamma\delta$& $\cdots$& \ref{fig:DS46}\\
 49& 12187& 2014-10-20& 03:00& 244.4& 150$\times$150& S09& E67& W67&  190$^{+ 80}_{- 20}$&  1& $\alpha$& $\beta$& $\beta$& $\beta$& $\beta$& $\beta$& $\alpha$& $\beta$& \textbf{1}(\textbf{1})[\textbf{1}]$\beta$& $\alpha$& $\beta$& $\beta$& \ref{fig:DS48}\\
 50& 12178& 2014-09-30& 22:00& 248.6& 210$\times$140& S02& E70& W67&  120$^{+ 70}_{- 40}$&  4& $\alpha$& $\beta$& \textbf{1}(\textbf{1})$\beta$& [\textbf{1}]$\beta\gamma$& $\beta\gamma$& $\beta\gamma$& $\beta$& $\beta$& \textbf{1}(\textbf{1})[\textbf{1}]$\beta$& \textbf{2}(\textbf{2})[\textbf{2}]$\alpha$& $\alpha$& $\cdots$& \ref{fig:DS50}\\
 51& 12175& 2014-09-26& 23:12& 147.8& 270$\times$180& N16& E04& W77&  360$^{+220}_{- 90}$&  6& \textbf{1}(\textbf{1})$\beta\gamma$& \textbf{1}(\textbf{1})[\textbf{2}]$\beta\gamma\delta$& \textbf{1}(\textbf{1})[\textbf{1}]$\beta\gamma\delta$& \textbf{3}(\textbf{2})[\textbf{2}]$\beta\gamma\delta$& (\textbf{1})[\textbf{1}]$\beta\gamma\delta$& $\beta\gamma$& $\beta\gamma$& $\cdots$& $\cdots$& $\cdots$& $\cdots$& $\cdots$& \ref{fig:DS51}\\
 52& 12172& 2014-09-22& 17:00& 199.8& 340$\times$190& S11& E55& W54&  460$^{+ 30}_{- 50}$&  4& \textbf{1}(\textbf{1})[\textbf{1}]$\beta\delta$& $\beta$& \textbf{1}(\textbf{1})[\textbf{1}]$\beta\gamma$& $\beta\gamma$& $\beta\gamma$& $\beta\gamma$& \textbf{2}(\textbf{2})[\textbf{1}]$\beta\gamma$& [\textbf{1}]$\beta\gamma$& $\beta\gamma$& $\cdots$& $\cdots$& $\cdots$& \ref{fig:DS52}\\
 53& 12158& 2014-09-06& 01:24& 253.8& 220$\times$210& N16& E70& W69&  380$^{+210}_{- 40}$&  3& \textbf{2}$\alpha$& (\textbf{2})[\textbf{1}]$\beta$& \textbf{1}[\textbf{1}]$\beta\delta$& (\textbf{1})[\textbf{1}]$\beta\delta$& $\beta\gamma\delta$& $\beta\gamma\delta$& $\beta\gamma$& $\beta\gamma$& $\beta\gamma$& $\beta$& $\beta\gamma$& $\beta\gamma$& \ref{fig:DS53}\\
 54& 12149& 2014-08-31& 01:00& 251.2& 220$\times$150& N10& E69& W69&  190$^{+ 70}_{- 90}$&  1& $\beta$& $\beta\gamma$& $\beta\gamma$& $\beta\gamma\delta$& $\beta\gamma$& $\beta$& $\beta$& $\beta$& \textbf{1}$\beta$& (\textbf{1})[\textbf{1}]$\beta$& $\beta$& $\cdots$& \ref{fig:DS55}\\
 55& 12146& 2014-08-17& 20:12& 254.4& 130$\times$120& N09& E68& W72&  150$^{+ 70}_{-140}$&  4& \textbf{2}(\textbf{2})[\textbf{1}]$\alpha$& \textbf{1}(\textbf{1})[\textbf{2}]$\alpha$& $\alpha$& $\alpha$& $\beta$& $\beta$& $\beta$& $\beta$& $\beta$& \textbf{1}$\beta\delta$& (\textbf{1})[\textbf{1}]$\beta\delta$& $\beta\gamma\delta$& \ref{fig:DS56}\\
 56& 12135& 2014-08-08& 01:48& 249.8& 260$\times$170& N12& E69& W68&  170$^{+ 80}_{- 70}$&  2& \textbf{1}$\beta$& (\textbf{1})[\textbf{1}]$\beta$& \textbf{1}(\textbf{1})[\textbf{1}]$\beta\gamma$& $\beta\gamma$& $\beta$& $\beta$& $\beta$& $\beta$& $\beta$& $\beta$& $\cdots$& $\cdots$& \ref{fig:DS57}\\
 57& 12109& 2014-07-06& 03:00& 265.8& 250$\times$220& S08& E74& W73&  620$^{+310}_{-180}$&  5& $\alpha$& $\beta$& \textbf{1}$\beta\gamma$& \textbf{1}(\textbf{1})[\textbf{1}]$\beta\gamma$& (\textbf{1})$\beta\gamma\delta$& [\textbf{1}]$\beta\gamma\delta$& $\beta\gamma\delta$& \textbf{3}(\textbf{3})[\textbf{2}]$\beta\delta$& [\textbf{1}]$\beta$& $\beta$& $\beta$& $\beta$& \ref{fig:DS58}\\
 58& 12108& 2014-07-05& 19:48& 238.8& 280$\times$185& S08& E65& W67&  560$^{+470}_{-270}$&  9& $\beta$& $\beta$& $\beta$& \textbf{1}(\textbf{1})[\textbf{1}]$\beta\gamma$& $\beta\gamma$& $\beta\gamma\delta$& \textbf{2}(\textbf{2})[\textbf{2}]$\beta\gamma\delta$& \textbf{2}(\textbf{2})[\textbf{2}]$\beta\gamma\delta$& \textbf{3}(\textbf{2})[\textbf{1}]$\beta\gamma\delta$& \textbf{1}(\textbf{2})[\textbf{3}]$\beta\gamma$& $\beta\gamma$& $\cdots$& \ref{fig:DS59}\\
 59& 12085& 2014-06-08& 02:36& 170.8& 300$\times$120& S20& E28& W64&  460$^{+280}_{- 30}$&  6& \textbf{1}$\beta\gamma$& (\textbf{1})[\textbf{1}]$\beta\gamma$& \textbf{1}(\textbf{1})[\textbf{1}]$\beta\gamma$& \textbf{1}(\textbf{1})[\textbf{1}]$\beta\gamma\delta$& $\beta\gamma$& \textbf{2}(\textbf{2})[\textbf{1}]$\beta\gamma$& \textbf{1}(\textbf{1})[\textbf{2}]$\beta\gamma\delta$& $\cdots$& $\cdots$& $\cdots$& $\cdots$& $\cdots$& \ref{fig:DS62}\\
 60& 12082& 2014-06-06& 10:48& 197.8& 315$\times$210& N16& E47& W60&  220$^{+ 50}_{- 60}$&  6& \textbf{3}(\textbf{3})[\textbf{2}]$\beta$& \textbf{1}(\textbf{1})[\textbf{1}]$\beta\gamma$& [\textbf{1}]$\beta\gamma$& $\beta$& $\beta$& \textbf{1}(\textbf{1})[\textbf{1}]$\beta$& $\beta$& \textbf{1}(\textbf{1})[\textbf{1}]$\beta$& $\cdots$& $\cdots$& $\cdots$& $\cdots$& \ref{fig:DS63}\\
 61& 12080& 2014-06-06& 03:24& 210.8& 300$\times$140& S12& E46& W69&  300$^{+270}_{- 40}$&  5& $\beta$& \textbf{2}$\beta$& (\textbf{2})[\textbf{2}]$\beta\gamma$& $\beta\gamma\delta$& $\beta\gamma\delta$& $\beta\gamma\delta$& \textbf{3}(\textbf{3})[\textbf{3}]$\beta\gamma\delta$& $\beta\gamma\delta$& $\beta\gamma\delta$& $\beta\gamma\delta$& $\cdots$& $\cdots$& \ref{fig:DS64}\\
 62& 12056& 2014-05-08& 02:24& 251.2& 300$\times$200& N05& E71& W67&  300$^{+110}_{- 50}$&  1& \textbf{1}$\beta$& (\textbf{1})$\beta\gamma$& [\textbf{1}]$\beta\gamma\delta$& $\beta\gamma$& $\beta\gamma$& $\beta\gamma$& $\beta$& $\beta$& $\beta\gamma$& $\beta\gamma$& $\beta\gamma$& $\cdots$& \ref{fig:DS65}\\
 63& 12055& 2014-05-08& 09:24& 244.8& 280$\times$140& N12& E69& W65&  330$^{+130}_{-120}$&  5& \textbf{1}$\beta$& $\beta$& \textbf{1}(\textbf{2})[\textbf{2}]$\beta$& \textbf{3}(\textbf{2})[\textbf{2}]$\beta$& (\textbf{1})[\textbf{1}]$\beta\gamma$& $\beta\gamma$& $\beta$& $\beta$& $\beta$& $\beta$& $\beta$& $\cdots$& \ref{fig:DS66}\\
 64& 12049& 2014-04-30& 13:24& 260.8& 270$\times$180& S07& E74& W69&  300$^{+120}_{-100}$&  2& $\beta$& $\beta$& \textbf{1}(\textbf{1})[\textbf{1}]$\beta\gamma$& $\beta\gamma$& $\beta\gamma$& $\beta\gamma$& $\beta$& $\beta\gamma$& $\beta\gamma$& \textbf{1}(\textbf{1})[\textbf{1}]$\beta$& $\alpha$& $\cdots$& \ref{fig:DS67}\\
 65& 12034& 2014-04-11& 23:36& 230.8& 300$\times$145& N04& E63& W64&  250$^{+120}_{-150}$&  3& \textbf{1}(\textbf{1})$\beta$& [\textbf{1}]$\beta$& $\beta$& \textbf{1}(\textbf{1})[\textbf{1}]$\beta\gamma$& $\beta$& $\beta\gamma$& $\beta$& \textbf{1}(\textbf{1})[\textbf{1}]$\beta$& $\beta$& $\beta\gamma$& $\cdots$& $\cdots$& \ref{fig:DS68}\\
 66& 12005& 2014-03-16& 11:36& 251.8& 190$\times$160& N13& E71& W67&  220$^{+ 50}_{- 30}$&  1& $\alpha$& $\alpha$& $\alpha$& \textbf{1}(\textbf{1})[\textbf{1}]$\alpha$& $\beta$& $\beta$& $\beta$& $\alpha$& $\alpha$& $\alpha$& $\alpha$& $\cdots$& \ref{fig:DS70}\\
 67& 11976& 2014-02-10& 01:12& 273.8& 345$\times$150& S15& E78& W71&  240$^{+ 60}_{-130}$&  2& \textbf{1}$\beta$& \textbf{1}(\textbf{2})[\textbf{2}]$\beta\gamma$& $\beta\gamma$& $\beta\gamma$& $\beta\gamma$& $\beta\gamma$& $\beta$& $\beta$& $\beta$& $\alpha$& $\beta$& $\alpha$& \ref{fig:DS72}\\
 68& 11960& 2014-01-26& 13:00& 274.4& 185$\times$170& S15& E74& W75&  200$^{+110}_{- 60}$&  3& $\alpha$& $\alpha$& $\alpha$& $\alpha$& $\alpha$& $\alpha$& $\beta$& \textbf{2}(\textbf{2})[\textbf{2}]$\beta$& $\beta$& \textbf{1}(\textbf{1})$\beta$& [\textbf{1}]$\beta$& $\alpha$& \ref{fig:DS74}\\
 69& 11944& 2014-01-04& 08:48& 199.6& 420$\times$280& S09& E53& W56& 1420$^{+140}_{-120}$& 12& $\beta\gamma$& \textbf{3}(\textbf{2})[\textbf{2}]$\beta\gamma$& \textbf{2}(\textbf{3})[\textbf{3}]$\beta\gamma\delta$& \textbf{1}(\textbf{1})[\textbf{1}]$\beta\gamma\delta$& \textbf{1}$\beta\gamma\delta$& \textbf{1}(\textbf{1})$\beta\gamma\delta$& \textbf{2}(\textbf{2})[\textbf{3}]$\beta\gamma\delta$& \textbf{2}(\textbf{3})[\textbf{3}]$\beta\gamma\delta$& $\beta\gamma\delta$& $\beta\gamma\delta$& $\cdots$& $\cdots$& \ref{fig:DS75}\\
 70& 11899& 2013-11-15& 04:12& 259.8& 260$\times$240& N06& E71& W72&  560$^{+200}_{- 50}$&  3& $\alpha$& \textbf{1}$\alpha$& (\textbf{1})[\textbf{1}]$\beta$& $\beta\gamma$& $\beta\gamma$& $\beta\gamma$& $\alpha$& $\beta$& $\beta$& \textbf{2}(\textbf{2})[\textbf{2}]$\beta$& $\beta$& $\alpha$& \ref{fig:DS78}\\
 71& 11893& 2013-11-10& 14:12& 251.8& 260$\times$150& S13& E71& W67&  240$^{+100}_{- 50}$&  3& $\alpha$& \textbf{1}(\textbf{1})$\alpha$& [\textbf{1}]$\beta$& $\beta$& $\beta$& $\beta$& $\beta$& $\beta\gamma$& \textbf{2}(\textbf{2})$\beta\gamma$& [\textbf{2}]$\beta\delta$& $\cdots$& $\cdots$& \ref{fig:DS79}\\
 72& 11890& 2013-11-05& 12:48& 155.8& 280$\times$240& S11& E57& W28&  910$^{+250}_{- 10}$&  5& $\beta\gamma$& \textbf{2}(\textbf{2})$\beta\gamma\delta$& \textbf{1}[\textbf{1}]$\beta\gamma\delta$& \textbf{1}(\textbf{2})[\textbf{2}]$\beta\gamma\delta$& [\textbf{1}]$\beta\gamma\delta$& $\beta\gamma\delta$& \textbf{1}(\textbf{1})[\textbf{1}]$\beta\gamma\delta$& $\beta\gamma\delta$& $\cdots$& $\cdots$& $\cdots$& $\cdots$& \ref{fig:DS80}\\
 73& 11877& 2013-10-19& 17:36& 252.8& 235$\times$190& S12& E69& W70&  330$^{+100}_{- 70}$&  5& \textbf{1}(\textbf{1})$\alpha$& \textbf{2}(\textbf{2})[\textbf{1}]$\beta$& [\textbf{2}]$\beta\gamma$& $\beta$& \textbf{1}(\textbf{1})$\beta$& \textbf{1}(\textbf{1})[\textbf{2}]$\beta\gamma\delta$& $\beta\gamma$& $\beta\gamma$& $\beta\gamma$& $\beta\gamma\delta$& $\beta\gamma$& $\beta\gamma$& \ref{fig:DS81}\\
 74& 11875& 2013-10-20& 18:48& 255.8& 375$\times$185& N07& E37& W70&  420$^{+270}_{-300}$&  4& $\beta$& $\beta\gamma$& \textbf{1}(\textbf{1})[\textbf{1}]$\beta\gamma$& $\beta\gamma$& $\beta\gamma\delta$& $\beta\gamma\delta$& \textbf{3}(\textbf{3})[\textbf{2}]$\beta\gamma\delta$& [\textbf{1}]$\beta\gamma\delta$& $\beta\gamma\delta$& $\beta\gamma\delta$& $\beta\gamma\delta$& $\cdots$& \ref{fig:DS82}\\
 75& 11861& 2013-10-10& 20:12& 194.0& 310$\times$150& S10& E33& W74&  350$^{+290}_{- 50}$&  5& $\beta$& \textbf{1}(\textbf{1})$\beta$& \textbf{2}(\textbf{2})[\textbf{2}]$\beta$& [\textbf{1}]$\beta\gamma\delta$& \textbf{1}(\textbf{1})$\beta\gamma$& [\textbf{1}]$\beta\gamma$& $\beta\gamma$& \textbf{1}$\beta$& (\textbf{1})[\textbf{1}]$\beta\gamma$& $\cdots$& $\cdots$& $\cdots$& \ref{fig:DS83}\\
 76& 11818& 2013-08-12& 00:36& 219.8& 245$\times$160& S07& E00& W66&  270$^{+210}_{- 60}$&  3& \textbf{1}$\alpha$& (\textbf{1})[\textbf{1}]$\beta\gamma$& $\beta\gamma$& $\beta\gamma$& $\beta$& $\beta$& \textbf{1}(\textbf{1})$\beta\gamma\delta$& \textbf{1}(\textbf{1})[\textbf{2}]$\beta\gamma\delta$& $\beta\delta$& $\alpha$& $\cdots$& $\cdots$& \ref{fig:DS84}\\
 77& 11734& 2013-05-01& 02:36& 240.8& 300$\times$180& S18& E57& W73&  400$^{+ 50}_{-160}$&  6& \textbf{1}(\textbf{1})[\textbf{1}]$\beta$& \textbf{1}$\beta$& \textbf{1}(\textbf{1})[\textbf{1}]$\beta$& (\textbf{1})$\beta\gamma$& [\textbf{1}]$\beta\gamma$& $\beta\gamma$& $\beta$& \textbf{2}(\textbf{1})[\textbf{1}]$\beta$& \textbf{1}(\textbf{1})[\textbf{1}]$\beta$& (\textbf{1})$\alpha$& [\textbf{1}]$\alpha$& $\cdots$& \ref{fig:DS85}\\
 78& 11711& 2013-04-01& 22:24& 256.0& 200$\times$200& S17& E76& W63&  440$^{+ 90}_{-120}$&  3& $\beta$& \textbf{1}(\textbf{1})$\beta$& \textbf{2}(\textbf{1})[\textbf{1}]$\beta$& (\textbf{1})[\textbf{2}]$\beta$& $\beta$& $\beta$& $\beta$& $\beta$& $\beta$& $\beta$& $\alpha$& $\cdots$& \ref{fig:DS87}\\
 79& 11654& 2013-01-09& 21:48& 239.8& 450$\times$250& N08& E79& W53&  770$^{+510}_{-220}$&  7& \textbf{1}$\beta\gamma$& \textbf{1}(\textbf{2})$\beta\gamma$& \textbf{1}(\textbf{1})[\textbf{1}]$\beta\gamma$& \textbf{3}(\textbf{2})[\textbf{2}]$\beta\gamma$& \textbf{1}(\textbf{2})[\textbf{3}]$\beta\gamma$& [\textbf{1}]$\beta\gamma$& $\beta\gamma$& $\beta\gamma$& $\beta\gamma\delta$& $\beta\gamma\delta$& $\beta\gamma$& $\cdots$& \ref{fig:DS90}\\
 80& 11640& 2013-01-01& 08:00& 184.2& 270$\times$175& N28& E30& W68&  220$^{+140}_{-100}$&  4& $\beta$& \textbf{1}(\textbf{1})[\textbf{1}]$\beta$& $\beta$& \textbf{1}(\textbf{1})[\textbf{1}]$\beta$& \textbf{2}(\textbf{2})[\textbf{2}]$\beta\gamma\delta$& $\beta\gamma\delta$& $\beta\gamma\delta$& $\cdots$& $\cdots$& $\cdots$& $\cdots$& $\cdots$& \ref{fig:DS91}\\
 81& 11623& 2012-12-01& 20:48& 265.8& 290$\times$120& N08& E74& W72&  150$^{+ 60}_{-200}$&  4& $\beta$& $\beta$& $\beta\gamma$& \textbf{1}(\textbf{1})$\beta\gamma$& \textbf{1}(\textbf{1})[\textbf{2}]$\beta\gamma$& $\beta$& \textbf{1}$\beta$& \textbf{1}(\textbf{2})[\textbf{2}]$\beta$& $\beta$& $\beta$& $\alpha$& $\alpha$& \ref{fig:DS92}\\
 82& 11598& 2012-10-22& 07:48& 265.8& 220$\times$170& S11& E71& W76&  330$^{+120}_{- 40}$&  3& \textbf{1}(\textbf{1})[\textbf{1}]$\beta$& $\beta$& $\beta\delta$& $\beta\delta$& $\beta\delta$& $\beta\delta$& \textbf{2}(\textbf{2})[\textbf{2}]$\beta\delta$& $\beta$& $\beta$& $\alpha$& $\alpha$& $\alpha$& \ref{fig:DS93}\\
 83& 11596& 2012-10-20& 16:12& 266.8& 310$\times$150& N07& E75& W72&  290$^{+120}_{- 40}$&  1& $\beta$& \textbf{1}(\textbf{1})[\textbf{1}]$\beta\gamma$& $\beta\gamma$& $\beta\gamma$& $\beta\gamma$& $\beta$& $\beta$& $\beta$& $\beta$& $\beta\gamma$& $\beta$& $\cdots$& \ref{fig:DS94}\\
 84& 11543& 2012-08-09& 21:48& 263.8& 230$\times$150& N21& E72& W71&  250$^{+ 20}_{- 70}$&  4& $\beta$& \textbf{1}(\textbf{1})$\beta$& \textbf{1}(\textbf{1})[\textbf{1}]$\beta\gamma$& \textbf{1}(\textbf{1})[\textbf{1}]$\beta$& [\textbf{1}]$\beta$& $\beta\gamma$& $\beta\gamma$& $\beta\gamma$& \textbf{1}(\textbf{1})[\textbf{1}]$\beta\gamma$& $\alpha$& $\beta$& $\cdots$& \ref{fig:DS96}\\
 85& 11520& 2012-07-08& 22:48& 191.6& 259$\times$181& S16& E50& W54& 1070$^{+150}_{-300}$& 15& \textbf{2}(\textbf{1})$\beta\gamma$& \textbf{1}(\textbf{2})[\textbf{3}]$\beta\gamma$& $\beta\gamma\delta$& \textbf{1}(\textbf{1})$\beta\gamma\delta$& \textbf{3}(\textbf{2})[\textbf{2}]$\beta\gamma\delta$& \textbf{2}(\textbf{3})[\textbf{2}]$\beta\gamma\delta$& \textbf{2}(\textbf{2})[\textbf{3}]$\beta\gamma\delta$& \textbf{4}(\textbf{4})[\textbf{5}]$\beta\gamma\delta$& $\beta\gamma\delta$& $\cdots$& $\cdots$& $\cdots$& \ref{fig:DS97}\\
 86& 11515& 2012-06-28& 08:48& 252.0& 390$\times$200& S17& E68& W69&  620$^{+420}_{-230}$&  3& \textbf{3}(\textbf{1})[\textbf{1}]$\beta$& (\textbf{2})[\textbf{2}]$\beta$& $\beta\gamma$& $\beta\gamma$& $\beta\gamma$& $\beta\gamma$& $\beta\gamma\delta$& $\beta\gamma\delta$& $\beta\gamma\delta$& $\beta\gamma\delta$& $\beta\gamma$& $\cdots$& \ref{fig:DS98}\\
 87& 11513& 2012-07-03& 08:48& 272.0& 190$\times$180& N16& E74& W74&  120$^{+ 40}_{-120}$&  1& $\alpha$& $\beta$& $\beta\gamma$& $\beta\gamma$& $\beta\gamma$& $\beta\gamma$& \textbf{1}(\textbf{1})[\textbf{1}]$\beta\gamma$& $\beta\gamma$& $\beta\gamma$& $\beta$& $\alpha$& $\cdots$& \ref{fig:DS99}\\
 88& 11504& 2012-06-12& 08:00& 271.8& 300$\times$165& S17& E73& W75&  570$^{+510}_{-170}$&  7& $\beta$& $\beta$& \textbf{1}$\beta$& \textbf{1}(\textbf{2})[\textbf{2}]$\beta$& \textbf{1}(\textbf{1})[\textbf{1}]$\beta$& $\beta\gamma\delta$& $\beta\gamma\delta$& \textbf{1}(\textbf{1})[\textbf{1}]$\beta\gamma$& \textbf{1}(\textbf{1})[\textbf{1}]$\beta\gamma$& \textbf{1}(\textbf{1})$\beta$& [\textbf{1}]$\beta$& \textbf{1}(\textbf{1})[\textbf{1}]$\beta$& \ref{fig:DS100}\\
 89& 11476& 2012-05-08& 10:48& 192.8& 365$\times$200& N10& E48& W57&  940$^{+340}_{-100}$&  7& \textbf{2}(\textbf{2})[\textbf{1}]$\beta\gamma$& \textbf{2}(\textbf{1})[\textbf{1}]$\beta\gamma\delta$& (\textbf{1})[\textbf{2}]$\beta\gamma\delta$& $\beta\gamma\delta$& $\beta\gamma\delta$& \textbf{2}(\textbf{1})[\textbf{1}]$\beta\gamma\delta$& \textbf{1}(\textbf{2})[\textbf{1}]$\beta\gamma\delta$& [\textbf{1}]$\beta\gamma$& $\beta$& $\cdots$& $\cdots$& $\cdots$& \ref{fig:DS101}\\
 90& 11460& 2012-04-21& 09:00& 203.8& 225$\times$130& N16& E42& W69&  300$^{+250}_{-150}$&  5& $\beta$& $\beta$& $\beta$& \textbf{1}$\beta$& \textbf{3}(\textbf{4})[\textbf{3}]$\beta$& [\textbf{1}]$\beta$& \textbf{1}(\textbf{1})[\textbf{1}]$\beta$& $\beta$& $\cdots$& $\cdots$& $\cdots$& $\cdots$& \ref{fig:DS103}\\
 91& 11384& 2011-12-22& 02:24& 273.8& 270$\times$185& N12& E76& W74&  330$^{+240}_{-150}$&  5& $\beta$& \textbf{1}$\beta$& \textbf{1}(\textbf{1})[\textbf{1}]$\beta$& (\textbf{1})[\textbf{1}]$\beta$& $\beta$& $\beta$& $\beta$& \textbf{2}(\textbf{1})[\textbf{1}]$\beta$& \textbf{1}(\textbf{2})[\textbf{2}]$\beta$& $\beta$& $\beta$& $\beta$& \ref{fig:DS105}\\
 92& 11339& 2011-11-03& 02:00& 277.8& 340$\times$220& N19& E79& W72& 1030$^{+440}_{-370}$&  5& (\textbf{1})[\textbf{1}]$\beta\gamma$& \textbf{1}(\textbf{1})[\textbf{1}]$\beta\gamma\delta$& \textbf{1}(\textbf{1})[\textbf{1}]$\beta\gamma\delta$& $\beta\gamma\delta$& \textbf{1}(\textbf{1})[\textbf{1}]$\beta\gamma\delta$& \textbf{1}$\beta\gamma\delta$& (\textbf{1})[\textbf{1}]$\beta\gamma$& $\beta\gamma$& $\beta\gamma$& $\beta\gamma$& $\beta\gamma$& $\cdots$& \ref{fig:DS106}\\
 93& 11330& 2011-10-23& 23:48& 241.8& 360$\times$195& N08& E75& W58&  450$^{+240}_{- 50}$&  6& \textbf{2}(\textbf{1})$\beta$& \textbf{1}(\textbf{2})[\textbf{3}]$\beta$& \textbf{2}(\textbf{2})[\textbf{1}]$\beta\gamma$& \textbf{1}(\textbf{1})[\textbf{2}]$\beta\gamma$& $\beta\gamma$& $\beta\gamma$& $\beta\gamma$& $\beta\gamma$& $\beta\gamma$& $\beta\gamma$& $\cdots$& $\cdots$& \ref{fig:DS107}\\
 94& 11314& 2011-10-12& 06:48& 207.8& 190$\times$150& N27& E54& W56&  280$^{+ 50}_{- 50}$&  4& $\alpha$& \textbf{2}(\textbf{1})[\textbf{2}]$\beta$& (\textbf{1})$\beta$& $\beta$& \textbf{1}(\textbf{1})[\textbf{1}]$\beta$& \textbf{1}$\beta$& (\textbf{1})[\textbf{1}]$\beta$& $\beta$& $\beta$& $\cdots$& $\cdots$& $\cdots$& \ref{fig:DS108}\\
 95& 11302& 2011-09-23& 22:12& 267.2& 360$\times$225& N13& E73& W74&  840$^{+430}_{-230}$&  7& \textbf{2}(\textbf{1})$\beta\gamma$& \textbf{1}(\textbf{2})[\textbf{2}]$\beta\gamma$& \textbf{2}(\textbf{1})[\textbf{2}]$\beta\gamma\delta$& (\textbf{1})$\beta\gamma\delta$& \textbf{1}(\textbf{1})[\textbf{2}]$\beta\gamma$& $\beta\gamma\delta$& $\beta\gamma\delta$& \textbf{1}(\textbf{1})$\beta\gamma\delta$& [\textbf{1}]$\beta\gamma\delta$& $\beta\gamma\delta$& $\beta\gamma\delta$& $\beta$& \ref{fig:DS110}\\
 96& 11289& 2011-09-09& 18:36& 233.6& 190$\times$160& N22& E60& W65&  370$^{+ 70}_{- 60}$&  5& \textbf{1}$\beta$& (\textbf{1})$\beta$& \textbf{1}(\textbf{1})[\textbf{1}]$\beta$& \textbf{1}[\textbf{1}]$\beta$& (\textbf{1})[\textbf{1}]$\beta$& $\beta$& $\beta$& \textbf{1}$\beta$& (\textbf{1})[\textbf{1}]$\beta$& \textbf{1}(\textbf{1})[\textbf{1}]$\beta$& $\cdots$& $\cdots$& \ref{fig:DS111}\\
 97& 11263& 2011-07-30& 10:12& 228.8& 310$\times$155& N17& E63& W61&  510$^{+ 70}_{- 90}$&  4& \textbf{1}(\textbf{1})[\textbf{1}]$\beta$& \textbf{2}(\textbf{2})[\textbf{1}]$\beta$& \textbf{1}(\textbf{1})[\textbf{2}]$\beta$& $\beta\gamma\delta$& $\beta\gamma\delta$& $\beta\gamma\delta$& $\beta\gamma\delta$& $\beta\gamma\delta$& $\beta\gamma\delta$& $\beta\gamma\delta$& $\cdots$& $\cdots$& \ref{fig:DS112}\\
\end{longtable}
\end{landscape}
}}

\begin{appendix}

\section{ARs without CEFs}\label{sec:arnocefs}

\begin{figure}%[tbhp]
\sidecaption
  \includegraphics[width=.48\textwidth]{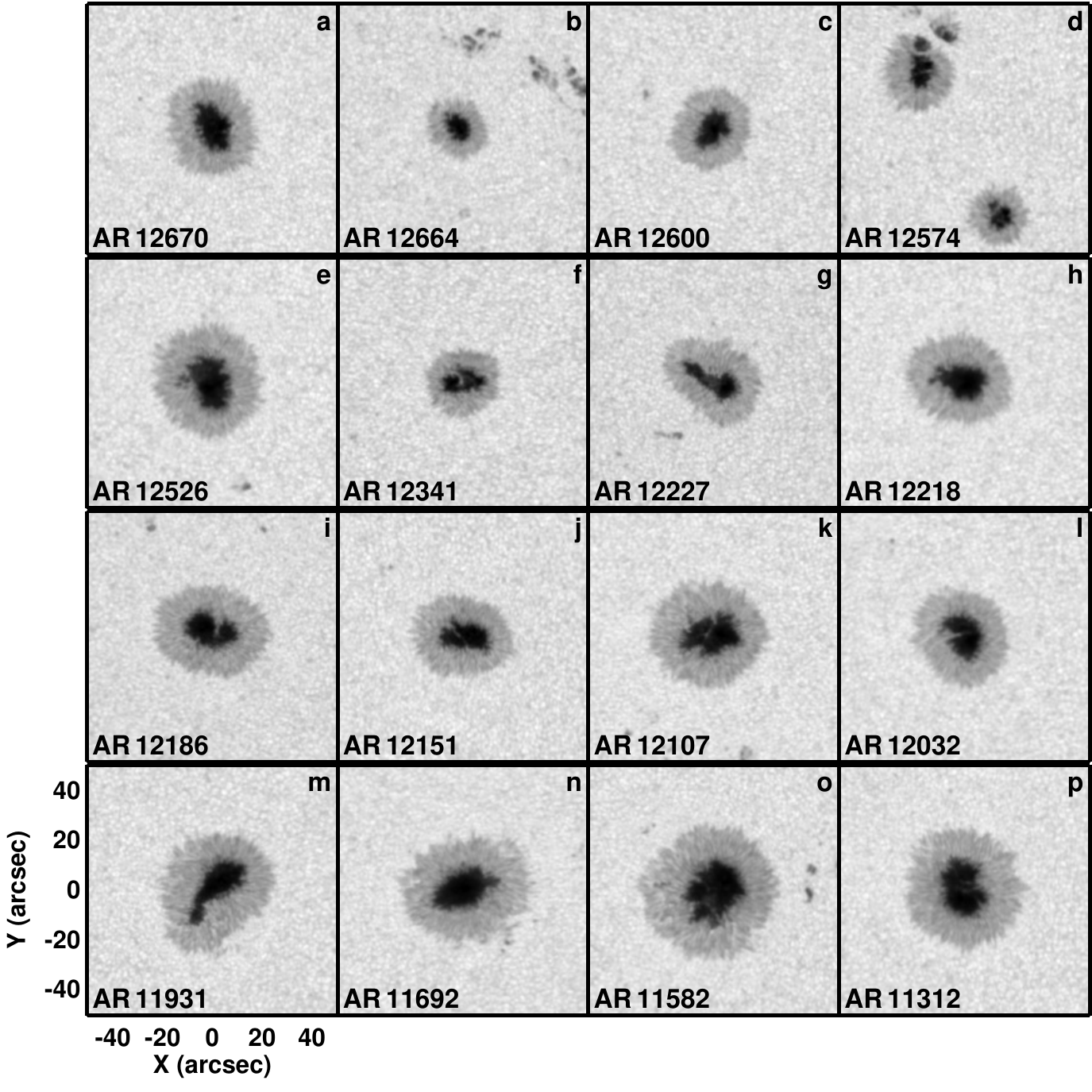}
 \caption{Continuum images of the ARs without CEFs. Images were taken when the ARs were close to disk center.}\label{fig:arswithoutcefs}
 \end{figure}

CEFs appear to be a more common phenomenon than was previously conceived. We could not detect CEFs in only \narswithoutcefper{} of the ARs in the sample. Continuum images of the \narswithoutcef{} ARs without CEFs are presented in Fig~\ref{fig:arswithoutcefs}. Most of these ARs are composed of just one sunspot.  The top part on Table~\ref{tab:OtherCEFs} summarizes the general characteristics of the ARs without CEFs.

%%%%% ALL PLOTS OF CEFS %%%%%
\onecolumn
\section{All CEFs}\label{sec:allcefs}
\begin{figure*}[htbp]
 \centering
 \includegraphics[width=.48\textwidth]{colorbars.pdf}
 \includegraphics[width=.48\textwidth]{colorbars.pdf}
 \includegraphics[width=.48\textwidth]{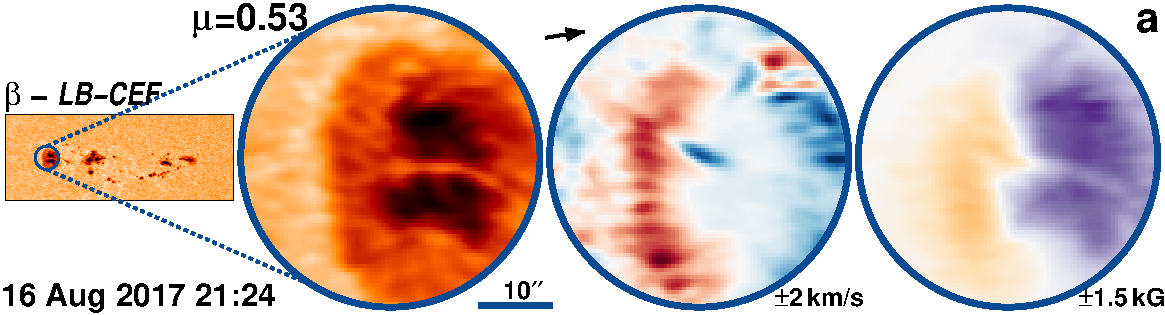}
 \includegraphics[width=.48\textwidth]{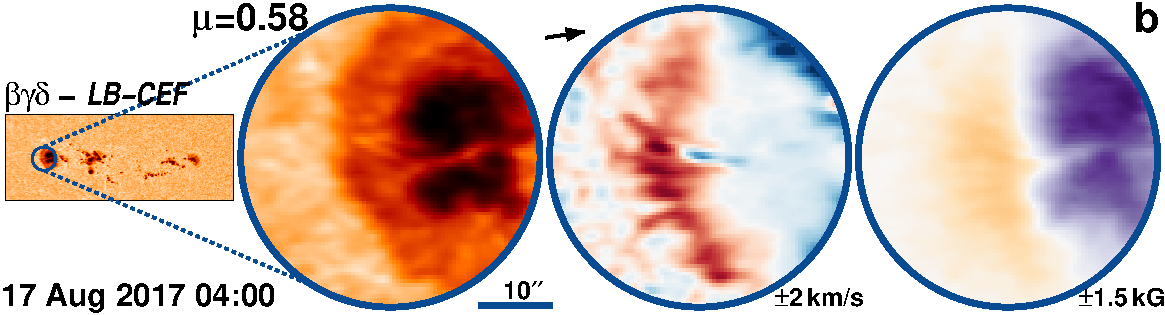}
 \includegraphics[width=.48\textwidth]{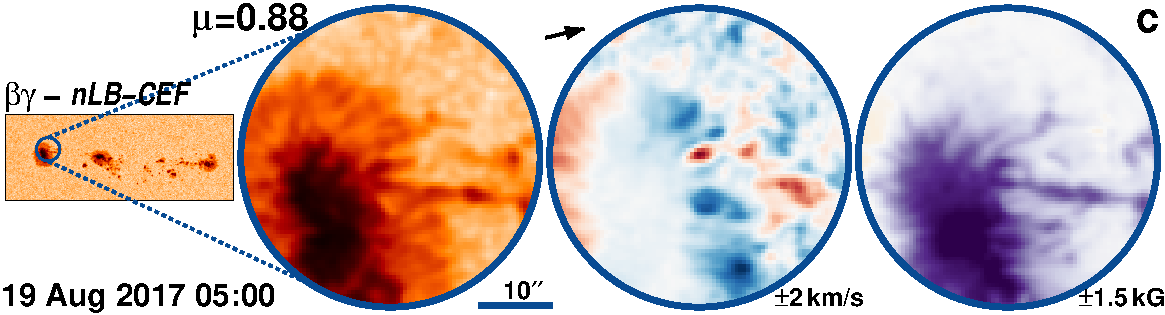}
 \includegraphics[width=.48\textwidth]{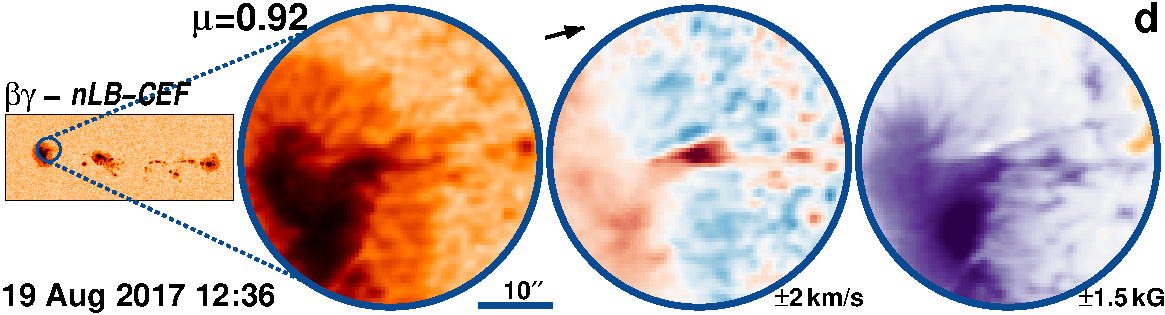}
 \caption{AR\,12671 followed for    8.6 days from 16-Aug-2017 starting at 06:12\,UT.\label{fig:DS00}}
 \end{figure*}

\begin{figure*}[htbp]
 \centering
 \includegraphics[width=.48\textwidth]{colorbars.pdf}
 \includegraphics[width=.48\textwidth]{colorbars.pdf}
 \includegraphics[width=.48\textwidth]{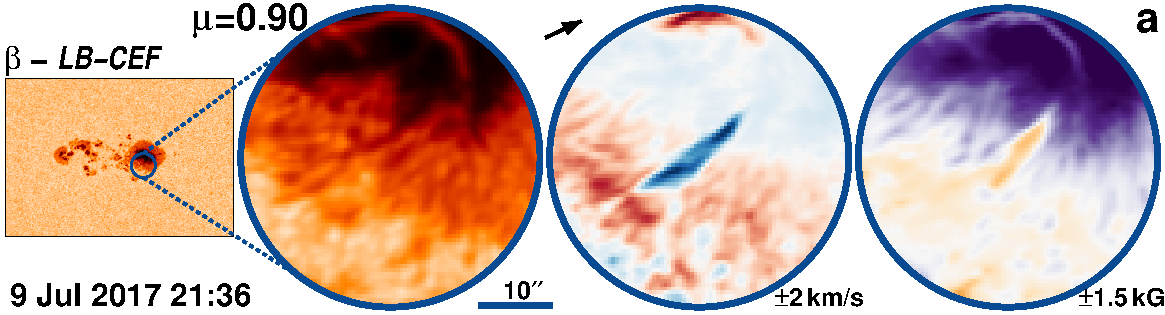}
 \includegraphics[width=.48\textwidth]{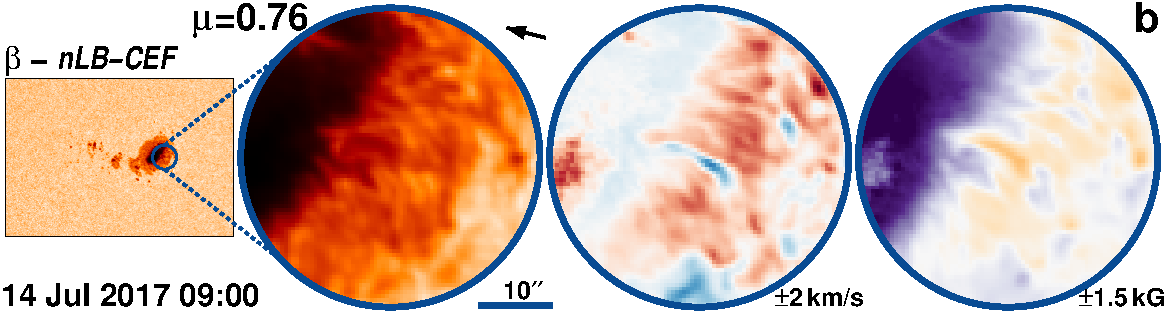}
 \includegraphics[width=.48\textwidth]{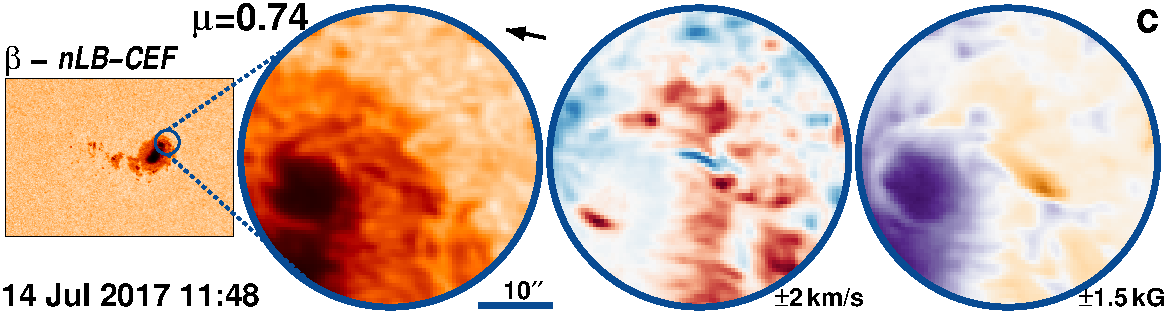}
 \includegraphics[width=.48\textwidth]{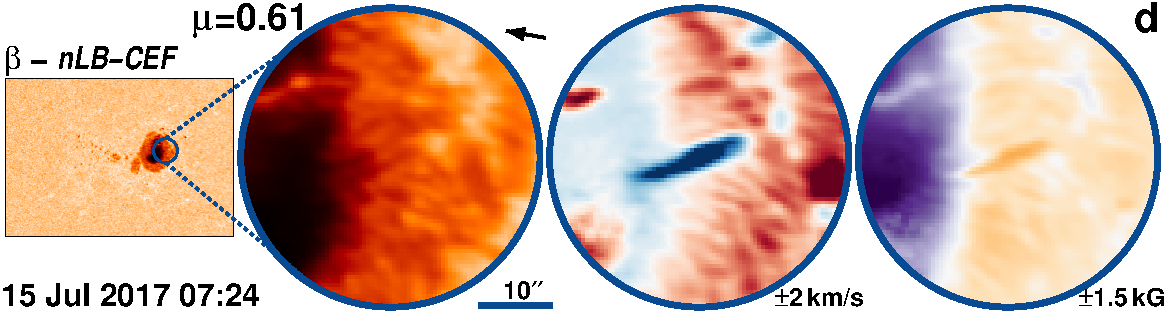}
 \includegraphics[width=.48\textwidth]{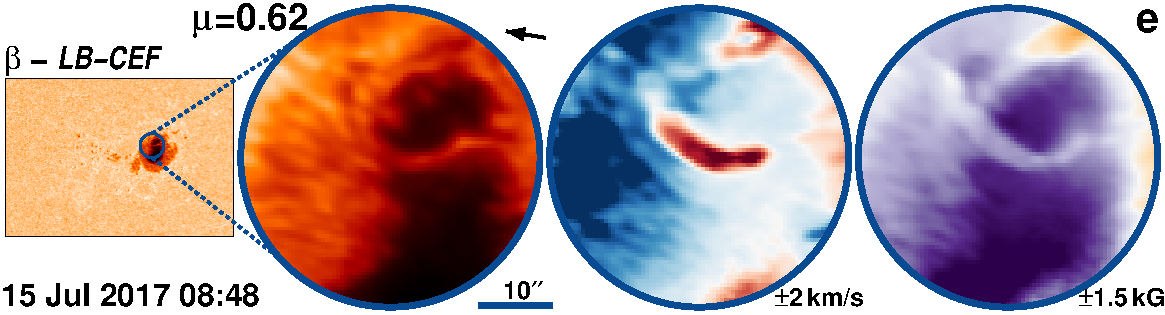}
 \includegraphics[width=.48\textwidth]{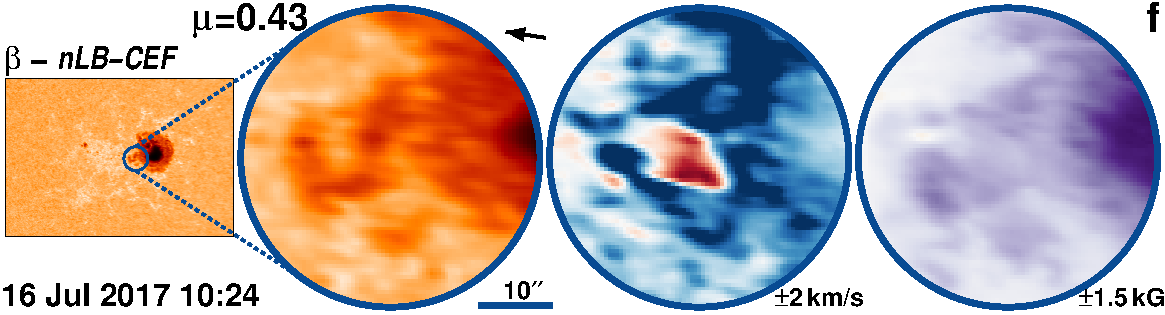}
 \caption{AR\,12665 followed for    9.4 days from  7-Jul-2017 starting at 12:00\,UT.\label{fig:DS02}}
 \end{figure*}

\begin{figure*}[htbp]
 \centering
 \includegraphics[width=.48\textwidth]{colorbars.pdf}
 \includegraphics[width=.48\textwidth]{colorbars.pdf}
 \includegraphics[width=.48\textwidth]{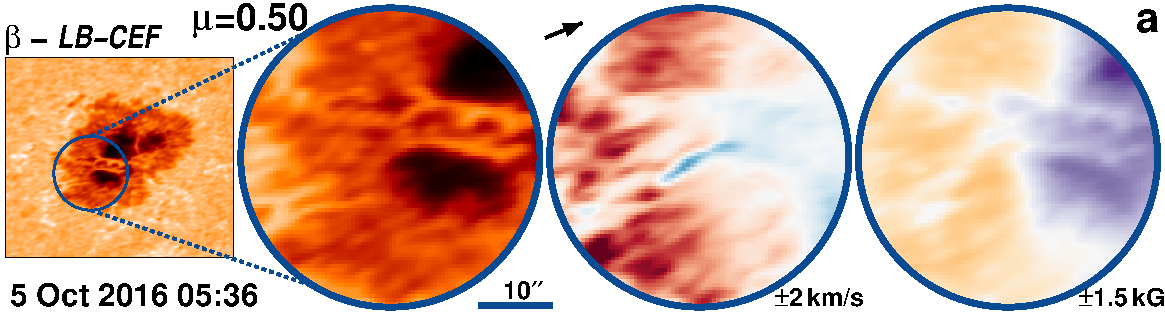}
 \includegraphics[width=.48\textwidth]{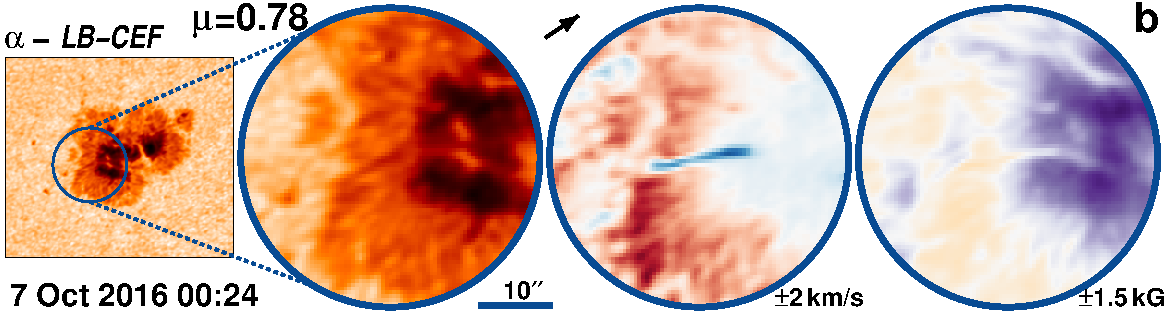}
 \includegraphics[width=.48\textwidth]{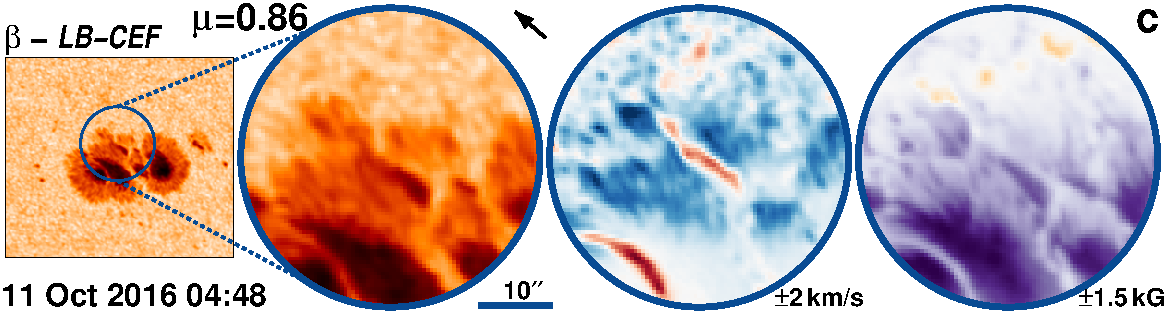}
 \includegraphics[width=.48\textwidth]{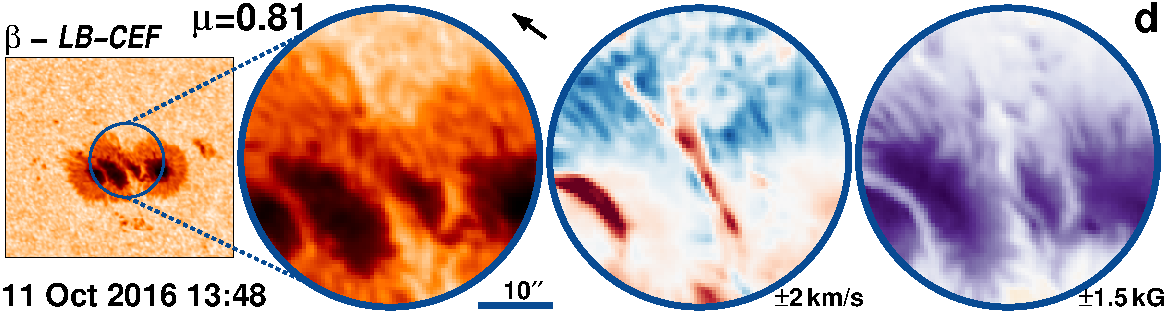}
 \includegraphics[width=.48\textwidth]{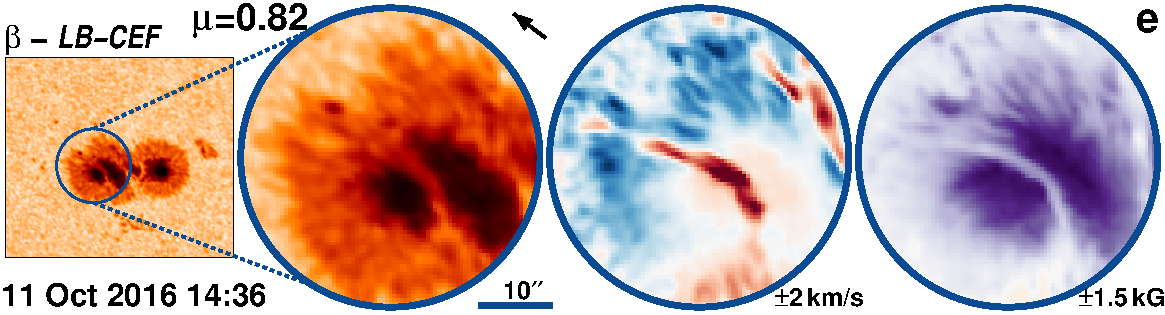}
 \includegraphics[width=.48\textwidth]{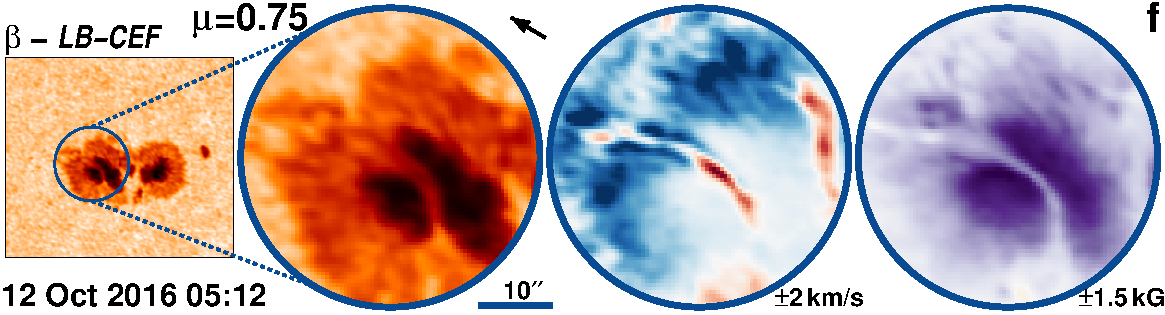}
 \includegraphics[width=.48\textwidth]{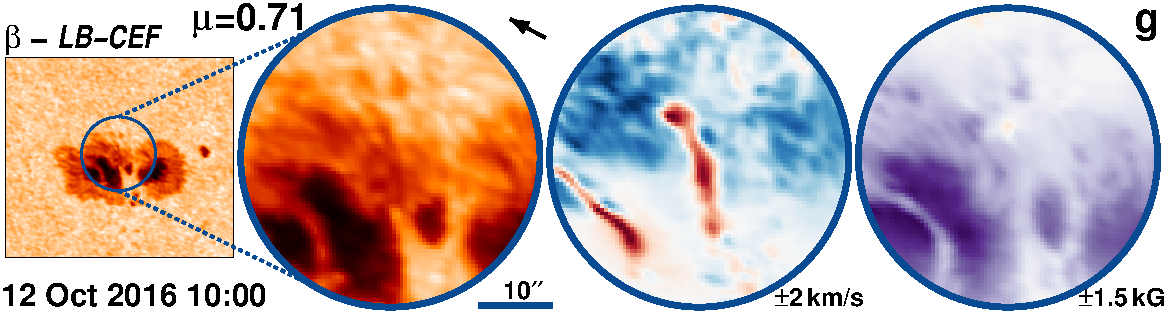}
 \includegraphics[width=.48\textwidth]{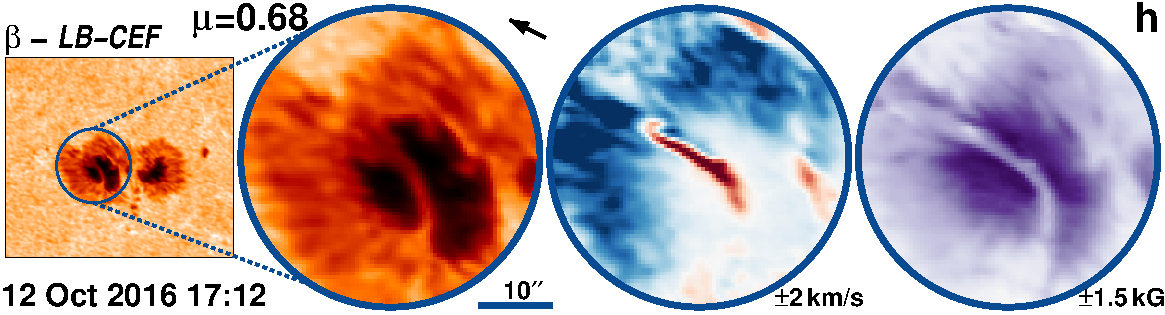}
 \caption{AR\,12599 followed for    9.2 days from  5-Oct-2016 starting at 00:00\,UT.\label{fig:DS05}}
 \end{figure*}

\begin{figure*}[htbp]
 \centering
 \includegraphics[width=.48\textwidth]{colorbars.pdf}
 \includegraphics[width=.48\textwidth]{colorbars.pdf}
 \includegraphics[width=.48\textwidth]{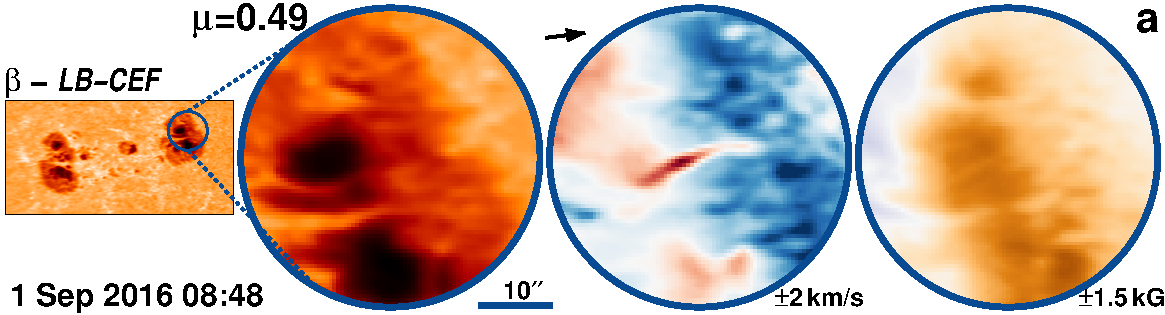}
 \includegraphics[width=.48\textwidth]{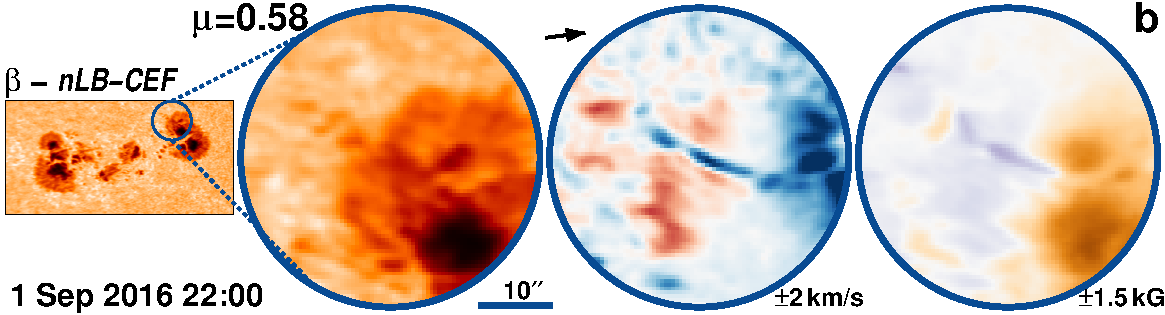}
 \includegraphics[width=.48\textwidth]{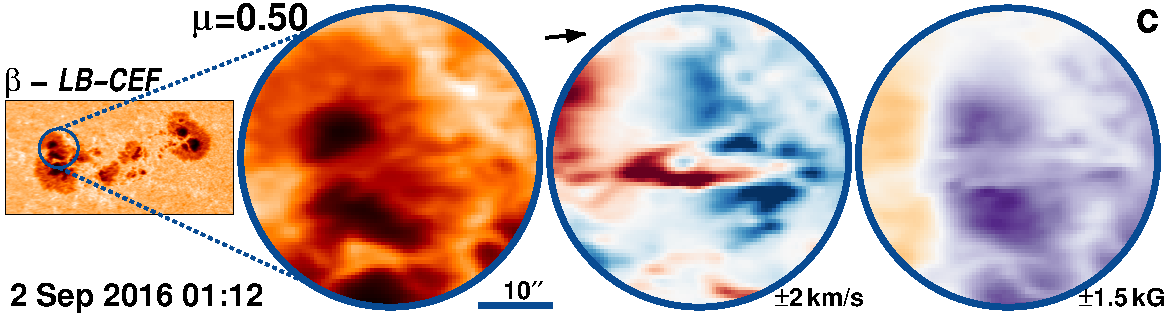}
 \includegraphics[width=.48\textwidth]{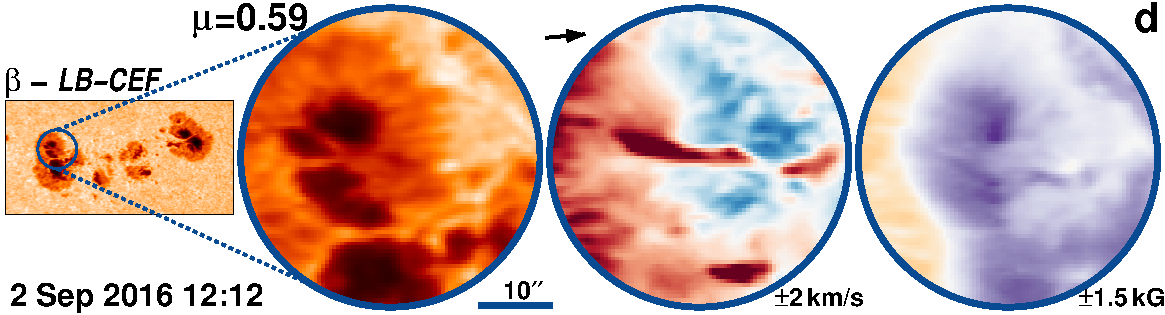}
 \includegraphics[width=.48\textwidth]{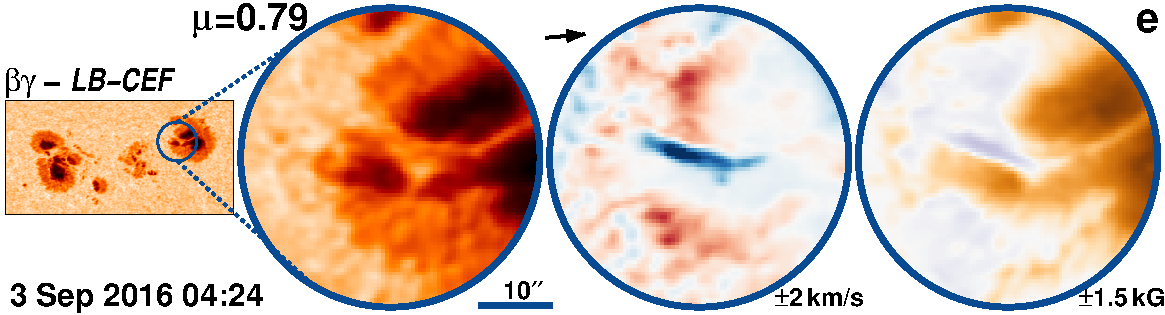}
 \includegraphics[width=.48\textwidth]{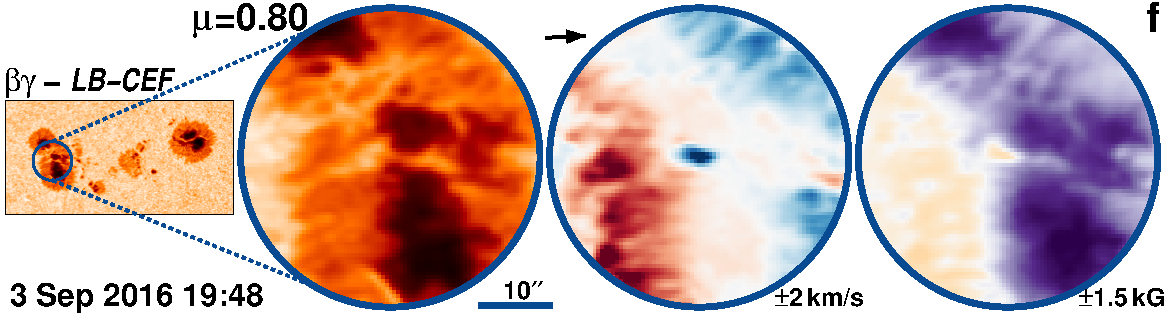}
 \includegraphics[width=.48\textwidth]{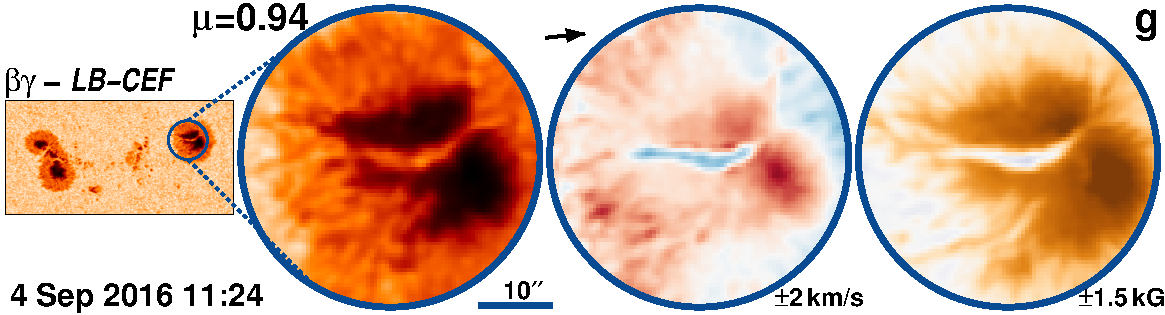}
 \includegraphics[width=.48\textwidth]{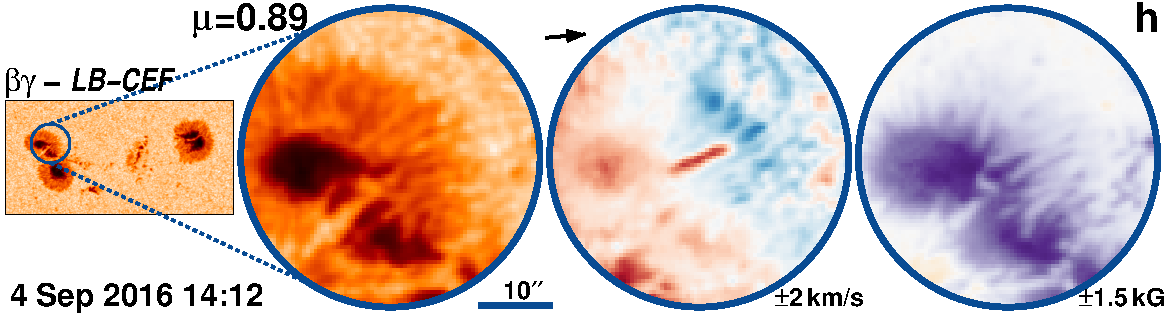}
 \includegraphics[width=.48\textwidth]{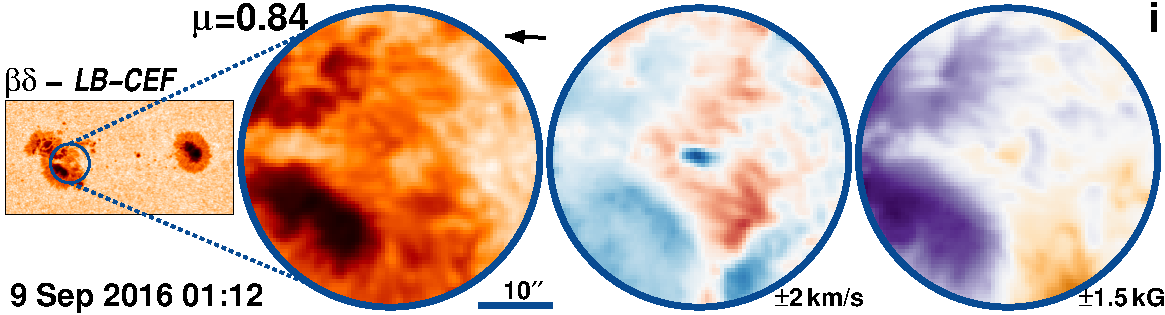}
 \includegraphics[width=.48\textwidth]{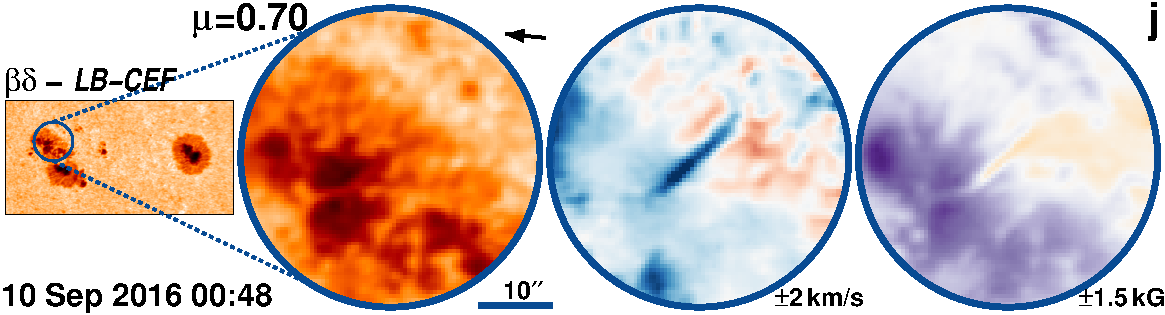}
 \caption{AR\,12585 followed for    9.4 days from  1-Sep-2016 starting at 00:36\,UT.\label{fig:DS06}}
 \end{figure*}

\begin{figure*}[htbp]
 \includegraphics[width=.48\textwidth]{colorbars.pdf}

 \includegraphics[width=.48\textwidth]{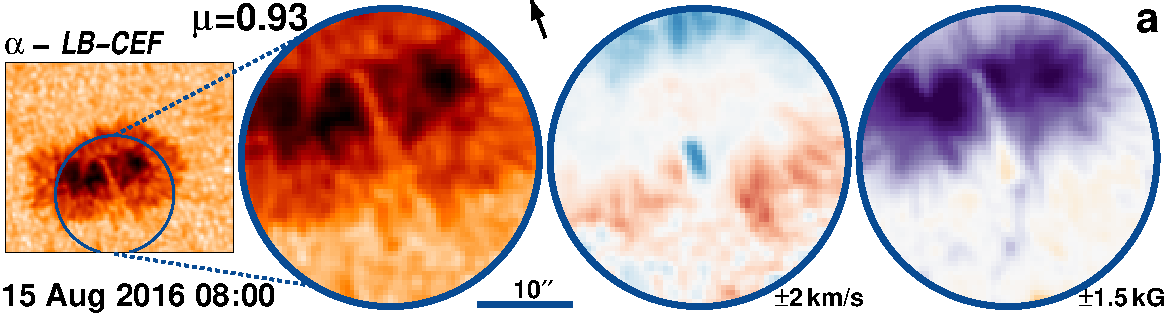}
 \caption{AR\,12576 followed for    6.4 days from 10-Aug-2016 starting at 14:36\,UT.\label{fig:DS07}}
 \end{figure*}

\begin{figure*}[htbp]
 \centering
 \includegraphics[width=.48\textwidth]{colorbars.pdf}
 \includegraphics[width=.48\textwidth]{colorbars.pdf}
 \includegraphics[width=.48\textwidth]{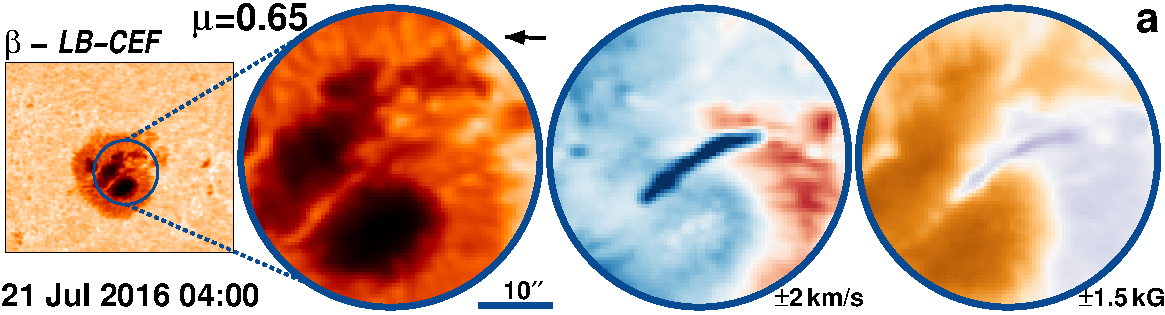}
 \includegraphics[width=.48\textwidth]{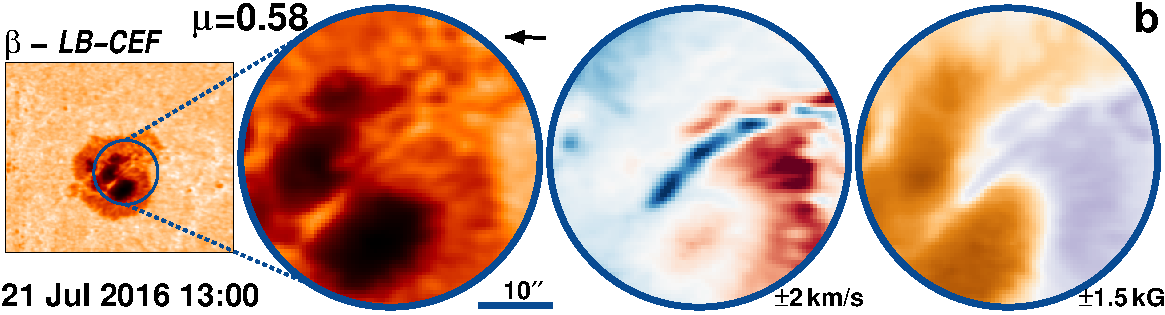}
 \caption{AR\,12565 followed for    8.7 days from 14-Jul-2016 starting at 17:00\,UT.\label{fig:DS09}}
 \end{figure*}

\begin{figure*}[htbp]
 \includegraphics[width=.48\textwidth]{colorbars.pdf}
 \includegraphics[width=.48\textwidth]{colorbars.pdf}
 \includegraphics[width=.48\textwidth]{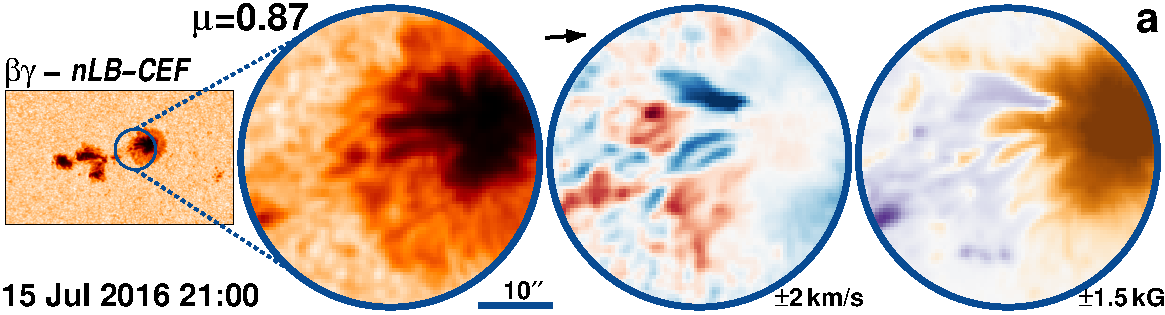}
 \includegraphics[width=.48\textwidth]{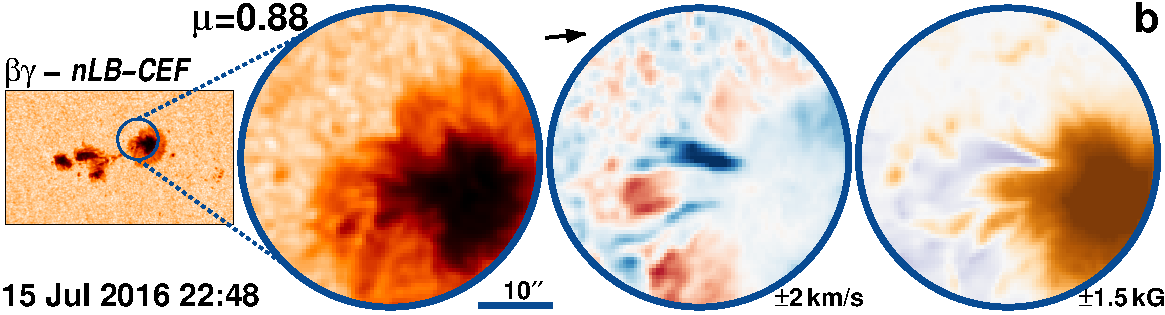}
 \includegraphics[width=.48\textwidth]{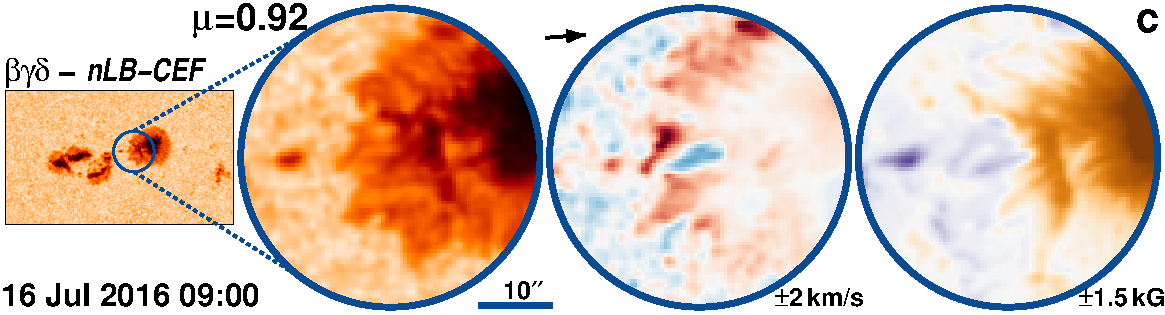}
 \includegraphics[width=.48\textwidth]{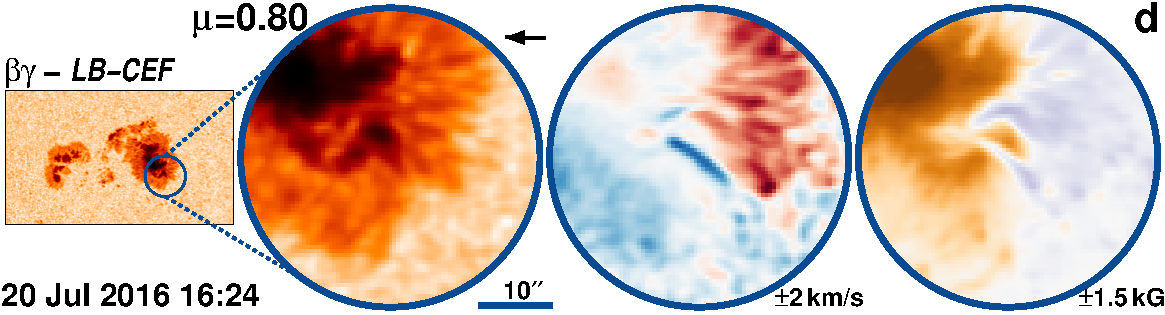}
 \includegraphics[width=.48\textwidth]{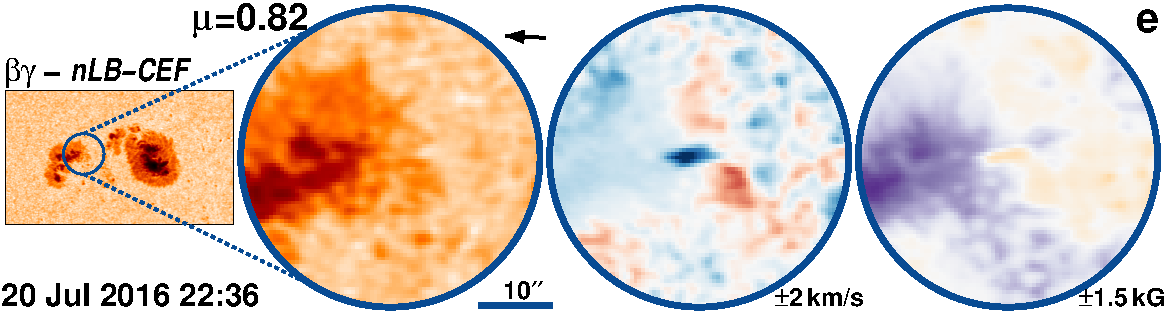}
 \includegraphics[width=.48\textwidth]{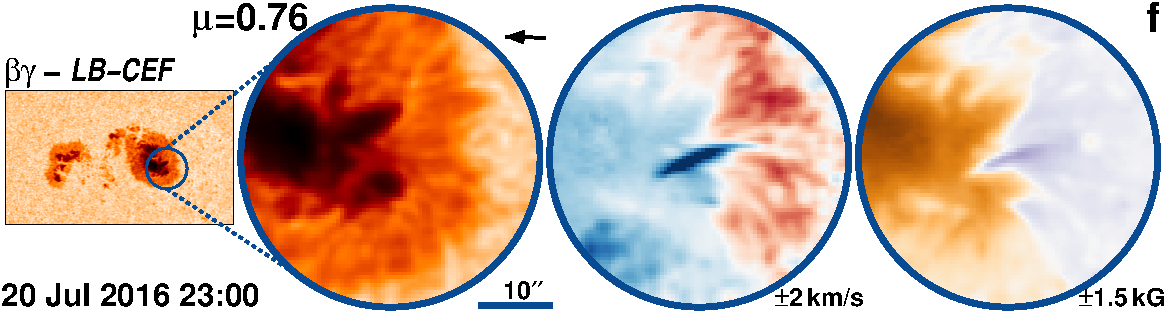}
 \includegraphics[width=.48\textwidth]{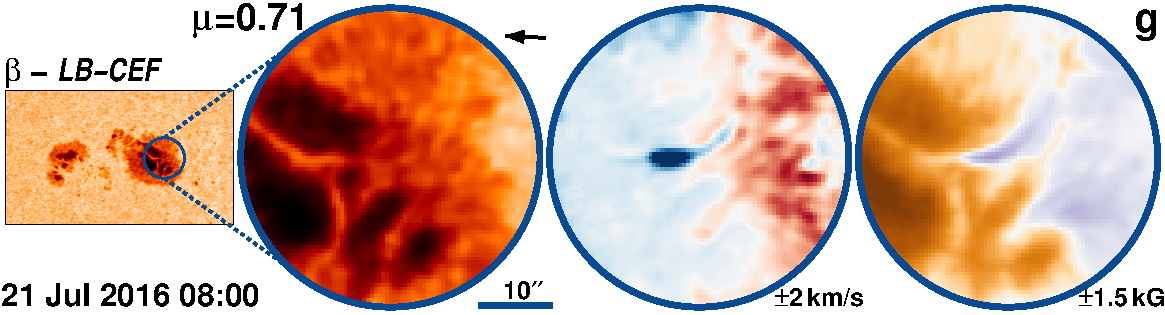}
 \includegraphics[width=.48\textwidth]{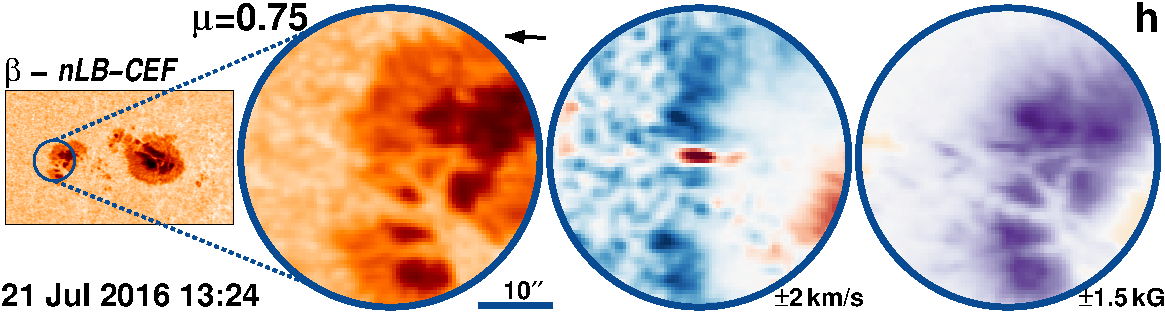}
 \includegraphics[width=.48\textwidth]{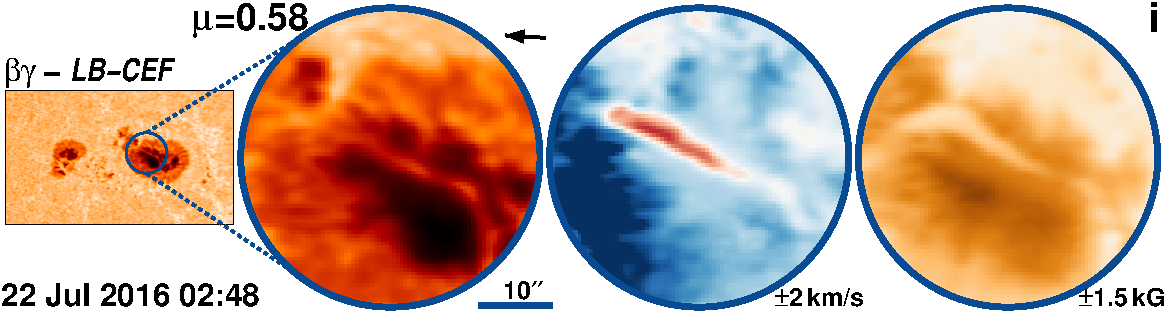}
 \caption{AR\,12567 followed for   10.6 days from 12-Jul-2016 starting at 19:00\,UT.\label{fig:DS10}}
 \end{figure*}

\begin{figure*}[htbp]
 \centering
 \includegraphics[width=.48\textwidth]{colorbars.pdf}
 \includegraphics[width=.48\textwidth]{colorbars.pdf}
 \includegraphics[width=.48\textwidth]{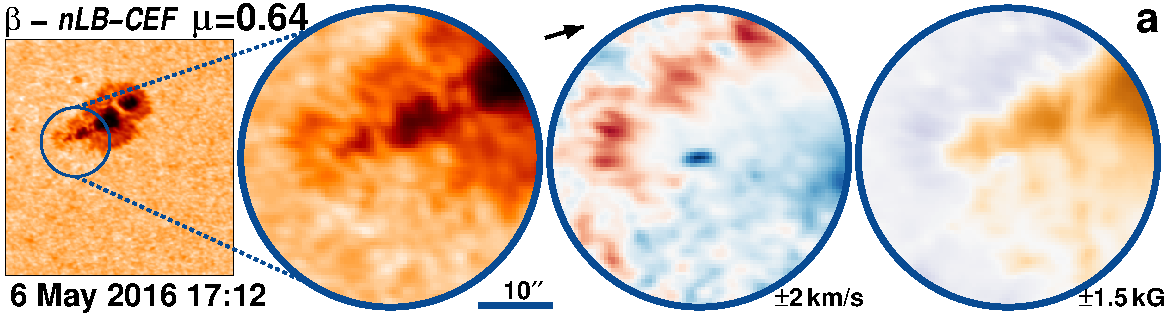}
 \includegraphics[width=.48\textwidth]{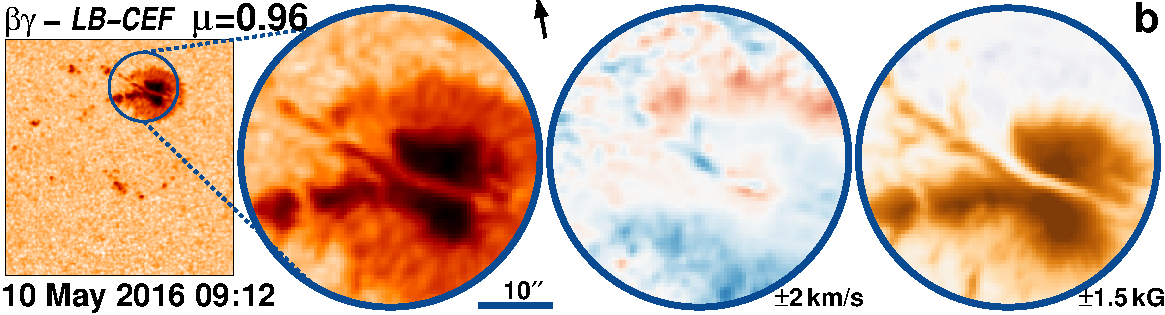}
 \caption{AR\,12542 followed for    8.1 days from  5-May-2016 starting at 21:00\,UT.\label{fig:DS13}}
 \end{figure*}

\begin{figure*}[htbp]
 \centering
 \includegraphics[width=.48\textwidth]{colorbars.pdf}
 \includegraphics[width=.48\textwidth]{colorbars.pdf}
 \includegraphics[width=.48\textwidth]{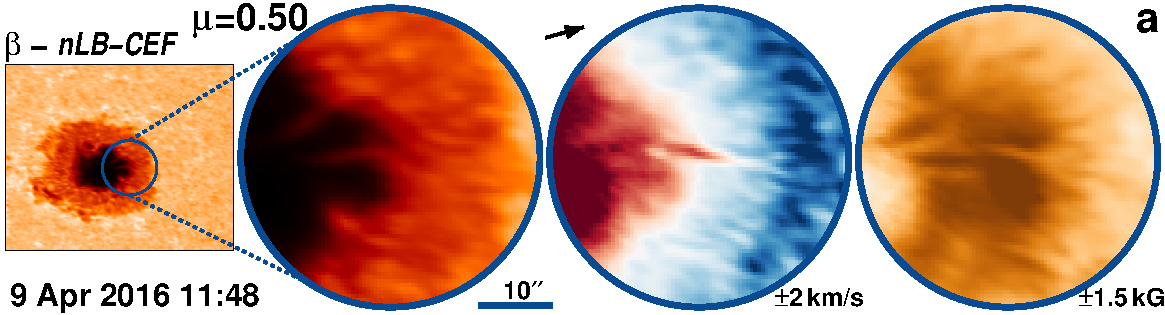}
 \includegraphics[width=.48\textwidth]{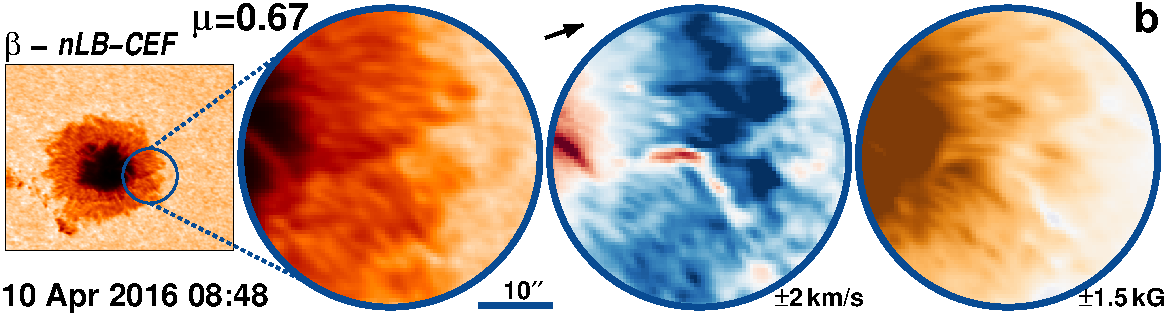}
 \includegraphics[width=.48\textwidth]{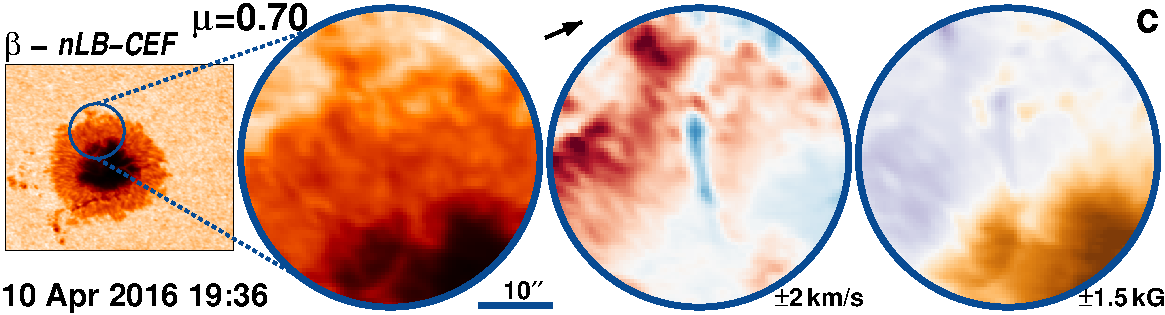}
 \includegraphics[width=.48\textwidth]{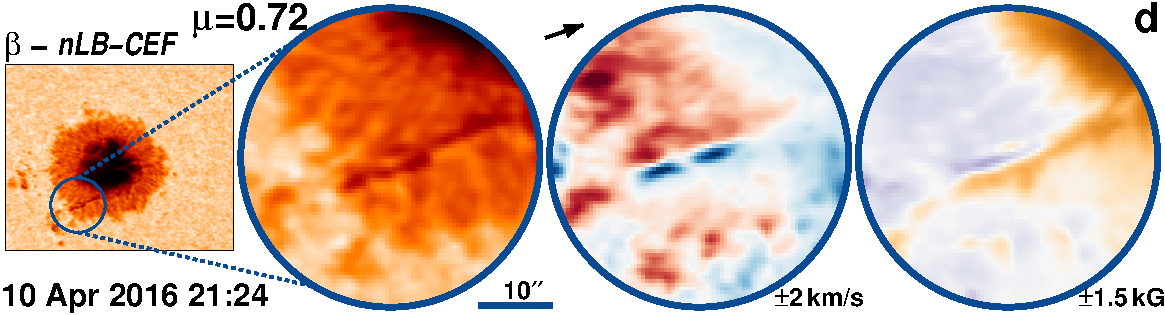}
 \includegraphics[width=.48\textwidth]{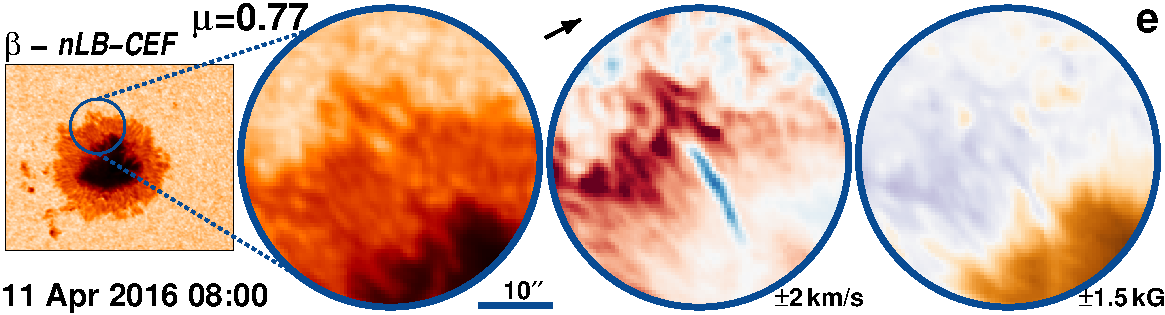}
 \includegraphics[width=.48\textwidth]{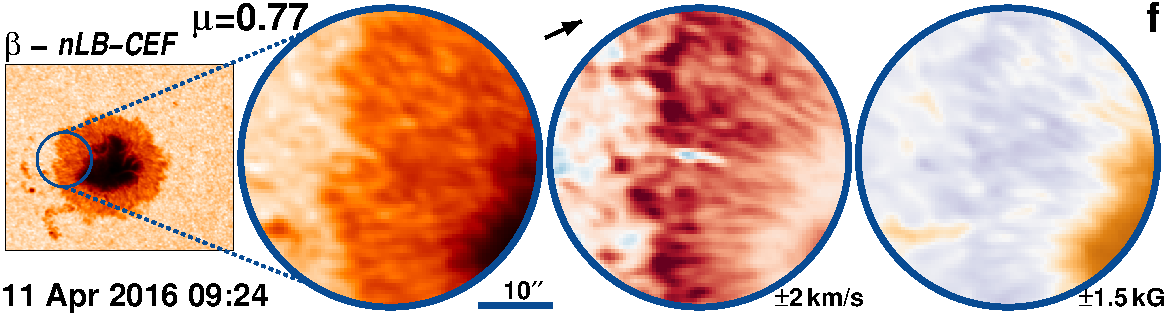}
 \includegraphics[width=.48\textwidth]{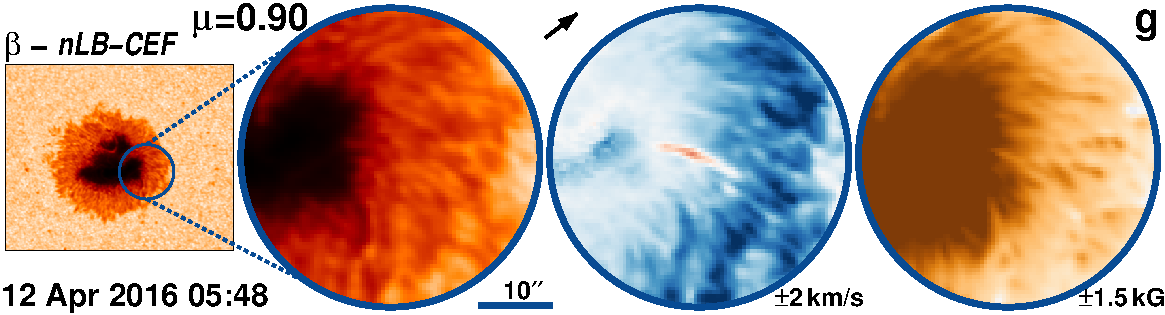}
 \includegraphics[width=.48\textwidth]{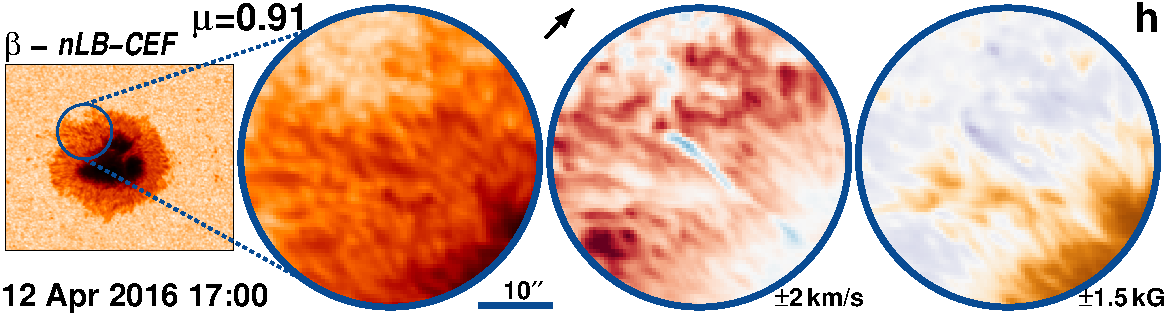}
 \includegraphics[width=.48\textwidth]{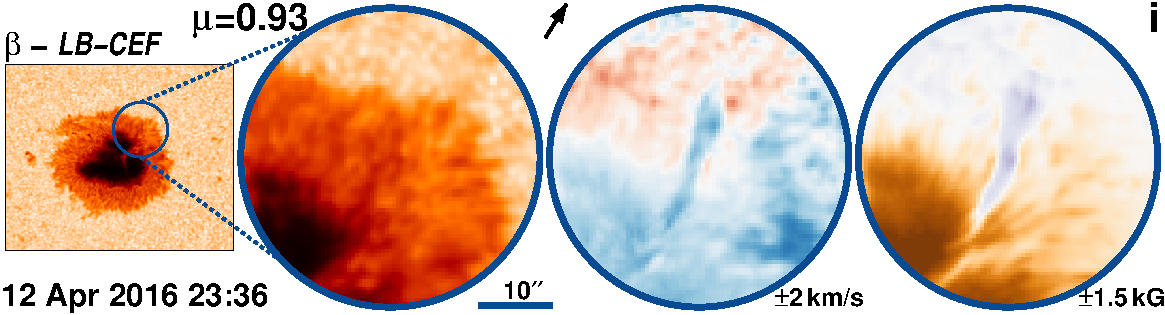}
 \includegraphics[width=.48\textwidth]{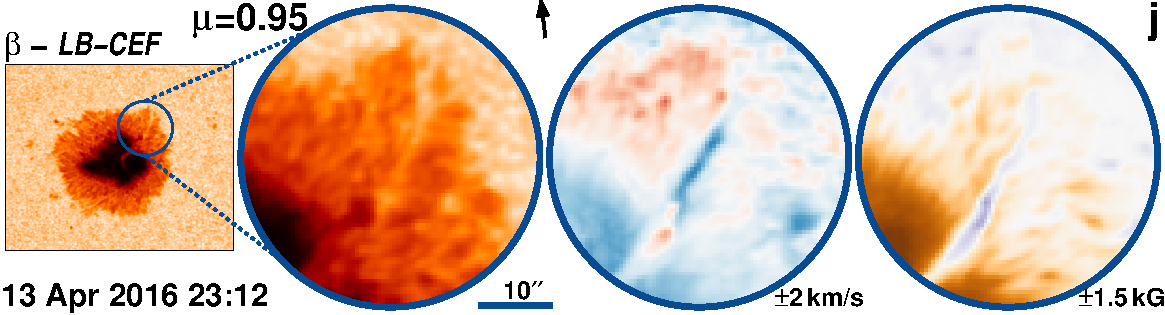}
 \includegraphics[width=.48\textwidth]{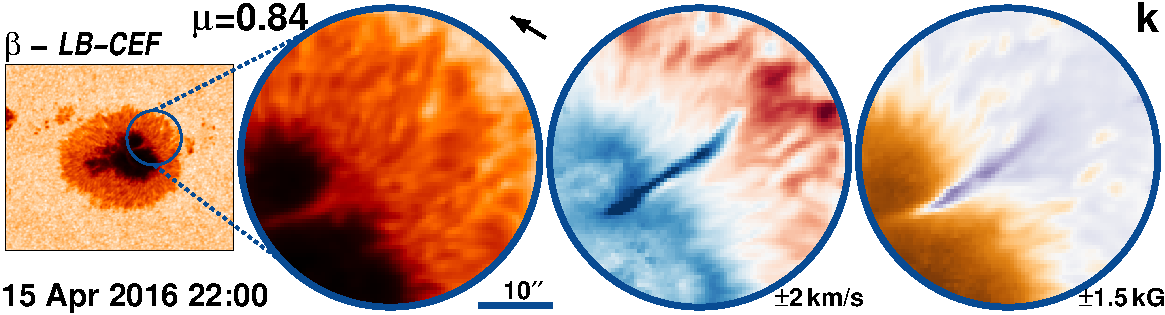}
 \includegraphics[width=.48\textwidth]{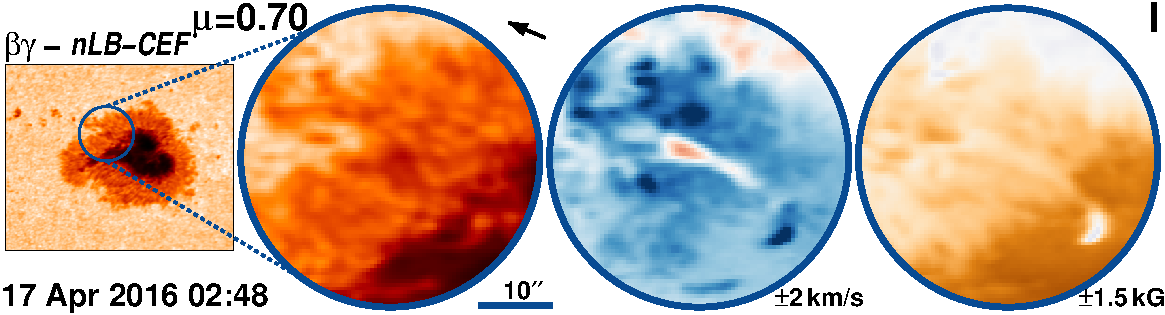}
 \includegraphics[width=.48\textwidth]{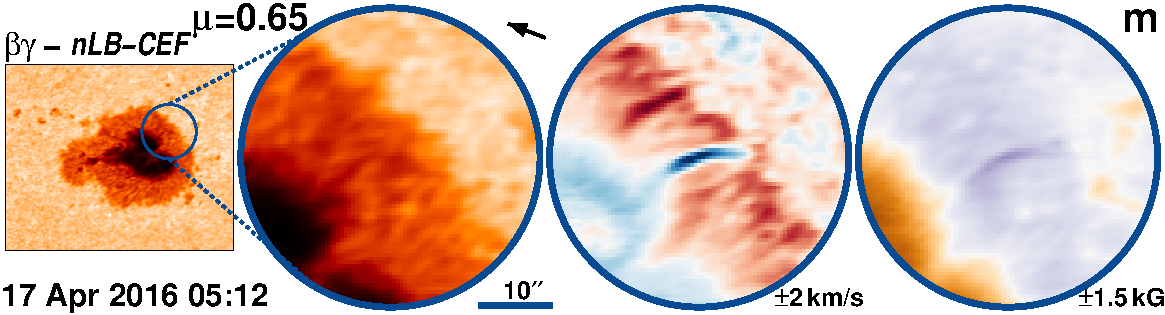}
 \includegraphics[width=.48\textwidth]{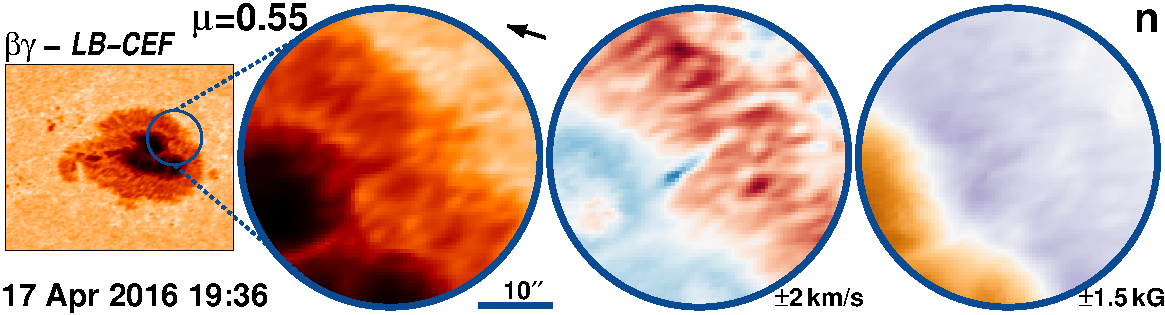}
 \caption{AR\,12529 followed for    9.2 days from  9-Apr-2016 starting at 08:00\,UT.\label{fig:DS14}}
 \end{figure*}

\begin{figure*}[htbp]
 \centering
 \includegraphics[width=.48\textwidth]{colorbars.pdf}
 \includegraphics[width=.48\textwidth]{colorbars.pdf}
 \includegraphics[width=.48\textwidth]{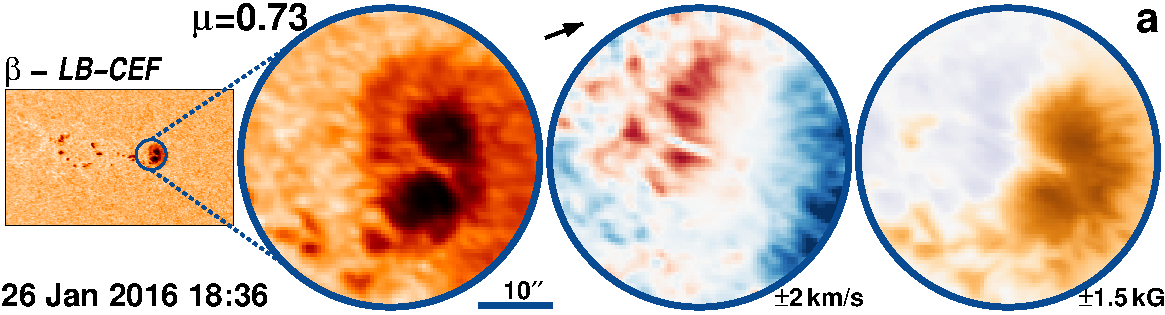}
 \includegraphics[width=.48\textwidth]{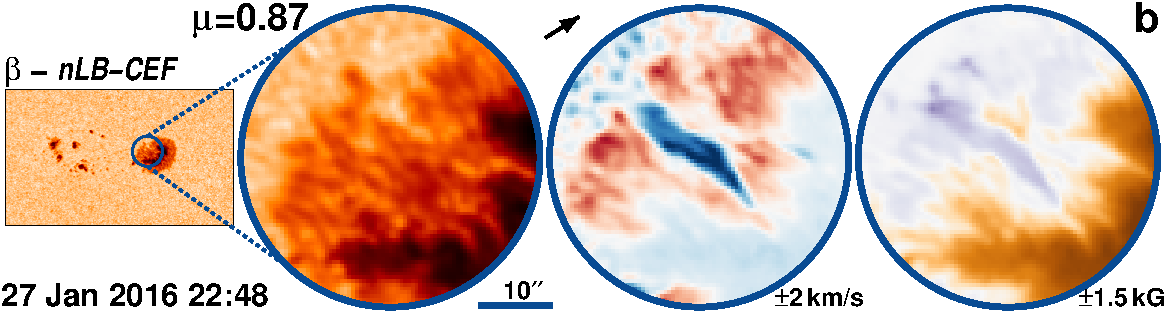}
 \includegraphics[width=.48\textwidth]{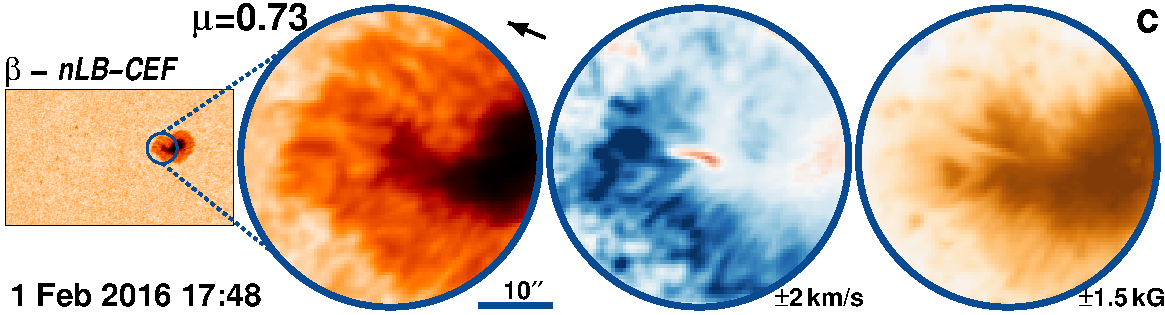}
 \includegraphics[width=.48\textwidth]{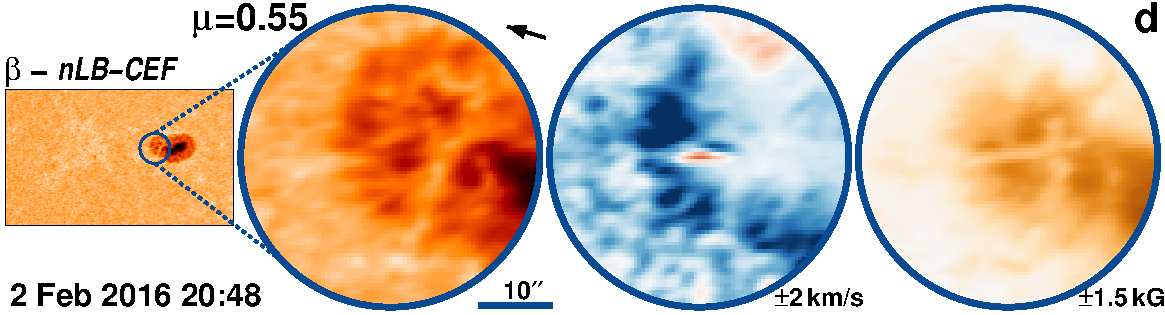}
 \caption{AR\,12489 followed for    8.9 days from 26-Jan-2016 starting at 00:00\,UT.\label{fig:DS16}}
 \end{figure*}

\begin{figure*}[htbp]
 \includegraphics[width=.48\textwidth]{colorbars.pdf}

 \includegraphics[width=.48\textwidth]{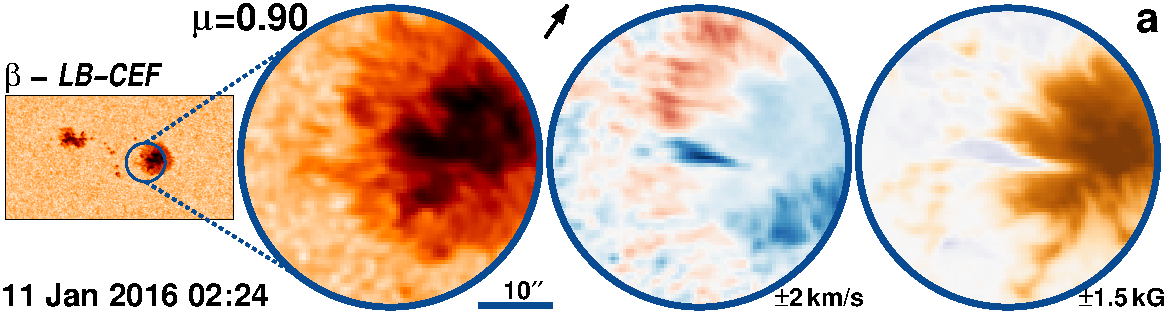}
 \caption{AR\,12483 followed for    8.6 days from  8-Jan-2016 starting at 10:36\,UT.\label{fig:DS17}}
 \end{figure*}

\begin{figure*}[htbp]
 \includegraphics[width=.48\textwidth]{colorbars.pdf}

 \includegraphics[width=.48\textwidth]{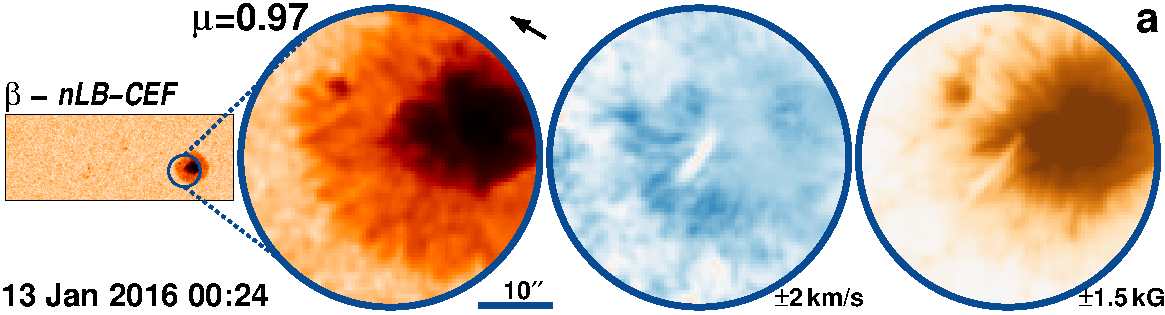}
 \caption{AR\,12480 followed for    9.4 days from  7-Jan-2016 starting at 13:36\,UT.\label{fig:DS18}}
 \end{figure*}

\begin{figure*}[htbp]
 \centering
 \includegraphics[width=.48\textwidth]{colorbars.pdf}
 \includegraphics[width=.48\textwidth]{colorbars.pdf}
 \includegraphics[width=.48\textwidth]{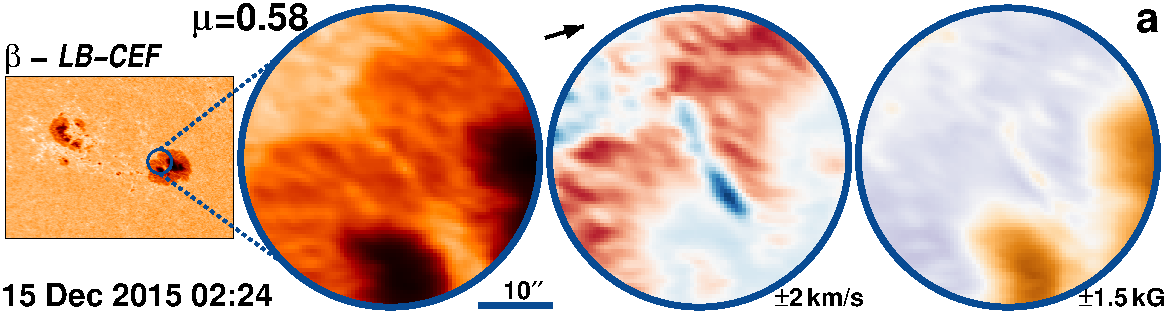}
 \includegraphics[width=.48\textwidth]{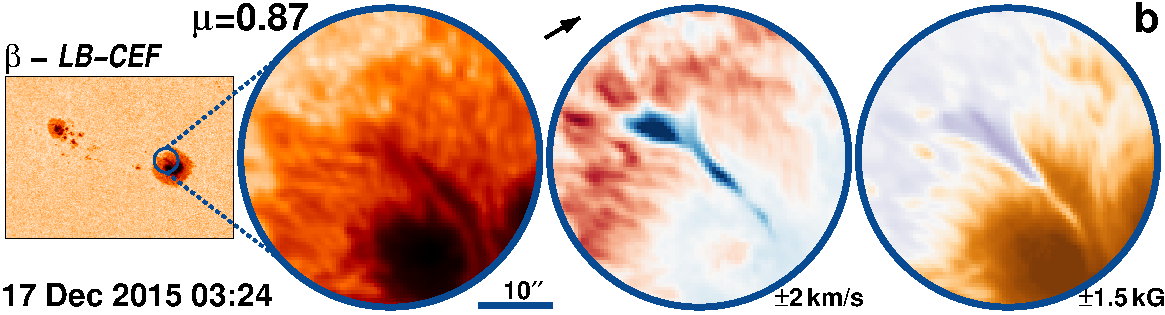}
 \caption{AR\,12470 followed for   10.9 days from 13-Dec-2015 starting at 21:36\,UT.\label{fig:DS19}}
 \end{figure*}

\begin{figure*}[htbp]
 \includegraphics[width=.48\textwidth]{colorbars.pdf}
 \includegraphics[width=.48\textwidth]{colorbars.pdf}
 \includegraphics[width=.48\textwidth]{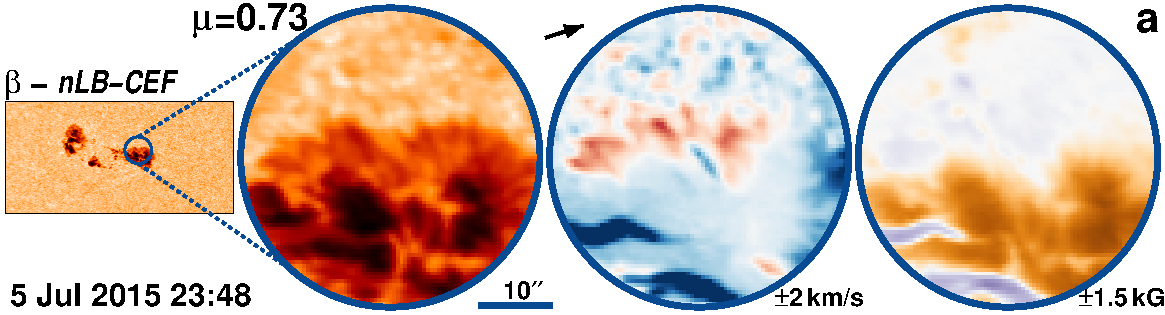}
 \includegraphics[width=.48\textwidth]{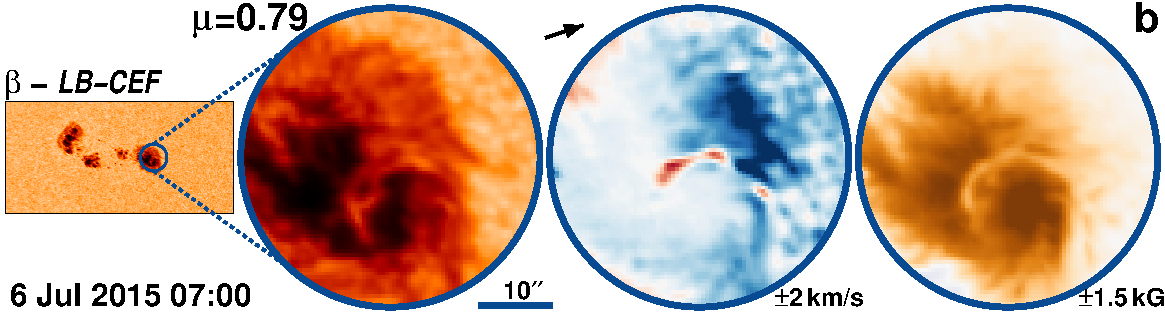}
 \includegraphics[width=.48\textwidth]{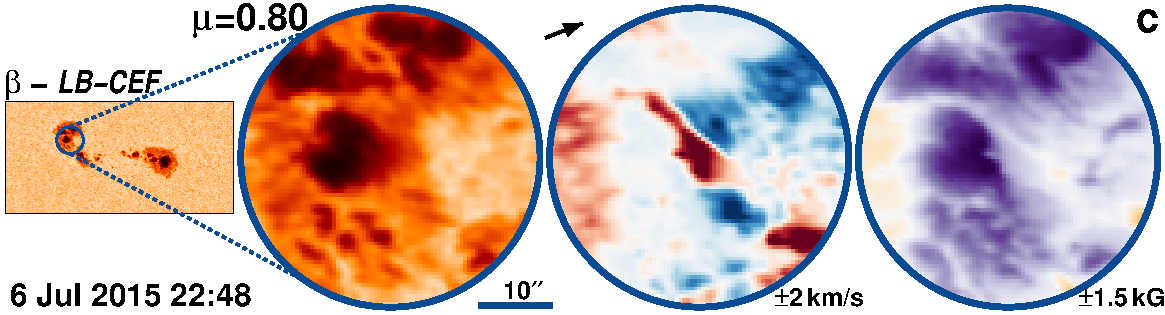}
 \includegraphics[width=.48\textwidth]{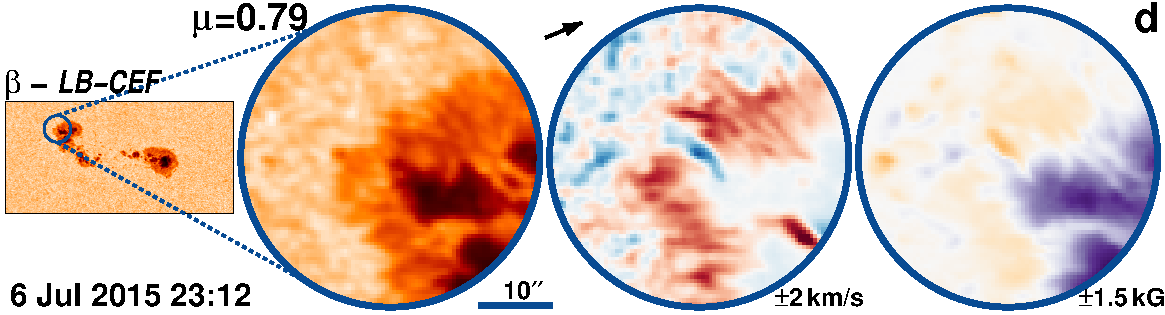}
 \includegraphics[width=.48\textwidth]{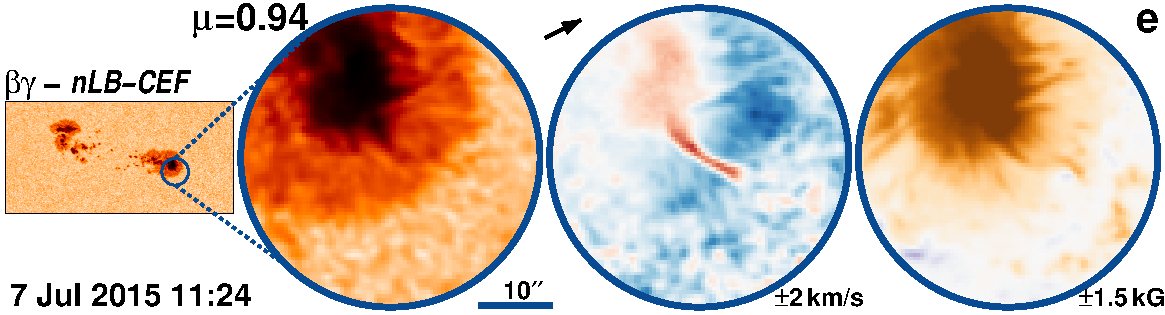}
 \includegraphics[width=.48\textwidth]{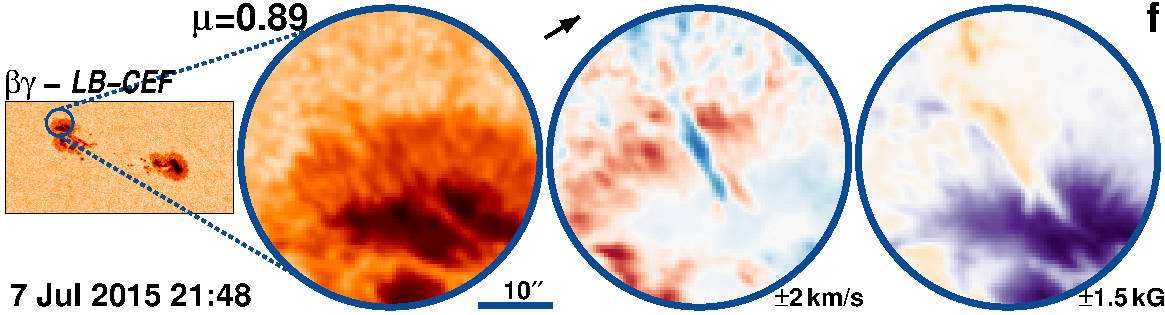}
 \includegraphics[width=.48\textwidth]{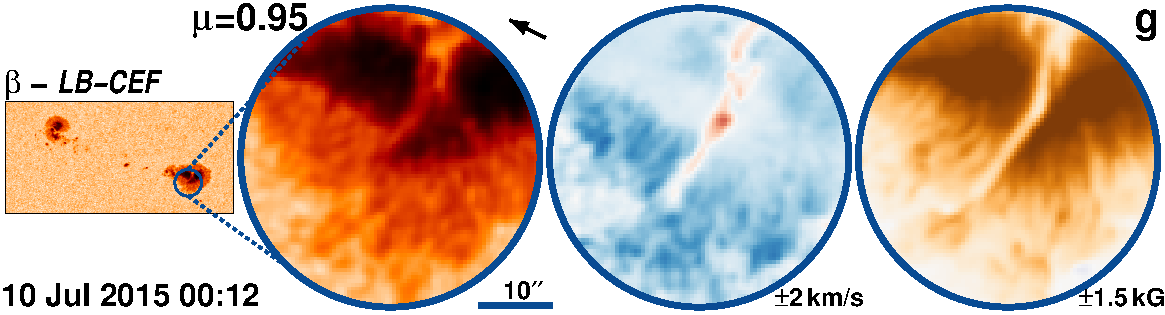}
 \includegraphics[width=.48\textwidth]{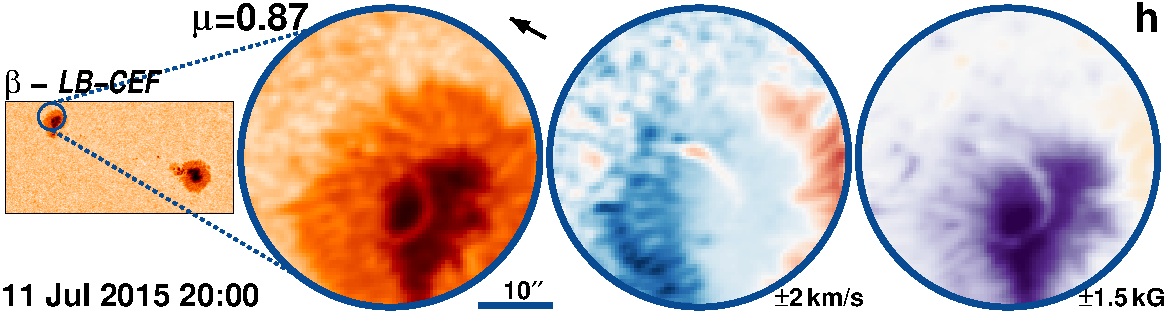}
 \includegraphics[width=.48\textwidth]{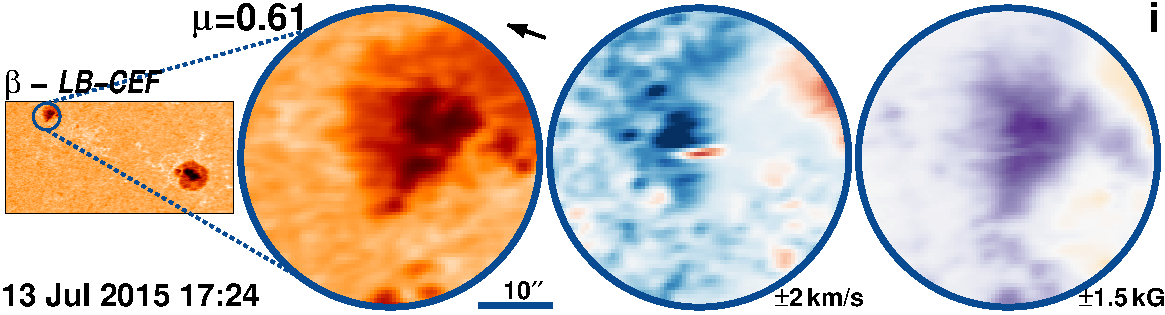}
 \caption{AR\,12381 followed for   10.7 days from  5-Jul-2015 starting at 06:12\,UT.\label{fig:DS23}}
 \end{figure*}

\begin{figure*}[htbp]
 \includegraphics[width=.48\textwidth]{colorbars.pdf}
 \includegraphics[width=.48\textwidth]{colorbars.pdf}
 \includegraphics[width=.48\textwidth]{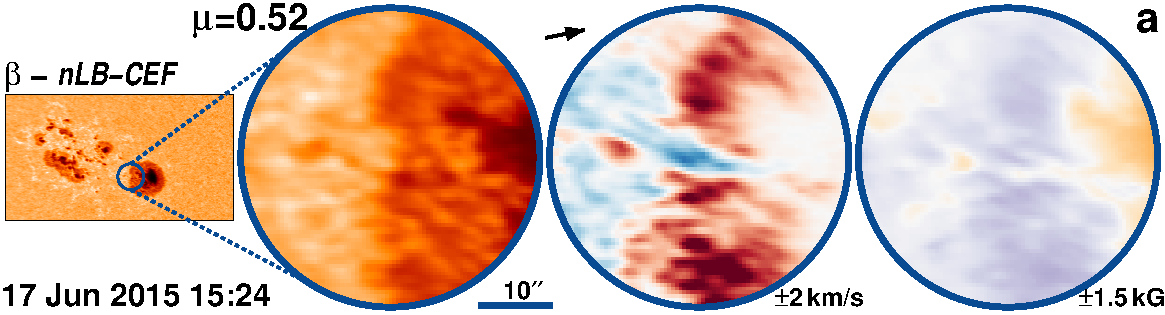}
 \includegraphics[width=.48\textwidth]{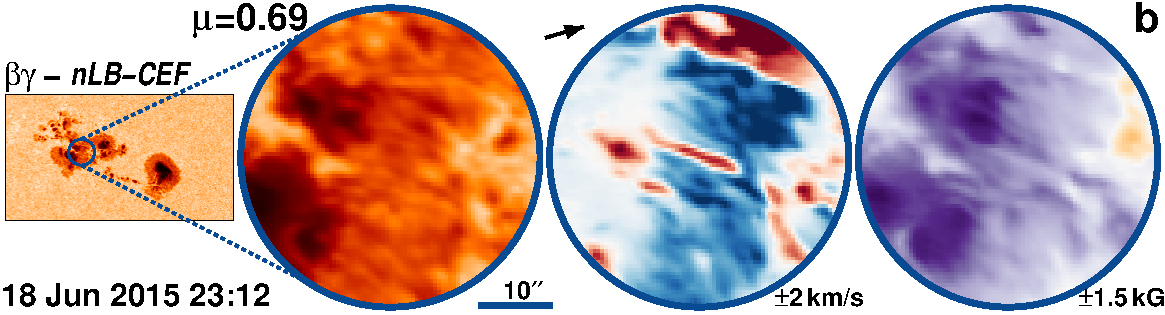}
 \includegraphics[width=.48\textwidth]{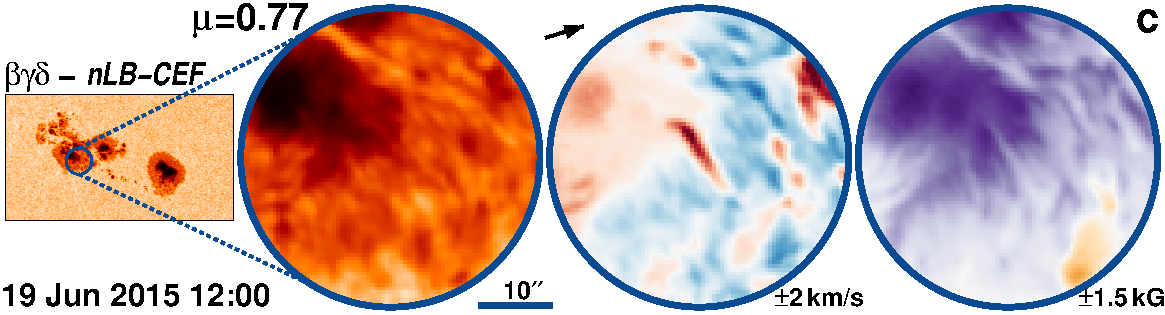}
 \includegraphics[width=.48\textwidth]{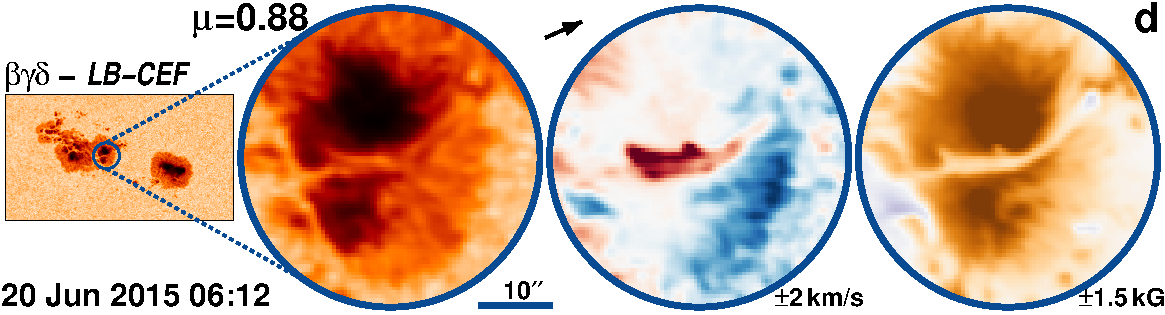}
 \includegraphics[width=.48\textwidth]{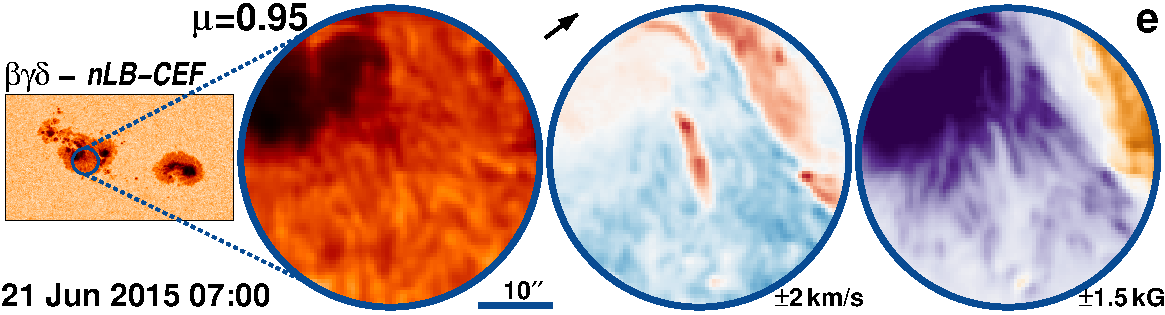}
 \includegraphics[width=.48\textwidth]{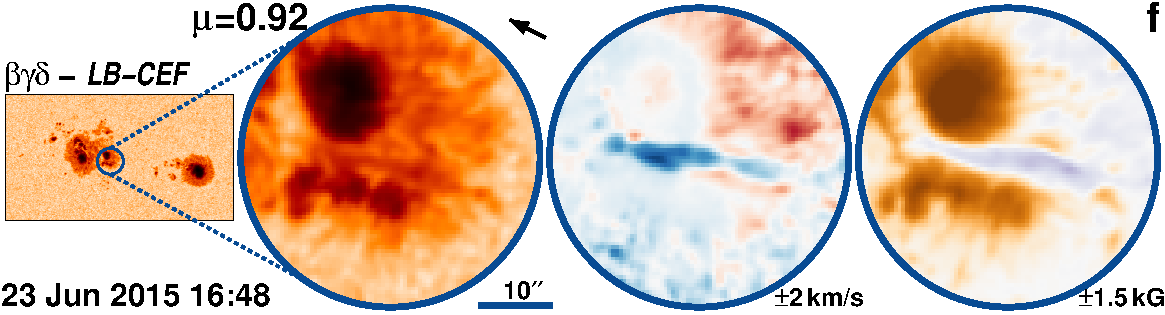}
 \includegraphics[width=.48\textwidth]{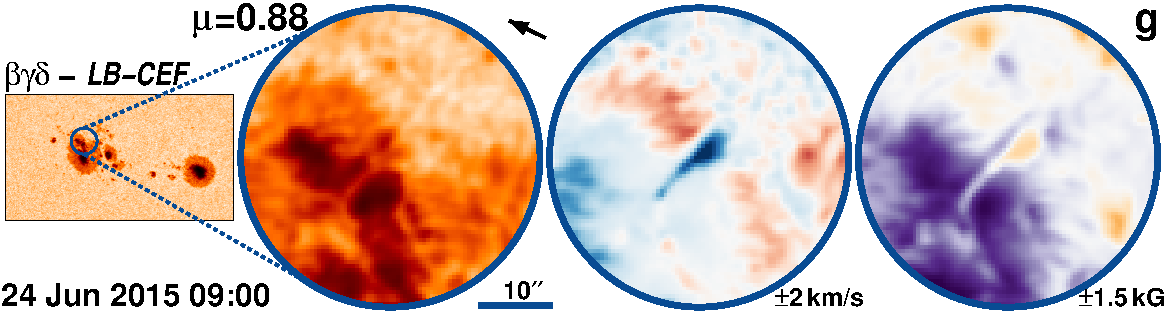}
 \includegraphics[width=.48\textwidth]{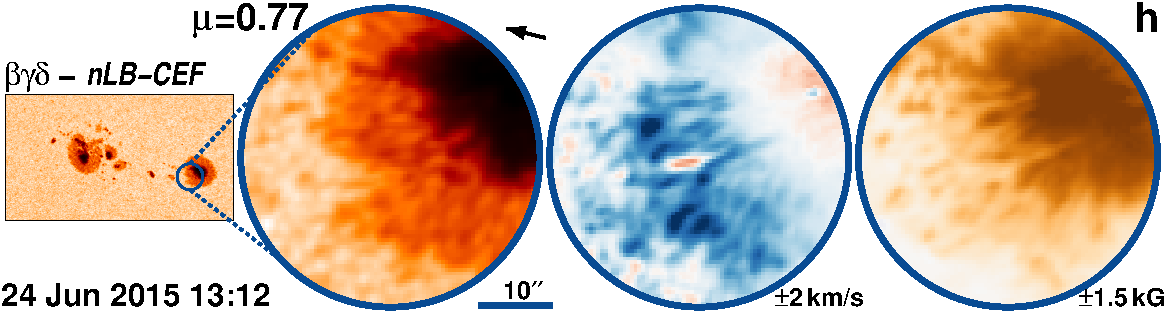}
 \includegraphics[width=.48\textwidth]{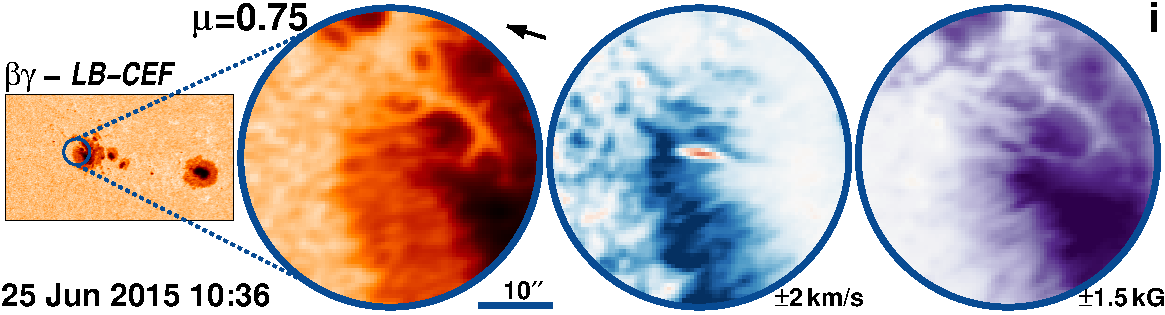}
 \caption{AR\,12371 followed for    9.6 days from 17-Jun-2015 starting at 05:00\,UT.\label{fig:DS24}}
 \end{figure*}

\begin{figure*}[htbp]
 \centering
 \includegraphics[width=.48\textwidth]{colorbars.pdf}
 \includegraphics[width=.48\textwidth]{colorbars.pdf}
 \includegraphics[width=.48\textwidth]{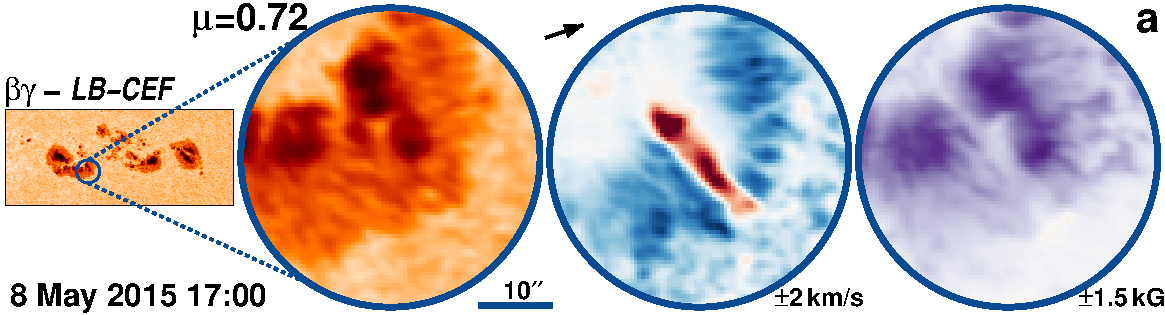}
 \includegraphics[width=.48\textwidth]{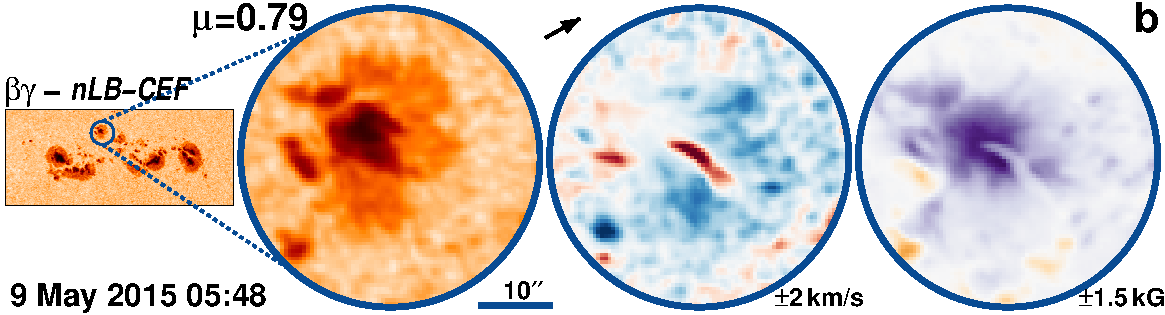}
 \includegraphics[width=.48\textwidth]{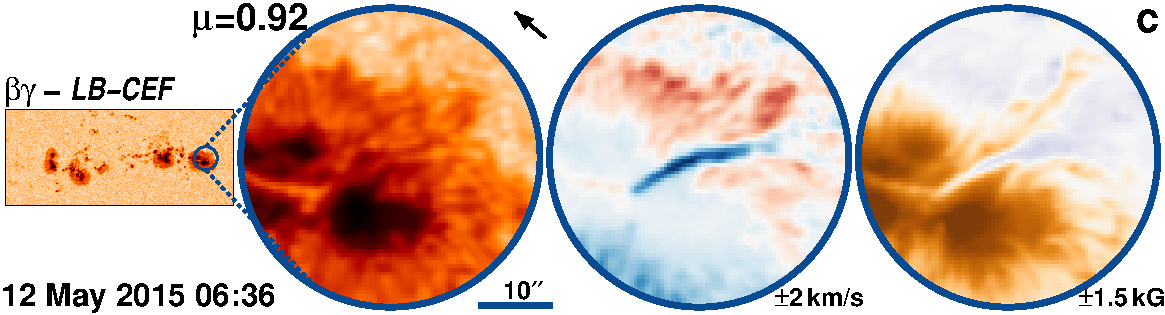}
 \includegraphics[width=.48\textwidth]{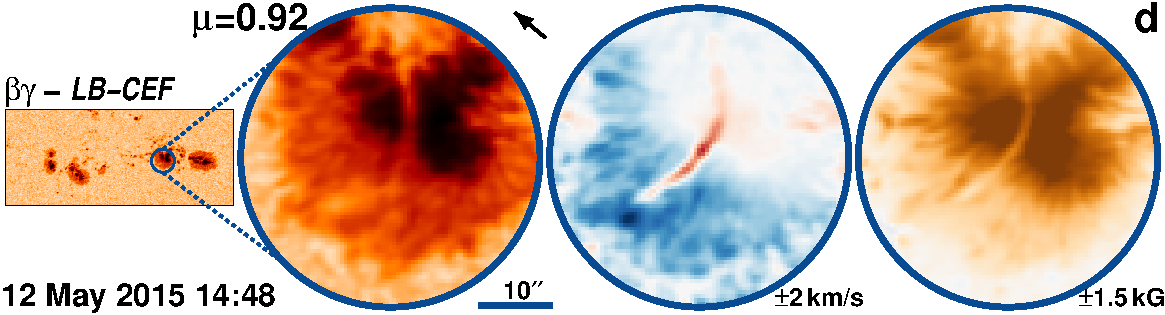}
 \caption{AR\,12339 followed for   10.3 days from  6-May-2015 starting at 08:12\,UT.\label{fig:DS26}}
 \end{figure*}

\begin{figure*}[htbp]
 \includegraphics[width=.48\textwidth]{colorbars.pdf}
 \includegraphics[width=.48\textwidth]{colorbars.pdf}
 \includegraphics[width=.48\textwidth]{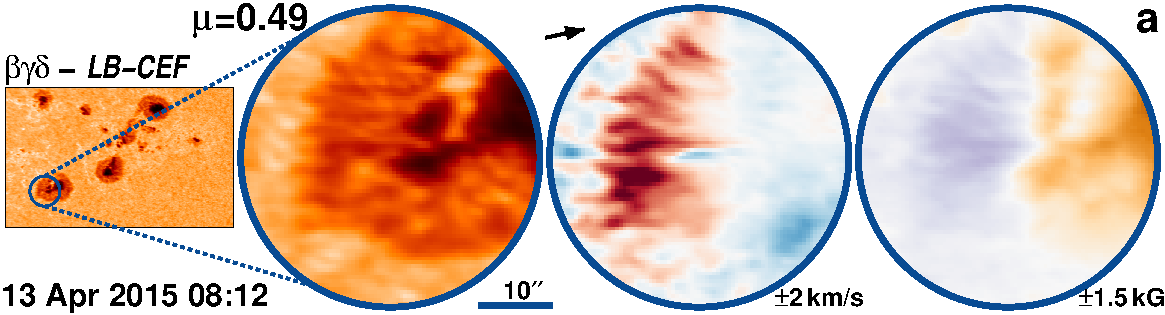}
 \includegraphics[width=.48\textwidth]{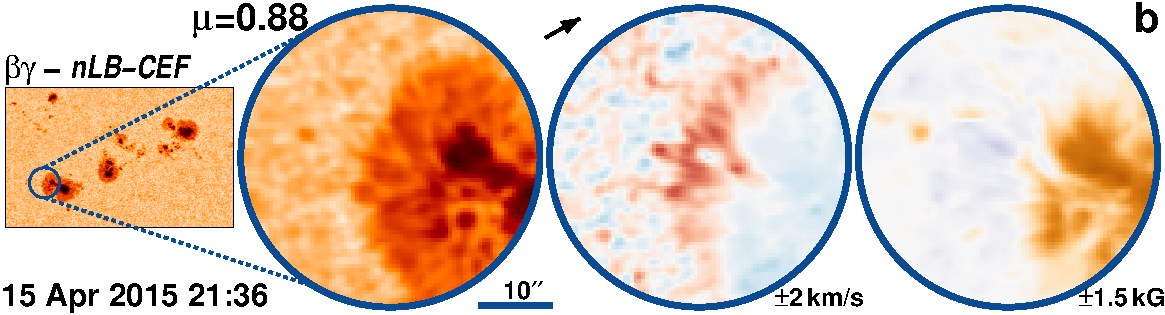}
 \includegraphics[width=.48\textwidth]{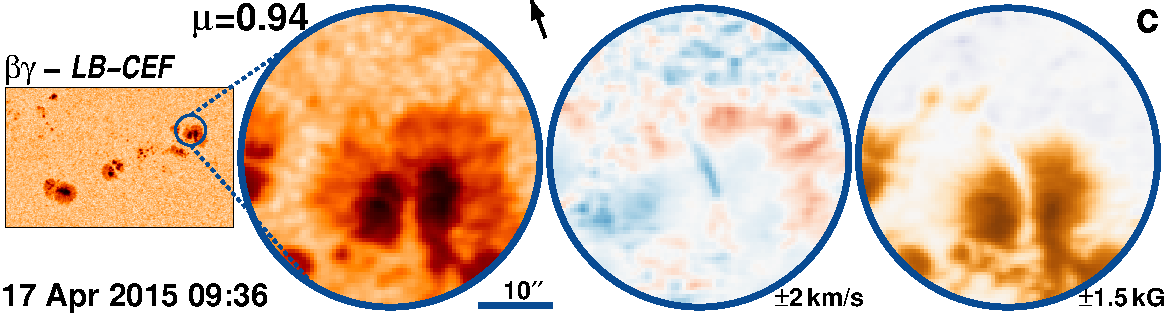}
 \caption{AR\,12321 followed for    9.1 days from 13-Apr-2015 starting at 05:12\,UT.\label{fig:DS27}}
 \end{figure*}

\begin{figure*}[htbp]
 \includegraphics[width=.48\textwidth]{colorbars.pdf}
 \includegraphics[width=.48\textwidth]{colorbars.pdf}
 \includegraphics[width=.48\textwidth]{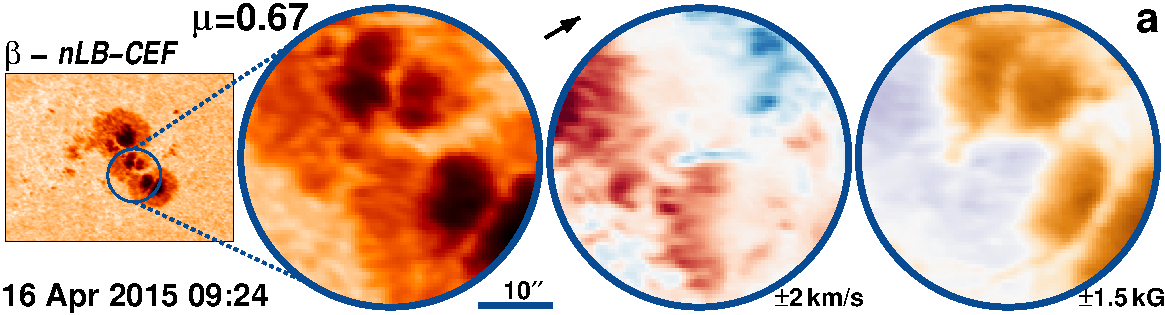}
 \includegraphics[width=.48\textwidth]{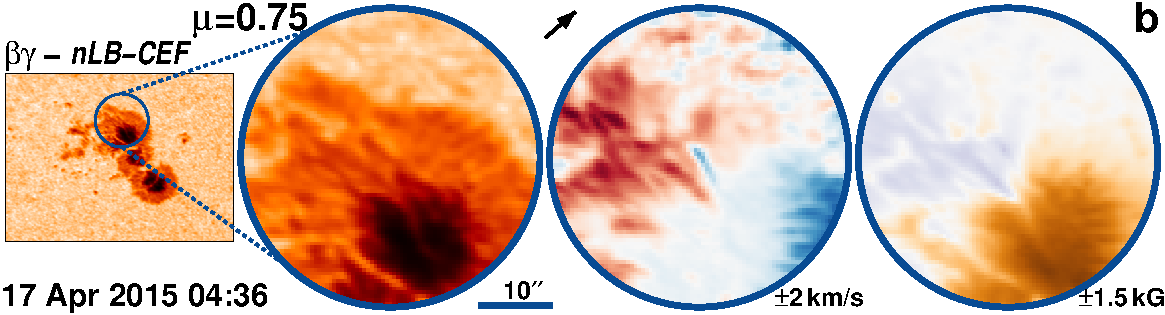}
 \includegraphics[width=.48\textwidth]{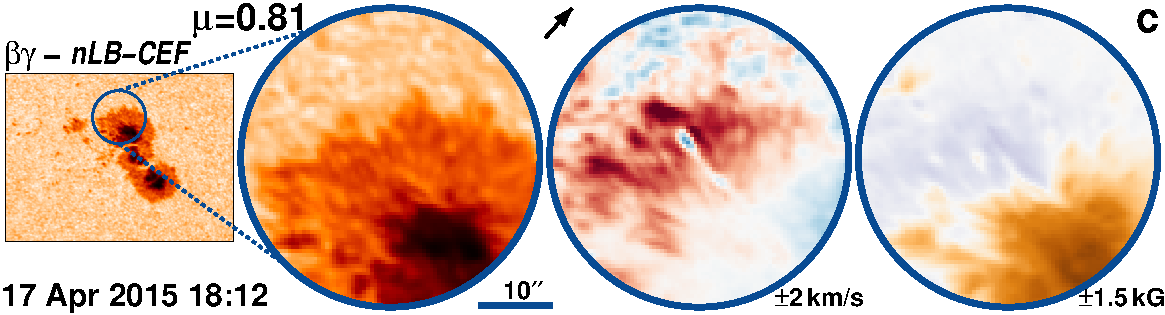}
 \caption{AR\,12324 followed for    9.0 days from 14-Apr-2015 starting at 23:12\,UT.\label{fig:DS28}}
 \end{figure*}

\begin{figure*}[htbp]
 \centering
 \includegraphics[width=.48\textwidth]{colorbars.pdf}
 \includegraphics[width=.48\textwidth]{colorbars.pdf}
 \includegraphics[width=.48\textwidth]{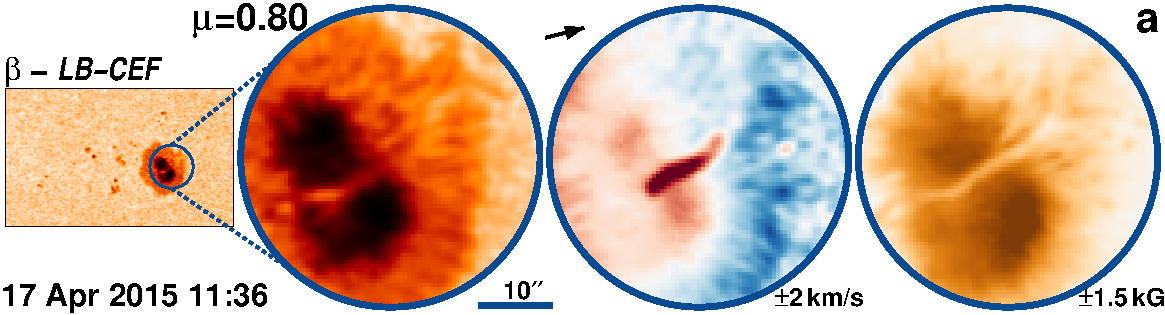}
 \includegraphics[width=.48\textwidth]{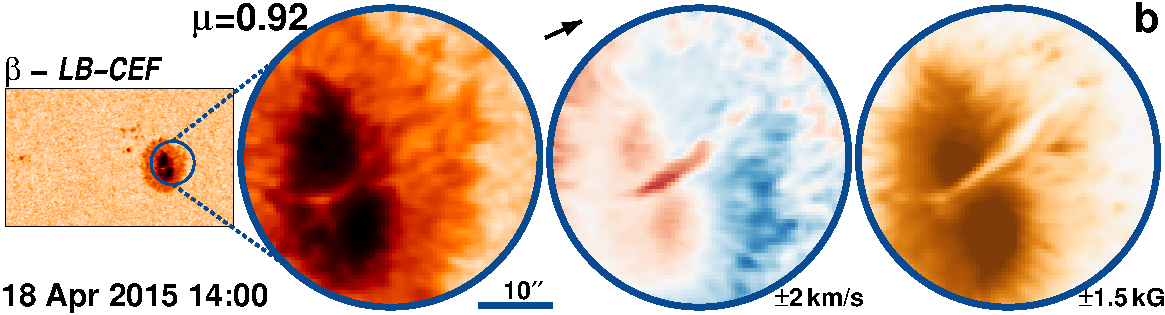}
 \includegraphics[width=.48\textwidth]{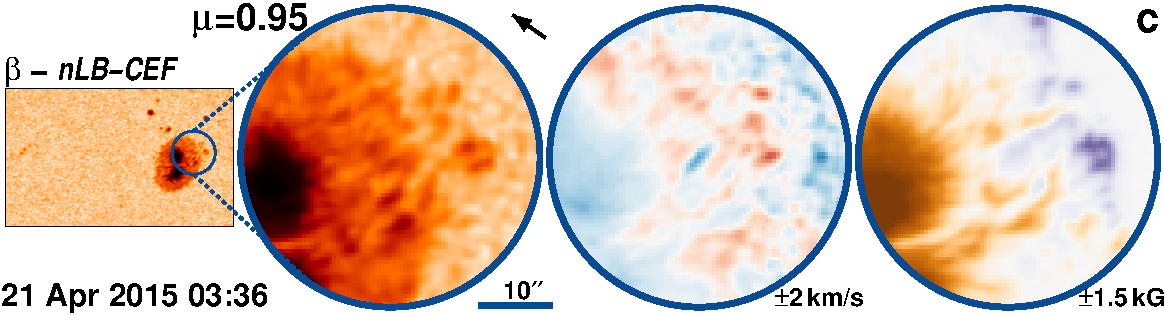}
 \includegraphics[width=.48\textwidth]{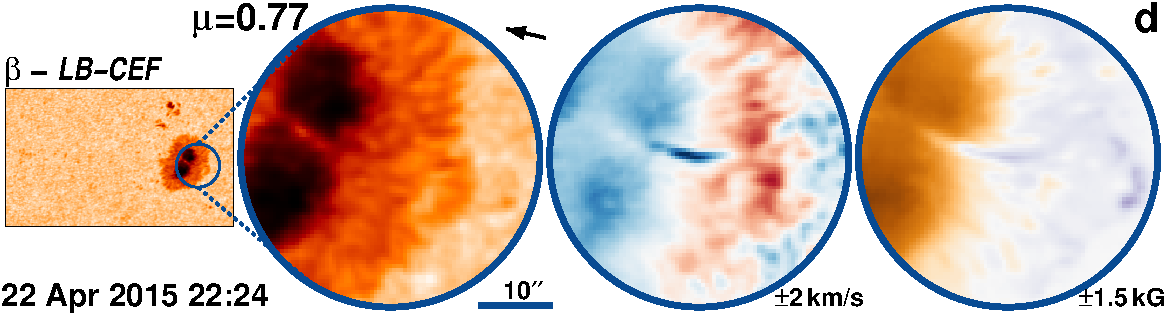}
 \caption{AR\,12325 followed for    9.4 days from 15-Apr-2015 starting at 06:24\,UT.\label{fig:DS29}}
 \end{figure*}

\begin{figure*}[htbp]
 \includegraphics[width=.48\textwidth]{colorbars.pdf}
 \includegraphics[width=.48\textwidth]{colorbars.pdf}
 \includegraphics[width=.48\textwidth]{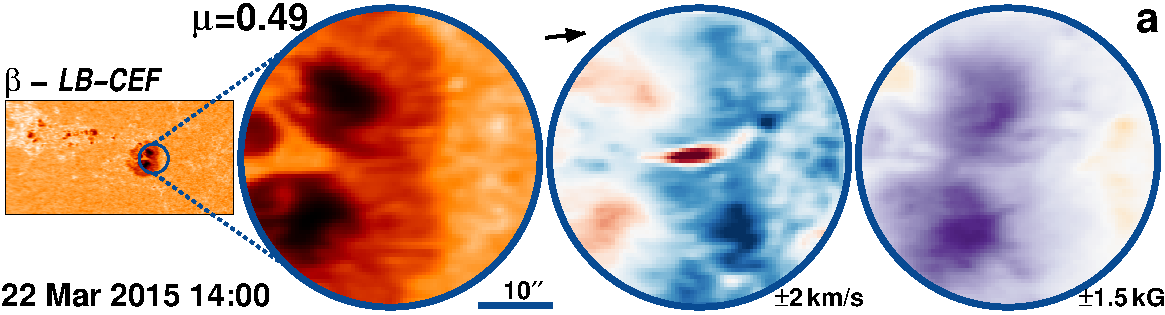}
 \includegraphics[width=.48\textwidth]{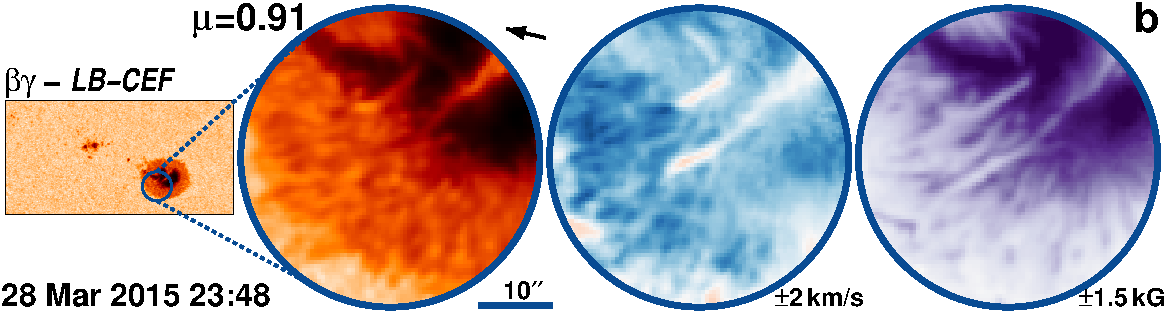}
 \includegraphics[width=.48\textwidth]{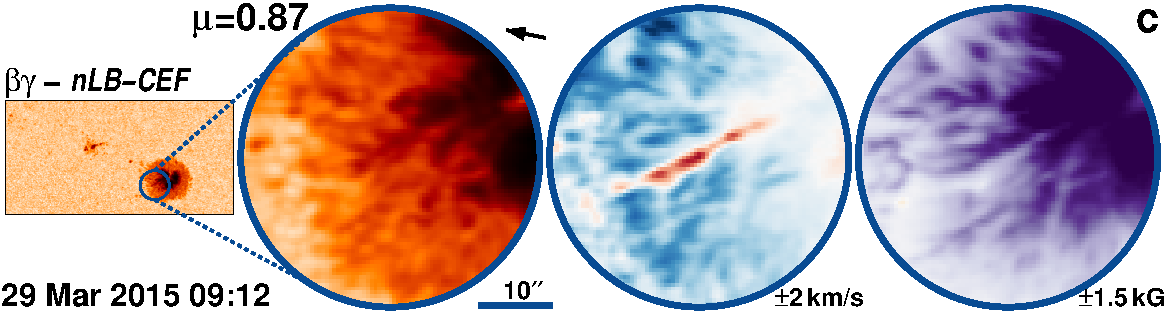}
 \includegraphics[width=.48\textwidth]{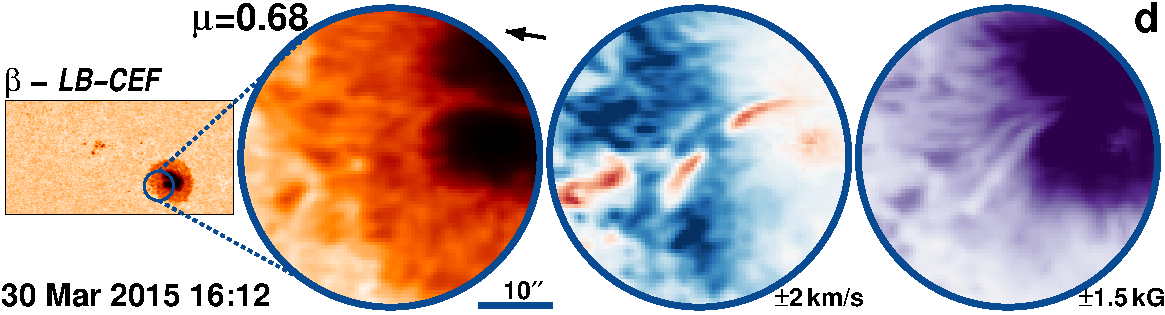}
 \includegraphics[width=.48\textwidth]{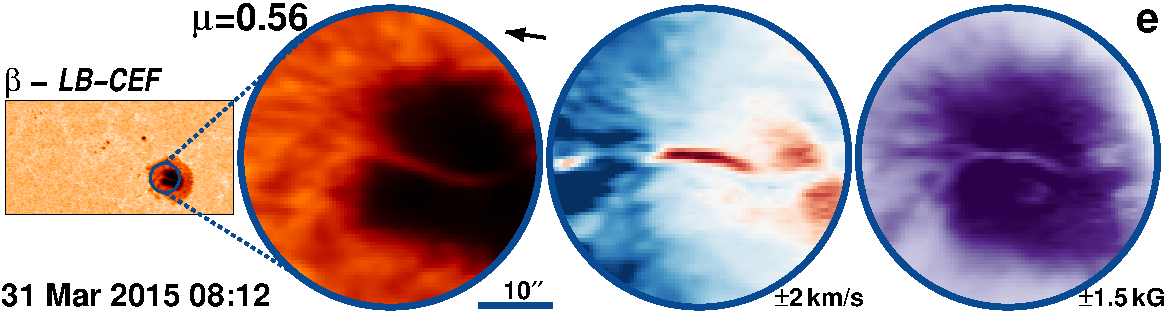}
 \caption{AR\,12305 followed for    9.9 days from 22-Mar-2015 starting at 07:00\,UT.\label{fig:DS30}}
 \end{figure*}

\begin{figure*}[htbp]
 \includegraphics[width=.48\textwidth]{colorbars.pdf}
 \includegraphics[width=.48\textwidth]{colorbars.pdf}
 \includegraphics[width=.48\textwidth]{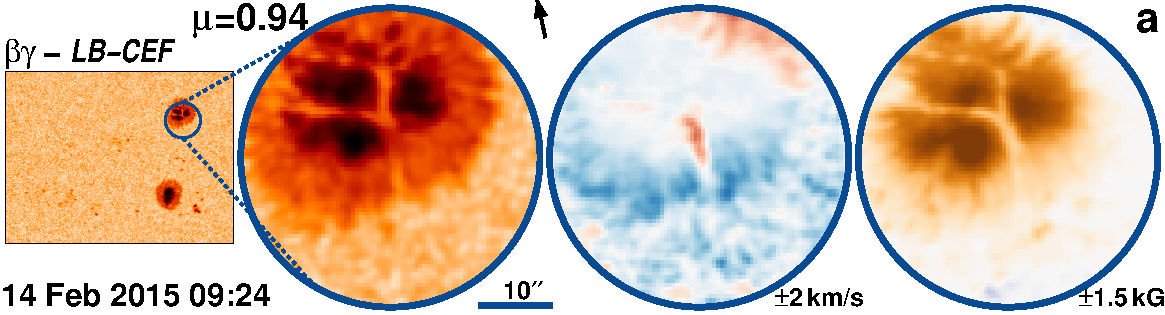}
 \includegraphics[width=.48\textwidth]{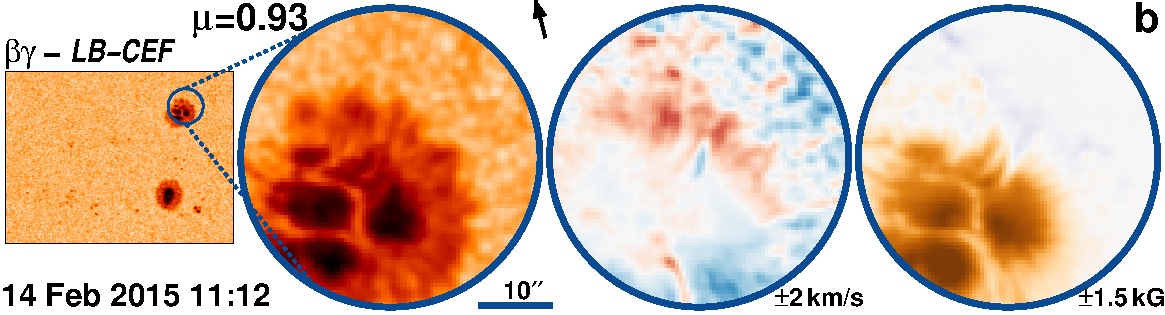}
 \includegraphics[width=.48\textwidth]{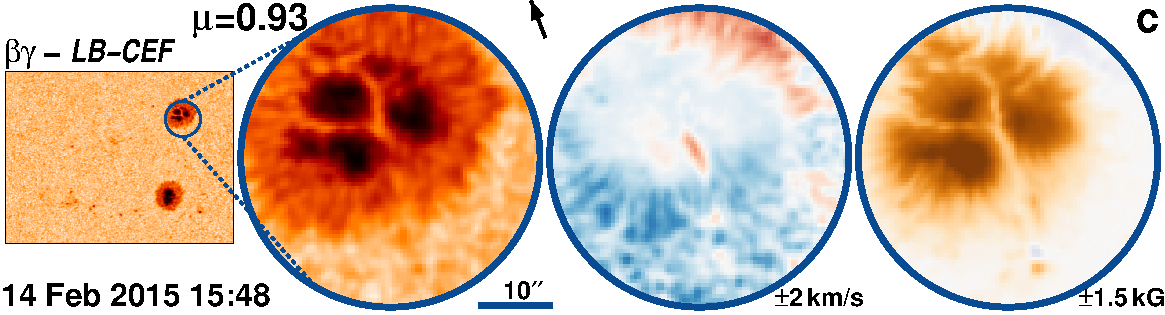}
 \includegraphics[width=.48\textwidth]{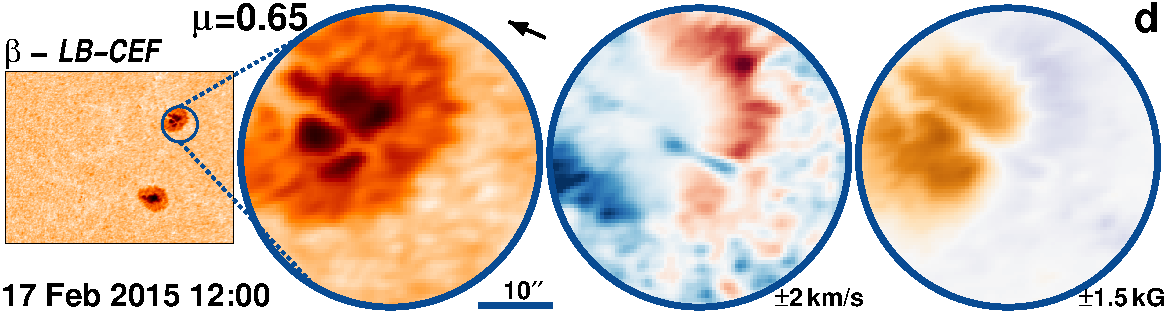}
 \includegraphics[width=.48\textwidth]{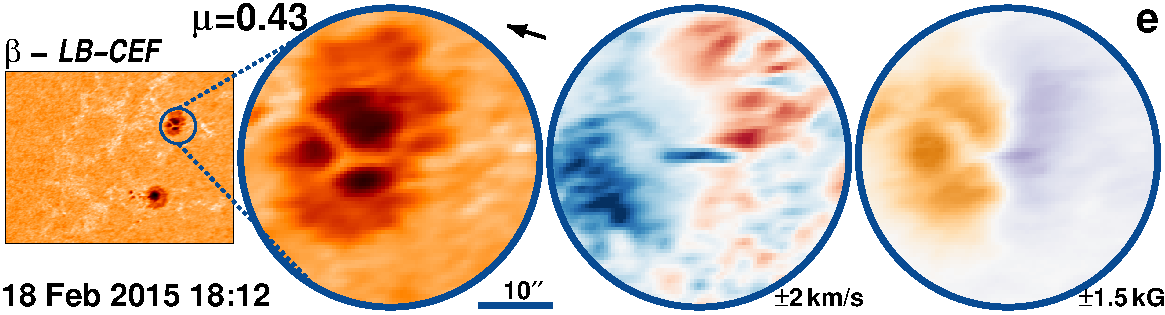}
 \caption{AR\,12282 followed for   10.5 days from  8-Feb-2015 starting at 23:24\,UT.\label{fig:DS32}}
 \end{figure*}

\begin{figure*}[htbp]
 \centering
 \includegraphics[width=.48\textwidth]{colorbars.pdf}
 \includegraphics[width=.48\textwidth]{colorbars.pdf}
 \includegraphics[width=.48\textwidth]{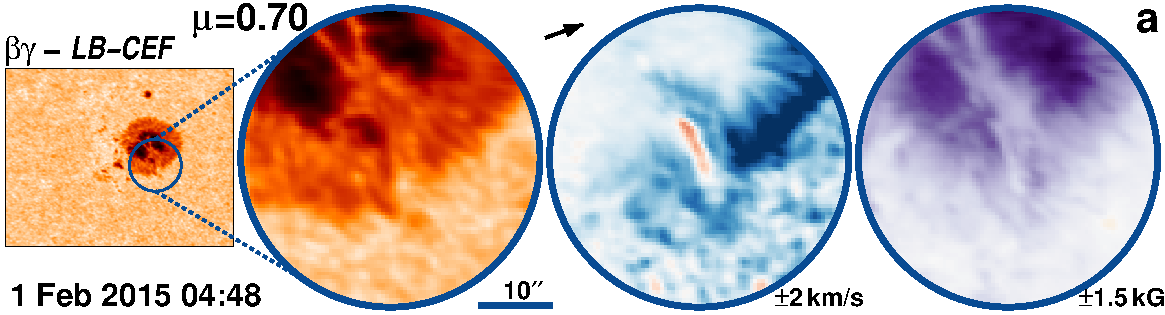}
 \includegraphics[width=.48\textwidth]{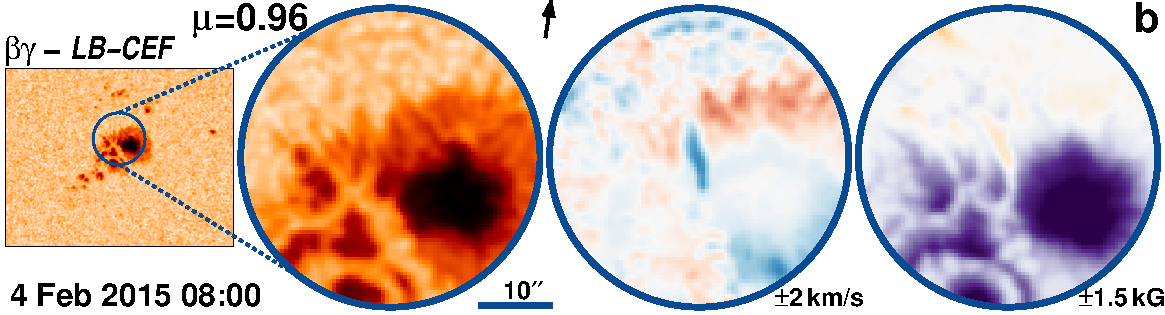}
 \caption{AR\,12277 followed for   10.2 days from 30-Jan-2015 starting at 02:48\,UT.\label{fig:DS33}}
 \end{figure*}

\begin{figure*}[htbp]
 \centering
 \includegraphics[width=.48\textwidth]{colorbars.pdf}
 \includegraphics[width=.48\textwidth]{colorbars.pdf}
 \includegraphics[width=.48\textwidth]{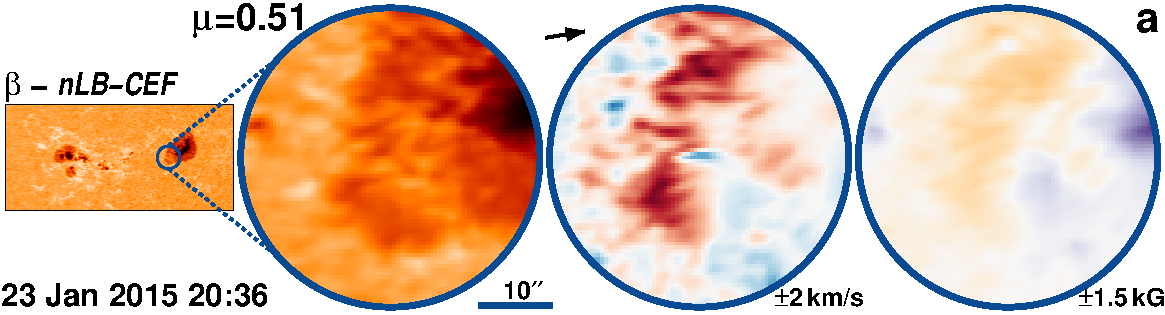}
 \includegraphics[width=.48\textwidth]{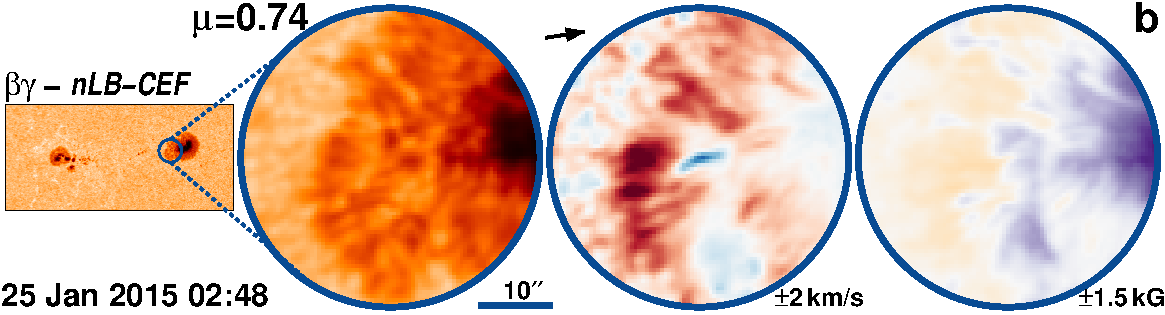}
 \includegraphics[width=.48\textwidth]{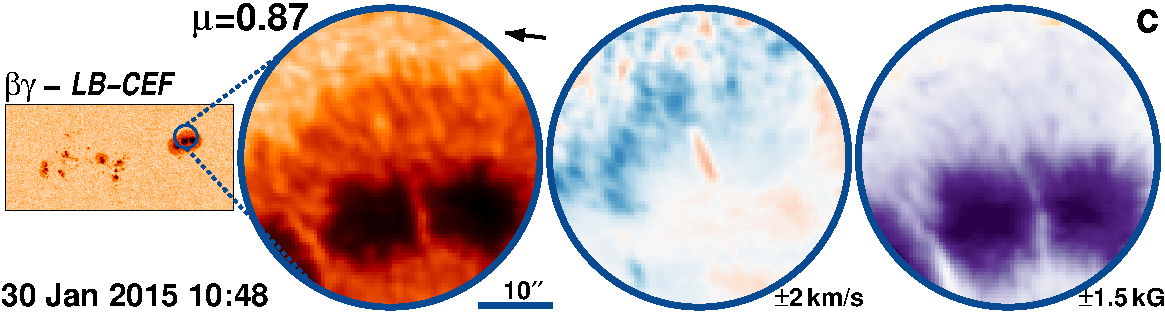}
 \includegraphics[width=.48\textwidth]{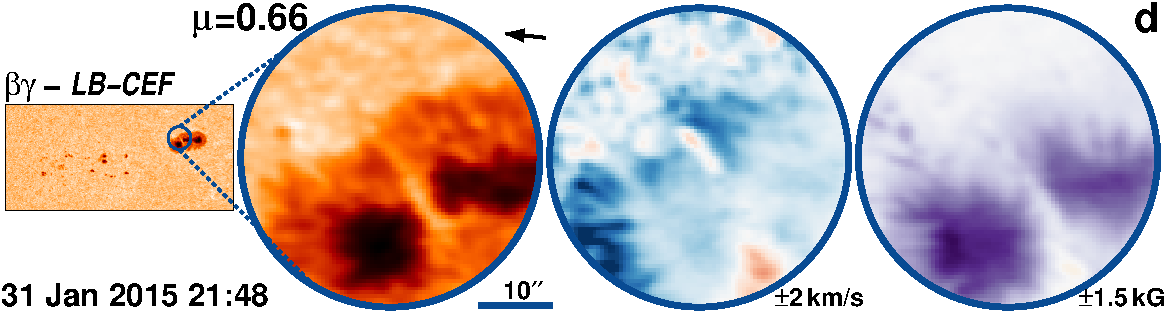}
 \caption{AR\,12268 followed for   10.6 days from 22-Jan-2015 starting at 21:48\,UT.\label{fig:DS34}}
 \end{figure*}

\begin{figure*}[htbp]
 \includegraphics[width=.48\textwidth]{colorbars.pdf}

 \includegraphics[width=.48\textwidth]{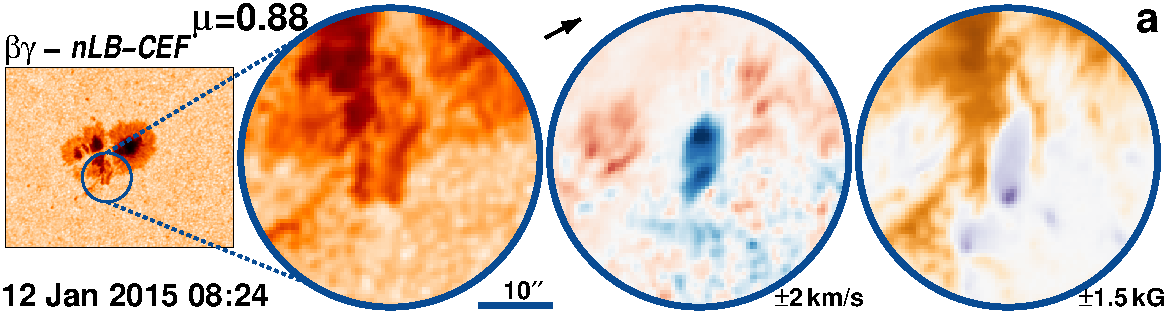}
 \caption{AR\,12259 followed for   11.6 days from  8-Jan-2015 starting at 13:24\,UT.\label{fig:DS36}}
 \end{figure*}

\begin{figure*}[htbp]
 \centering
 \includegraphics[width=.48\textwidth]{colorbars.pdf}
 \includegraphics[width=.48\textwidth]{colorbars.pdf}
 \includegraphics[width=.48\textwidth]{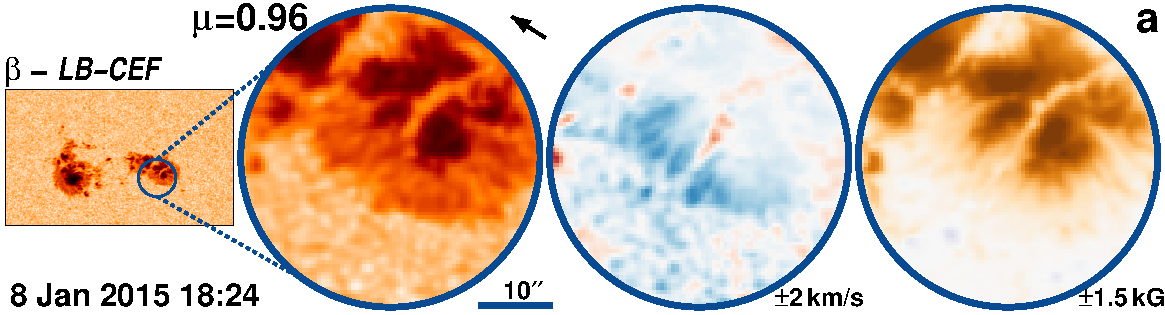}
 \includegraphics[width=.48\textwidth]{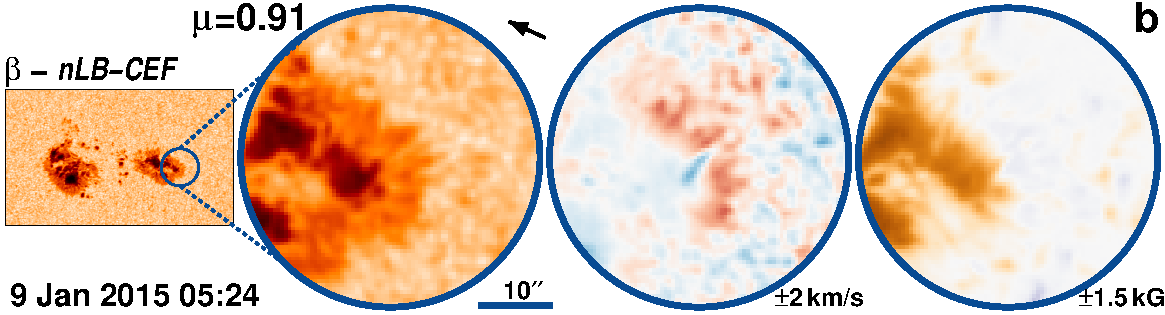}
 \includegraphics[width=.48\textwidth]{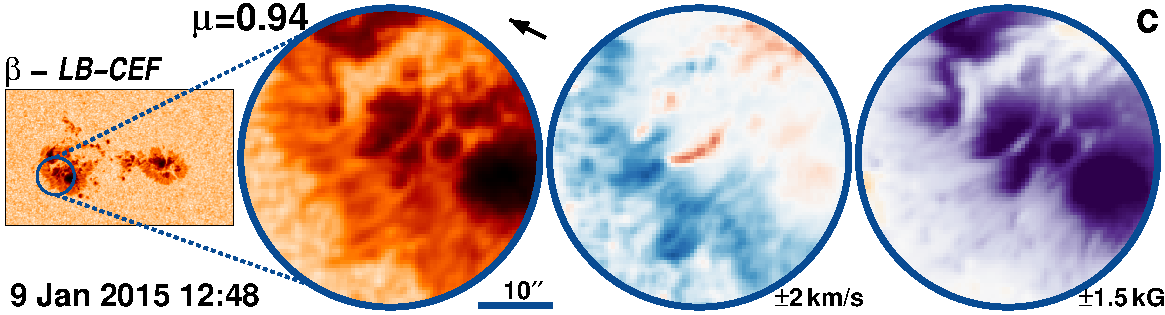}
 \includegraphics[width=.48\textwidth]{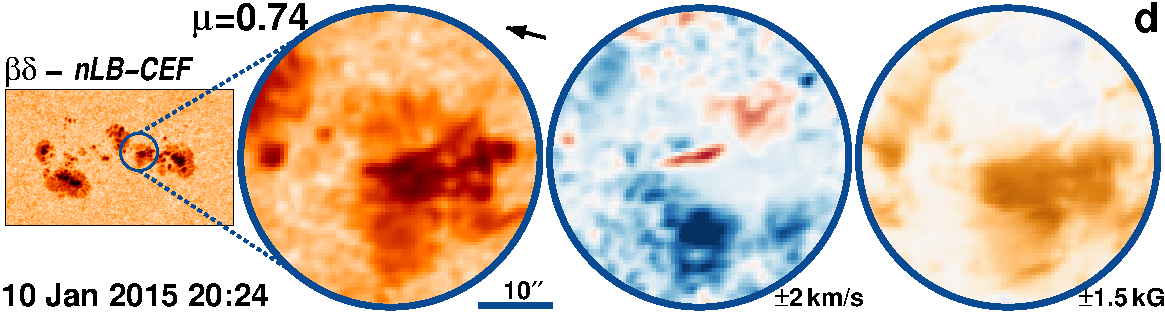}
 \caption{AR\,12257 followed for    6.8 days from  6-Jan-2015 starting at 13:36\,UT.\label{fig:DS37}}
 \end{figure*}

\begin{figure*}[htbp]
 \includegraphics[width=.48\textwidth]{colorbars.pdf}
 \includegraphics[width=.48\textwidth]{colorbars.pdf}
 \includegraphics[width=.48\textwidth]{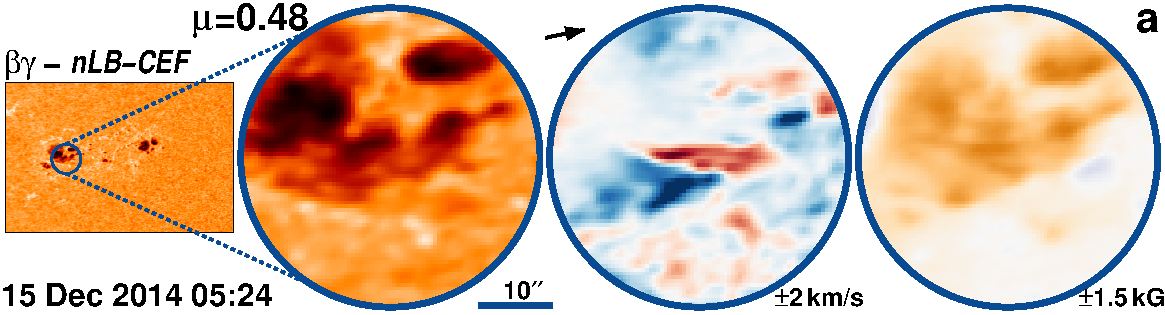}
 \includegraphics[width=.48\textwidth]{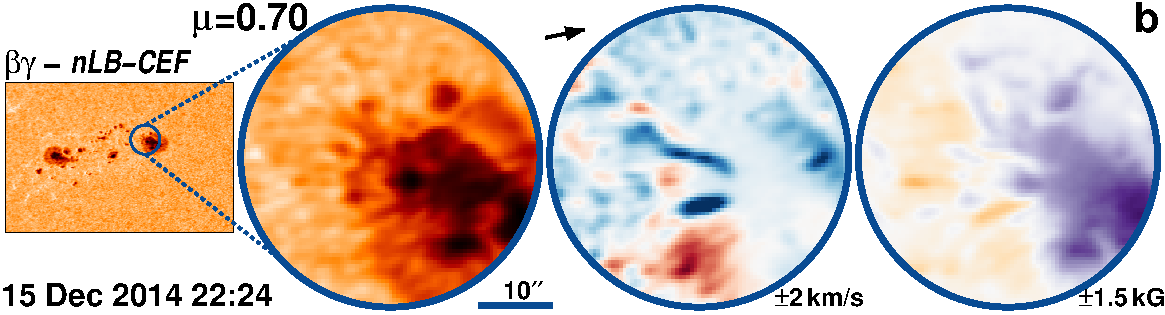}
 \includegraphics[width=.48\textwidth]{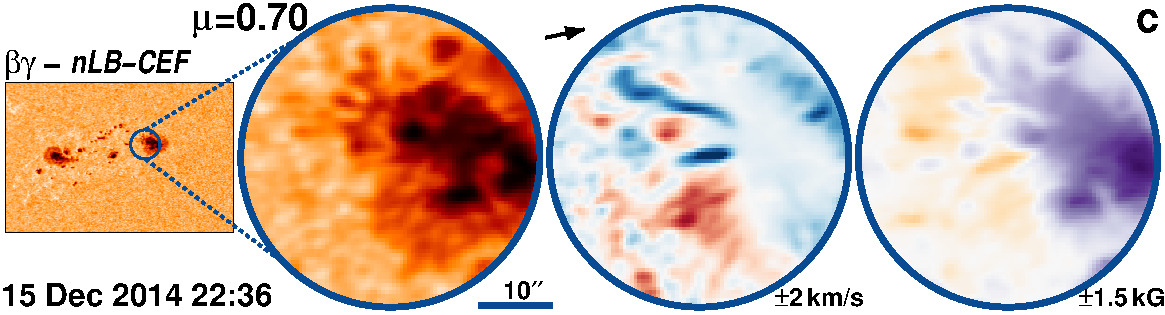}
 \includegraphics[width=.48\textwidth]{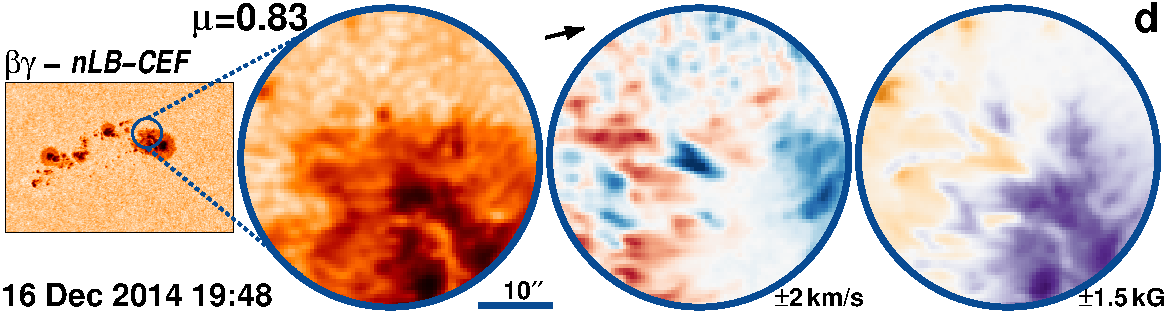}
 \includegraphics[width=.48\textwidth]{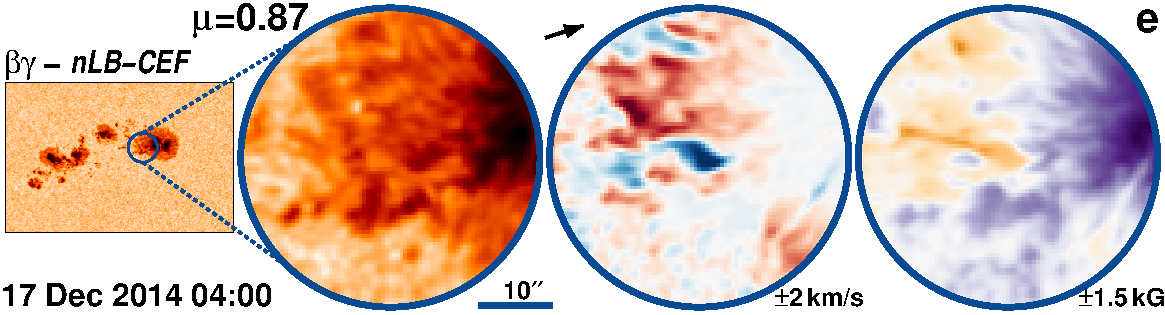}
 \includegraphics[width=.48\textwidth]{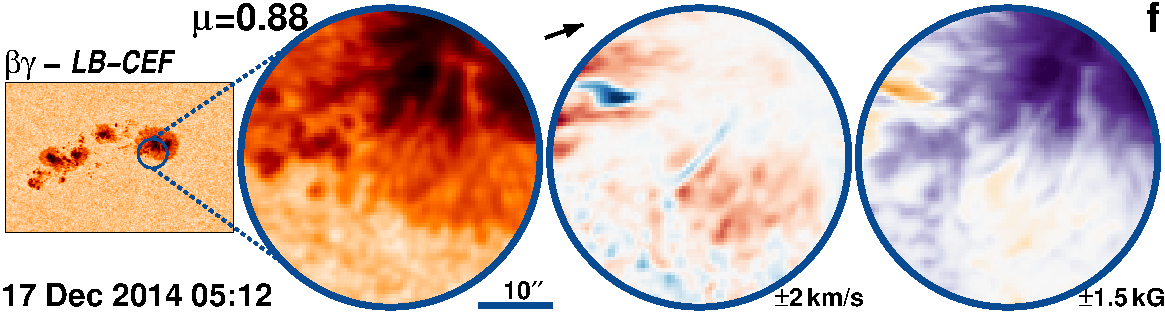}
 \includegraphics[width=.48\textwidth]{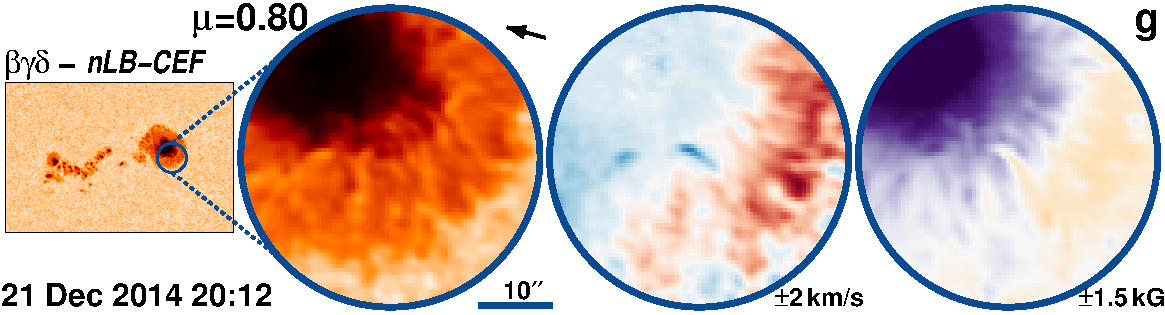}
 \caption{AR\,12241 followed for    8.8 days from 15-Dec-2014 starting at 03:12\,UT.\label{fig:DS38}}
 \end{figure*}

\begin{figure*}[htbp]
 \includegraphics[width=.48\textwidth]{colorbars.pdf}
 \includegraphics[width=.48\textwidth]{colorbars.pdf}
 \includegraphics[width=.48\textwidth]{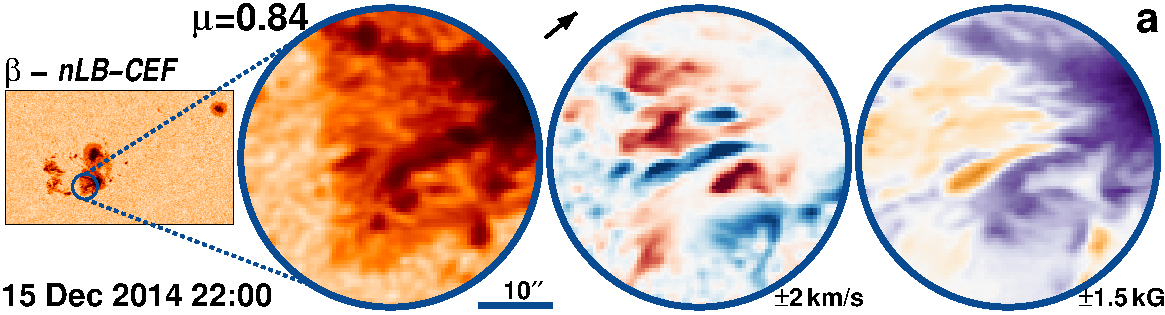}
 \includegraphics[width=.48\textwidth]{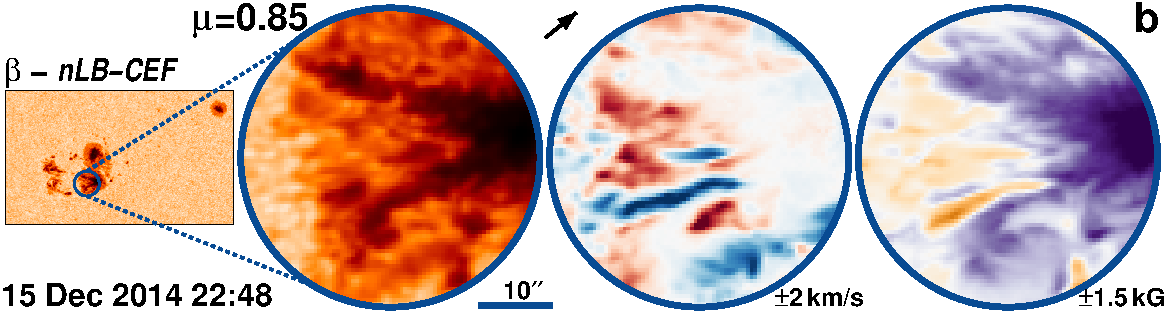}
 \includegraphics[width=.48\textwidth]{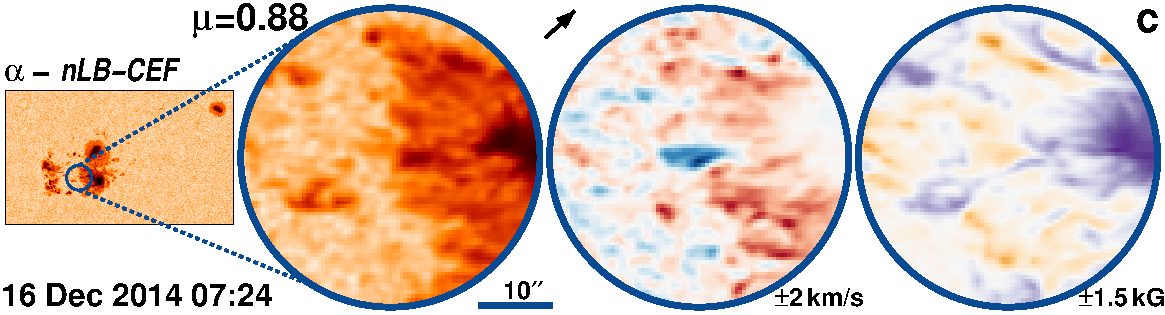}
 \includegraphics[width=.48\textwidth]{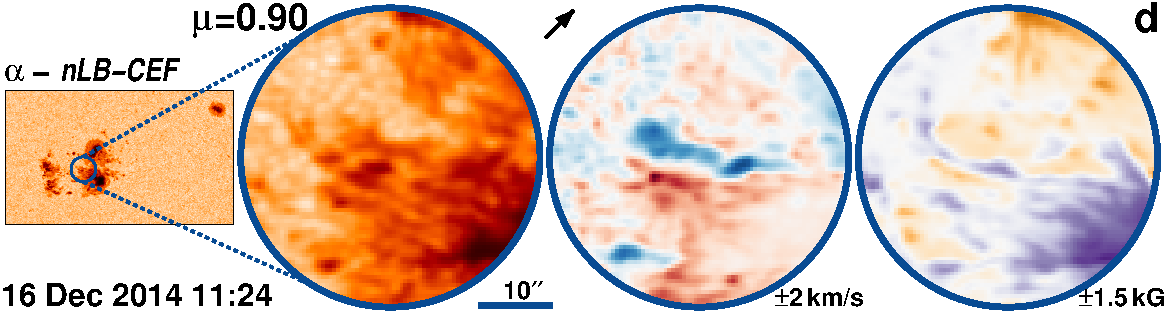}
 \includegraphics[width=.48\textwidth]{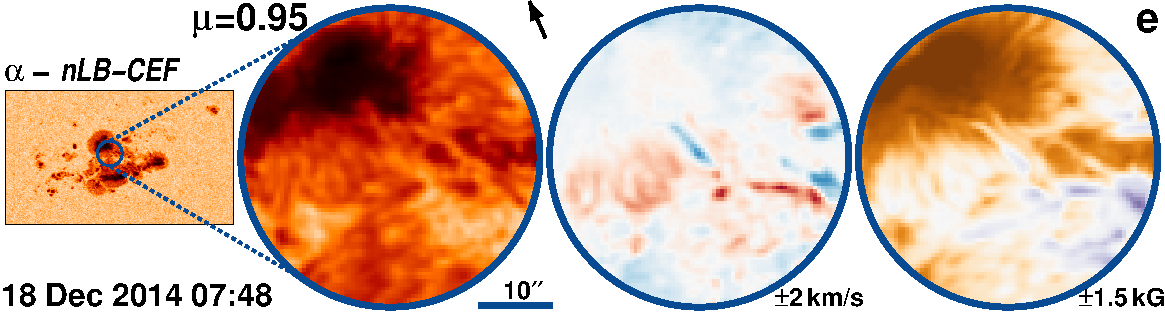}
 \includegraphics[width=.48\textwidth]{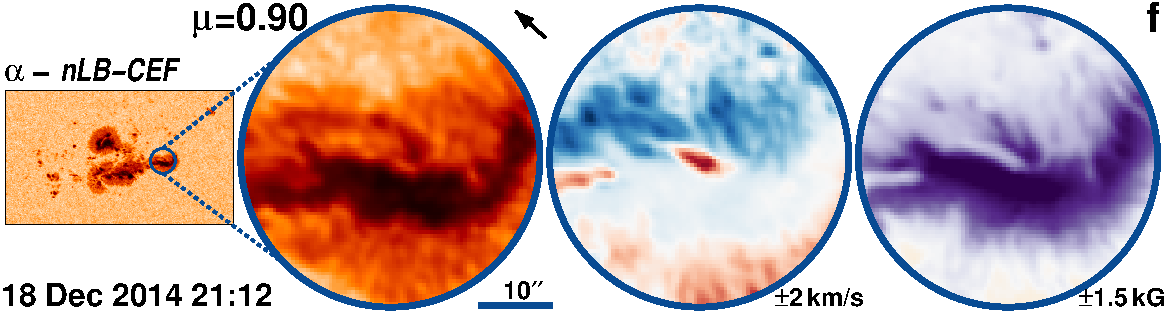}
 \includegraphics[width=.48\textwidth]{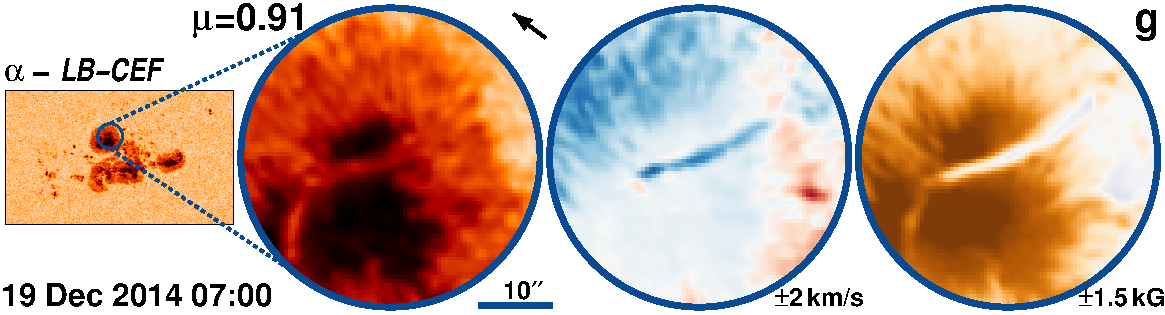}
 \caption{AR\,12237 followed for    8.2 days from 13-Dec-2014 starting at 23:00\,UT.\label{fig:DS39}}
 \end{figure*}

\begin{figure*}[htbp]
 \centering
 \includegraphics[width=.48\textwidth]{colorbars.pdf}
 \includegraphics[width=.48\textwidth]{colorbars.pdf}
 \includegraphics[width=.48\textwidth]{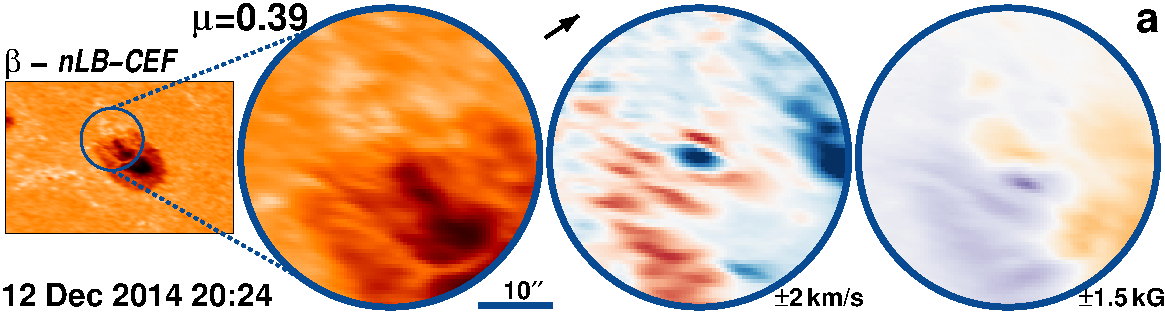}
 \includegraphics[width=.48\textwidth]{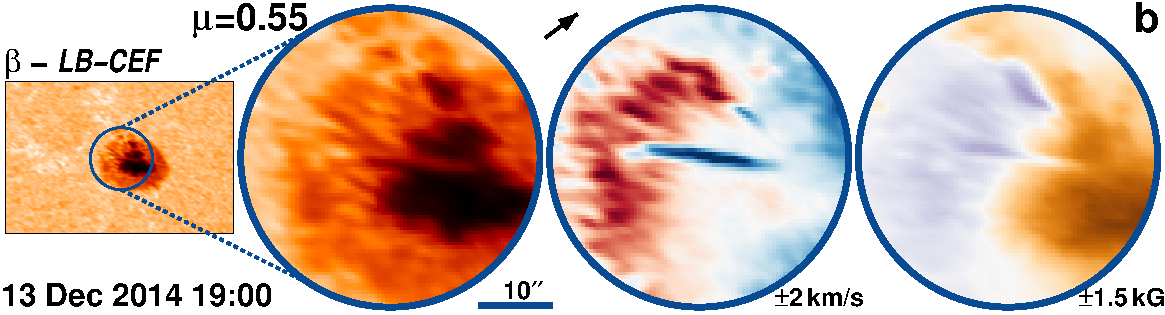}
 \caption{AR\,12236 followed for   10.0 days from 12-Dec-2014 starting at 02:00\,UT.\label{fig:DS40}}
 \end{figure*}

\begin{figure*}[htbp]
 \includegraphics[width=.48\textwidth]{colorbars.pdf}

 \includegraphics[width=.48\textwidth]{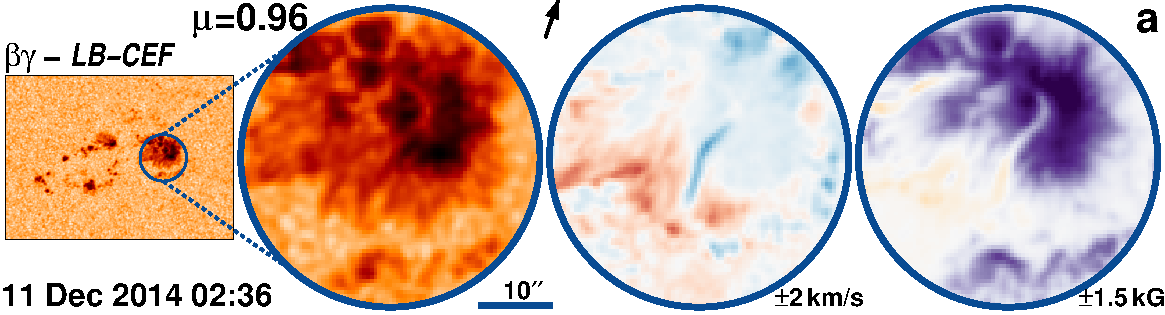}
 \caption{AR\,12230 followed for    7.4 days from  8-Dec-2014 starting at 15:00\,UT.\label{fig:DS41}}
 \end{figure*}

\begin{figure*}[htbp]
 \includegraphics[width=.48\textwidth]{colorbars.pdf}
 \includegraphics[width=.48\textwidth]{colorbars.pdf}
 \includegraphics[width=.48\textwidth]{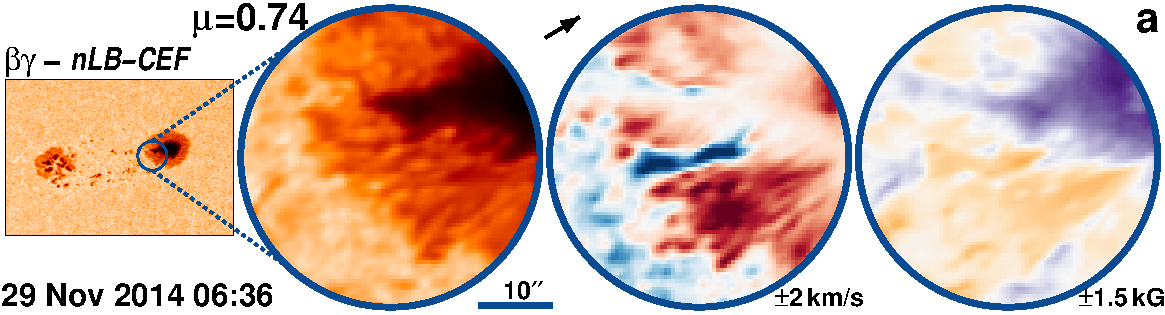}
 \includegraphics[width=.48\textwidth]{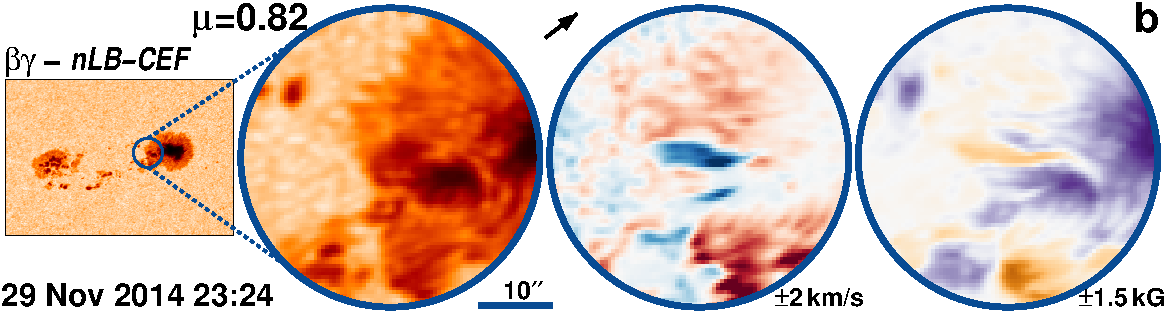}
 \includegraphics[width=.48\textwidth]{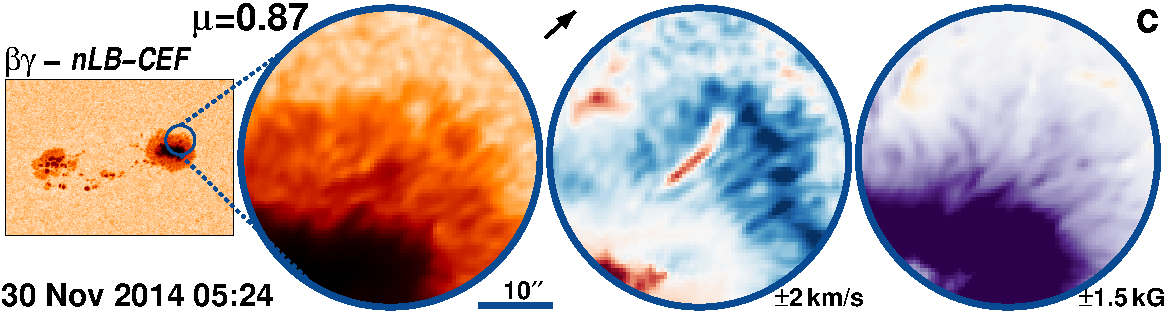}
 \includegraphics[width=.48\textwidth]{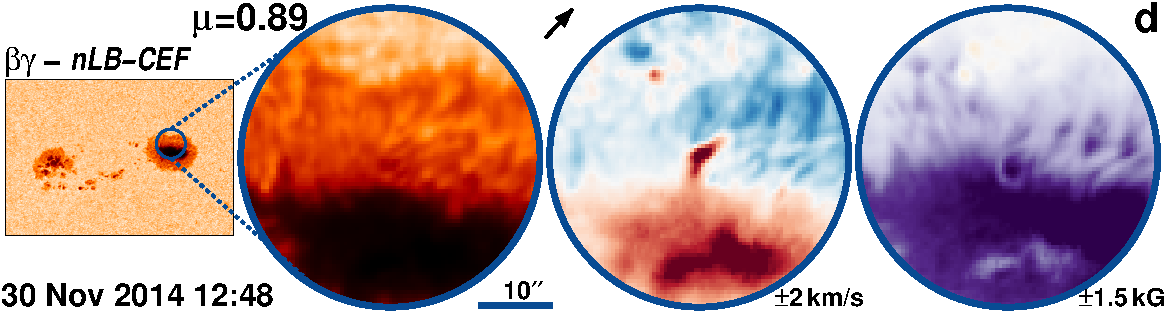}
 \includegraphics[width=.48\textwidth]{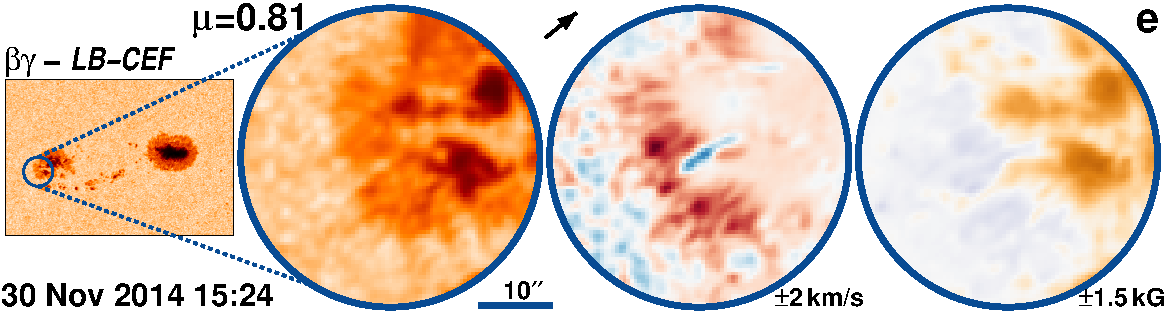}
 \includegraphics[width=.48\textwidth]{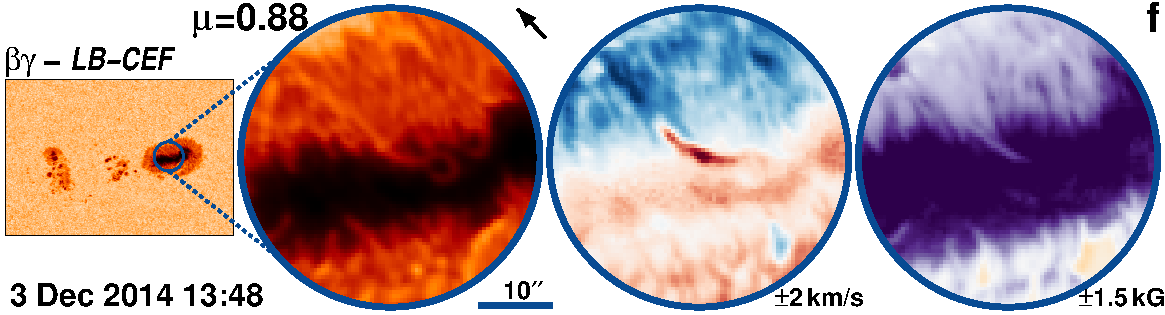}
 \includegraphics[width=.48\textwidth]{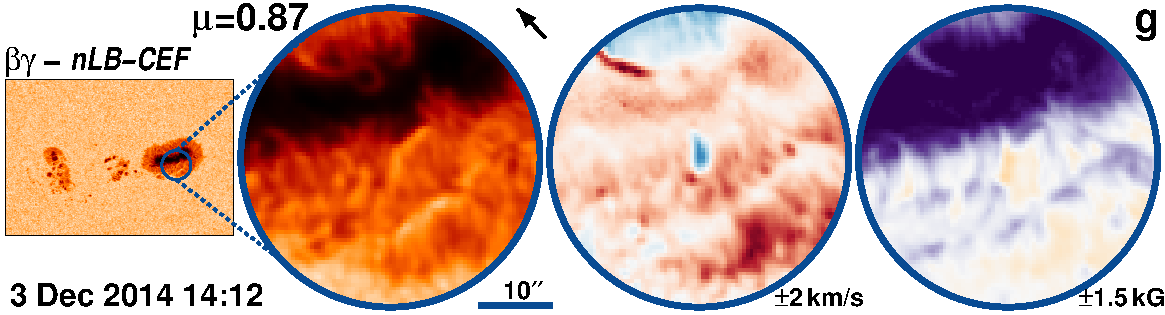}
 \includegraphics[width=.48\textwidth]{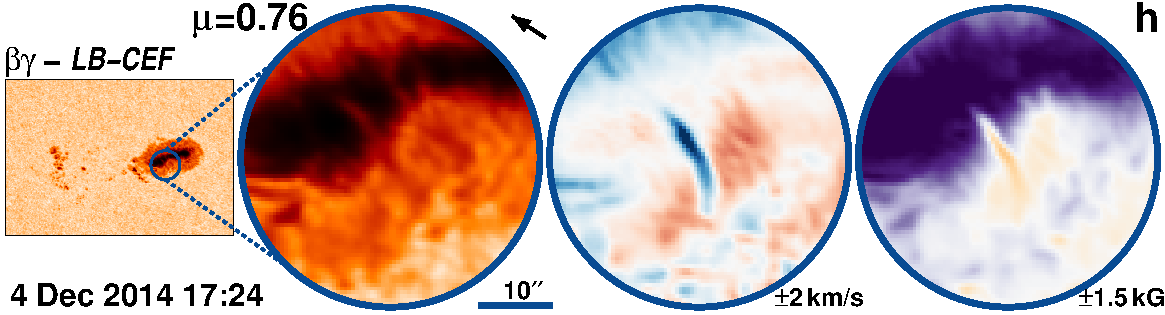}
 \includegraphics[width=.48\textwidth]{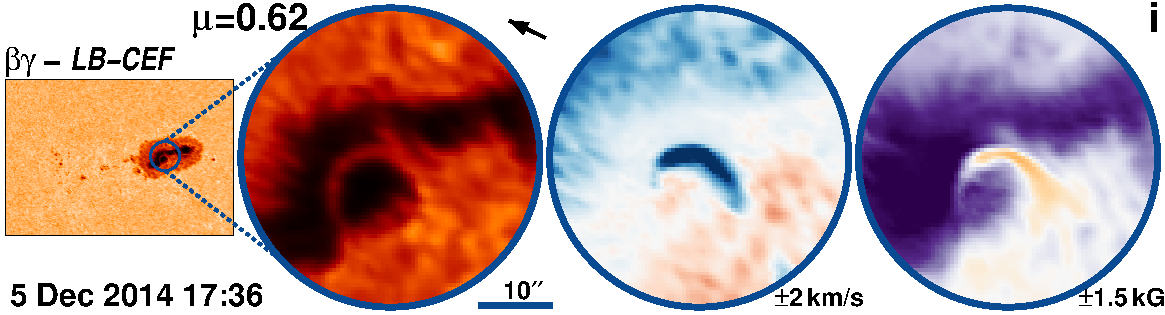}
 \includegraphics[width=.48\textwidth]{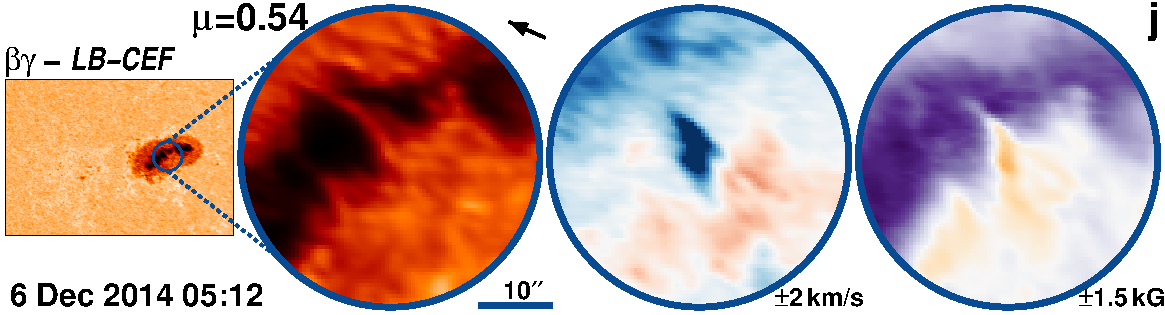}
 \includegraphics[width=.48\textwidth]{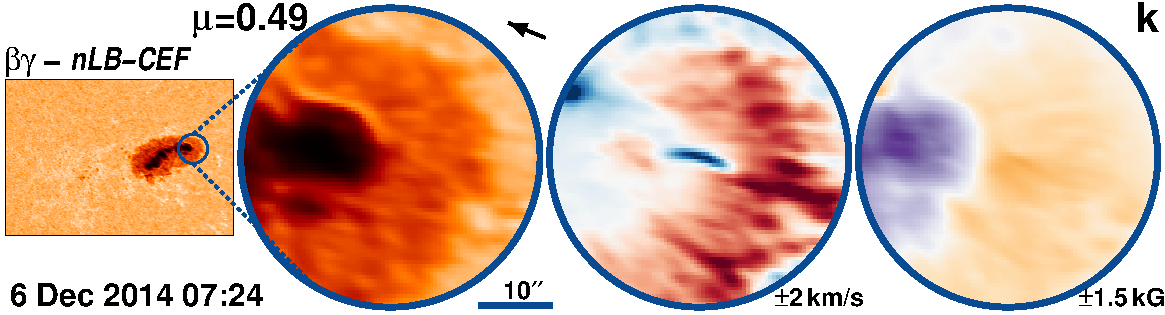}
 \caption{AR\,12222 followed for    8.8 days from 27-Nov-2014 starting at 20:00\,UT.\label{fig:DS43}}
 \end{figure*}

\begin{figure*}[htbp]
 \centering
 \includegraphics[width=.48\textwidth]{colorbars.pdf}
 \includegraphics[width=.48\textwidth]{colorbars.pdf}
 \includegraphics[width=.48\textwidth]{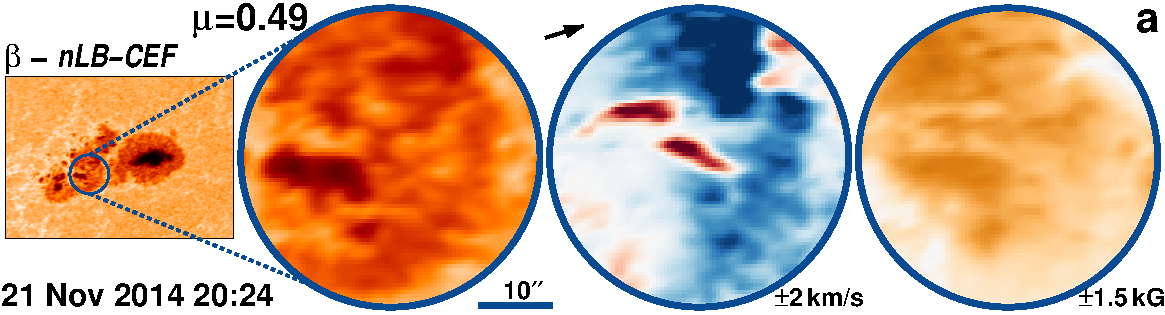}
 \includegraphics[width=.48\textwidth]{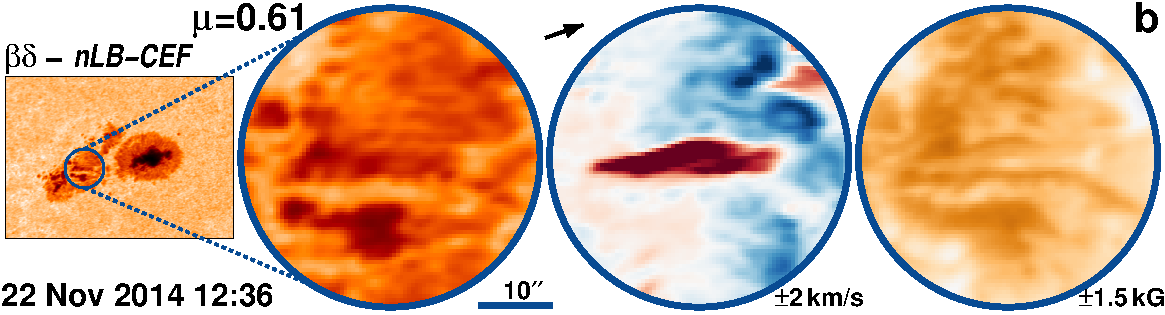}
 \includegraphics[width=.48\textwidth]{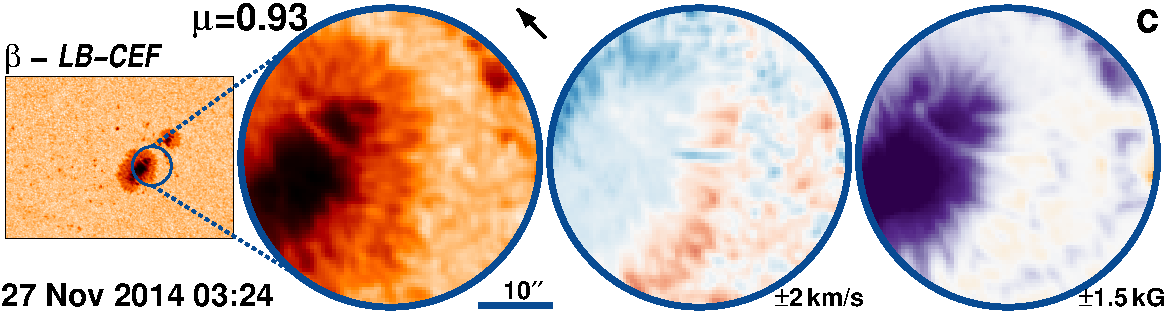}
 \includegraphics[width=.48\textwidth]{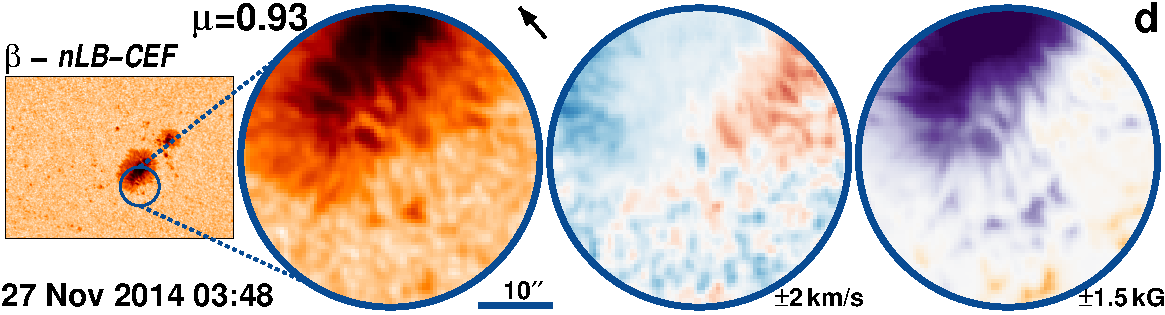}
 \includegraphics[width=.48\textwidth]{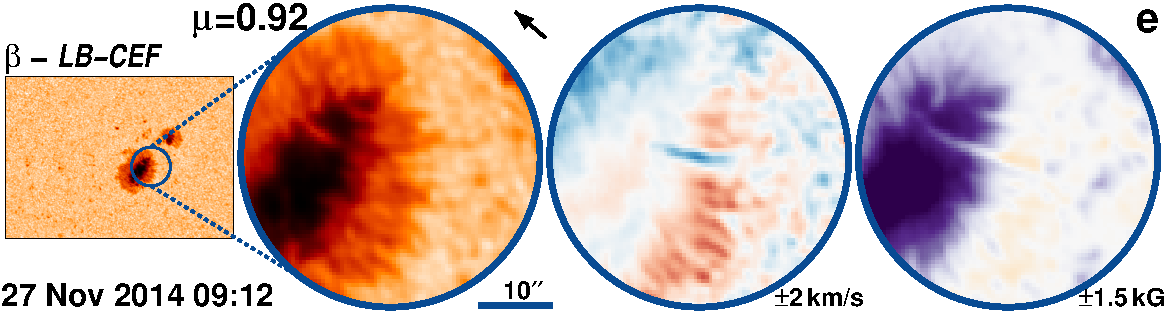}
 \includegraphics[width=.48\textwidth]{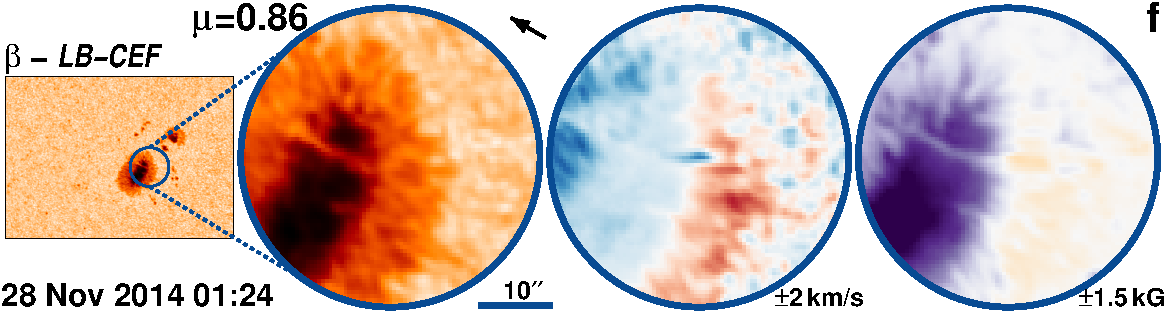}
 \includegraphics[width=.48\textwidth]{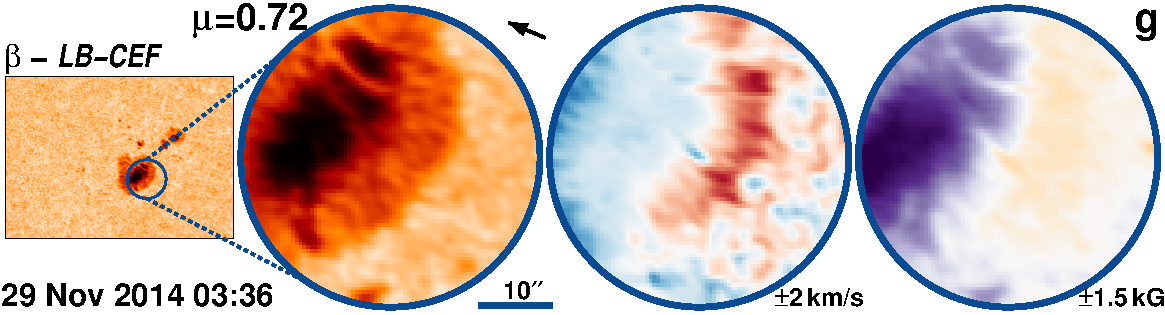}
 \includegraphics[width=.48\textwidth]{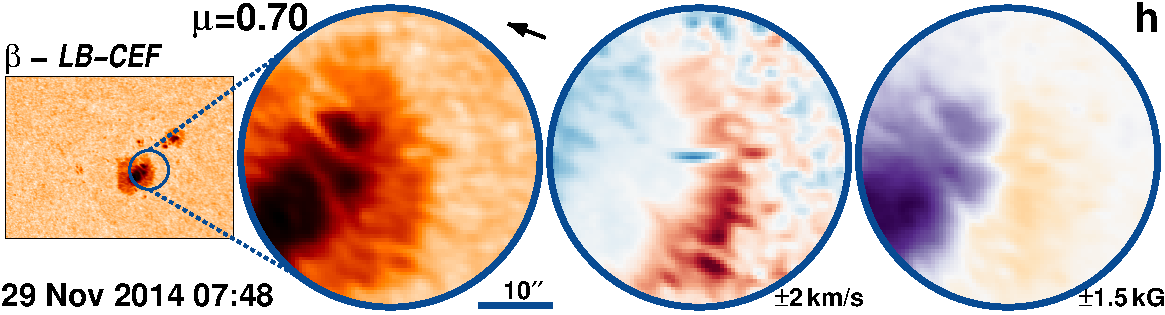}
 \caption{AR\,12216 followed for    9.6 days from 21-Nov-2014 starting at 00:24\,UT.\label{fig:DS45}}
 \end{figure*}

\begin{figure*}[htbp]
 \centering
 \includegraphics[width=.48\textwidth]{colorbars.pdf}
 \includegraphics[width=.48\textwidth]{colorbars.pdf}
 \includegraphics[width=.48\textwidth]{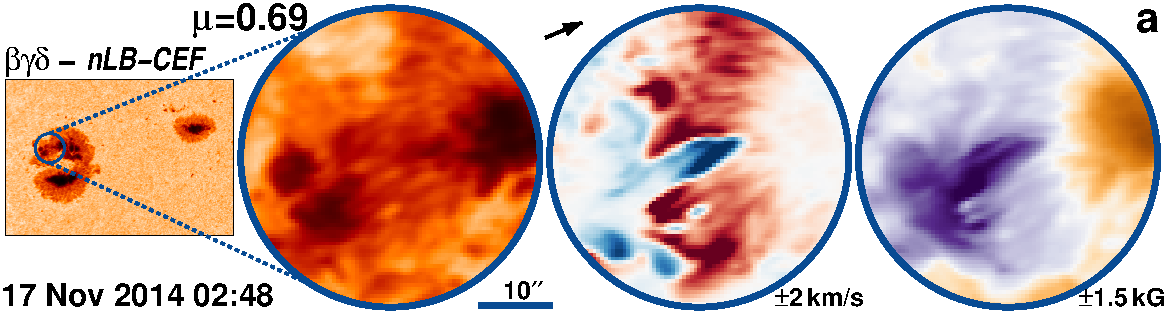}
 \includegraphics[width=.48\textwidth]{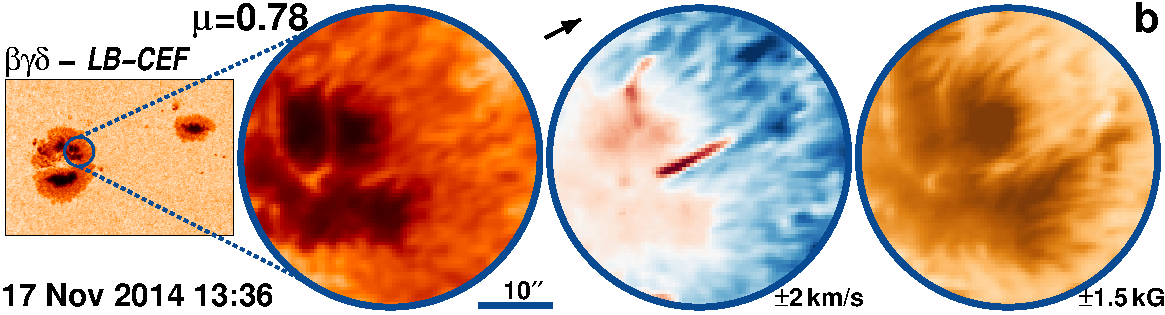}
 \includegraphics[width=.48\textwidth]{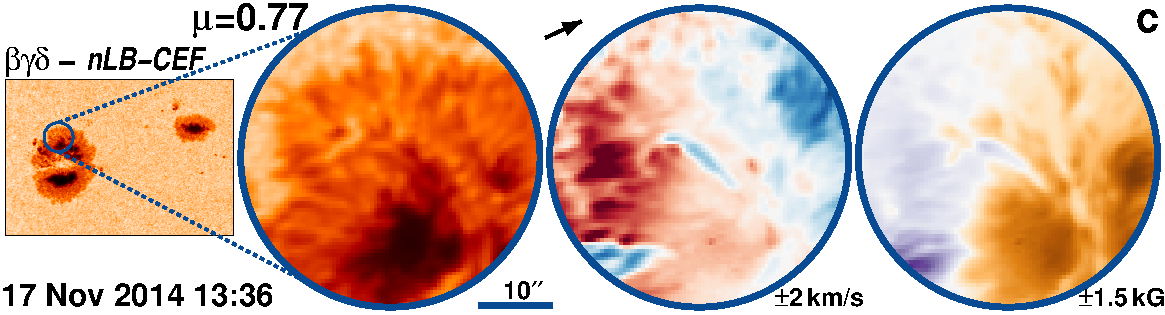}
 \includegraphics[width=.48\textwidth]{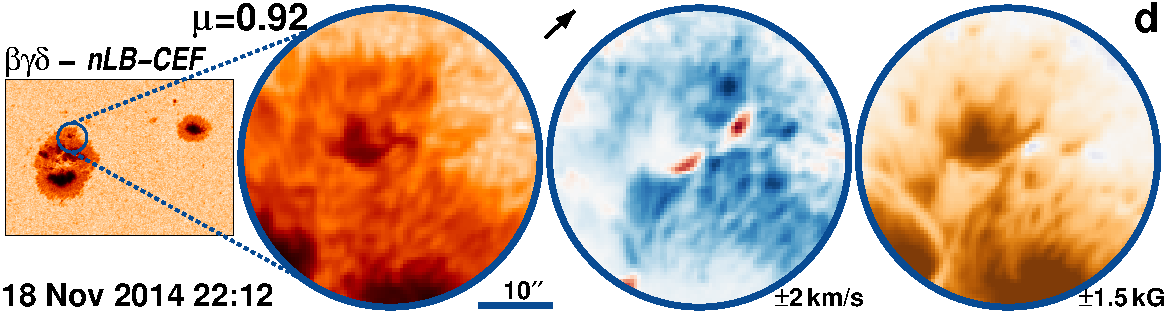}
 \includegraphics[width=.48\textwidth]{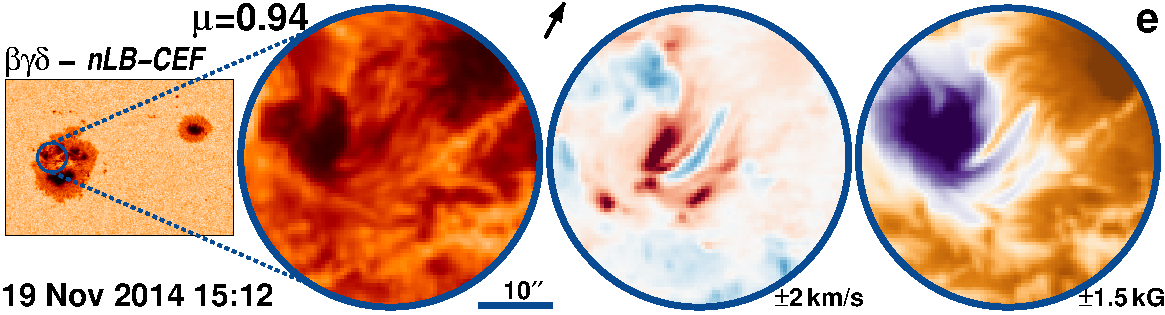}
 \includegraphics[width=.48\textwidth]{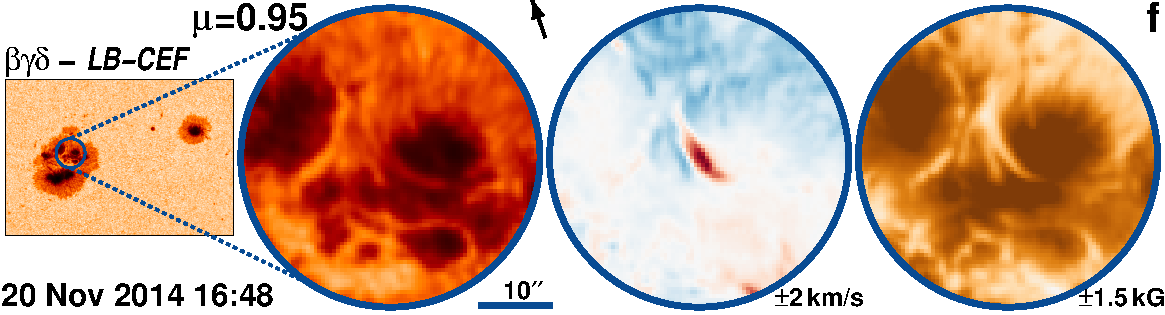}
 \includegraphics[width=.48\textwidth]{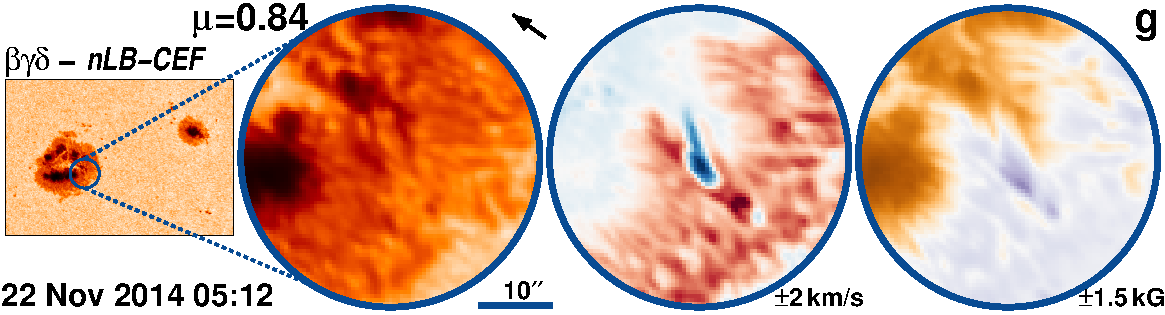}
 \includegraphics[width=.48\textwidth]{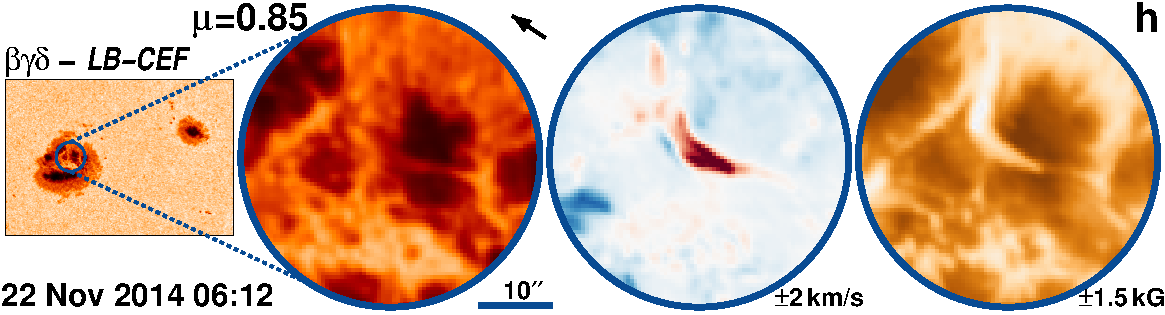}
 \includegraphics[width=.48\textwidth]{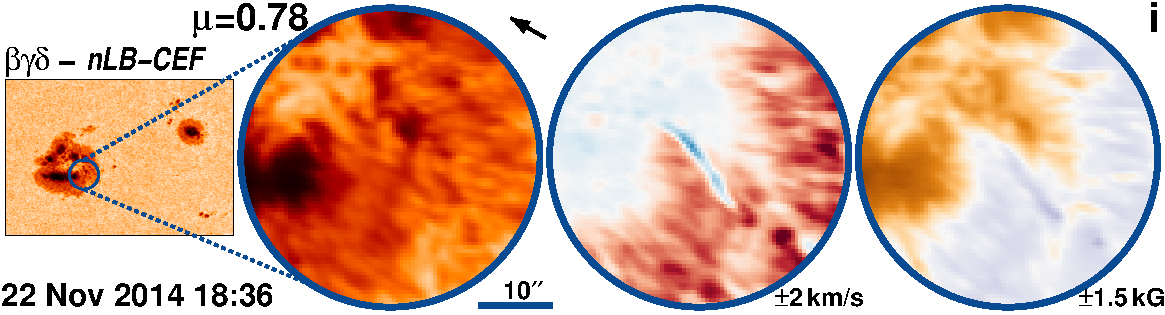}
 \includegraphics[width=.48\textwidth]{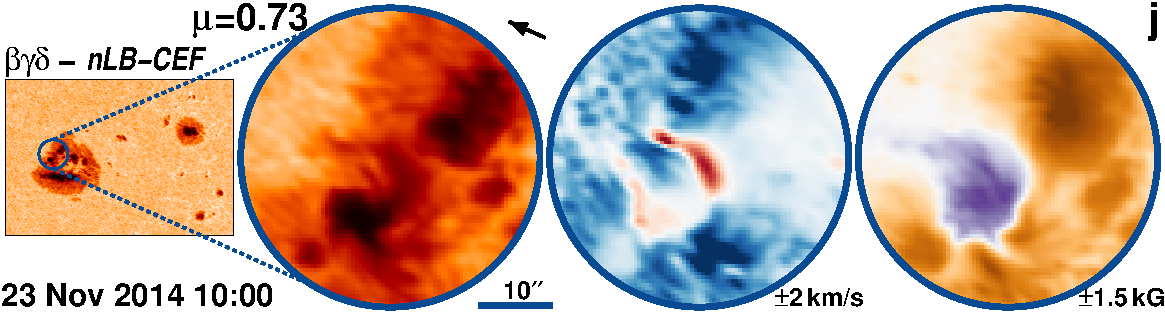}
 \includegraphics[width=.48\textwidth]{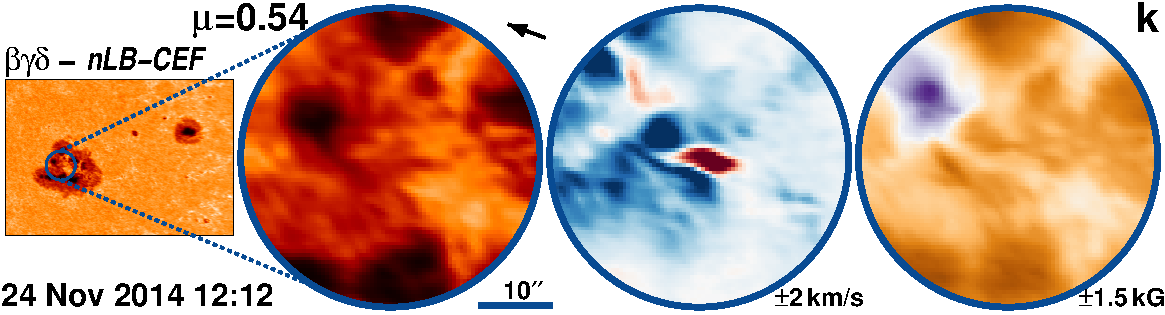}
 \includegraphics[width=.48\textwidth]{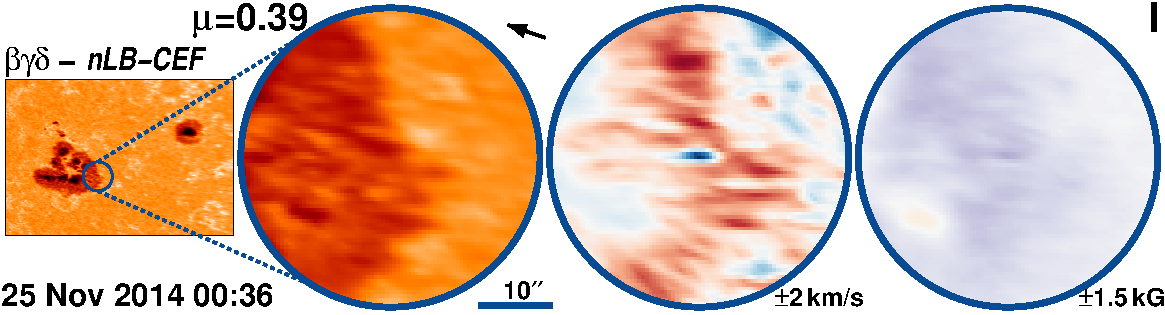}
 \caption{AR\,12209 followed for    9.9 days from 15-Nov-2014 starting at 18:12\,UT.\label{fig:DS46}}
 \end{figure*}

\begin{figure*}[htbp]
 \includegraphics[width=.48\textwidth]{colorbars.pdf}

 \includegraphics[width=.48\textwidth]{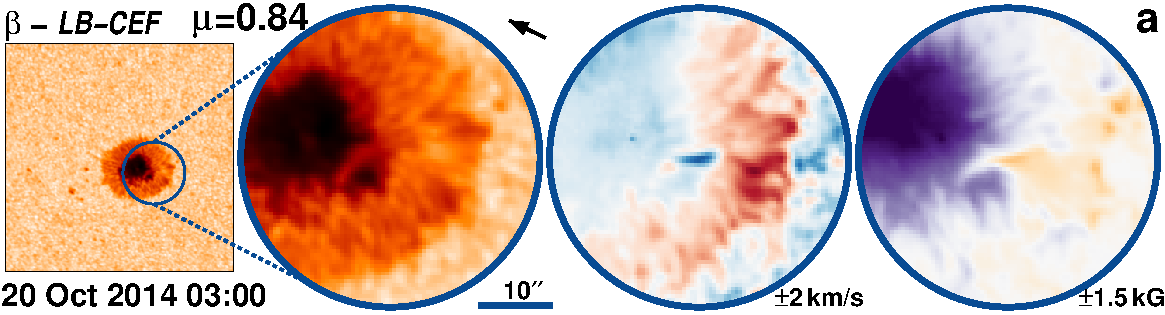}
 \caption{AR\,12187 followed for   10.2 days from 12-Oct-2014 starting at 21:00\,UT.\label{fig:DS48}}
 \end{figure*}

\begin{figure*}[htbp]
 \centering
 \includegraphics[width=.48\textwidth]{colorbars.pdf}
 \includegraphics[width=.48\textwidth]{colorbars.pdf}
 \includegraphics[width=.48\textwidth]{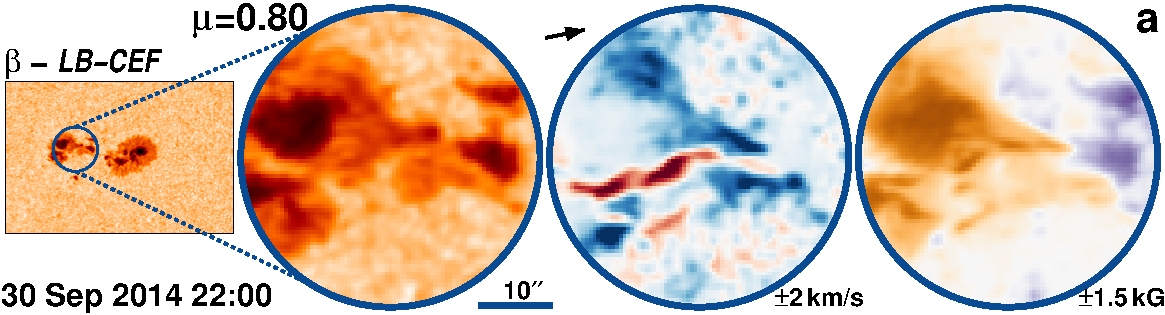}
 \includegraphics[width=.48\textwidth]{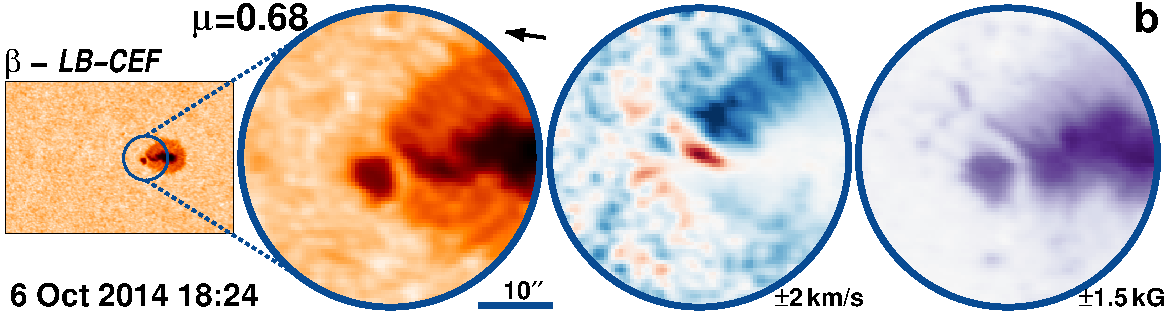}
 \includegraphics[width=.48\textwidth]{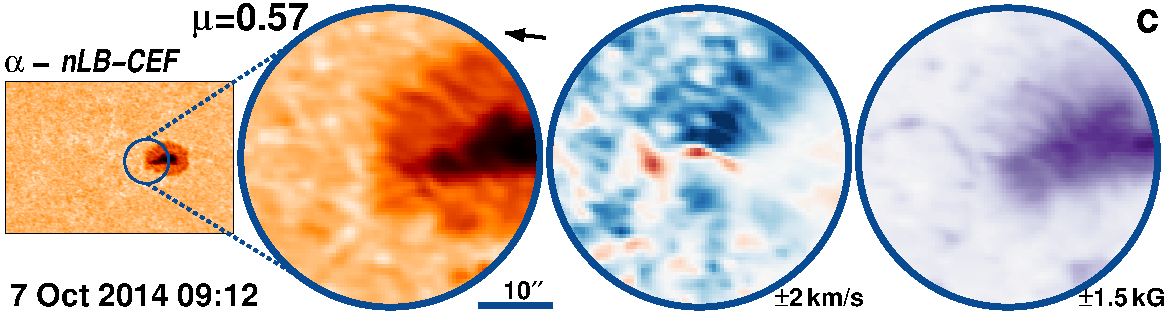}
 \includegraphics[width=.48\textwidth]{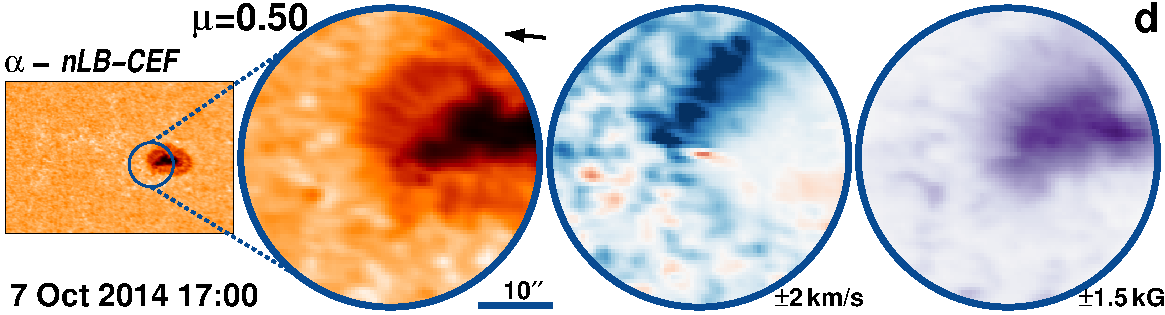}
 \caption{AR\,12178 followed for   10.4 days from 28-Sep-2014 starting at 02:36\,UT.\label{fig:DS50}}
 \end{figure*}

\begin{figure*}[htbp]
 \centering
 \includegraphics[width=.48\textwidth]{colorbars.pdf}
 \includegraphics[width=.48\textwidth]{colorbars.pdf}
 \includegraphics[width=.48\textwidth]{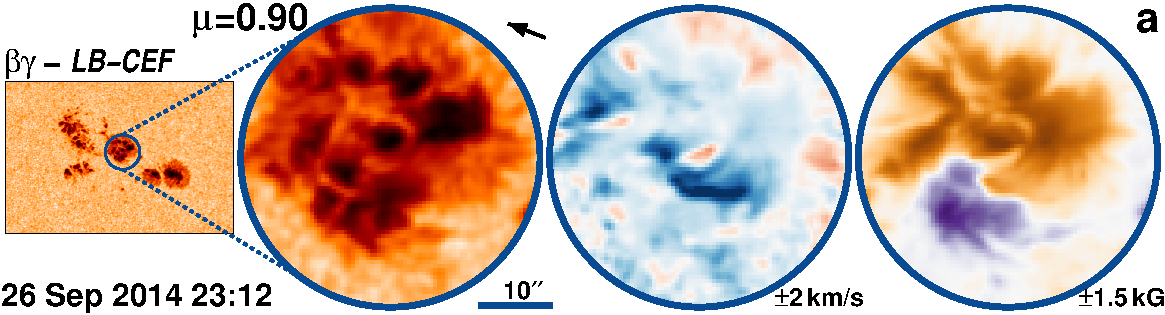}
 \includegraphics[width=.48\textwidth]{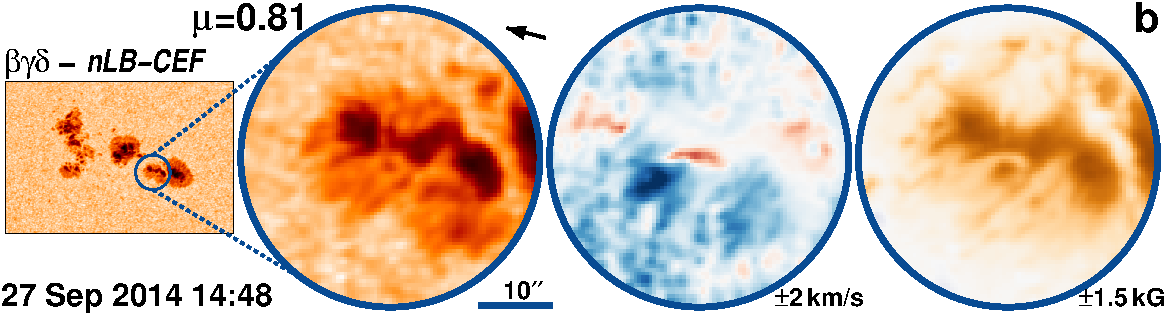}
 \includegraphics[width=.48\textwidth]{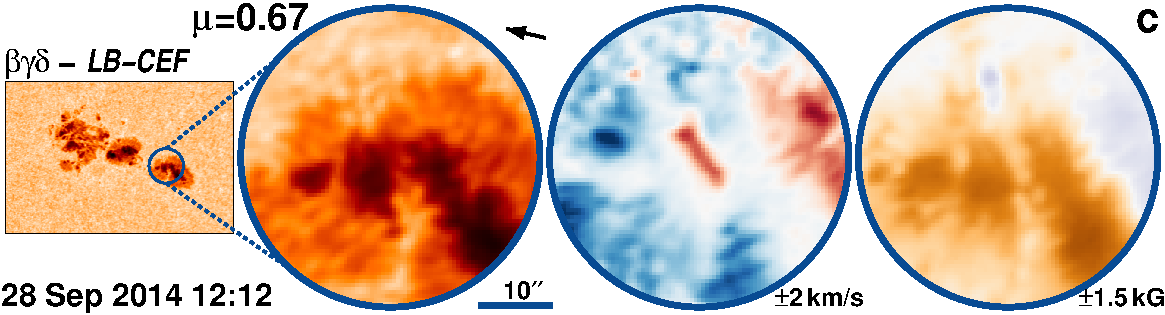}
 \includegraphics[width=.48\textwidth]{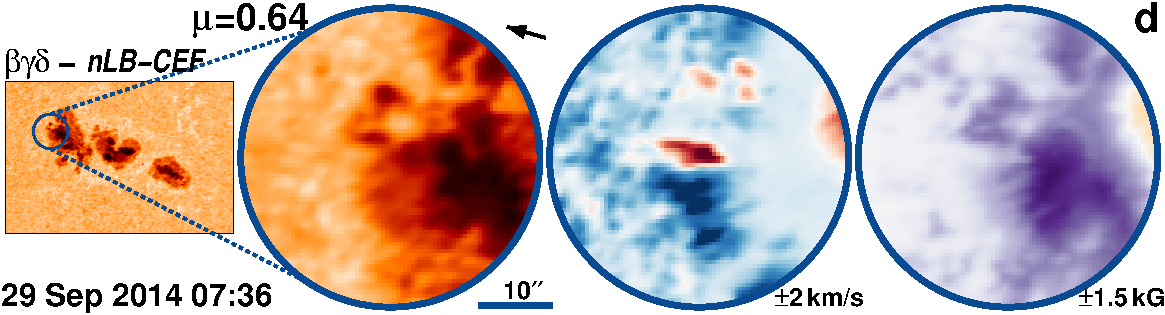}
 \includegraphics[width=.48\textwidth]{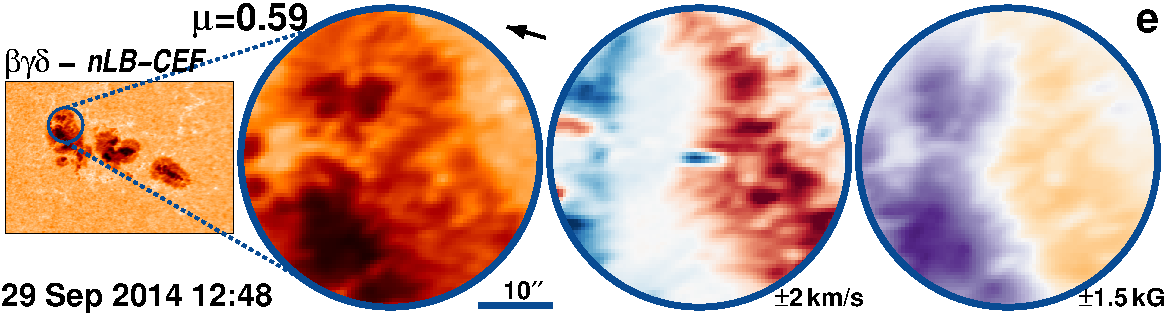}
 \includegraphics[width=.48\textwidth]{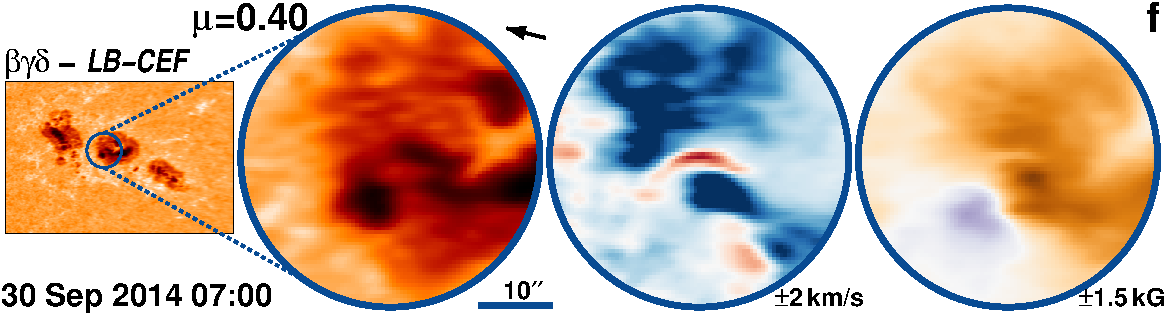}
 \caption{AR\,12175 followed for    6.2 days from 24-Sep-2014 starting at 19:48\,UT.\label{fig:DS51}}
 \end{figure*}

\begin{figure*}[htbp]
 \centering
 \includegraphics[width=.48\textwidth]{colorbars.pdf}
 \includegraphics[width=.48\textwidth]{colorbars.pdf}
 \includegraphics[width=.48\textwidth]{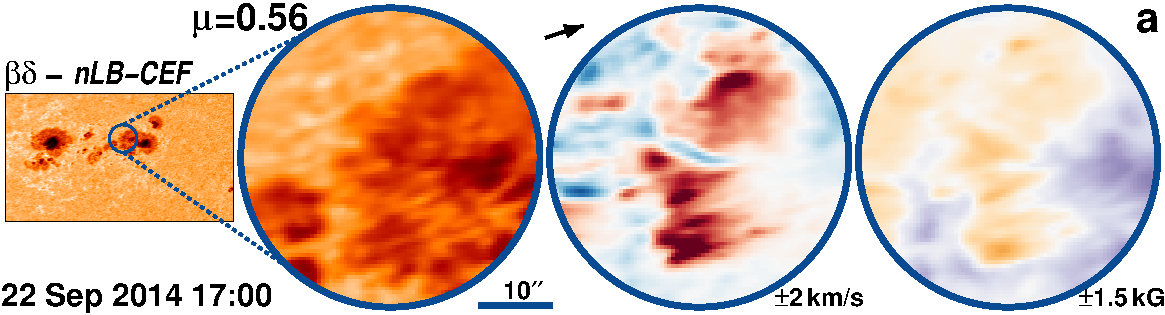}
 \includegraphics[width=.48\textwidth]{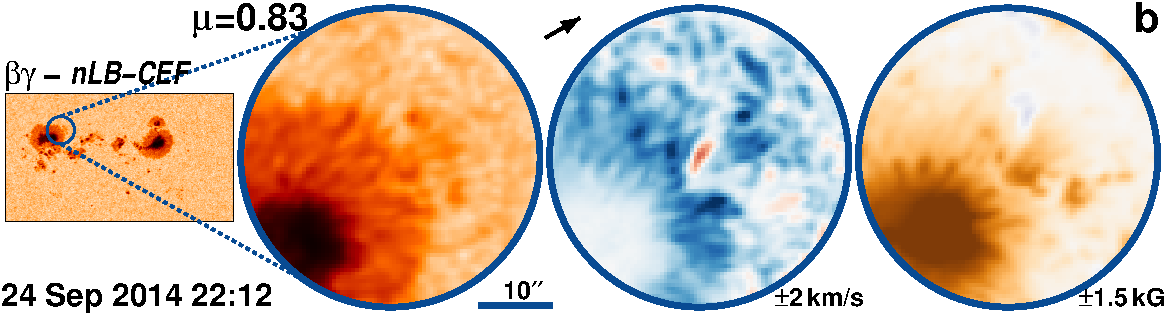}
 \includegraphics[width=.48\textwidth]{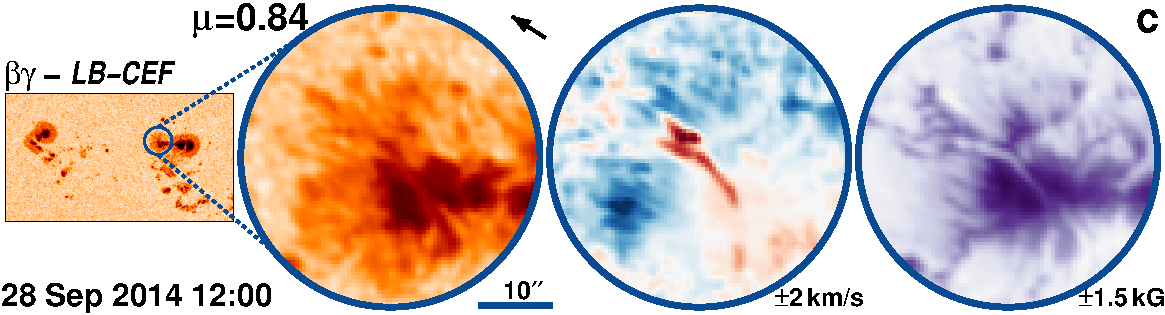}
 \includegraphics[width=.48\textwidth]{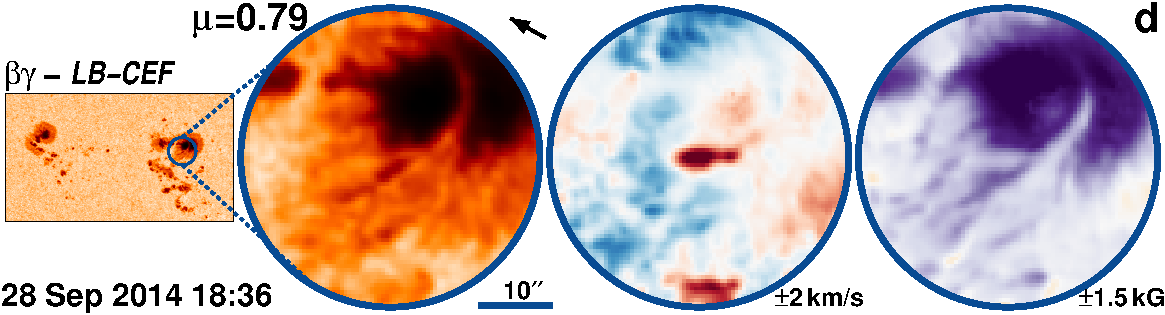}
 \caption{AR\,12172 followed for    8.3 days from 22-Sep-2014 starting at 12:00\,UT.\label{fig:DS52}}
 \end{figure*}

\begin{figure*}[htbp]
 \includegraphics[width=.48\textwidth]{colorbars.pdf}
 \includegraphics[width=.48\textwidth]{colorbars.pdf}
 \includegraphics[width=.48\textwidth]{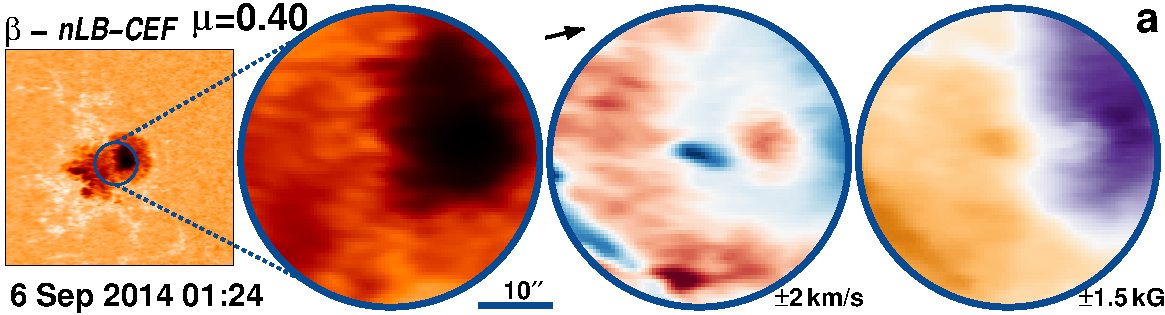}
 \includegraphics[width=.48\textwidth]{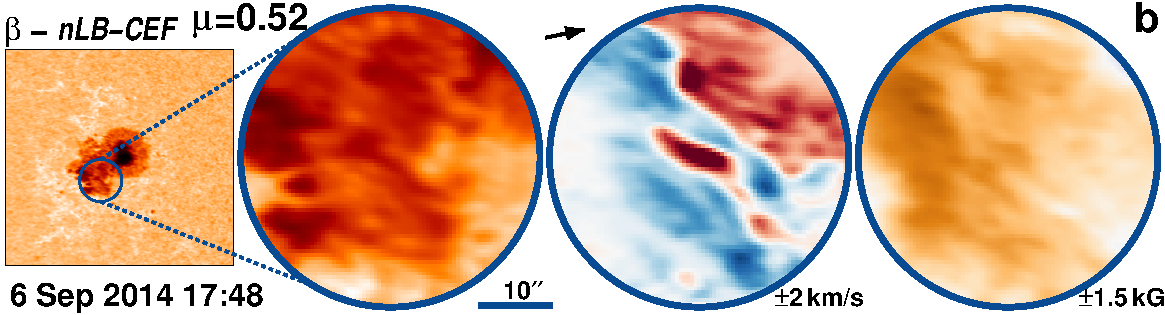}
 \includegraphics[width=.48\textwidth]{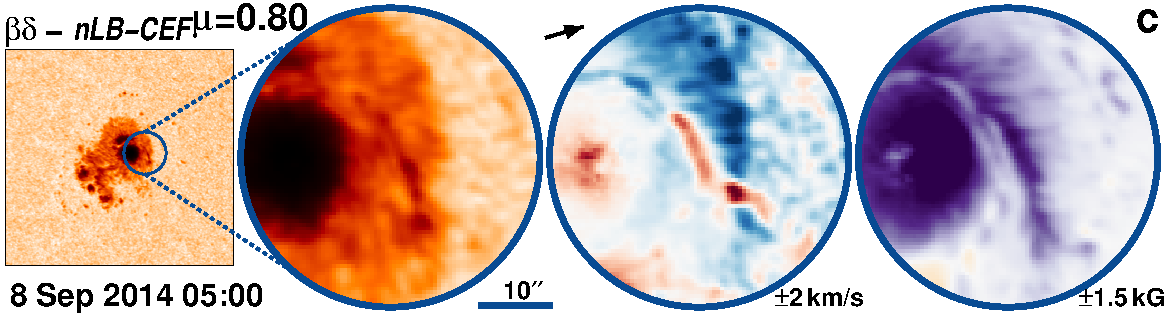}
 \caption{AR\,12158 followed for   10.6 days from  5-Sep-2014 starting at 18:36\,UT.\label{fig:DS53}}
 \end{figure*}

\begin{figure*}[htbp]
 \includegraphics[width=.48\textwidth]{colorbars.pdf}

 \includegraphics[width=.48\textwidth]{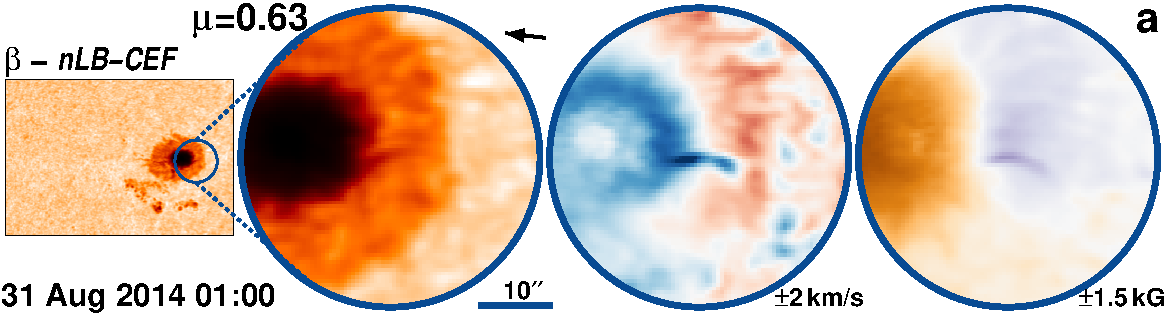}
 \caption{AR\,12149 followed for   10.5 days from 22-Aug-2014 starting at 07:12\,UT.\label{fig:DS55}}
 \end{figure*}

\begin{figure*}[htbp]
 \centering
 \includegraphics[width=.48\textwidth]{colorbars.pdf}
 \includegraphics[width=.48\textwidth]{colorbars.pdf}
 \includegraphics[width=.48\textwidth]{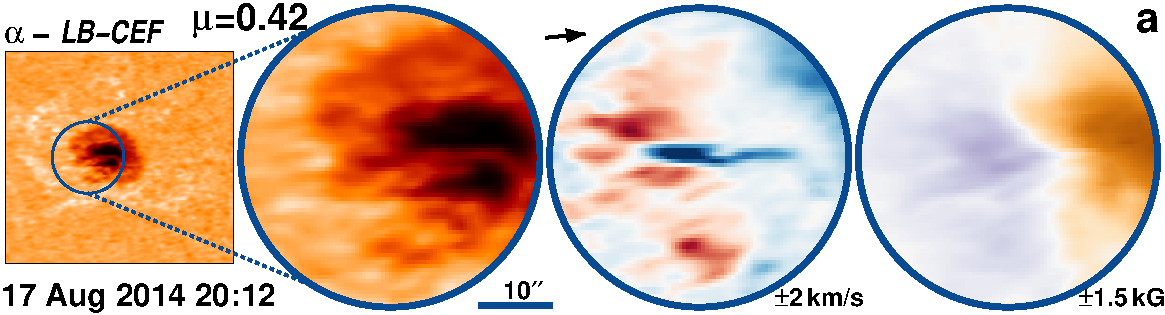}
 \includegraphics[width=.48\textwidth]{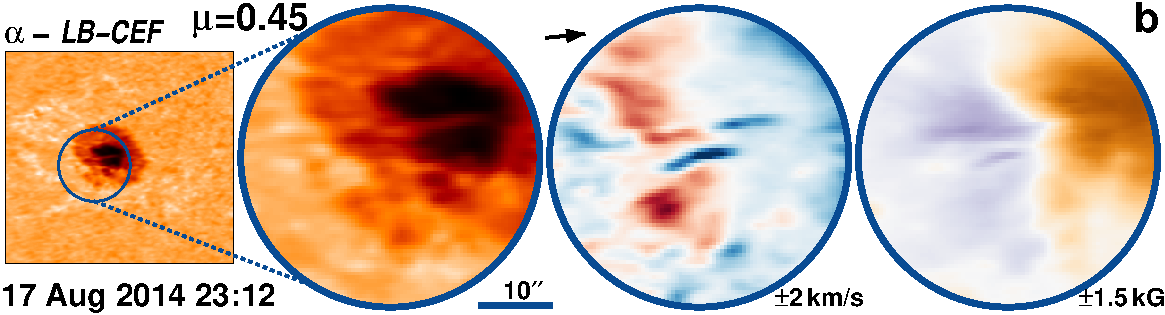}
 \includegraphics[width=.48\textwidth]{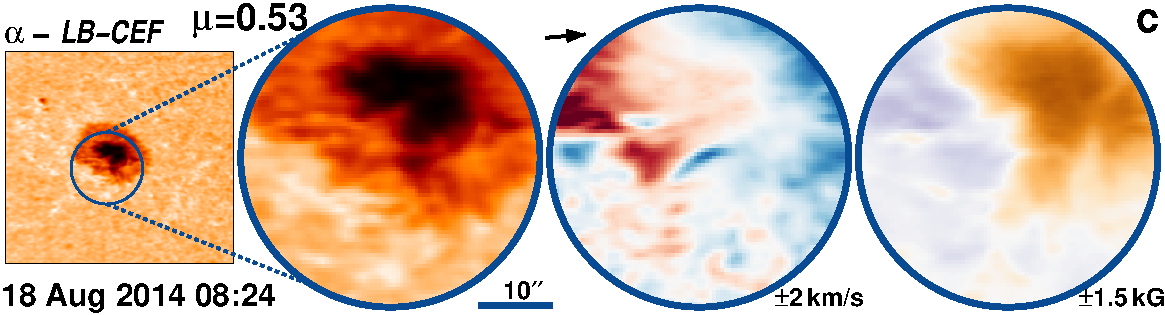}
 \includegraphics[width=.48\textwidth]{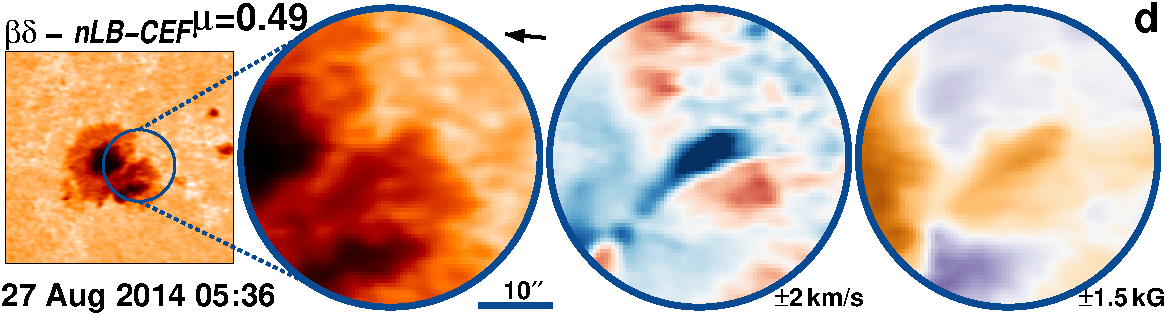}
 \caption{AR\,12146 followed for   10.6 days from 17-Aug-2014 starting at 12:00\,UT.\label{fig:DS56}}
 \end{figure*}

\begin{figure*}[htbp]
 \centering
 \includegraphics[width=.48\textwidth]{colorbars.pdf}
 \includegraphics[width=.48\textwidth]{colorbars.pdf}
 \includegraphics[width=.48\textwidth]{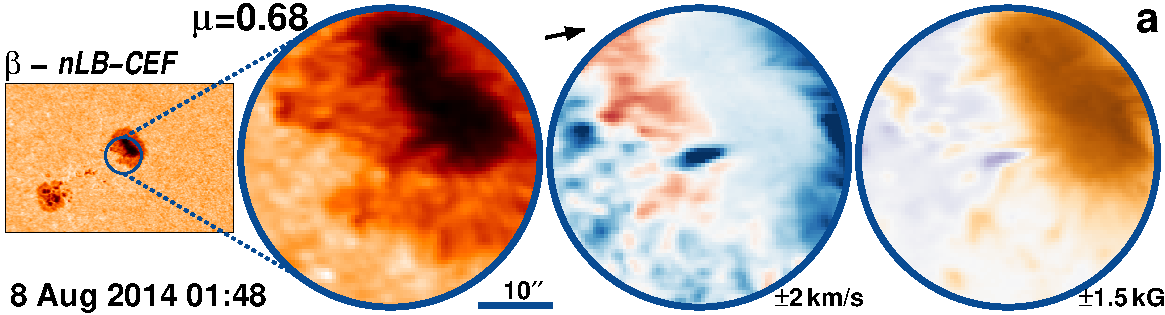}
 \includegraphics[width=.48\textwidth]{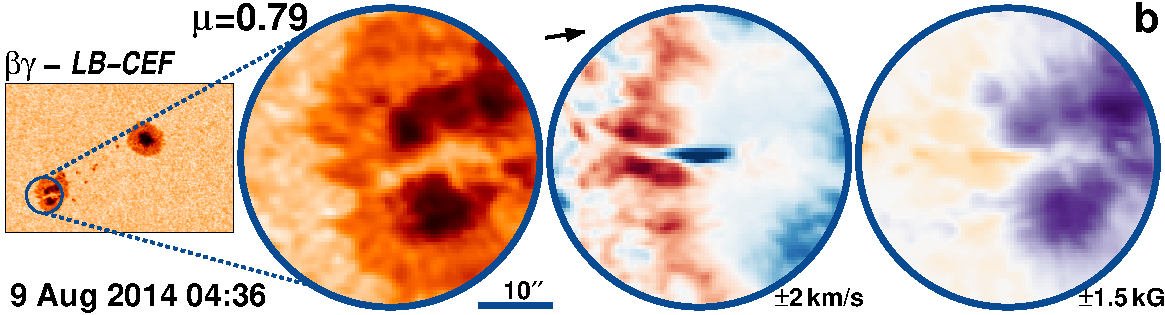}
 \caption{AR\,12135 followed for   10.4 days from  6-Aug-2014 starting at 10:48\,UT.\label{fig:DS57}}
 \end{figure*}

\begin{figure*}[htbp]
 \includegraphics[width=.48\textwidth]{colorbars.pdf}
 \includegraphics[width=.48\textwidth]{colorbars.pdf}
 \includegraphics[width=.48\textwidth]{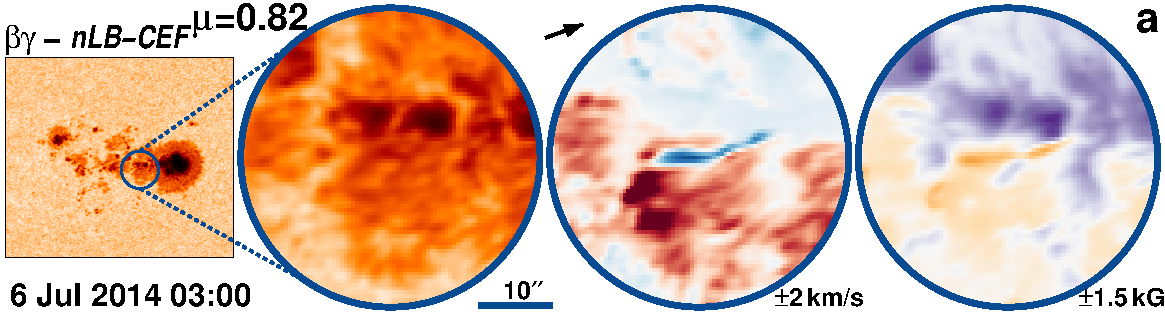}
 \includegraphics[width=.48\textwidth]{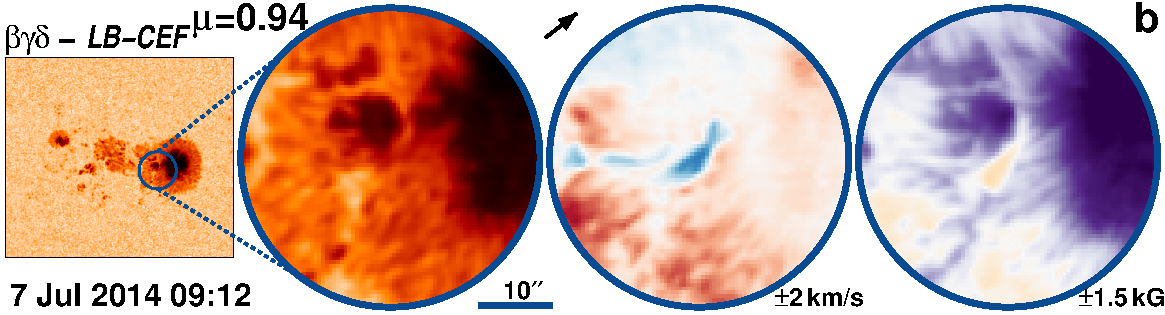}
 \includegraphics[width=.48\textwidth]{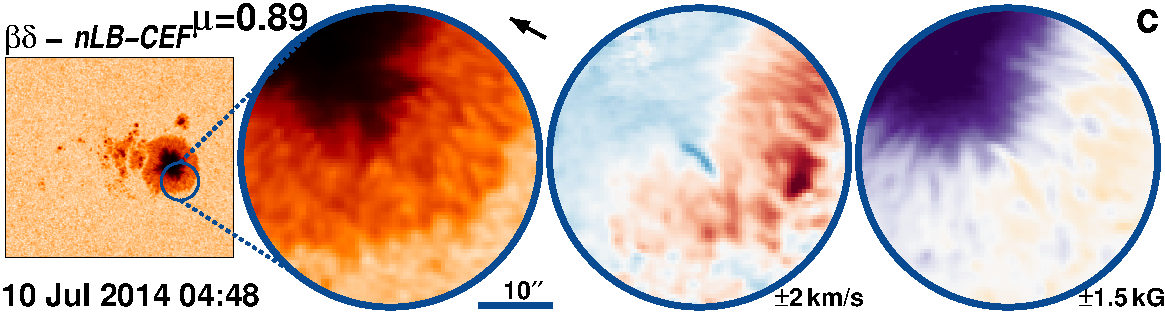}
 \includegraphics[width=.48\textwidth]{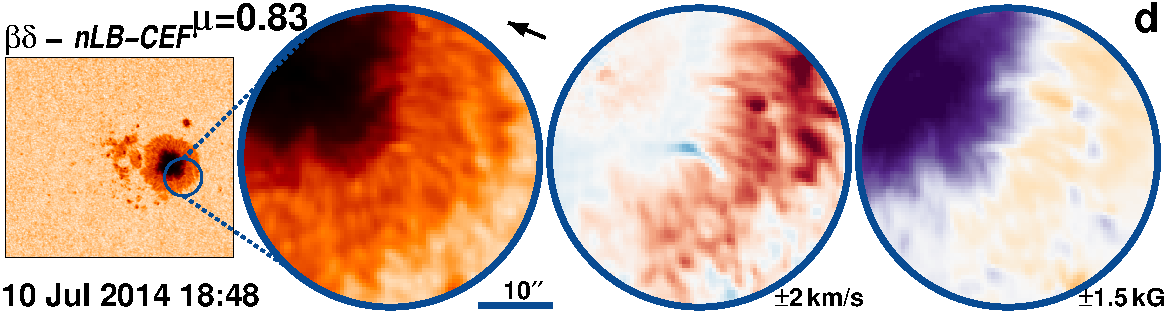}
 \includegraphics[width=.48\textwidth]{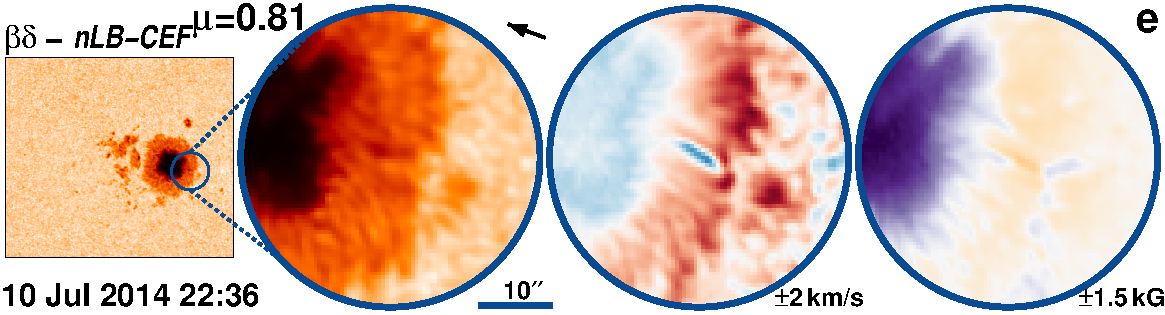}
 \caption{AR\,12109 followed for   11.1 days from  3-Jul-2014 starting at 03:24\,UT.\label{fig:DS58}}
 \end{figure*}

\begin{figure*}[htbp]
 \includegraphics[width=.48\textwidth]{colorbars.pdf}
 \includegraphics[width=.48\textwidth]{colorbars.pdf}
 \includegraphics[width=.48\textwidth]{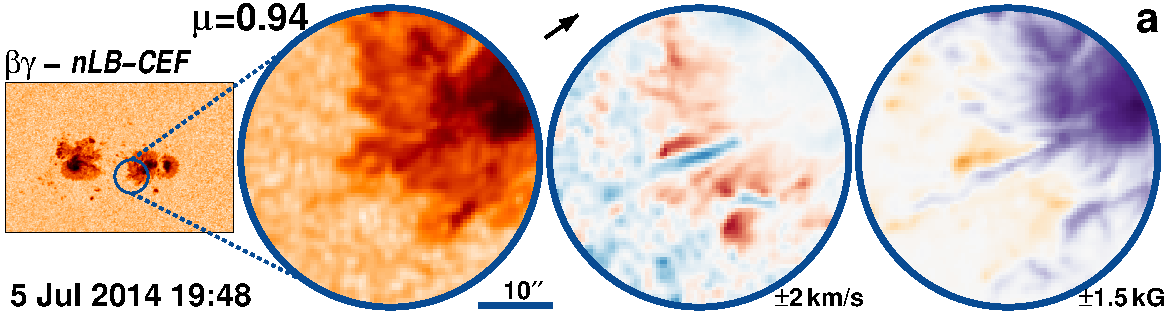}
 \includegraphics[width=.48\textwidth]{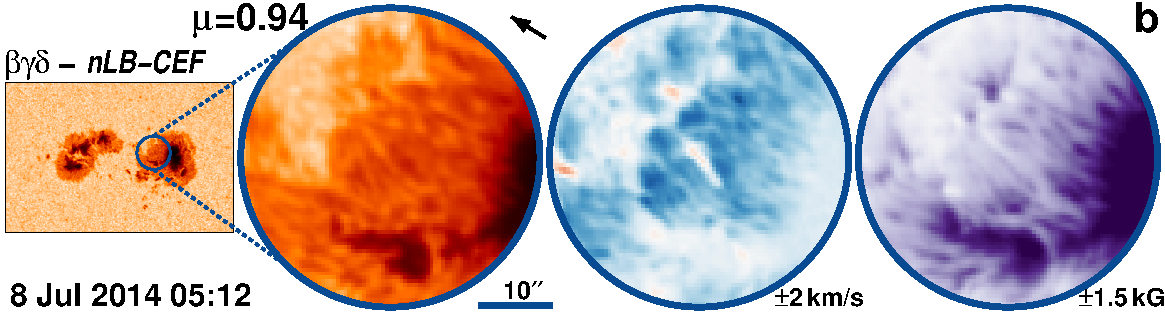}
 \includegraphics[width=.48\textwidth]{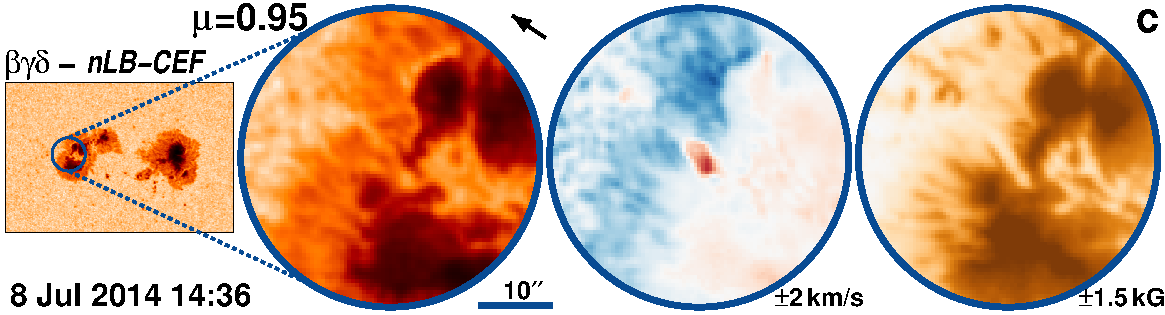}
 \includegraphics[width=.48\textwidth]{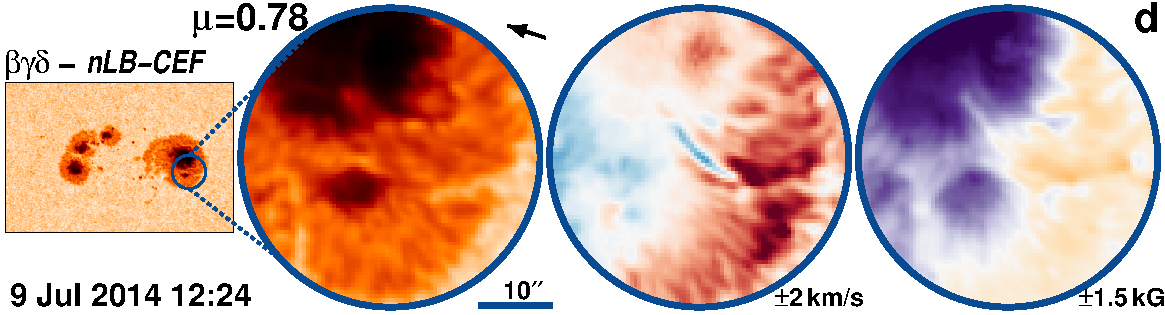}
 \includegraphics[width=.48\textwidth]{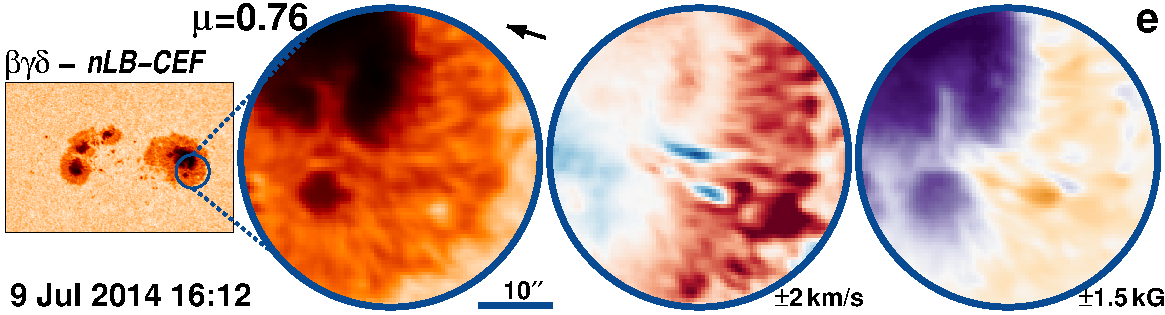}
 \includegraphics[width=.48\textwidth]{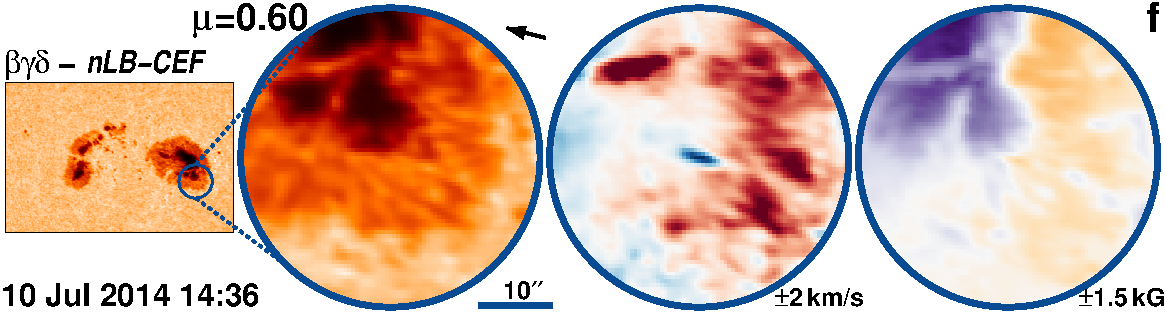}
 \includegraphics[width=.48\textwidth]{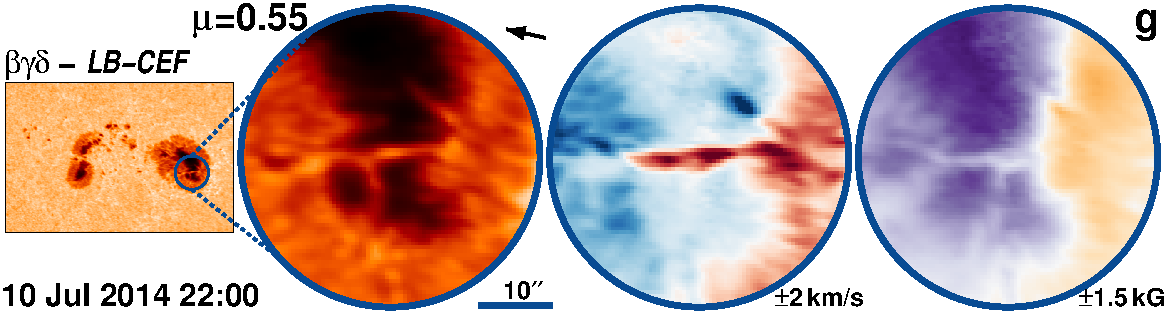}
 \includegraphics[width=.48\textwidth]{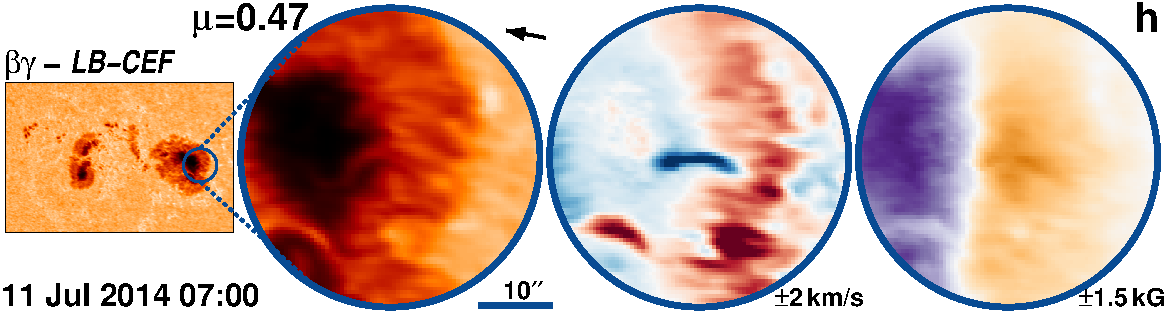}
 \includegraphics[width=.48\textwidth]{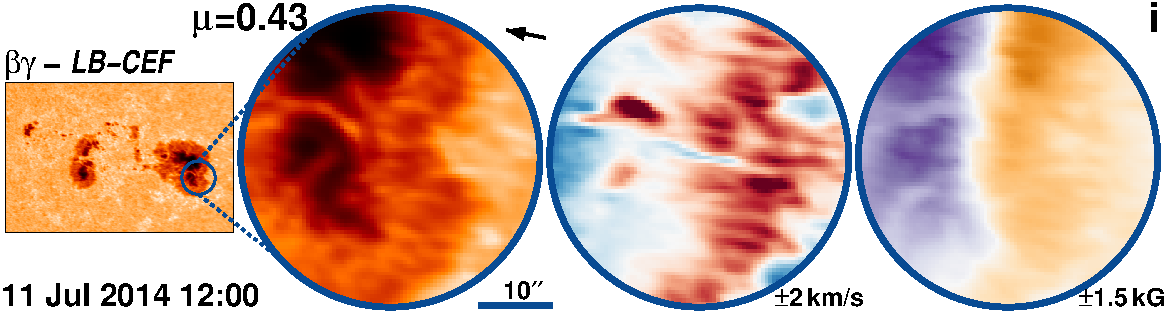}
 \caption{AR\,12108 followed for    9.9 days from  2-Jul-2014 starting at 06:24\,UT.\label{fig:DS59}}
 \end{figure*}

\begin{figure*}[htbp]
 \centering
 \includegraphics[width=.48\textwidth]{colorbars.pdf}
 \includegraphics[width=.48\textwidth]{colorbars.pdf}
 \includegraphics[width=.48\textwidth]{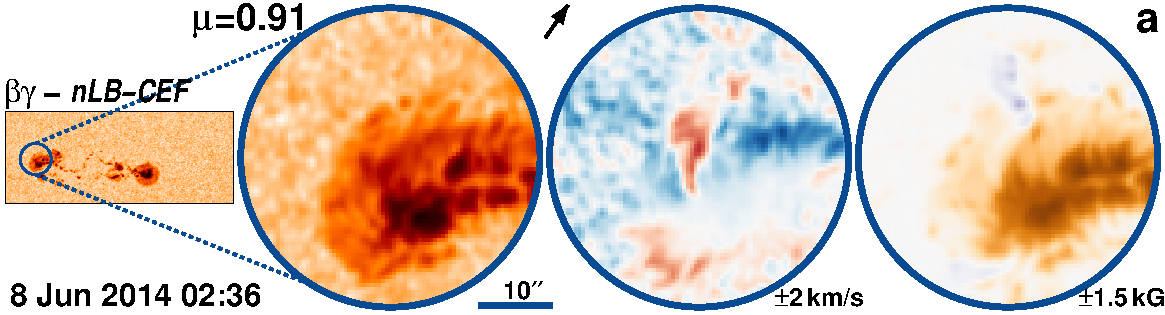}
 \includegraphics[width=.48\textwidth]{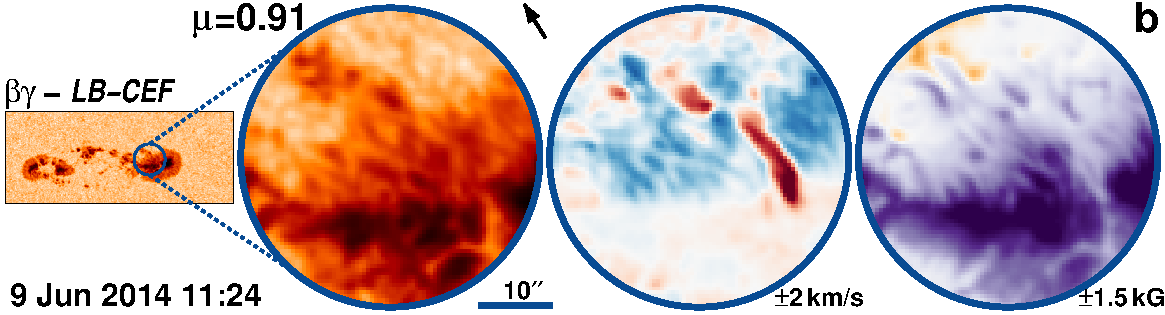}
 \includegraphics[width=.48\textwidth]{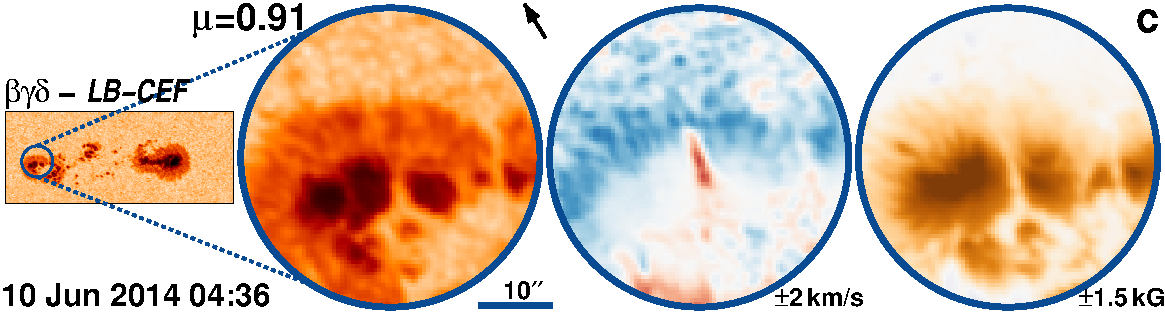}
 \includegraphics[width=.48\textwidth]{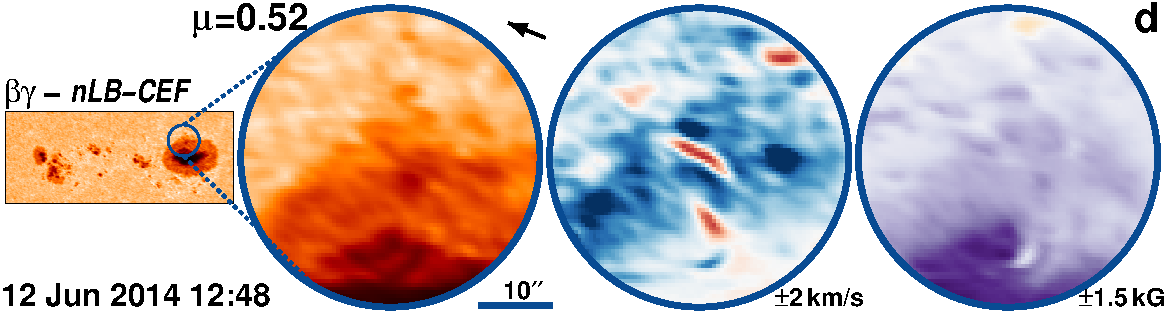}
 \includegraphics[width=.48\textwidth]{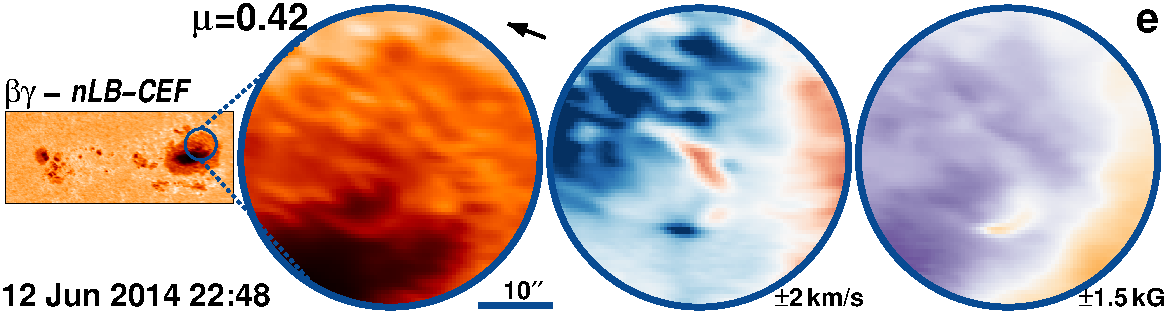}
 \includegraphics[width=.48\textwidth]{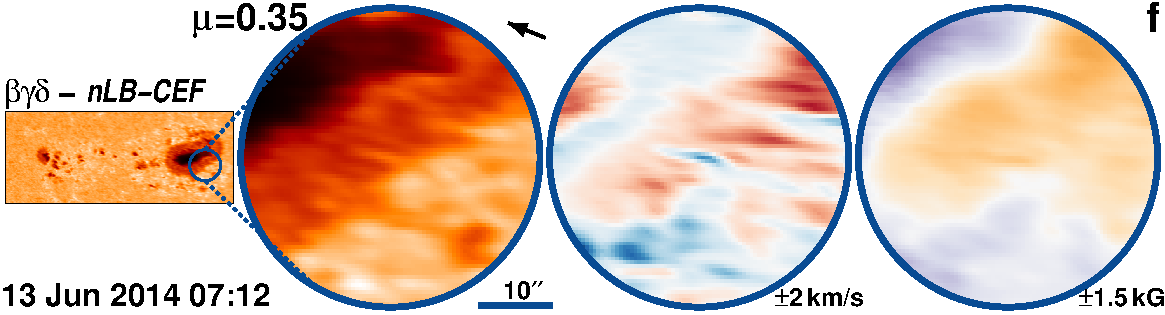}
 \caption{AR\,12085 followed for    7.1 days from  6-Jun-2014 starting at 11:24\,UT.\label{fig:DS62}}
 \end{figure*}

\begin{figure*}[htbp]
 \centering
 \includegraphics[width=.48\textwidth]{colorbars.pdf}
 \includegraphics[width=.48\textwidth]{colorbars.pdf}
 \includegraphics[width=.48\textwidth]{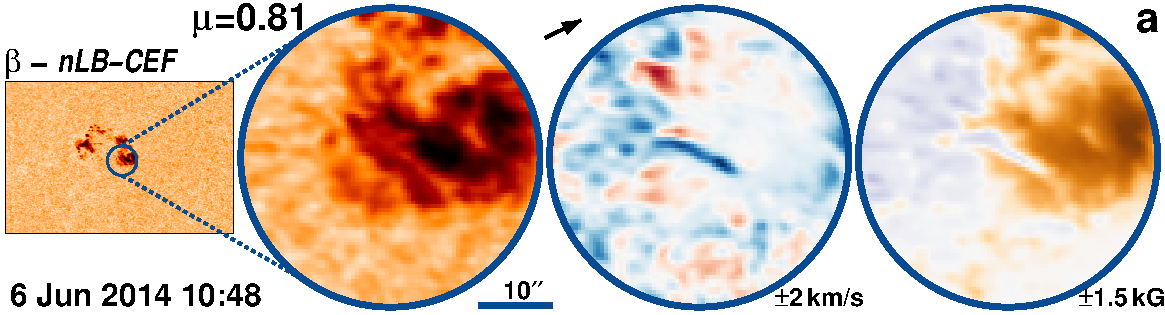}
 \includegraphics[width=.48\textwidth]{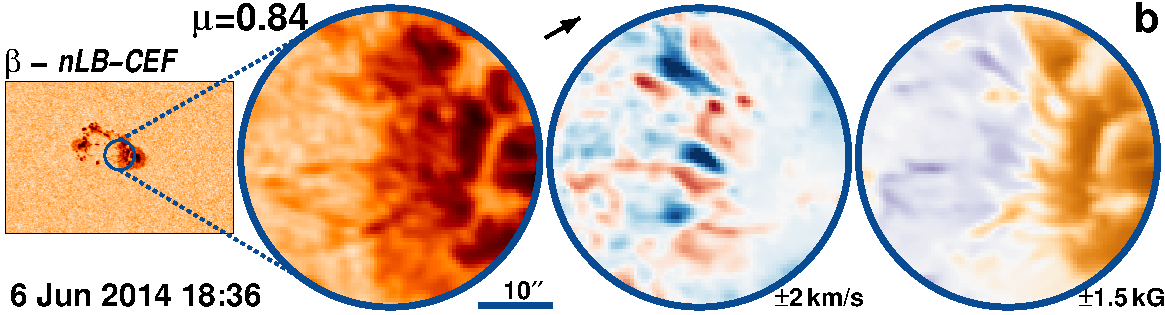}
 \includegraphics[width=.48\textwidth]{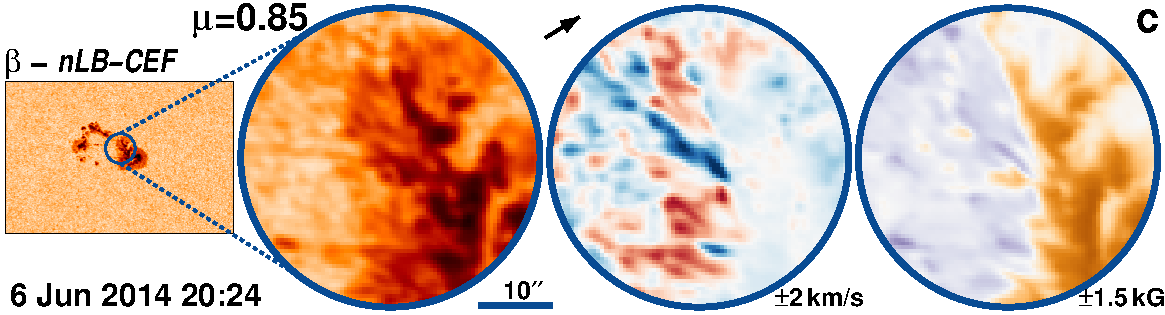}
 \includegraphics[width=.48\textwidth]{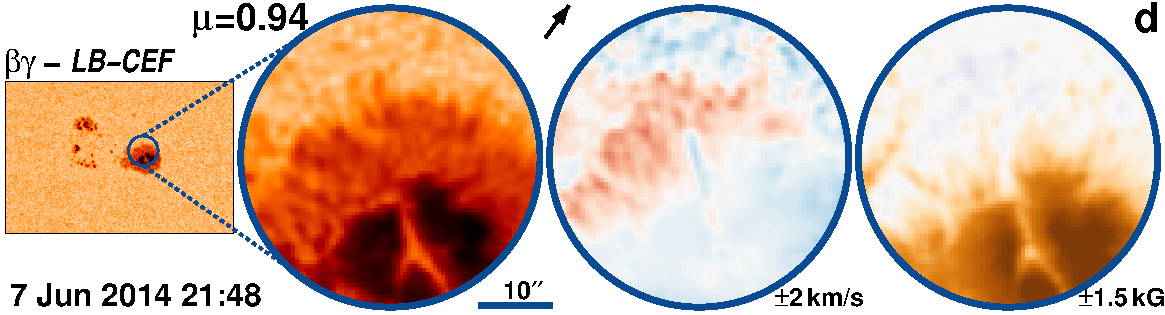}
 \includegraphics[width=.48\textwidth]{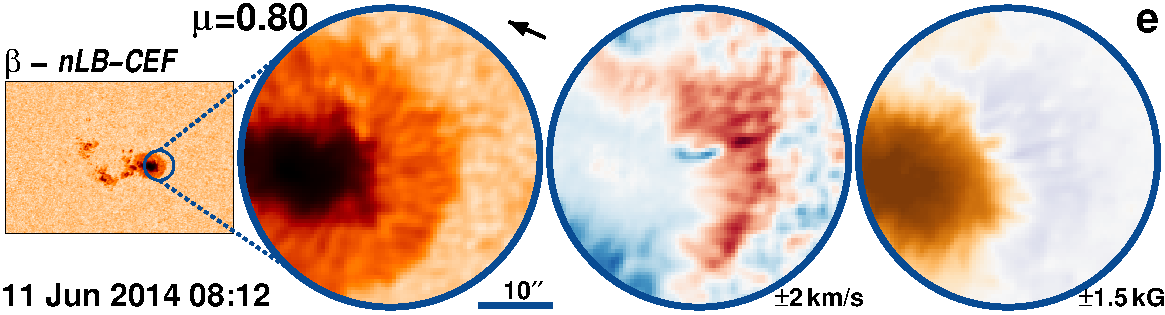}
 \includegraphics[width=.48\textwidth]{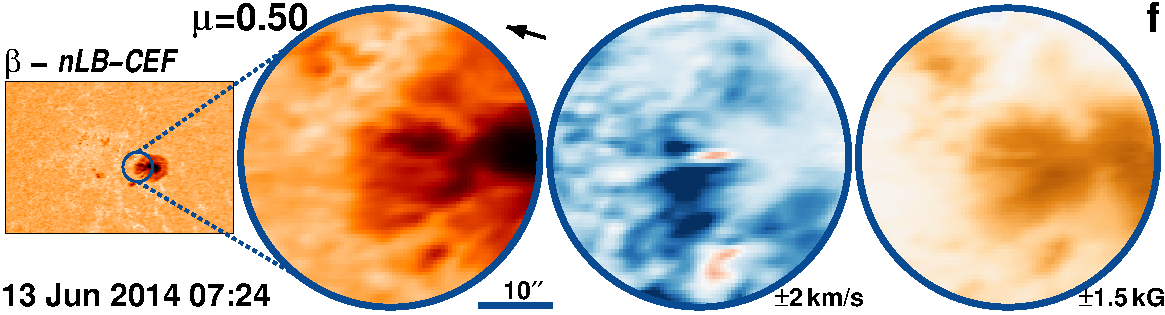}
 \caption{AR\,12082 followed for    8.2 days from  5-Jun-2014 starting at 08:24\,UT.\label{fig:DS63}}
 \end{figure*}

\begin{figure*}[htbp]
 \includegraphics[width=.48\textwidth]{colorbars.pdf}
 \includegraphics[width=.48\textwidth]{colorbars.pdf}
 \includegraphics[width=.48\textwidth]{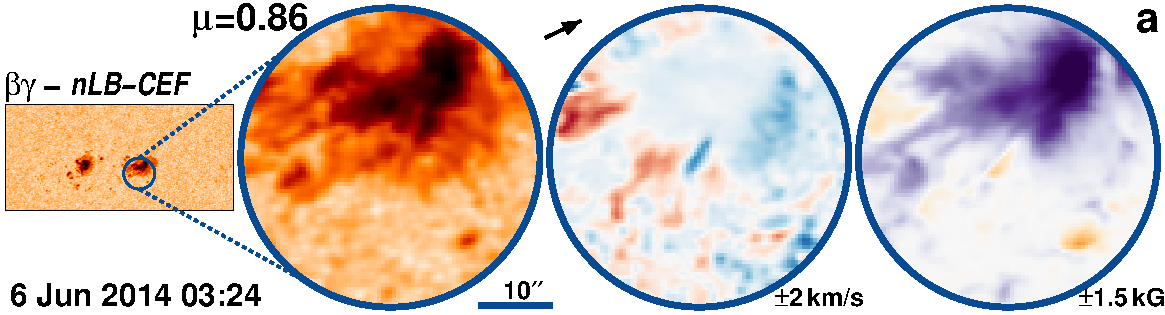}
 \includegraphics[width=.48\textwidth]{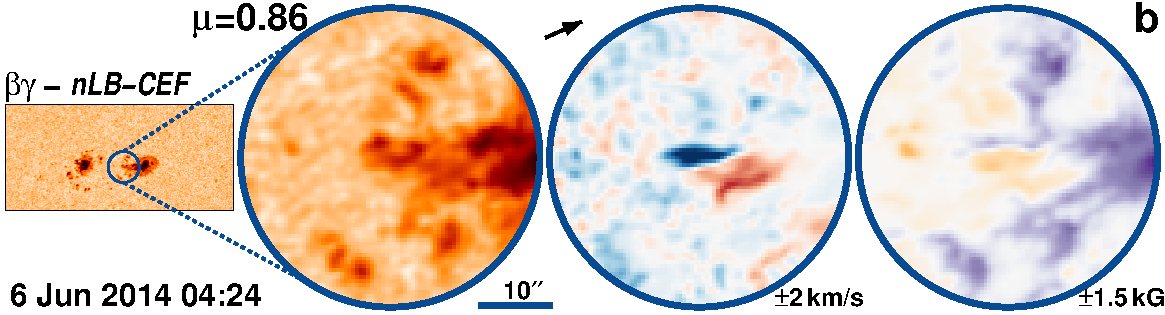}
 \includegraphics[width=.48\textwidth]{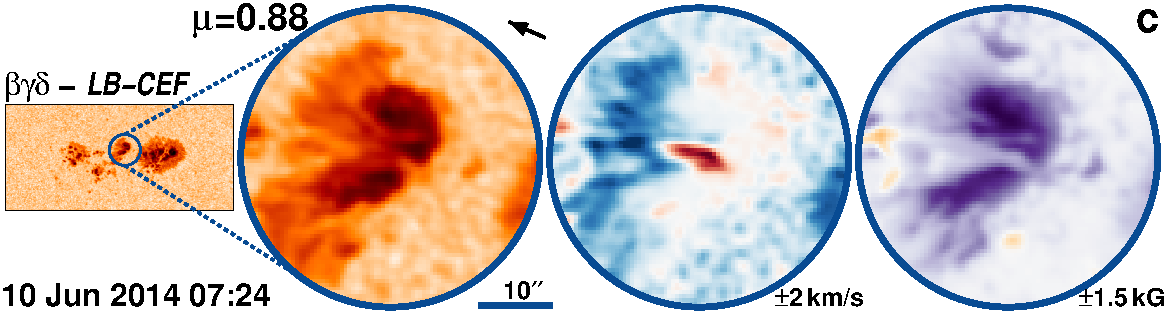}
 \includegraphics[width=.48\textwidth]{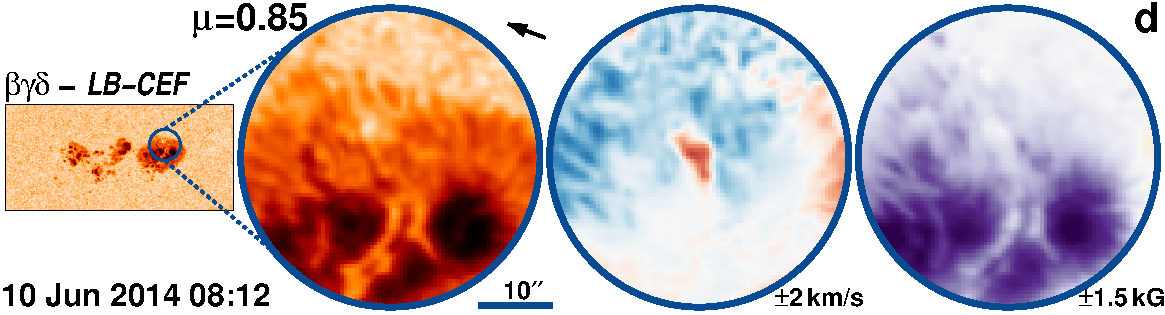}
 \includegraphics[width=.48\textwidth]{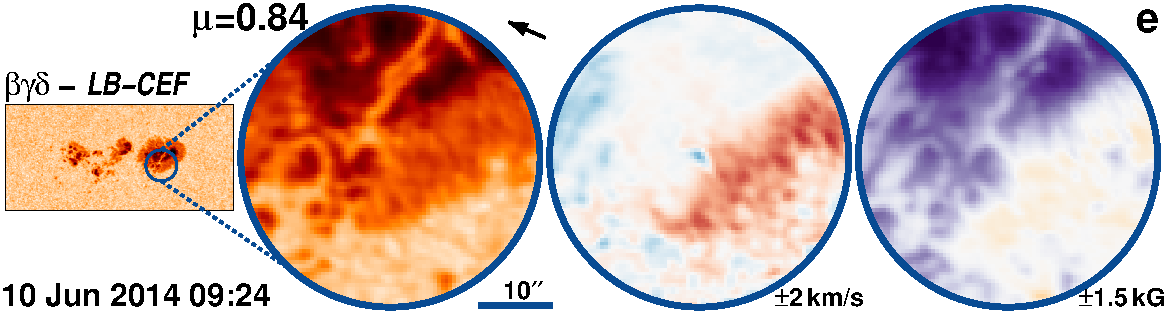}
 \caption{AR\,12080 followed for    8.8 days from  4-Jun-2014 starting at 19:24\,UT.\label{fig:DS64}}
 \end{figure*}

\begin{figure*}[htbp]
 \includegraphics[width=.48\textwidth]{colorbars.pdf}

 \includegraphics[width=.48\textwidth]{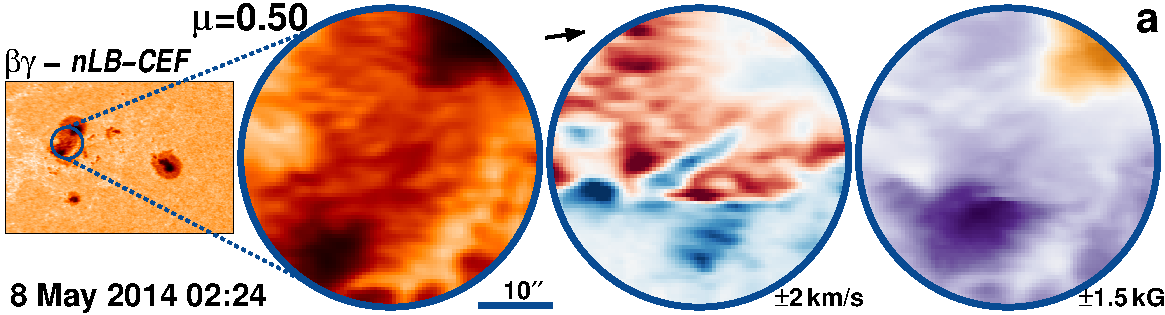}
 \caption{AR\,12056 followed for   10.5 days from  6-May-2014 starting at 21:48\,UT.\label{fig:DS65}}
 \end{figure*}

\begin{figure*}[htbp]
 \includegraphics[width=.48\textwidth]{colorbars.pdf}
 \includegraphics[width=.48\textwidth]{colorbars.pdf}
 \includegraphics[width=.48\textwidth]{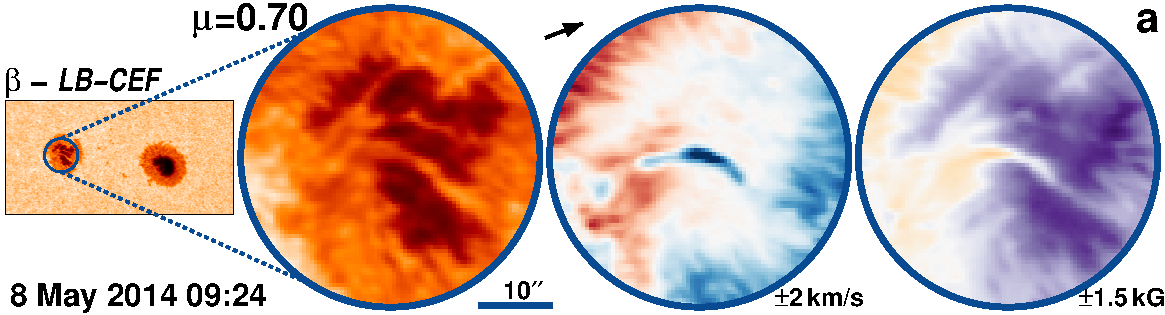}
 \includegraphics[width=.48\textwidth]{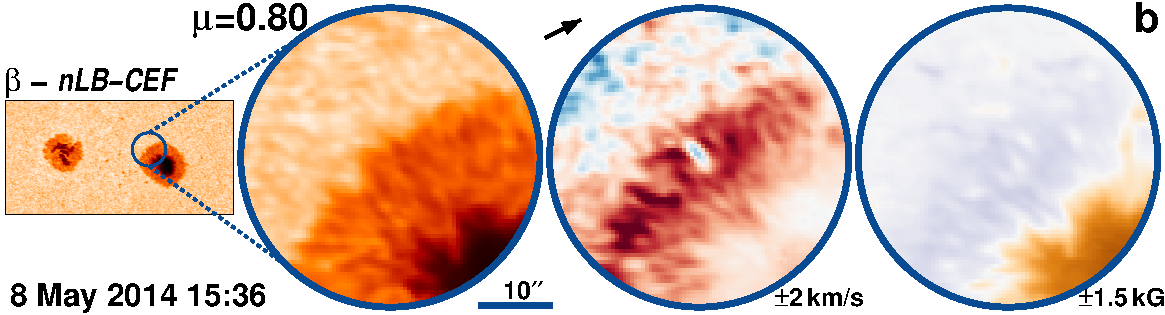}
 \includegraphics[width=.48\textwidth]{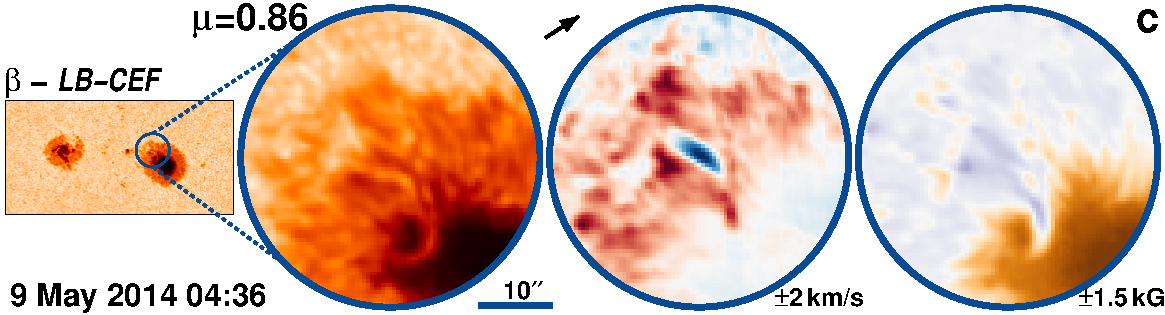}
 \includegraphics[width=.48\textwidth]{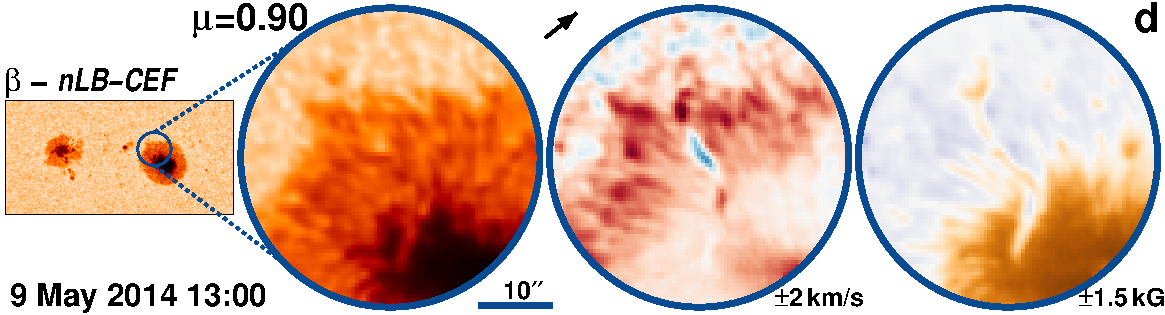}
 \includegraphics[width=.48\textwidth]{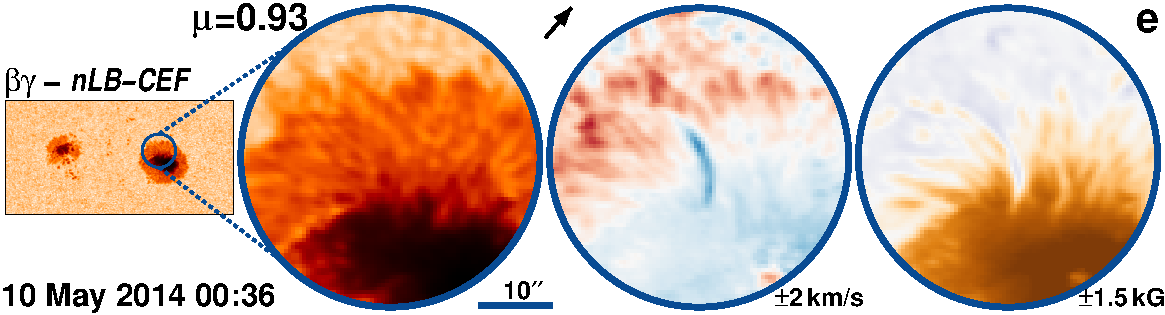}
 \caption{AR\,12055 followed for   10.2 days from  6-May-2014 starting at 01:48\,UT.\label{fig:DS66}}
 \end{figure*}

\begin{figure*}[htbp]
 \centering
 \includegraphics[width=.48\textwidth]{colorbars.pdf}
 \includegraphics[width=.48\textwidth]{colorbars.pdf}
 \includegraphics[width=.48\textwidth]{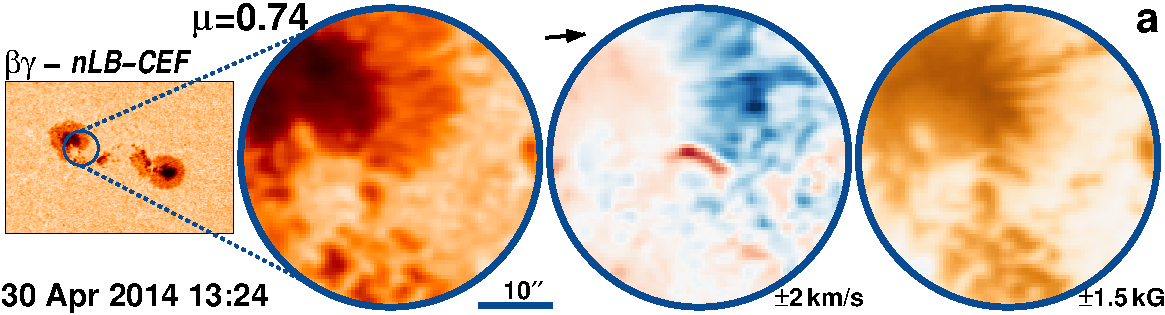}
 \includegraphics[width=.48\textwidth]{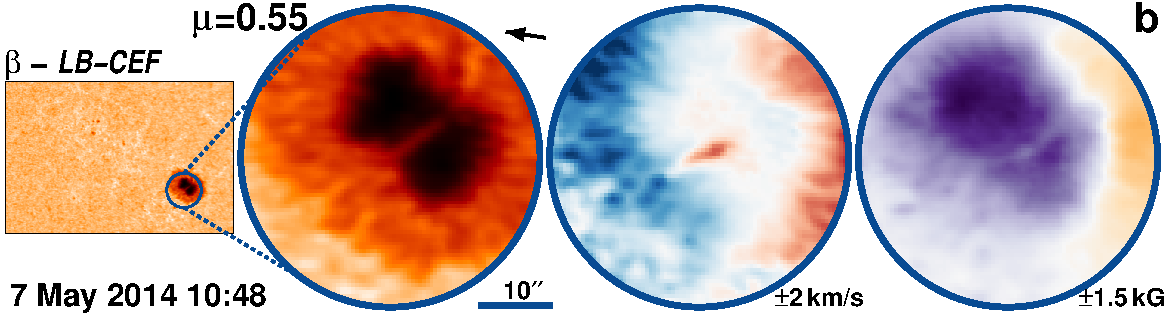}
 \caption{AR\,12049 followed for   10.9 days from 27-Apr-2014 starting at 22:12\,UT.\label{fig:DS67}}
 \end{figure*}

\begin{figure*}[htbp]
 \includegraphics[width=.48\textwidth]{colorbars.pdf}
 \includegraphics[width=.48\textwidth]{colorbars.pdf}
 \includegraphics[width=.48\textwidth]{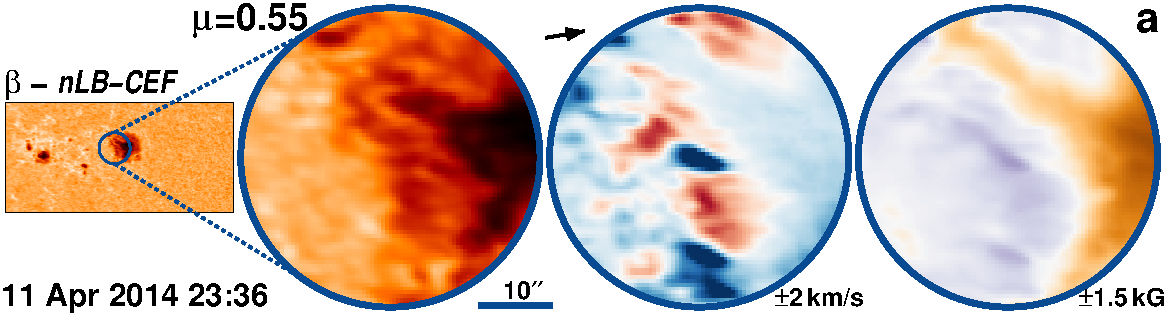}
 \includegraphics[width=.48\textwidth]{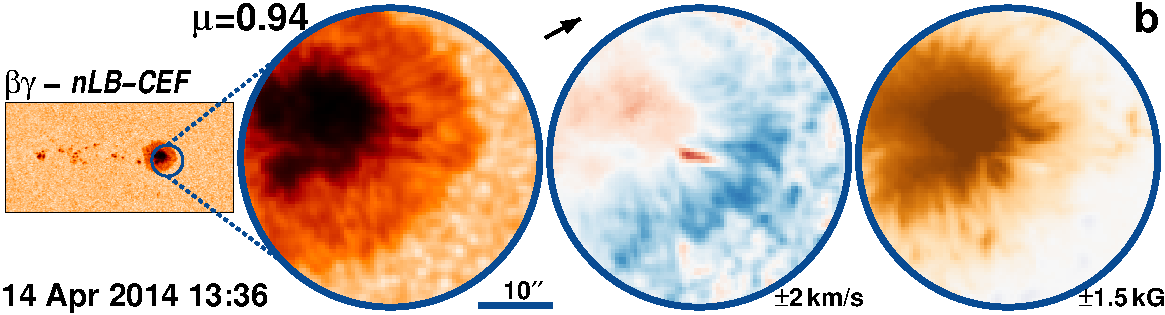}
 \includegraphics[width=.48\textwidth]{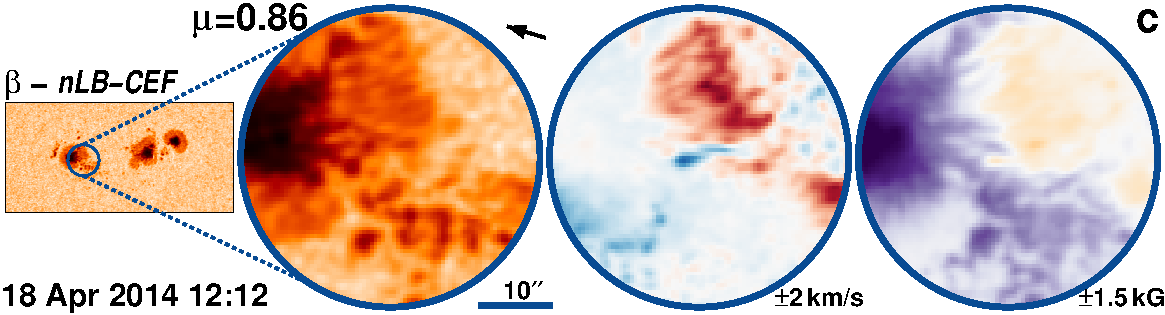}
 \caption{AR\,12034 followed for    9.6 days from 11-Apr-2014 starting at 09:00\,UT.\label{fig:DS68}}
 \end{figure*}

\begin{figure*}[htbp]
 \includegraphics[width=.48\textwidth]{colorbars.pdf}

 \includegraphics[width=.48\textwidth]{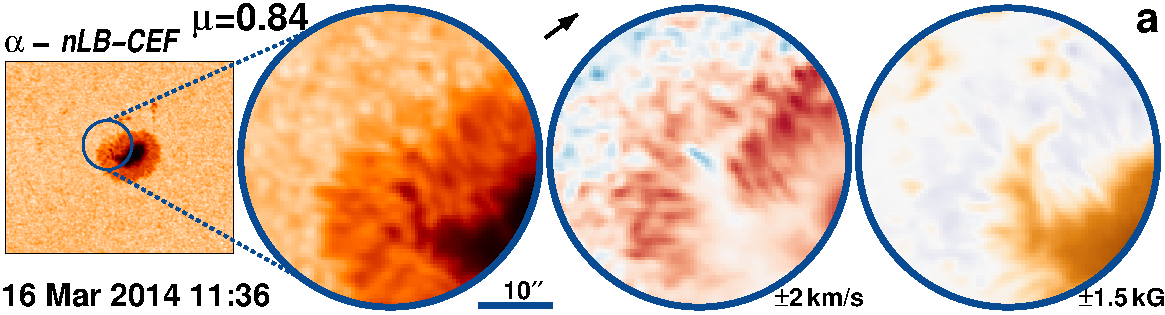}
 \caption{AR\,12005 followed for   10.5 days from 13-Mar-2014 starting at 02:00\,UT.\label{fig:DS70}}
 \end{figure*}

\begin{figure*}[htbp]
 \centering
 \includegraphics[width=.48\textwidth]{colorbars.pdf}
 \includegraphics[width=.48\textwidth]{colorbars.pdf}
 \includegraphics[width=.48\textwidth]{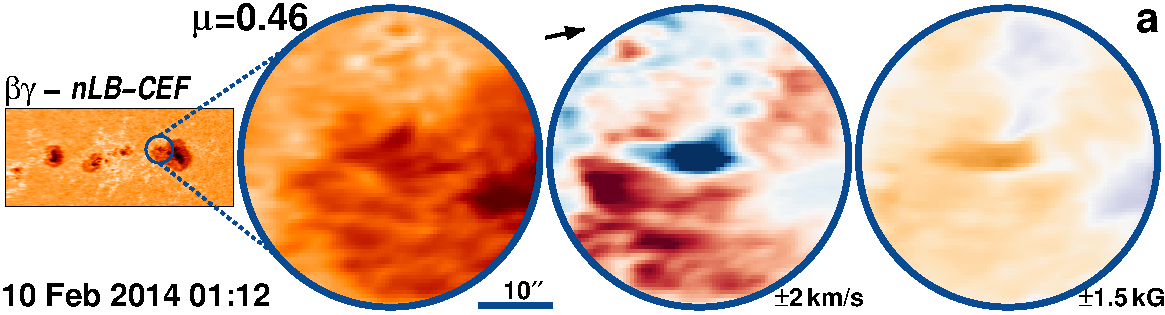}
 \includegraphics[width=.48\textwidth]{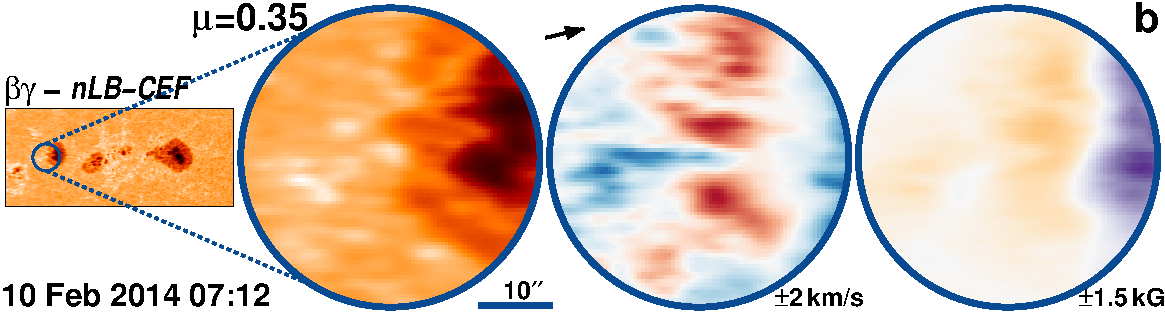}
 \caption{AR\,11976 followed for   11.4 days from  9-Feb-2014 starting at 03:12\,UT.\label{fig:DS72}}
 \end{figure*}

All the CEFs found in this study are presented in Figs.~\ref{fig:DS00}-\ref{fig:DS112}. Each Figure groups all the CEFs found per AR.  Figures are in chronological order starting from 2017 to 2011. The layout is the same as in Fig.~\ref{fig:cefs}. We remark that CEFs are displayed at the time closest to their maximum contrast with respect to their surroundings in the line-of-sight velocity maps. Hence, for few cases, the association with light bridges and the LB-CEFs is buried.

\clearpage

\begin{figure*}[htbp]
 \includegraphics[width=.48\textwidth]{colorbars.pdf}
 \includegraphics[width=.48\textwidth]{colorbars.pdf}
 \includegraphics[width=.48\textwidth]{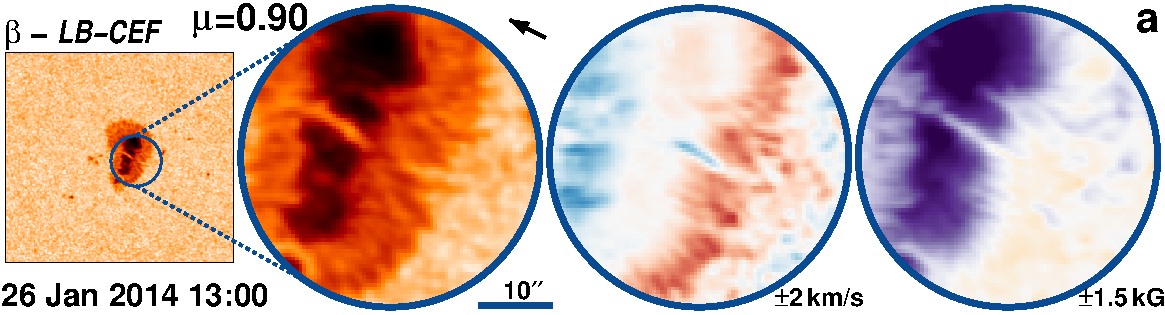}
 \includegraphics[width=.48\textwidth]{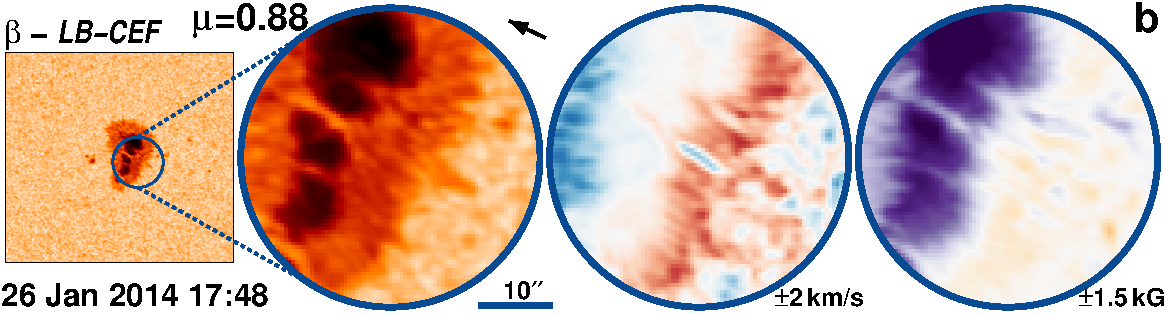}
 \includegraphics[width=.48\textwidth]{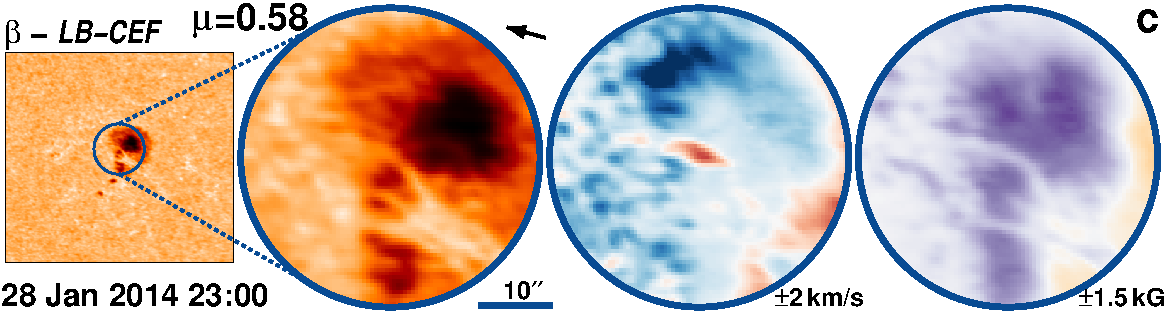}
 \caption{AR\,11960 followed for   11.4 days from 19-Jan-2014 starting at 03:00\,UT.\label{fig:DS74}}
 \end{figure*}

\begin{figure*}[htbp]
 \centering
 \includegraphics[width=.48\textwidth]{colorbars.pdf}
 \includegraphics[width=.48\textwidth]{colorbars.pdf}
 \includegraphics[width=.48\textwidth]{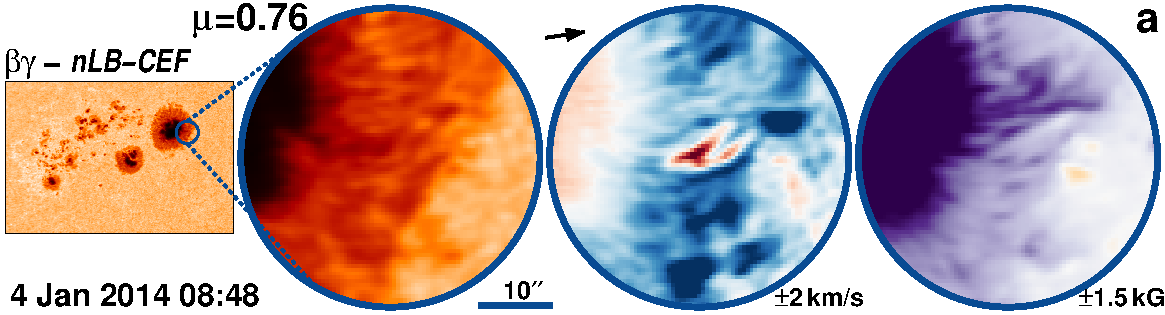}
 \includegraphics[width=.48\textwidth]{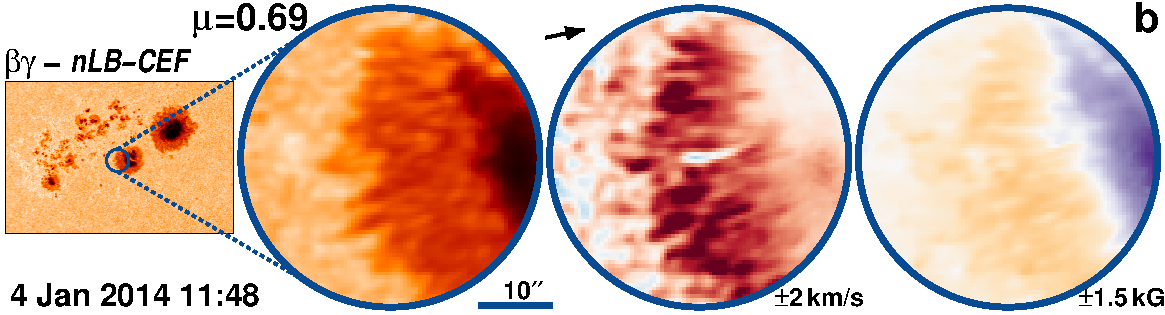}
 \includegraphics[width=.48\textwidth]{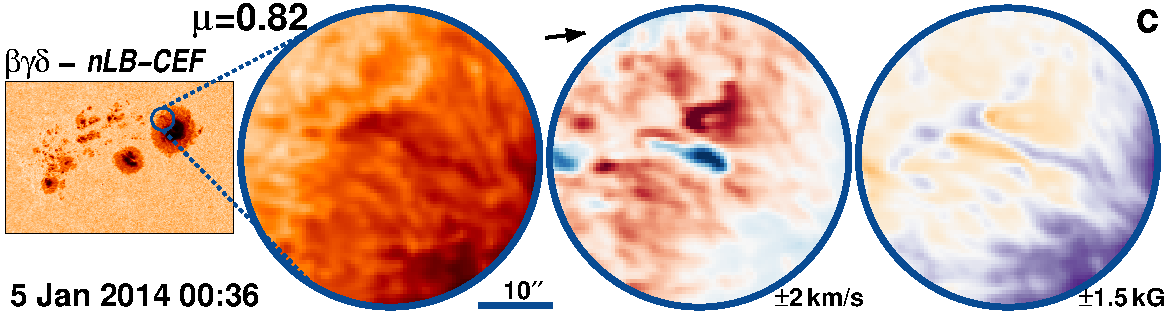}
 \includegraphics[width=.48\textwidth]{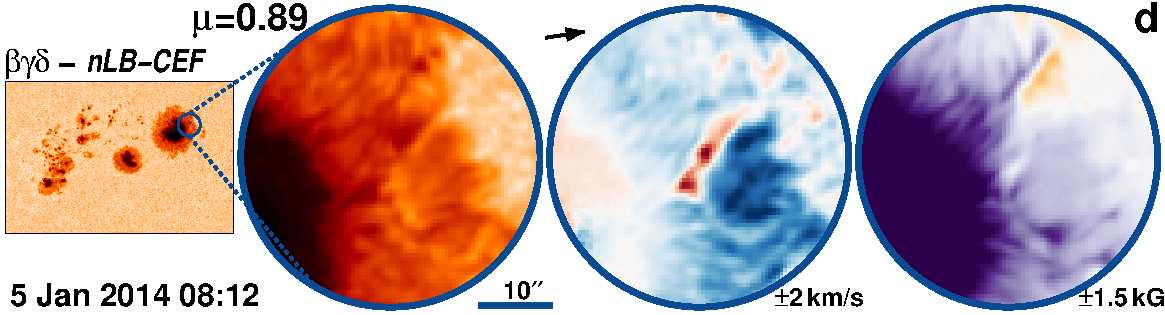}
 \includegraphics[width=.48\textwidth]{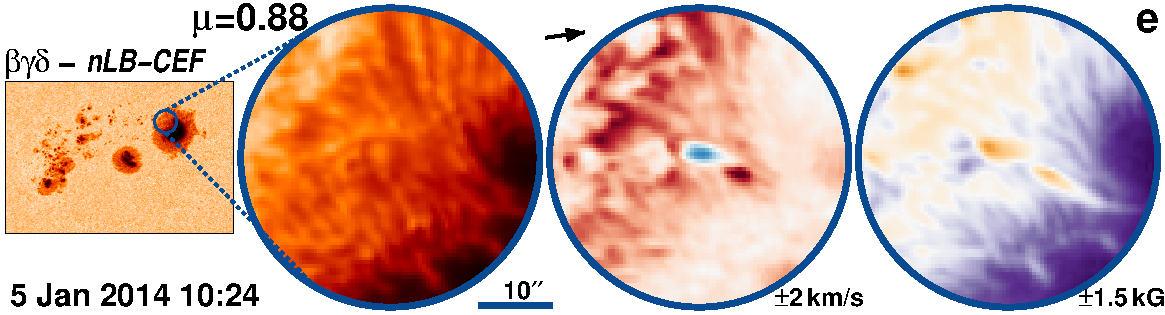}
 \includegraphics[width=.48\textwidth]{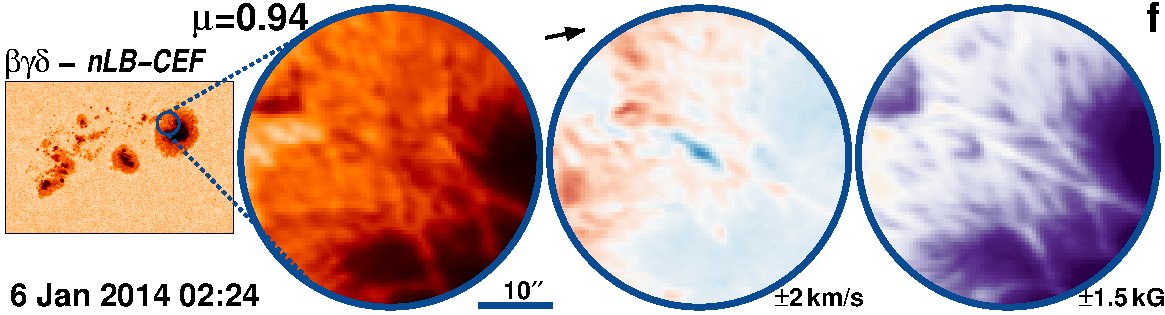}
 \includegraphics[width=.48\textwidth]{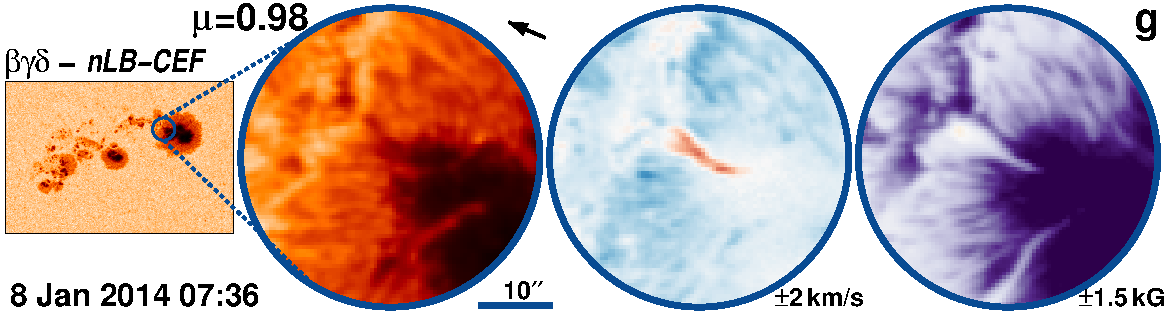}
 \includegraphics[width=.48\textwidth]{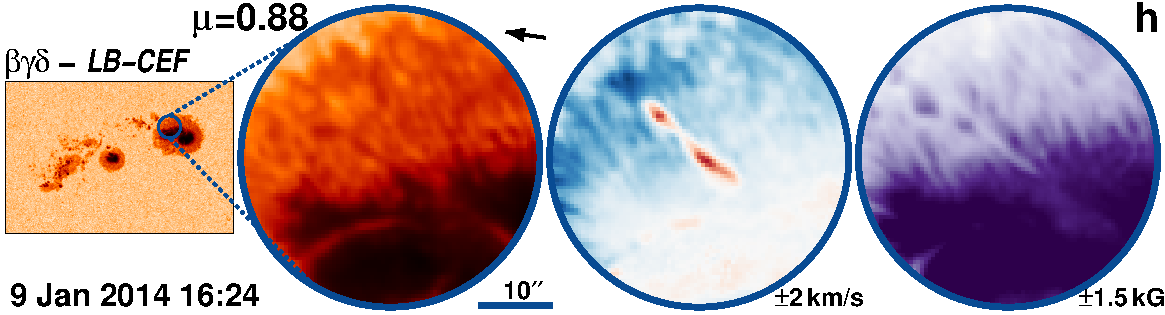}
 \includegraphics[width=.48\textwidth]{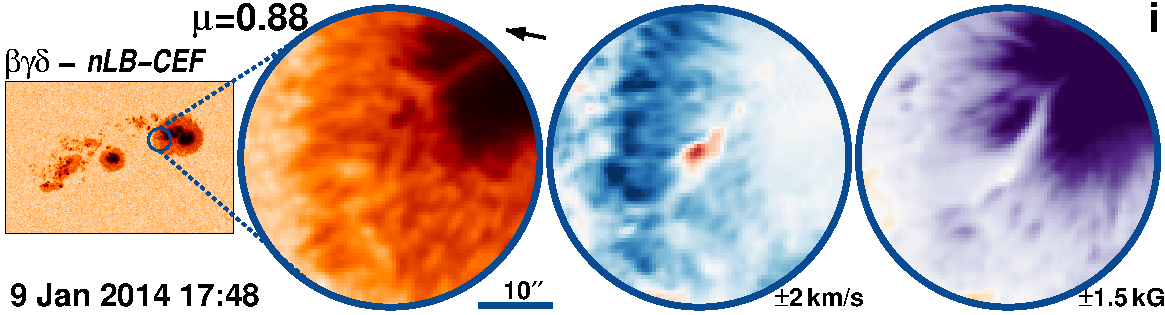}
 \includegraphics[width=.48\textwidth]{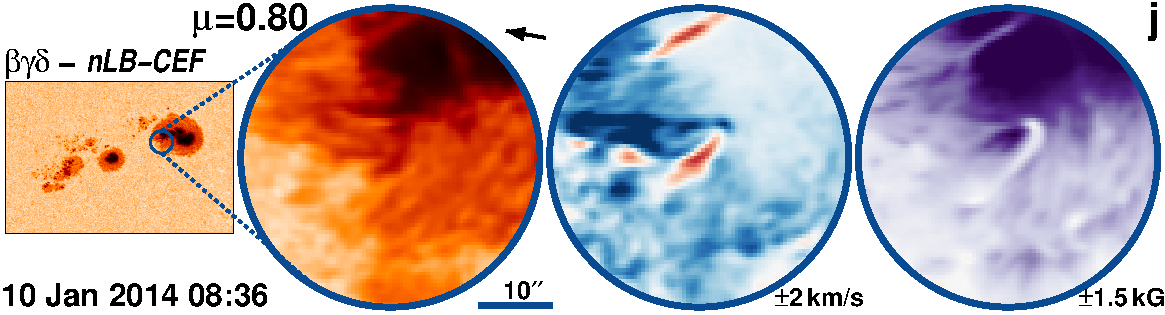}
 \includegraphics[width=.48\textwidth]{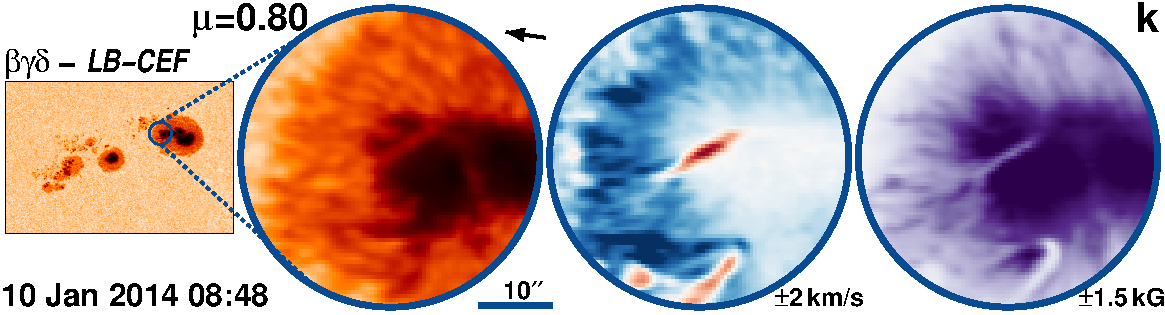}
 \includegraphics[width=.48\textwidth]{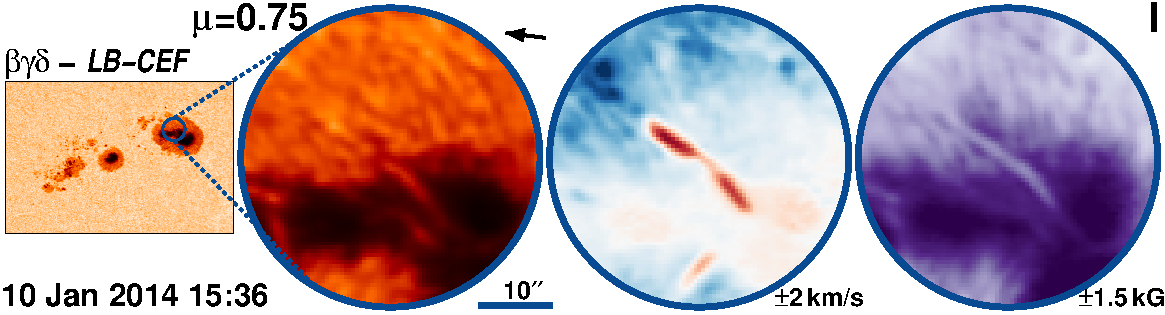}
 \caption{AR\,11944 followed for    8.3 days from  3-Jan-2014 starting at 22:12\,UT.\label{fig:DS75}}
 \end{figure*}

\begin{figure*}[htbp]
 \includegraphics[width=.48\textwidth]{colorbars.pdf}
 \includegraphics[width=.48\textwidth]{colorbars.pdf}
 \includegraphics[width=.48\textwidth]{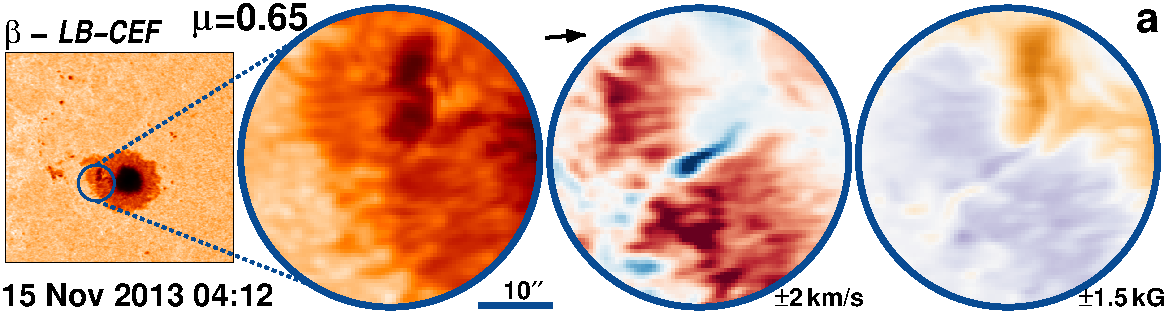}
 \includegraphics[width=.48\textwidth]{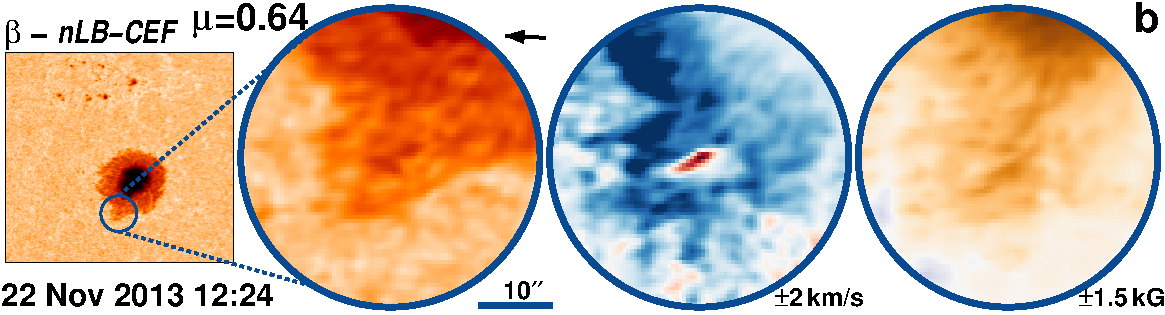}
 \includegraphics[width=.48\textwidth]{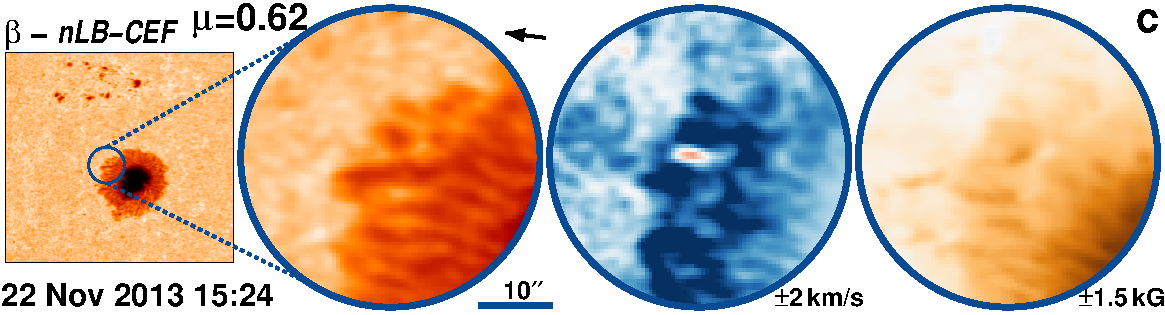}
 \caption{AR\,11899 followed for   10.8 days from 13-Nov-2013 starting at 07:48\,UT.\label{fig:DS78}}
 \end{figure*}

\begin{figure*}[htbp]
 \includegraphics[width=.48\textwidth]{colorbars.pdf}
 \includegraphics[width=.48\textwidth]{colorbars.pdf}
 \includegraphics[width=.48\textwidth]{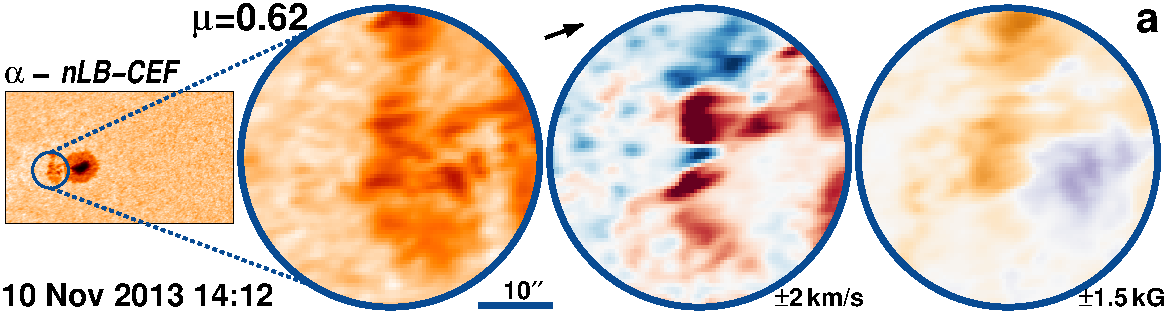}
 \includegraphics[width=.48\textwidth]{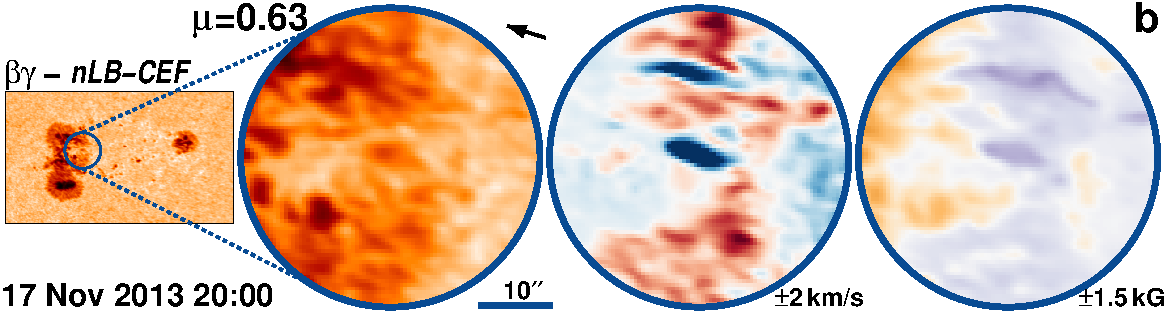}
 \includegraphics[width=.48\textwidth]{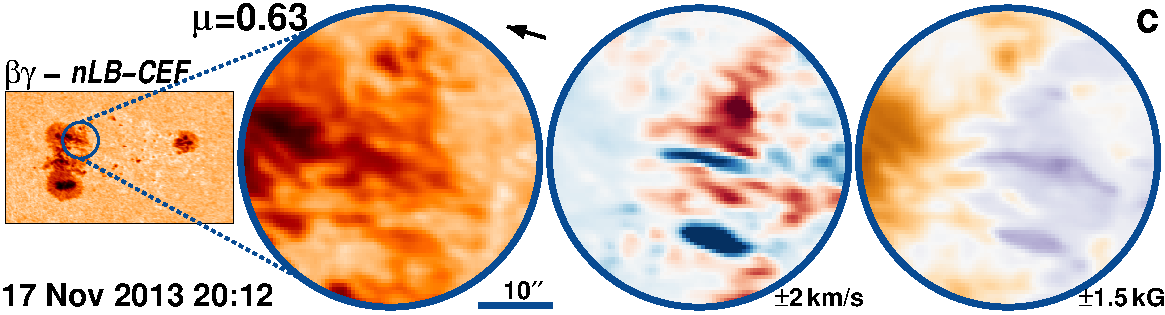}
 \caption{AR\,11893 followed for   10.5 days from  8-Nov-2013 starting at 12:36\,UT.\label{fig:DS79}}
 \end{figure*}

\begin{figure*}[htbp]
 \includegraphics[width=.48\textwidth]{colorbars.pdf}
 \includegraphics[width=.48\textwidth]{colorbars.pdf}
 \includegraphics[width=.48\textwidth]{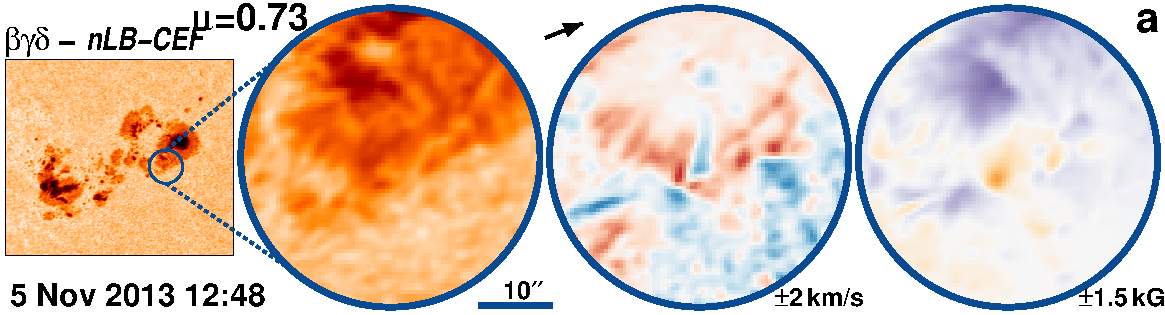}
 \includegraphics[width=.48\textwidth]{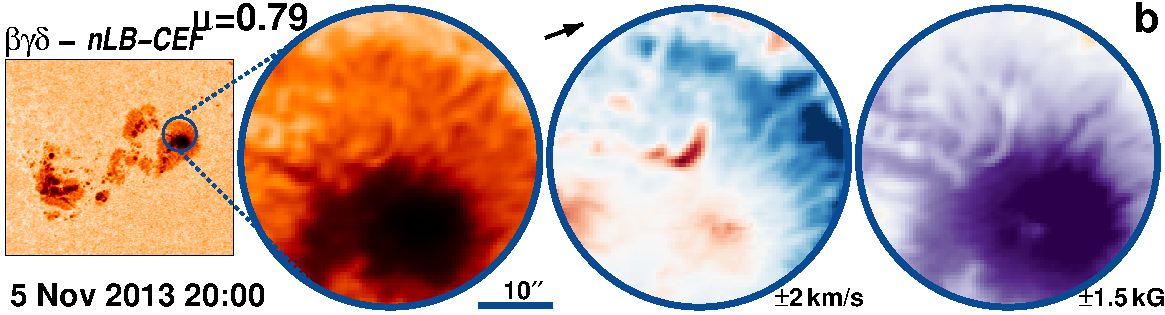}
 \includegraphics[width=.48\textwidth]{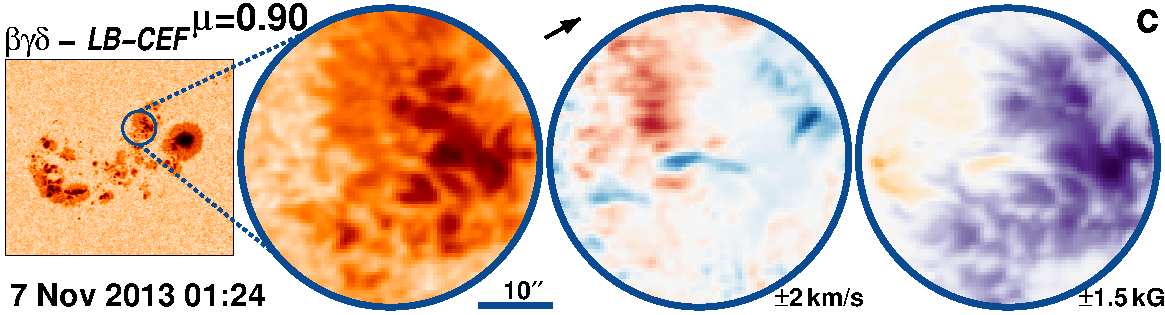}
 \includegraphics[width=.48\textwidth]{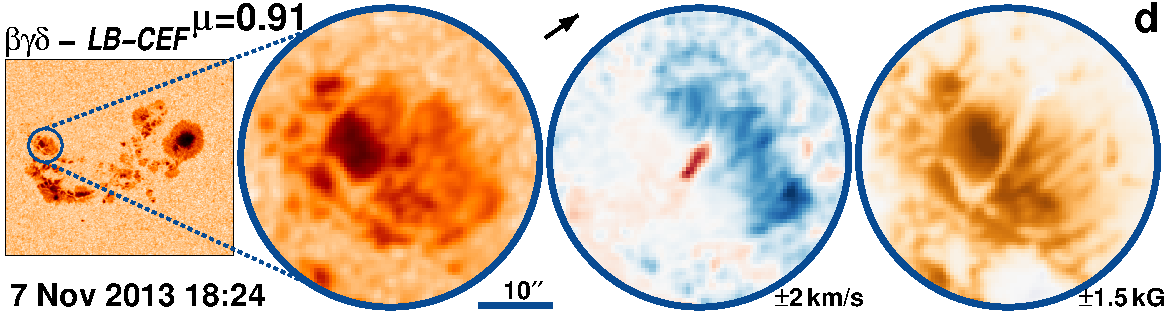}
 \includegraphics[width=.48\textwidth]{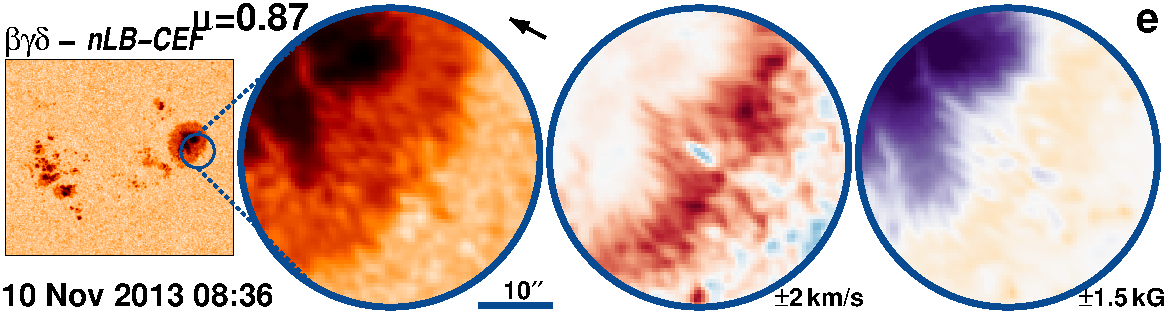}
 \caption{AR\,11890 followed for    6.5 days from  4-Nov-2013 starting at 12:00\,UT.\label{fig:DS80}}
 \end{figure*}

\begin{figure*}[htbp]
 \includegraphics[width=.48\textwidth]{colorbars.pdf}
 \includegraphics[width=.48\textwidth]{colorbars.pdf}
 \includegraphics[width=.48\textwidth]{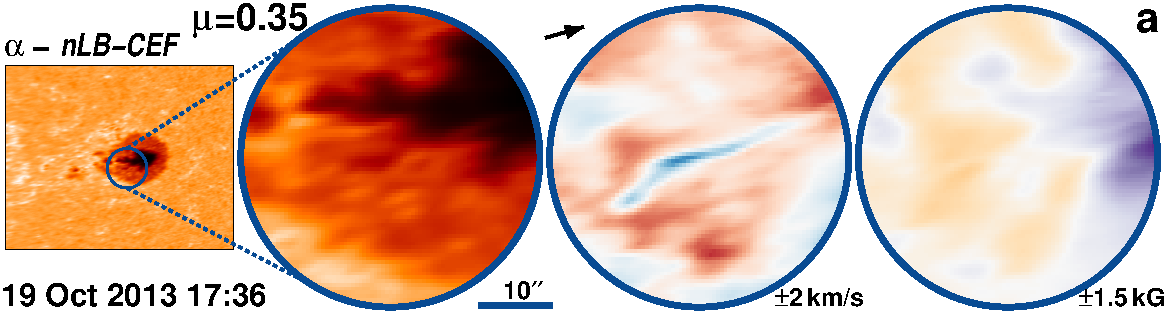}
 \includegraphics[width=.48\textwidth]{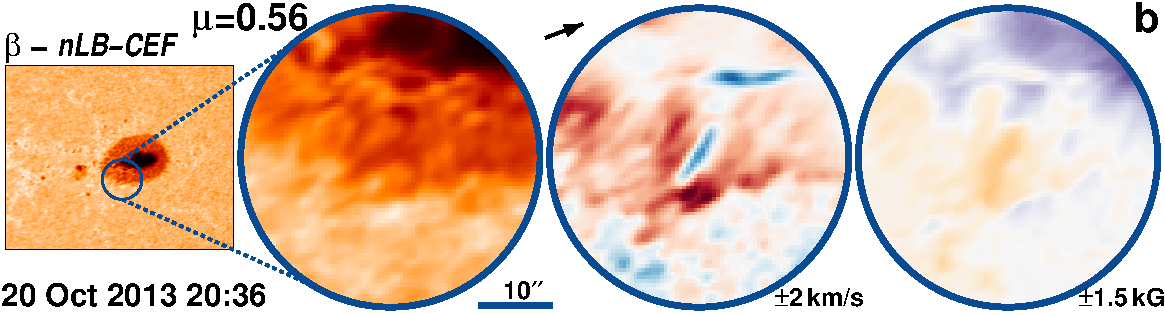}
 \includegraphics[width=.48\textwidth]{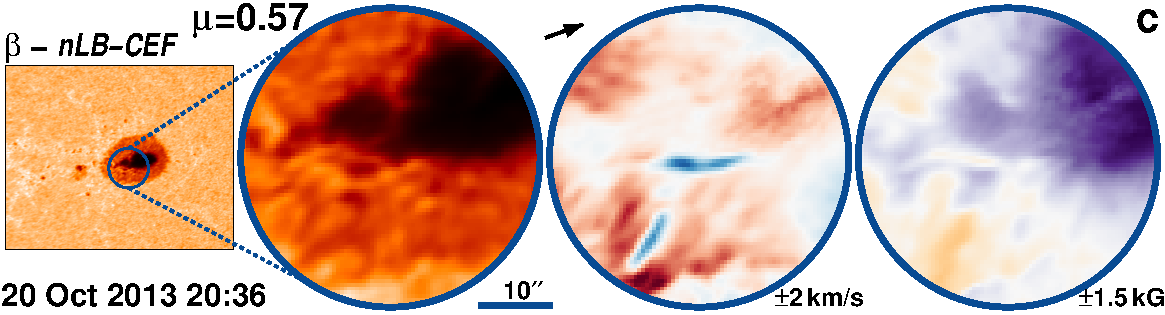}
 \includegraphics[width=.48\textwidth]{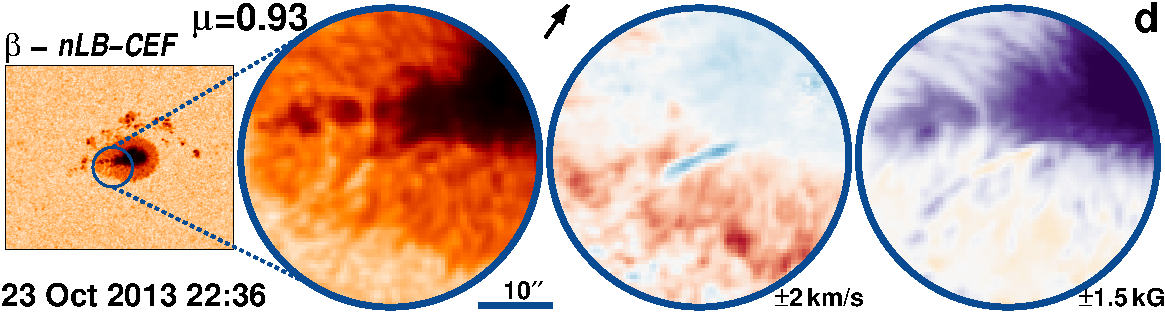}
 \includegraphics[width=.48\textwidth]{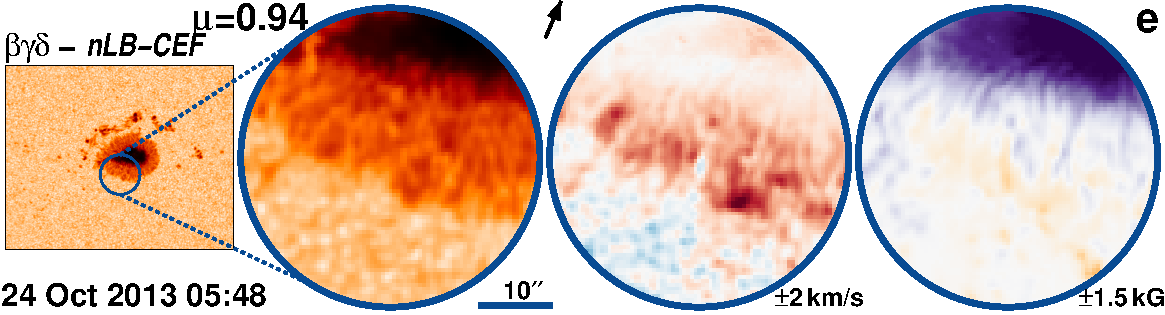}
 \caption{AR\,11877 followed for   10.5 days from 19-Oct-2013 starting at 15:00\,UT.\label{fig:DS81}}
 \end{figure*}

\begin{figure*}[htbp]
 \centering
 \includegraphics[width=.48\textwidth]{colorbars.pdf}
 \includegraphics[width=.48\textwidth]{colorbars.pdf}
 \includegraphics[width=.48\textwidth]{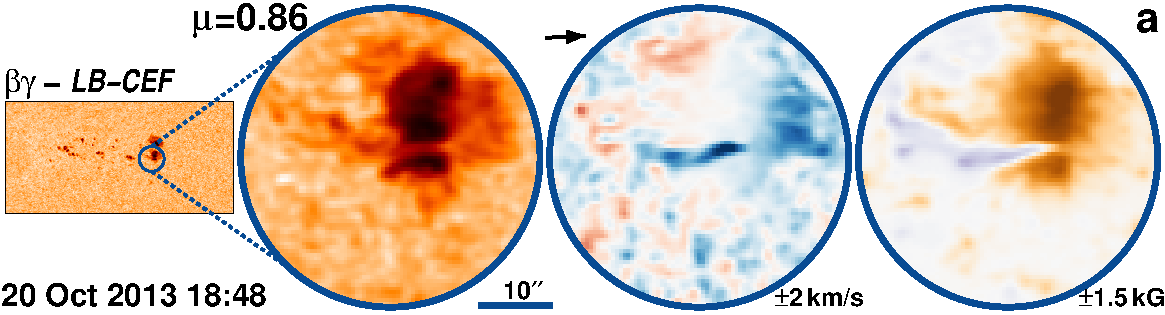}
 \includegraphics[width=.48\textwidth]{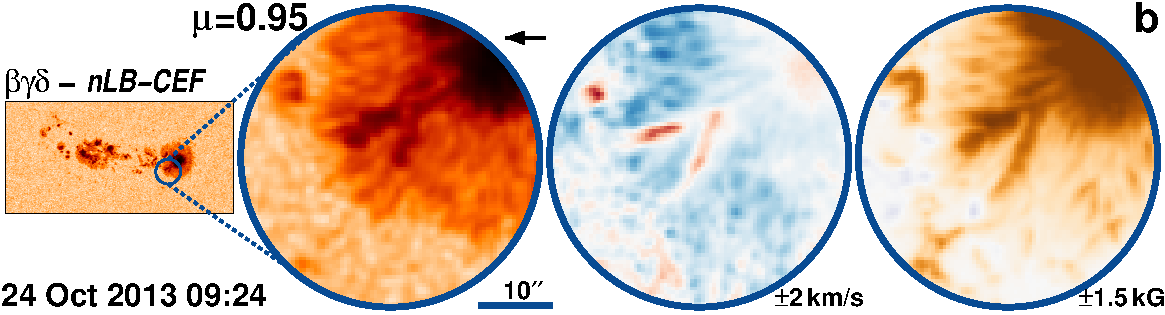}
 \includegraphics[width=.48\textwidth]{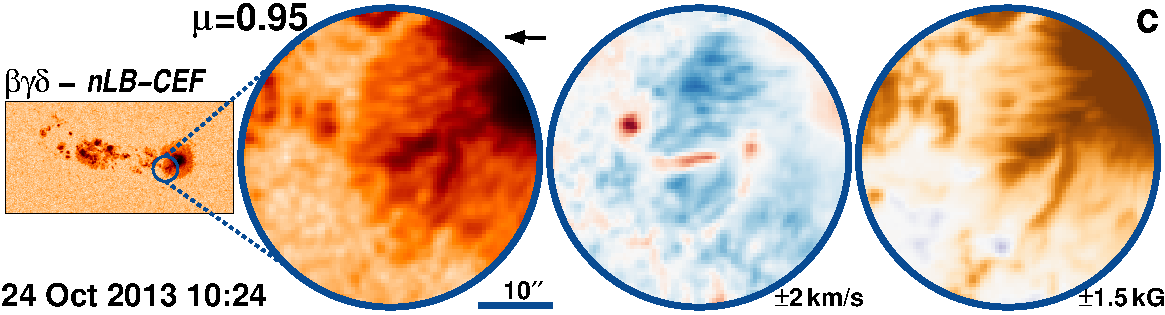}
 \includegraphics[width=.48\textwidth]{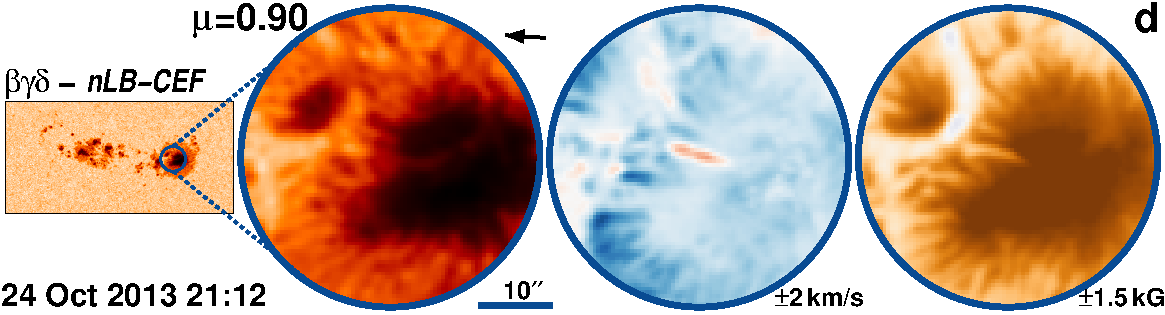}
 \caption{AR\,11875 followed for   10.7 days from 18-Oct-2013 starting at 00:00\,UT.\label{fig:DS82}}
 \end{figure*}

\begin{figure*}[htbp]
 \includegraphics[width=.48\textwidth]{colorbars.pdf}
 \includegraphics[width=.48\textwidth]{colorbars.pdf}
 \includegraphics[width=.48\textwidth]{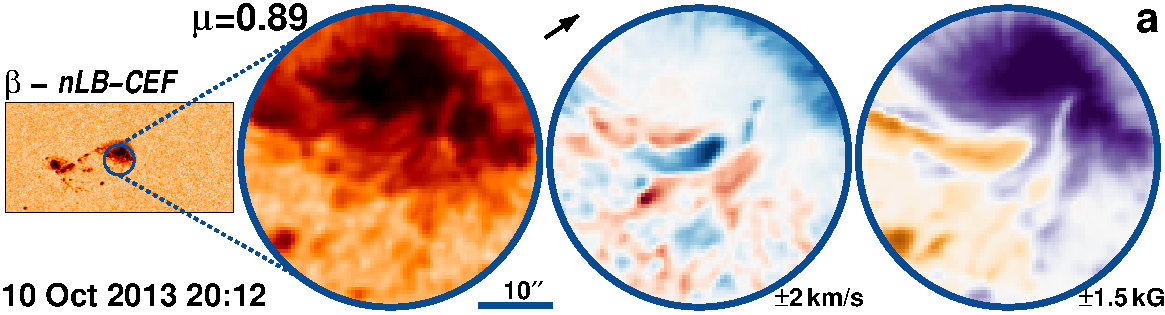}
 \includegraphics[width=.48\textwidth]{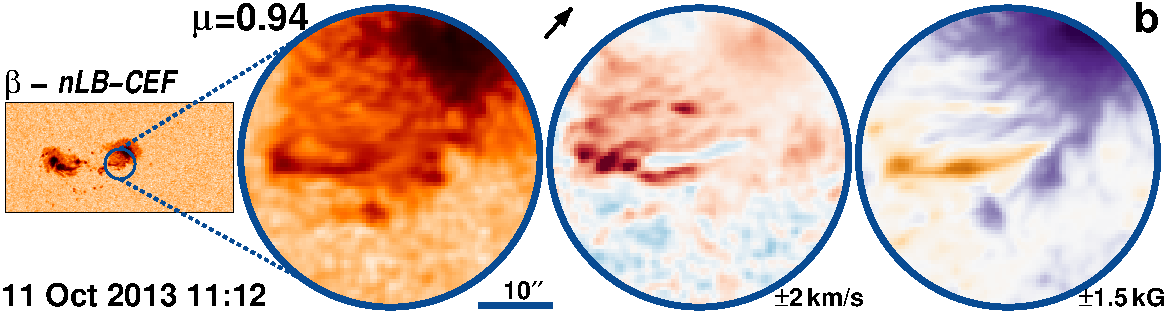}
 \includegraphics[width=.48\textwidth]{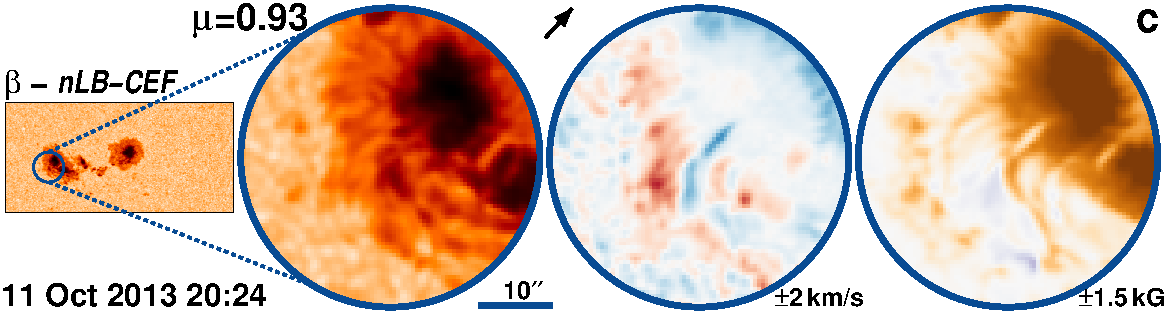}
 \includegraphics[width=.48\textwidth]{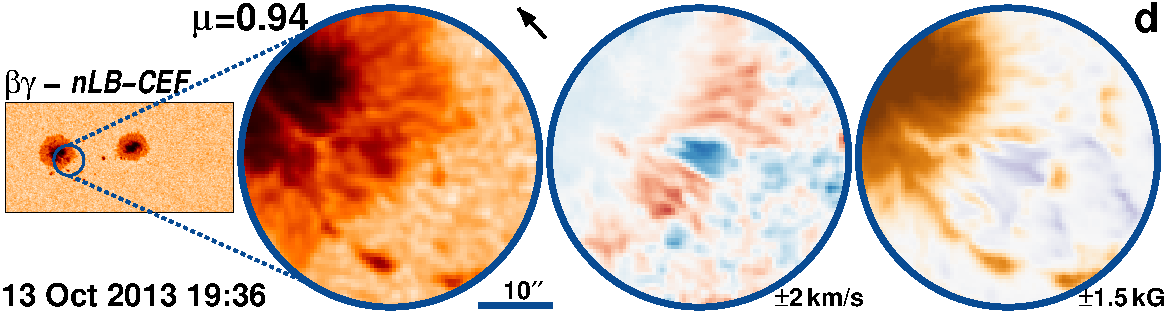}
 \includegraphics[width=.48\textwidth]{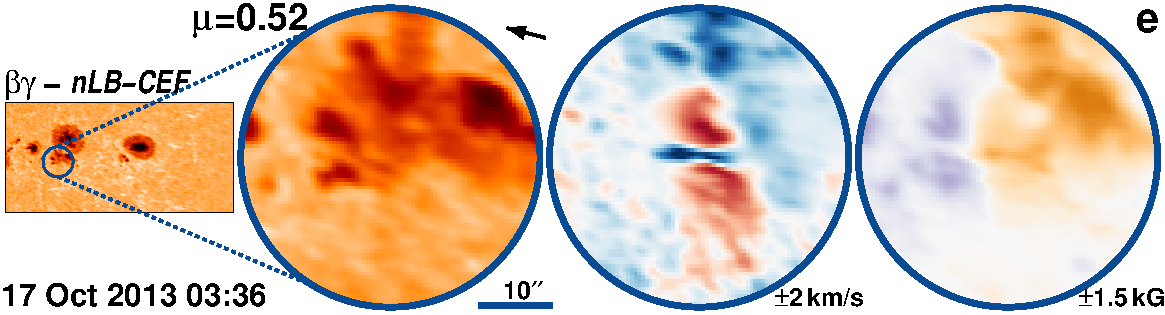}
 \caption{AR\,11861 followed for    8.1 days from  9-Oct-2013 starting at 22:48\,UT.\label{fig:DS83}}
 \end{figure*}

\begin{figure*}[htbp]
 \includegraphics[width=.48\textwidth]{colorbars.pdf}
 \includegraphics[width=.48\textwidth]{colorbars.pdf}
 \includegraphics[width=.48\textwidth]{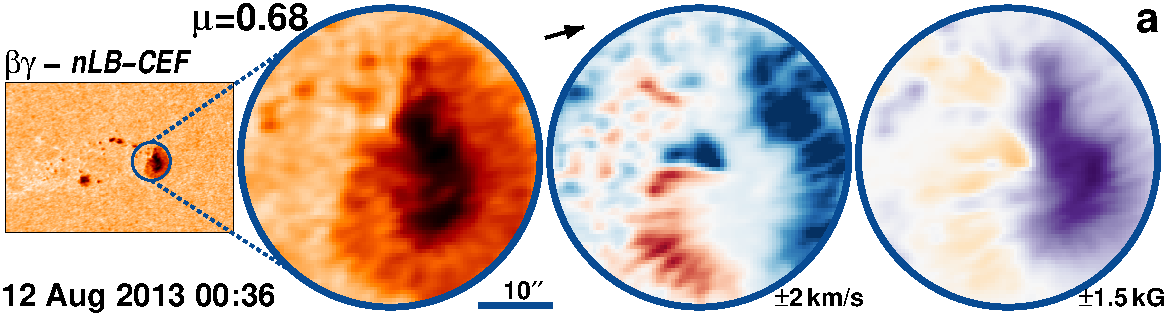}
 \includegraphics[width=.48\textwidth]{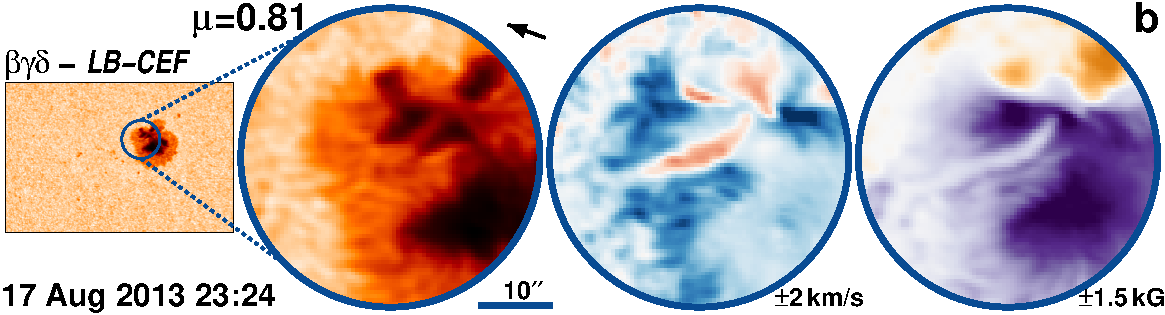}
 \includegraphics[width=.48\textwidth]{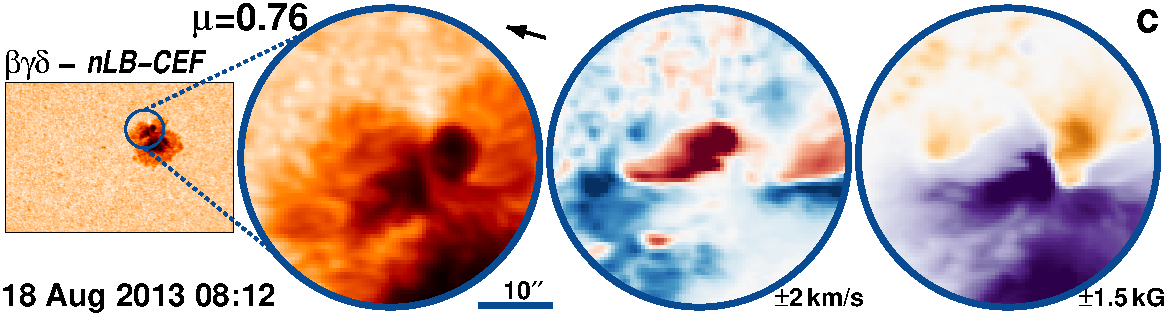}
 \caption{AR\,11818 followed for    9.2 days from 11-Aug-2013 starting at 10:00\,UT.\label{fig:DS84}}
 \end{figure*}

\begin{figure*}[htbp]
 \centering
 \includegraphics[width=.48\textwidth]{colorbars.pdf}
 \includegraphics[width=.48\textwidth]{colorbars.pdf}
 \includegraphics[width=.48\textwidth]{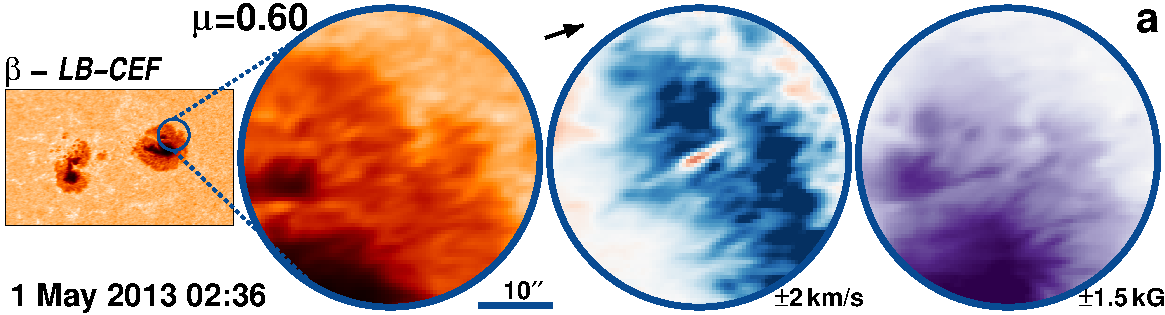}
 \includegraphics[width=.48\textwidth]{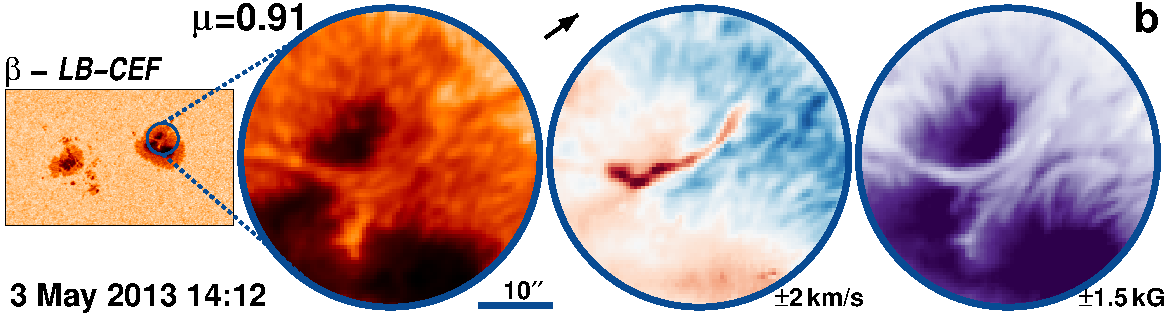}
 \includegraphics[width=.48\textwidth]{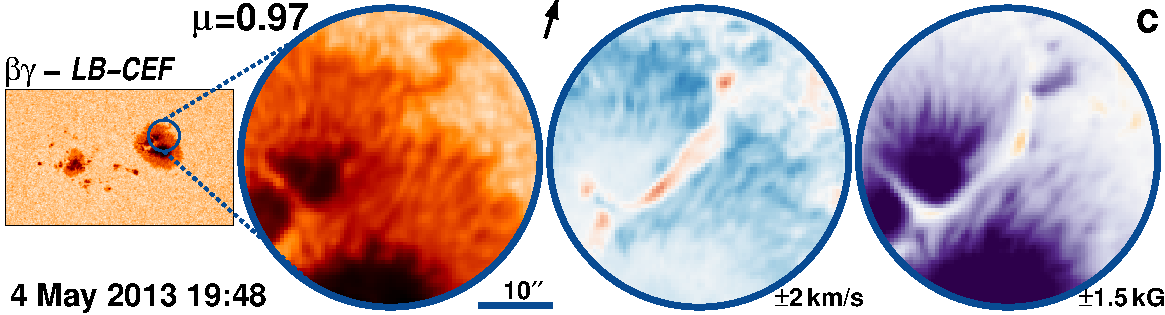}
 \includegraphics[width=.48\textwidth]{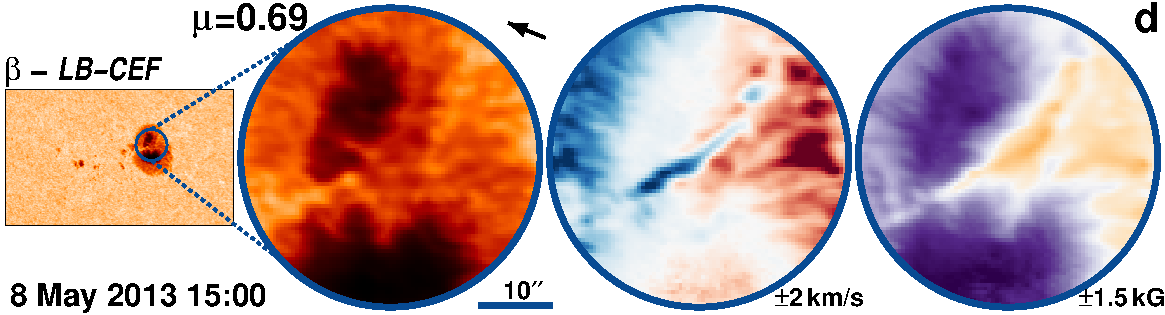}
 \includegraphics[width=.48\textwidth]{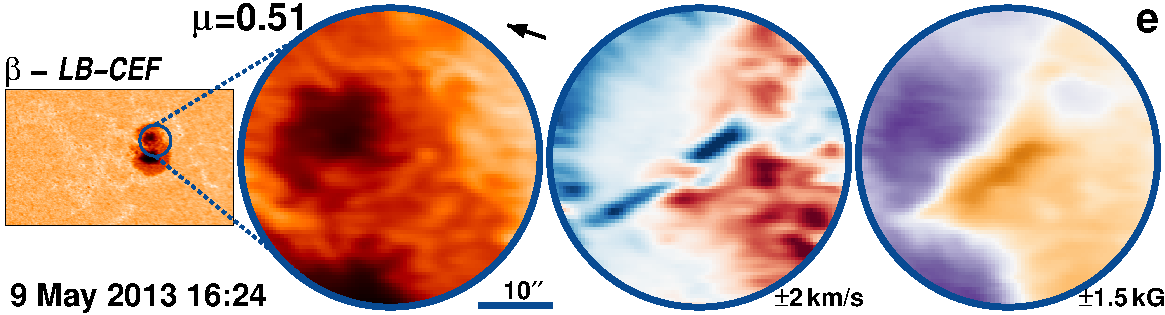}
 \includegraphics[width=.48\textwidth]{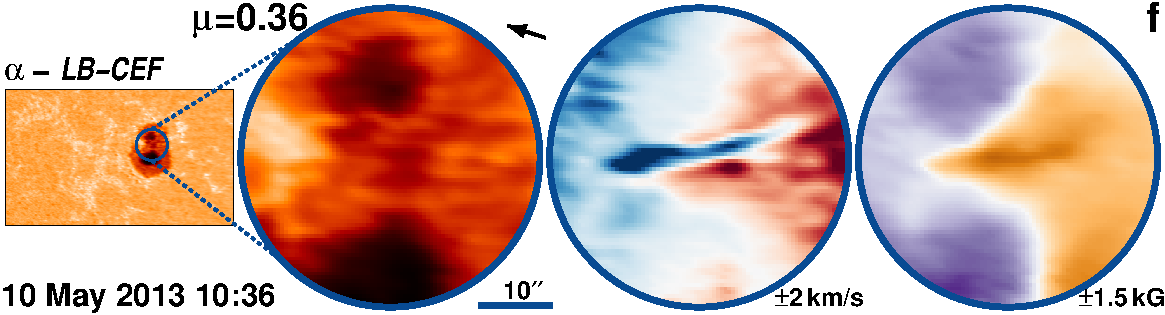}
 \caption{AR\,11734 followed for   10.0 days from  1-May-2013 starting at 01:00\,UT.\label{fig:DS85}}
 \end{figure*}

\begin{figure*}[htbp]
 \includegraphics[width=.48\textwidth]{colorbars.pdf}
 \includegraphics[width=.48\textwidth]{colorbars.pdf}
 \includegraphics[width=.48\textwidth]{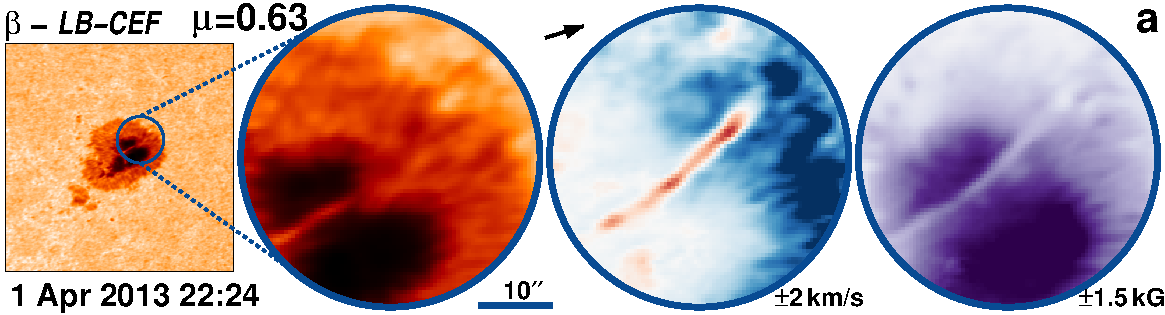}
 \includegraphics[width=.48\textwidth]{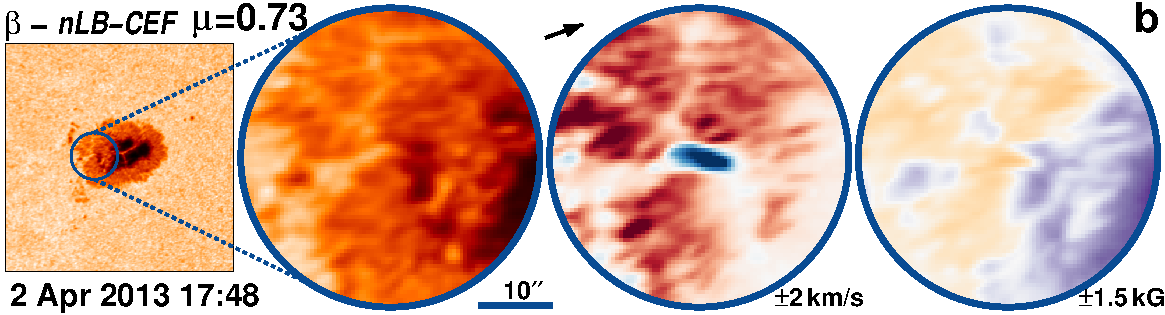}
 \includegraphics[width=.48\textwidth]{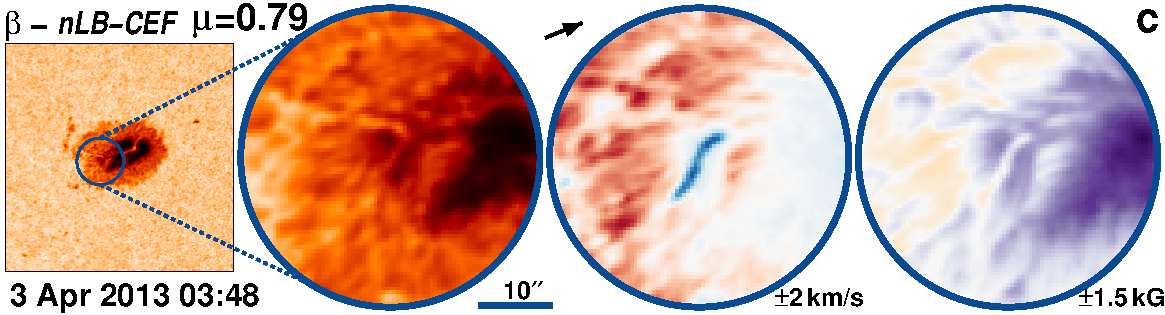}
 \caption{AR\,11711 followed for   10.7 days from 31-Mar-2013 starting at 01:48\,UT.\label{fig:DS87}}
 \end{figure*}

\begin{figure*}[htbp]
 \includegraphics[width=.48\textwidth]{colorbars.pdf}
 \includegraphics[width=.48\textwidth]{colorbars.pdf}
 \includegraphics[width=.48\textwidth]{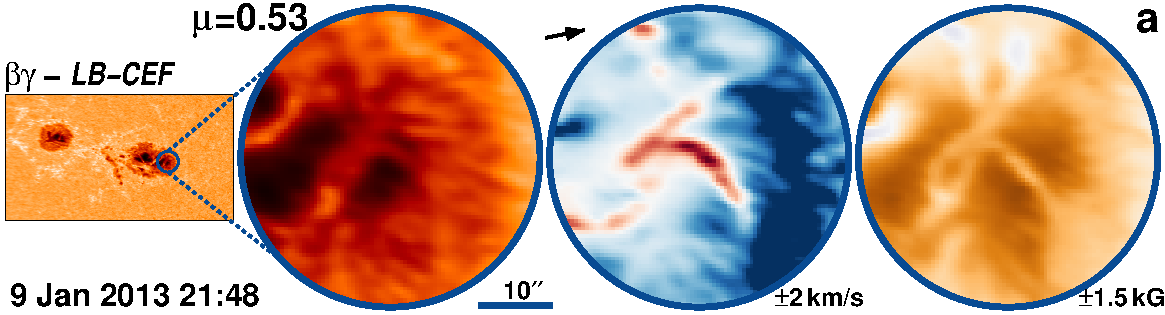}
 \includegraphics[width=.48\textwidth]{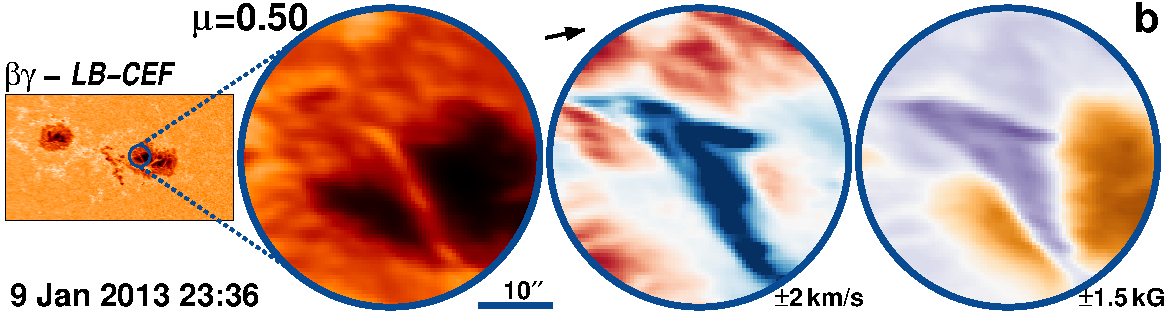}
 \includegraphics[width=.48\textwidth]{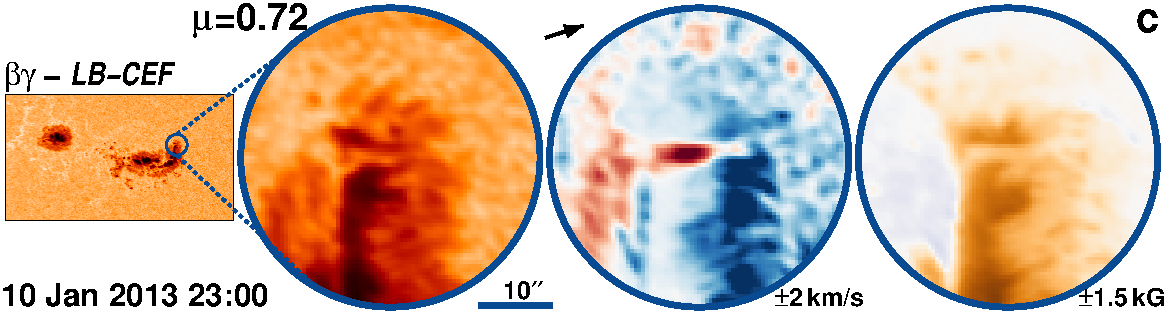}
 \includegraphics[width=.48\textwidth]{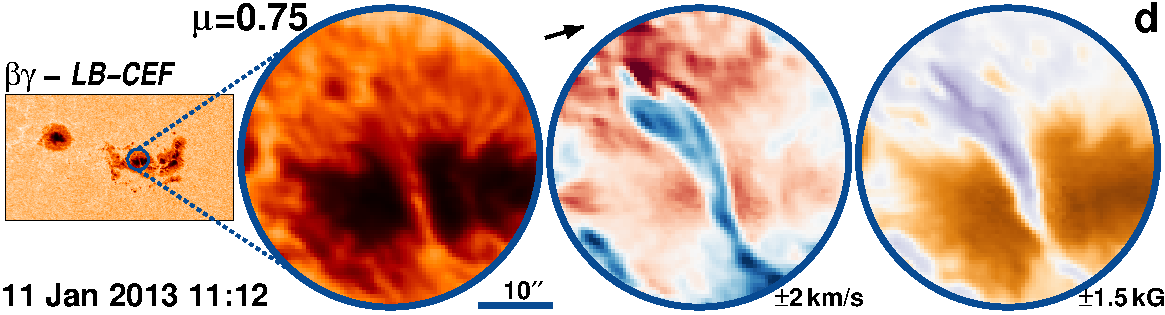}
 \includegraphics[width=.48\textwidth]{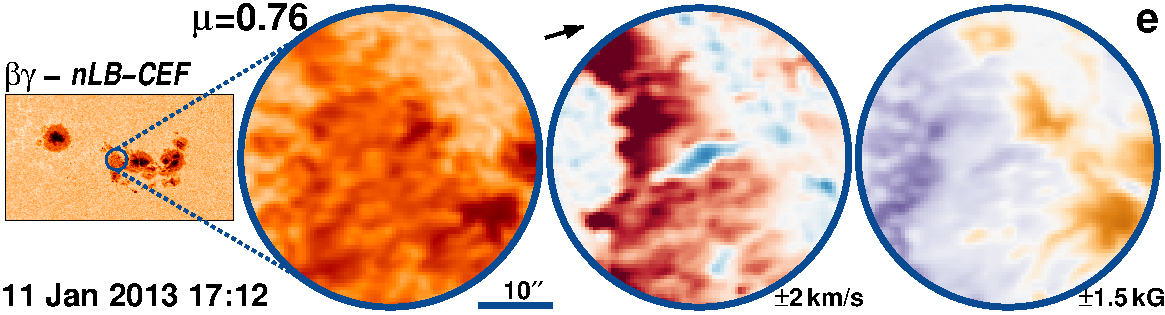}
 \includegraphics[width=.48\textwidth]{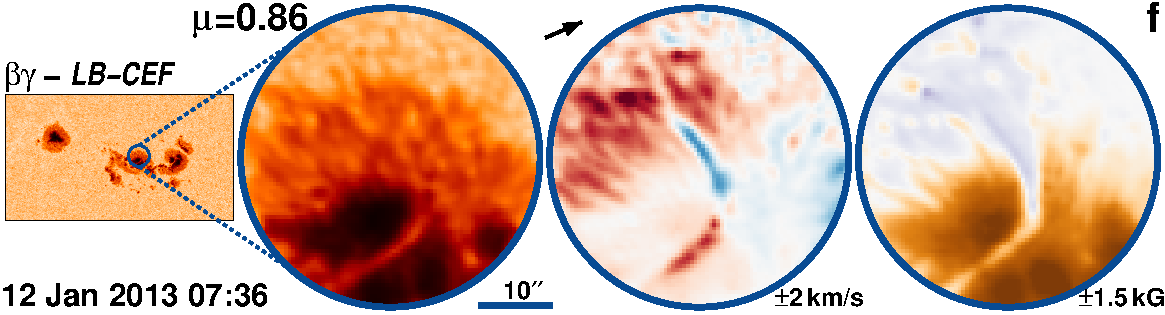}
 \includegraphics[width=.48\textwidth]{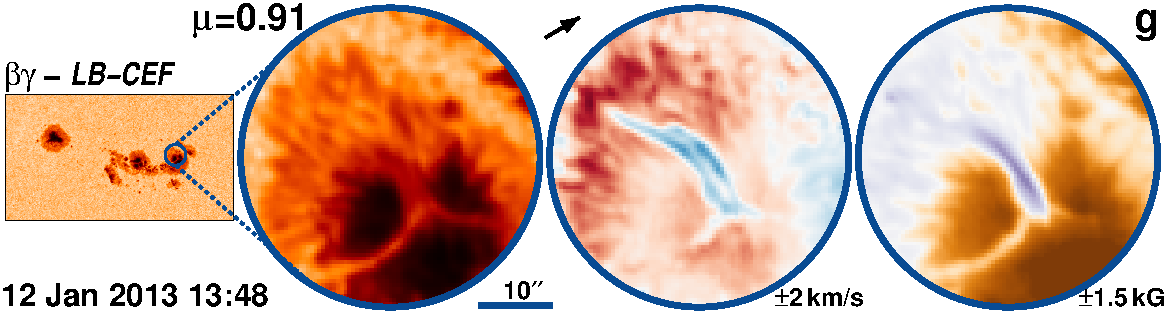}
 \caption{AR\,11654 followed for   10.0 days from  8-Jan-2013 starting at 15:00\,UT.\label{fig:DS90}}
 \end{figure*}

\begin{figure*}[htbp]
 \centering
 \includegraphics[width=.48\textwidth]{colorbars.pdf}
 \includegraphics[width=.48\textwidth]{colorbars.pdf}
 \includegraphics[width=.48\textwidth]{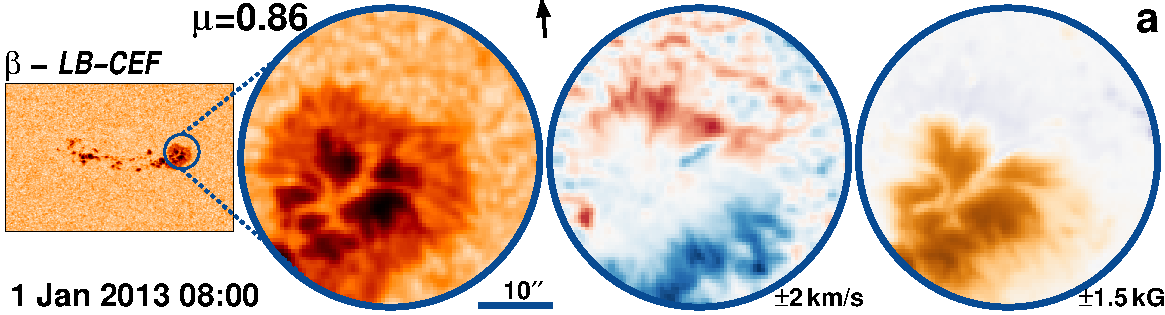}
 \includegraphics[width=.48\textwidth]{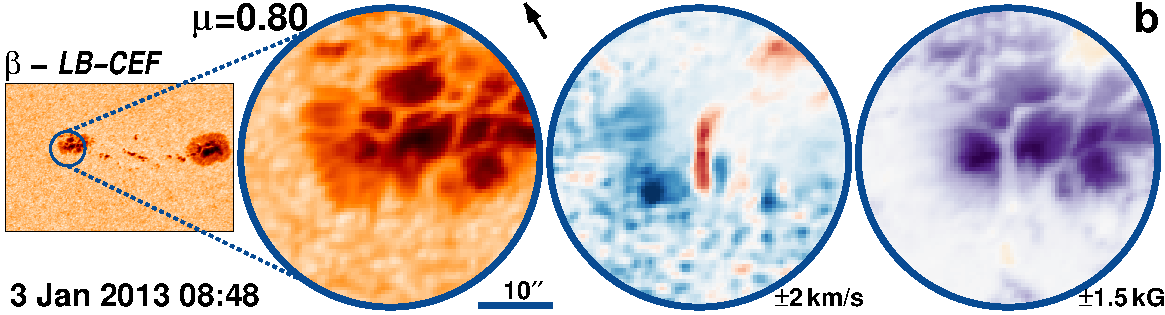}
 \includegraphics[width=.48\textwidth]{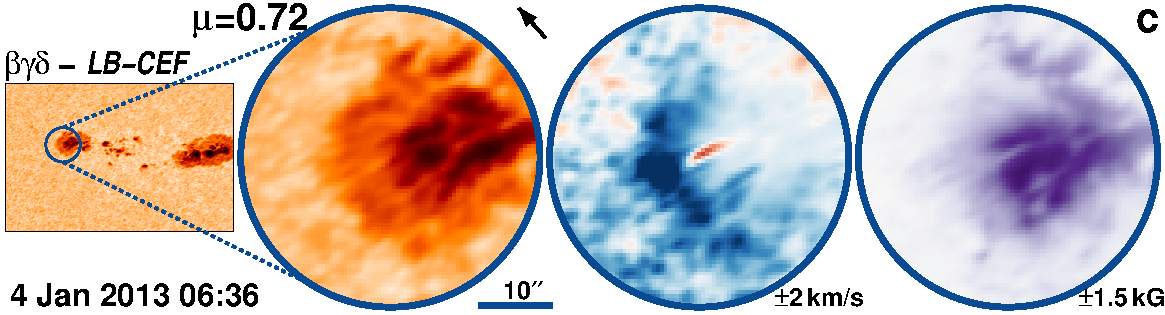}
 \includegraphics[width=.48\textwidth]{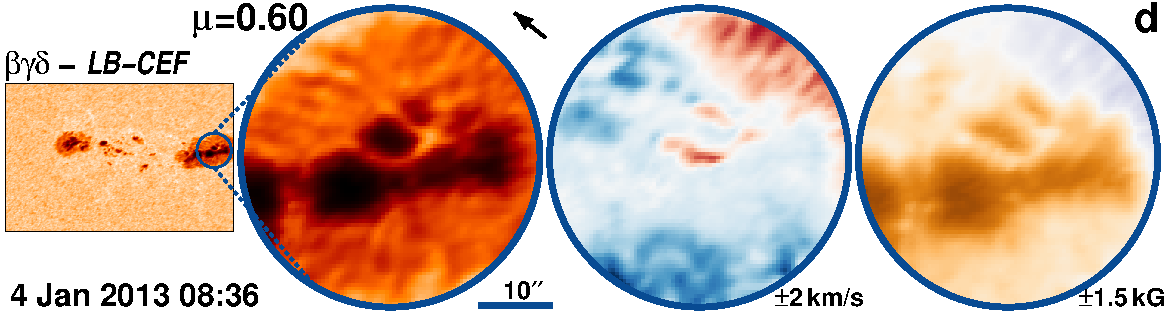}
 \caption{AR\,11640 followed for    7.7 days from 30-Dec-2012 starting at 02:00\,UT.\label{fig:DS91}}
 \end{figure*}

\begin{figure*}[htbp]
 \centering
 \includegraphics[width=.48\textwidth]{colorbars.pdf}
 \includegraphics[width=.48\textwidth]{colorbars.pdf}
 \includegraphics[width=.48\textwidth]{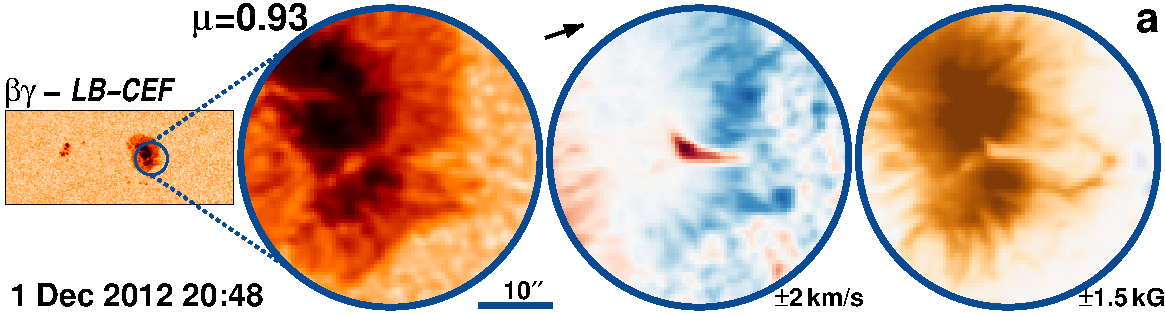}
 \includegraphics[width=.48\textwidth]{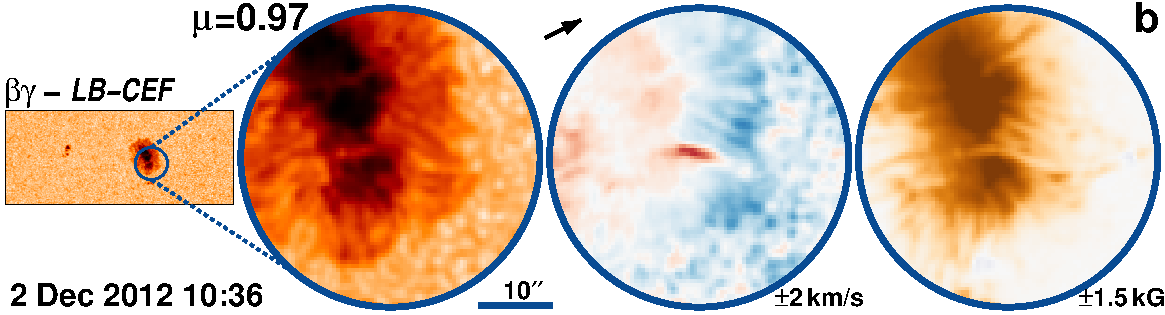}
 \includegraphics[width=.48\textwidth]{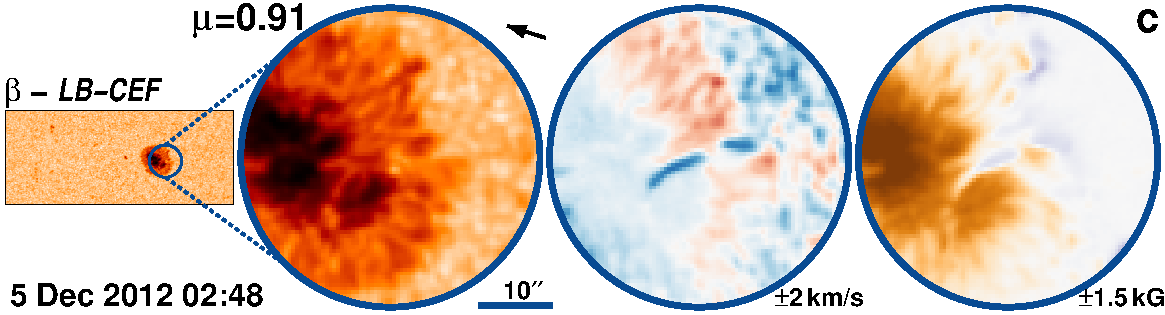}
 \includegraphics[width=.48\textwidth]{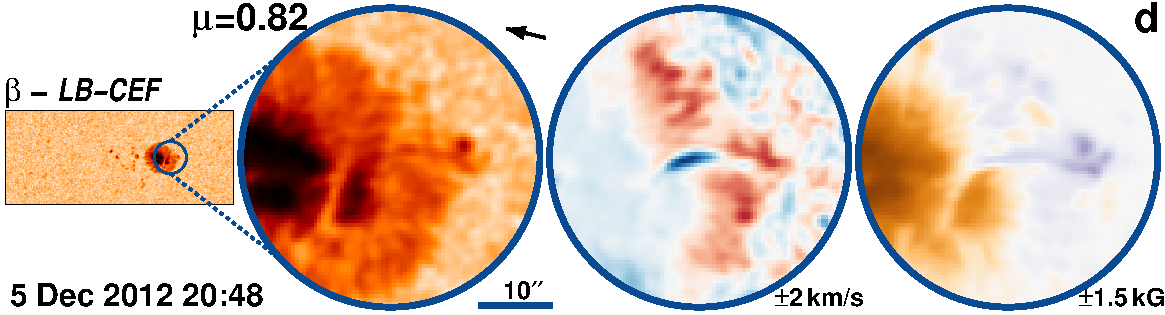}
 \caption{AR\,11623 followed for   11.1 days from 28-Nov-2012 starting at 00:12\,UT.\label{fig:DS92}}
 \end{figure*}

\clearpage

\begin{figure*}[htbp]
 \includegraphics[width=.48\textwidth]{colorbars.pdf}
 \includegraphics[width=.48\textwidth]{colorbars.pdf}
 \includegraphics[width=.48\textwidth]{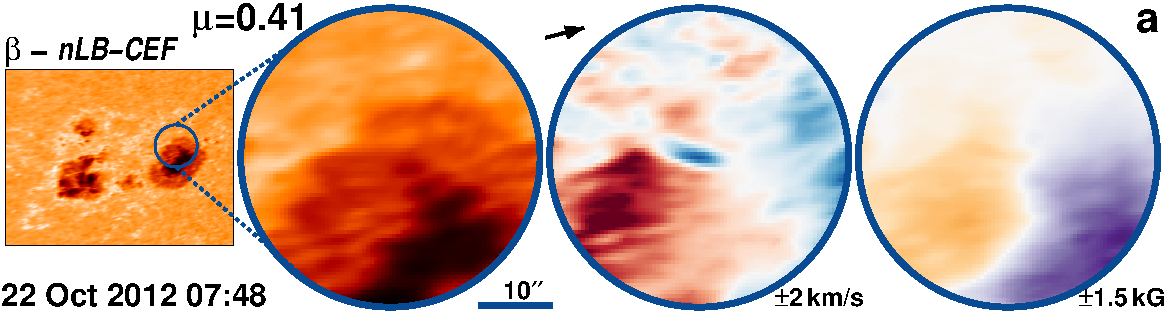}
 \includegraphics[width=.48\textwidth]{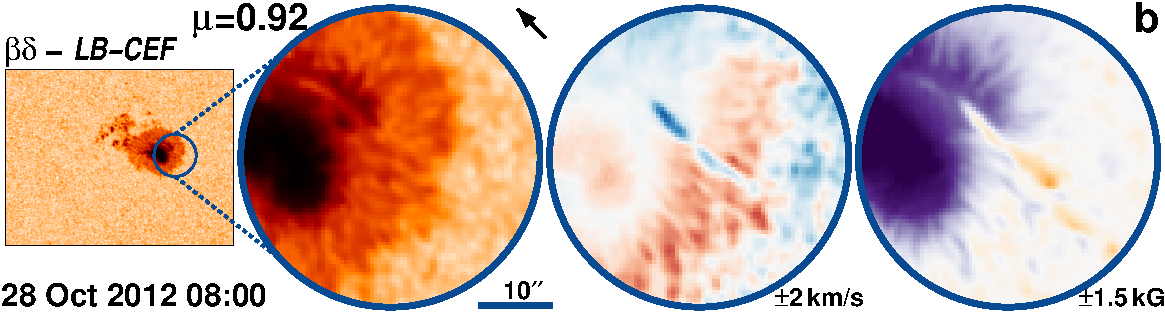}
 \includegraphics[width=.48\textwidth]{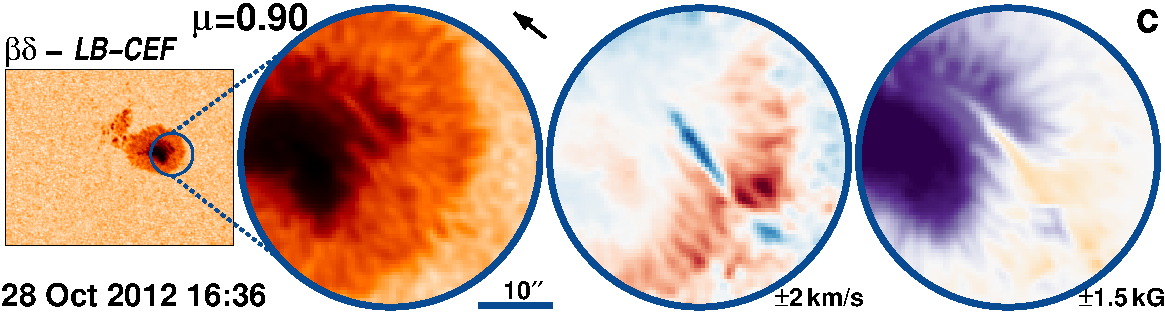}
 \caption{AR\,11598 followed for   11.1 days from 22-Oct-2012 starting at 00:00\,UT.\label{fig:DS93}}
 \end{figure*}

\begin{figure*}[htbp]
 \includegraphics[width=.48\textwidth]{colorbars.pdf}

 \includegraphics[width=.48\textwidth]{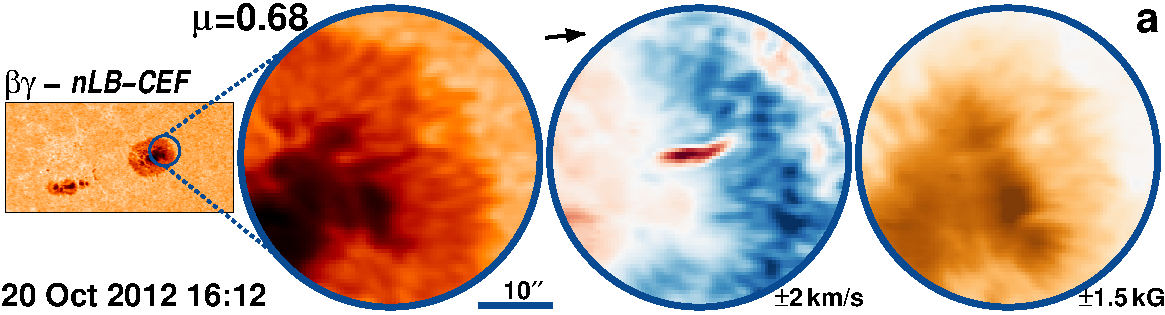}
 \caption{AR\,11596 followed for   11.1 days from 18-Oct-2012 starting at 18:48\,UT.\label{fig:DS94}}
 \end{figure*}

\begin{figure*}[htbp]
 \centering
 \includegraphics[width=.48\textwidth]{colorbars.pdf}
 \includegraphics[width=.48\textwidth]{colorbars.pdf}
 \includegraphics[width=.48\textwidth]{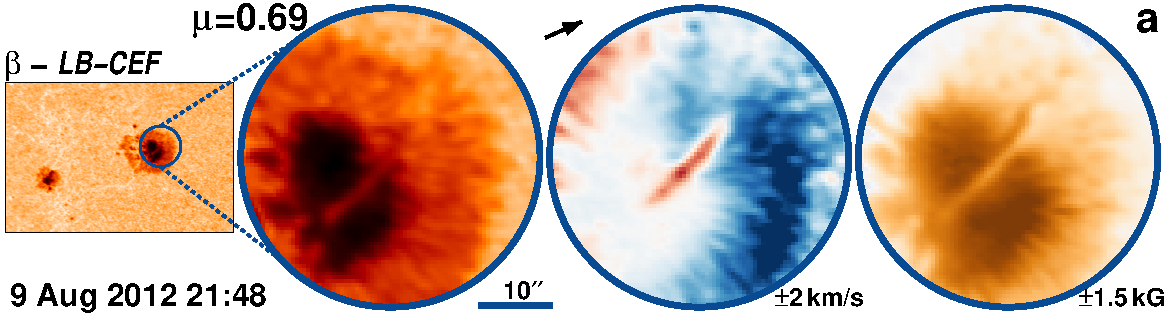}
 \includegraphics[width=.48\textwidth]{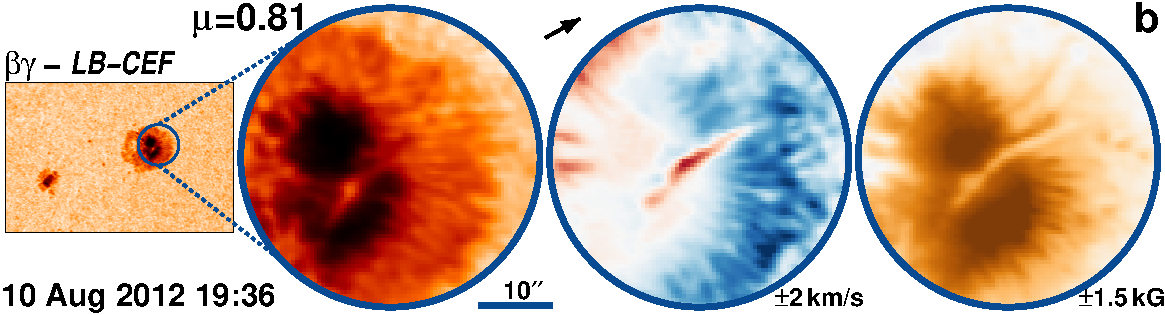}
 \includegraphics[width=.48\textwidth]{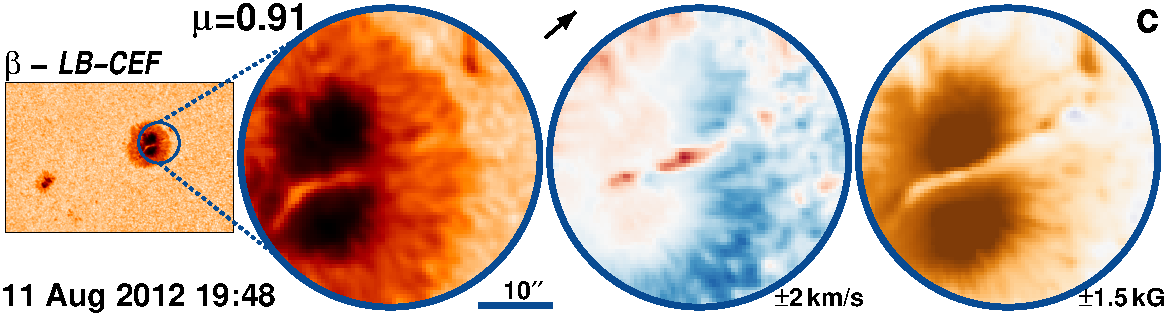}
 \includegraphics[width=.48\textwidth]{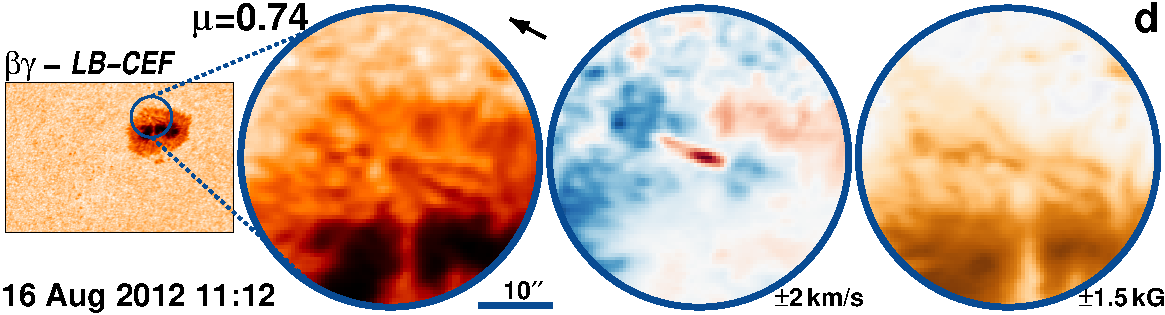}
 \caption{AR\,11543 followed for   11.0 days from  8-Aug-2012 starting at 00:24\,UT.\label{fig:DS96}}
 \end{figure*}

\begin{figure*}[htbp]
 \includegraphics[width=.48\textwidth]{colorbars.pdf}
 \includegraphics[width=.48\textwidth]{colorbars.pdf}
 \includegraphics[width=.48\textwidth]{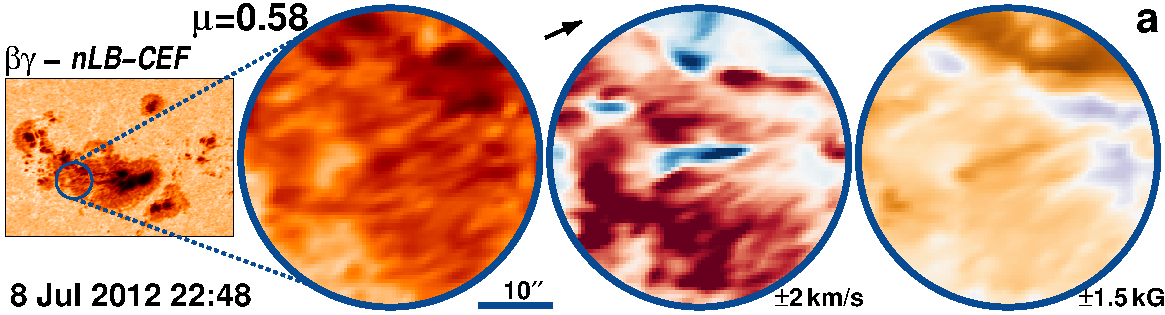}
 \includegraphics[width=.48\textwidth]{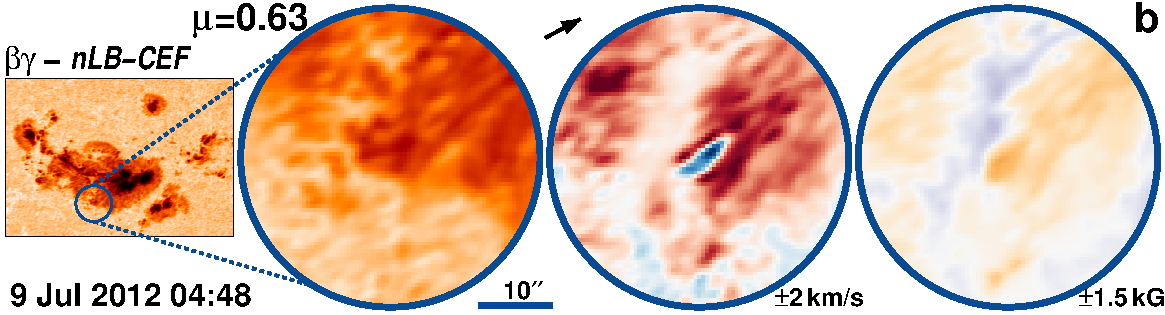}
 \includegraphics[width=.48\textwidth]{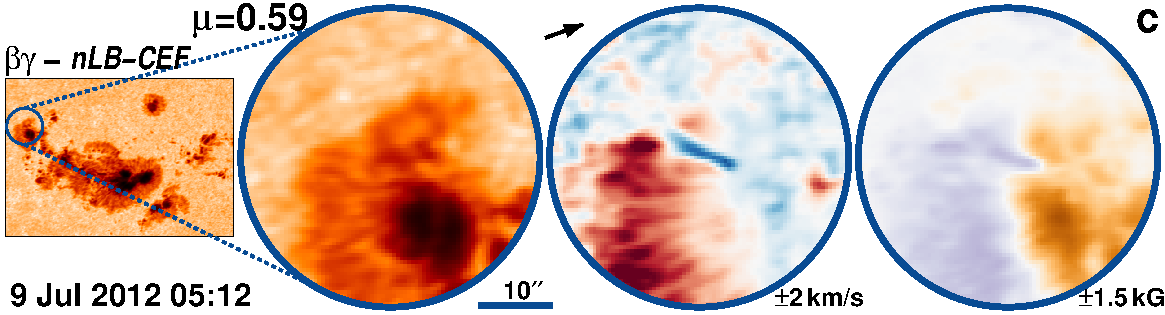}
 \includegraphics[width=.48\textwidth]{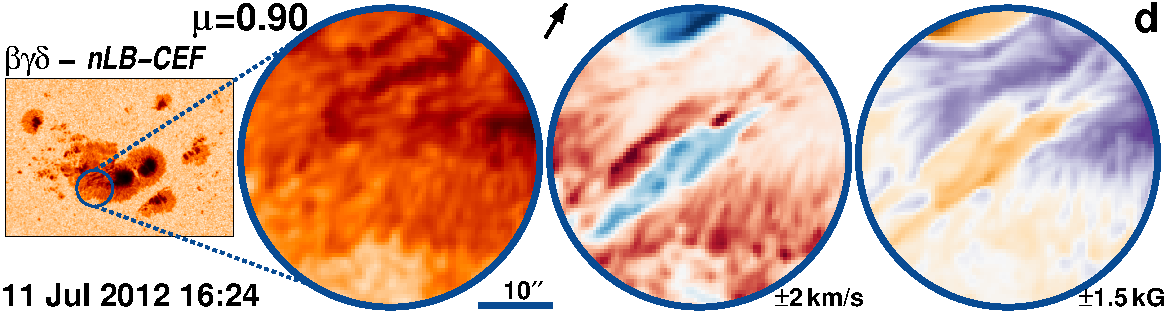}
 \includegraphics[width=.48\textwidth]{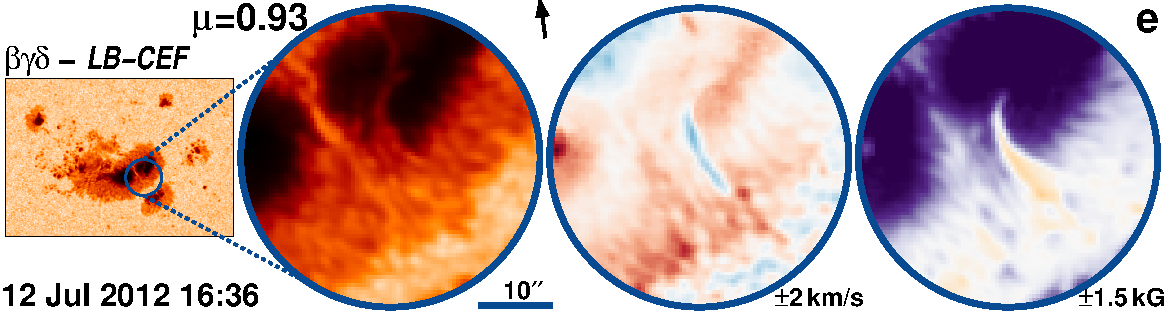}
 \includegraphics[width=.48\textwidth]{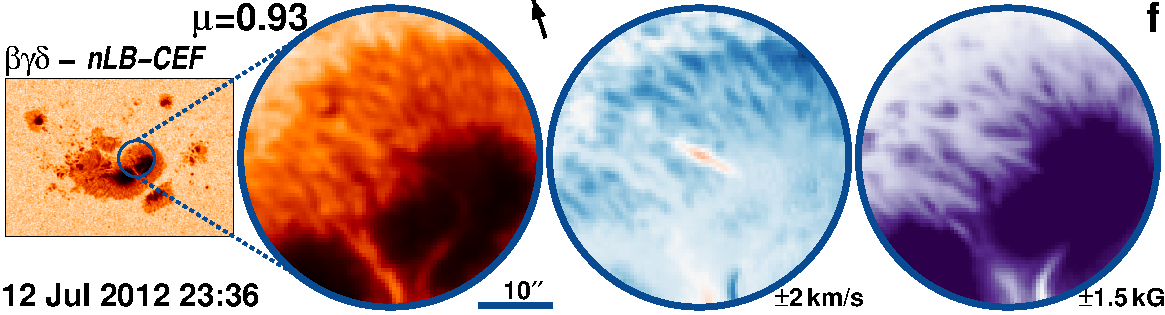}
 \includegraphics[width=.48\textwidth]{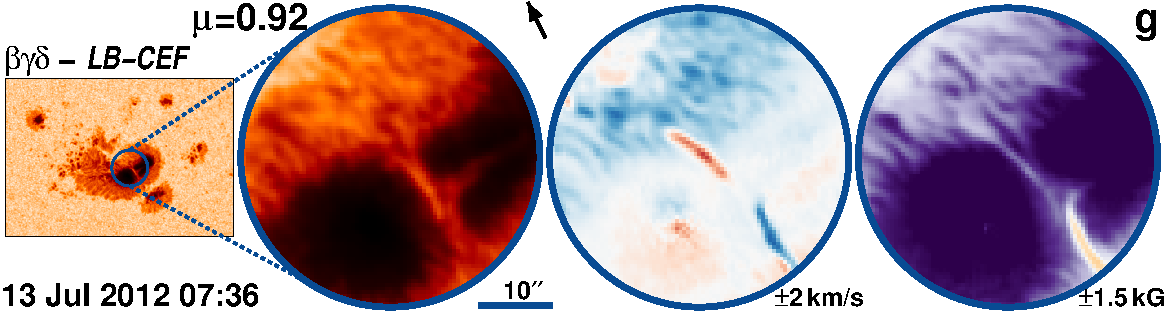}
 \includegraphics[width=.48\textwidth]{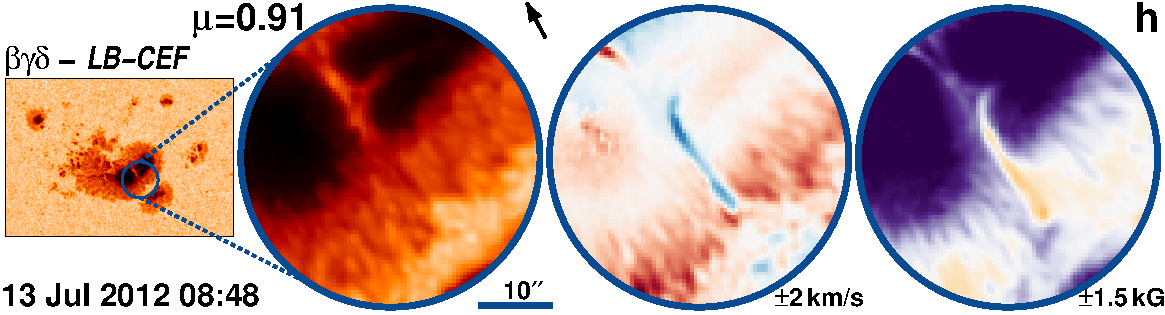}
 \includegraphics[width=.48\textwidth]{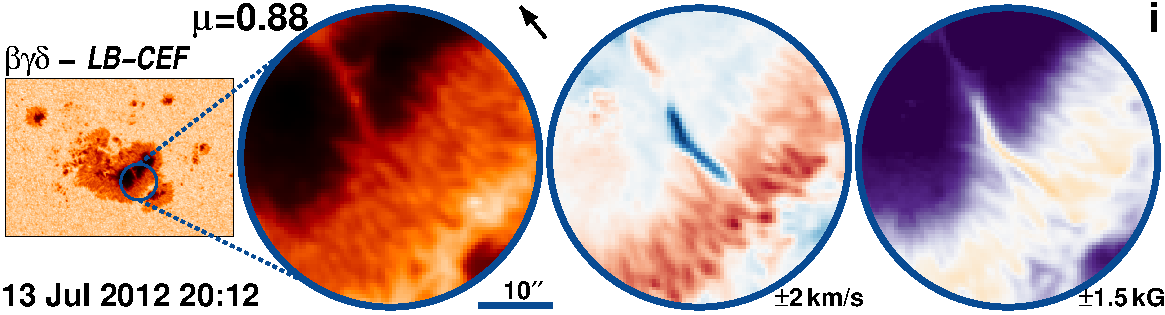}
 \includegraphics[width=.48\textwidth]{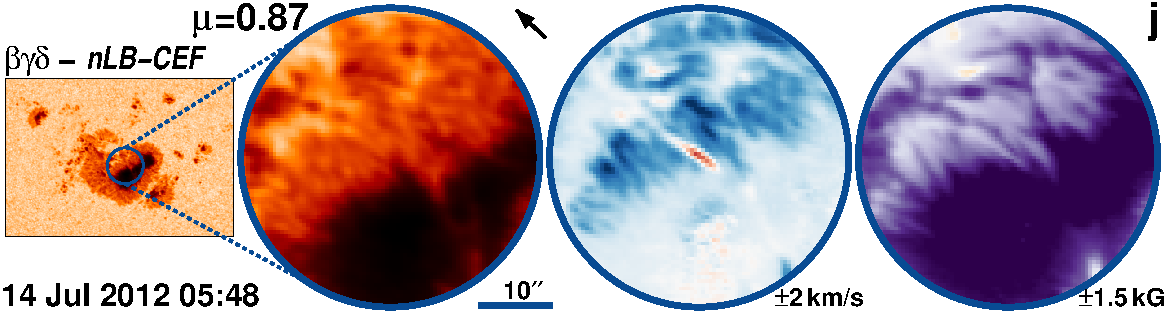}
 \includegraphics[width=.48\textwidth]{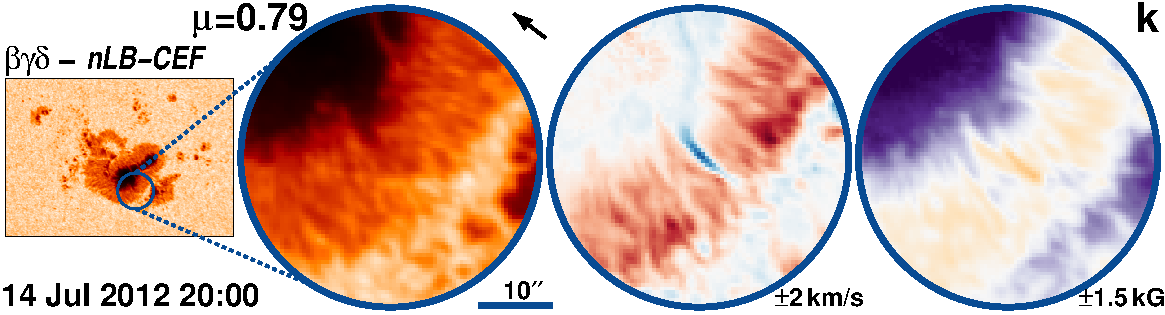}
 \includegraphics[width=.48\textwidth]{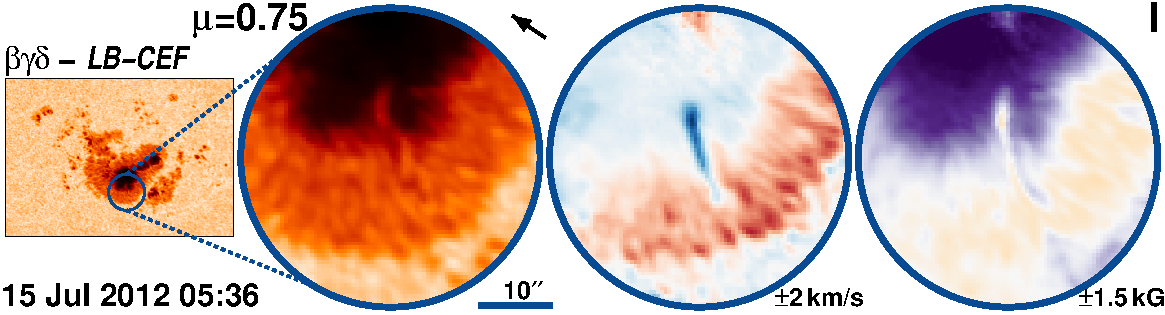}
 \includegraphics[width=.48\textwidth]{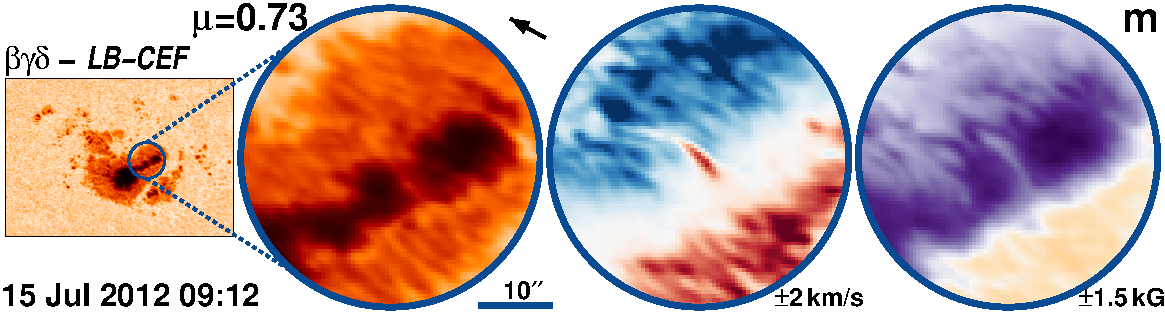}
 \includegraphics[width=.48\textwidth]{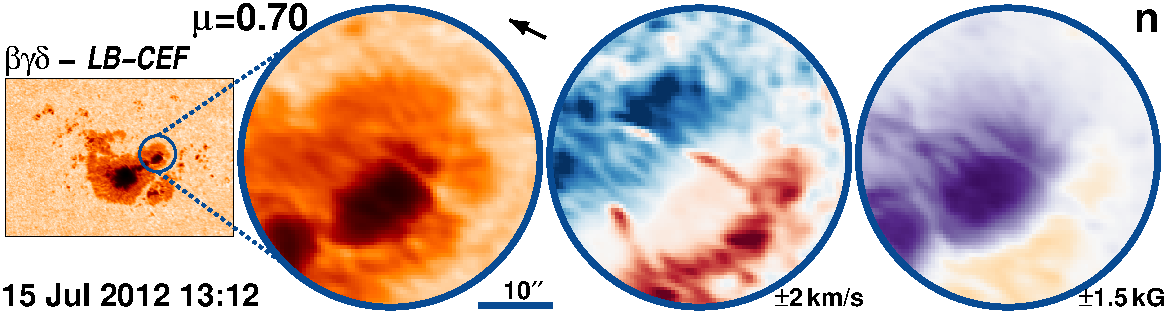}
 \includegraphics[width=.48\textwidth]{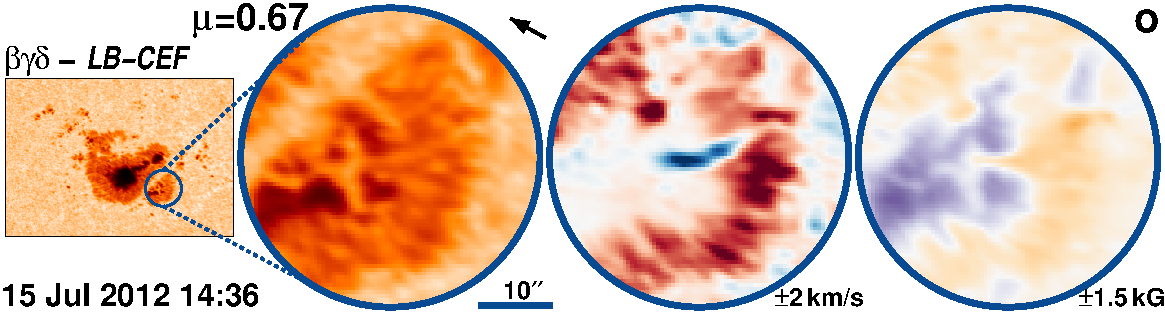}
 \caption{AR\,11520 followed for    8.0 days from  8-Jul-2012 starting at 18:12\,UT.\label{fig:DS97}}
 \end{figure*}

\begin{figure*}[htbp]
 \includegraphics[width=.48\textwidth]{colorbars.pdf}
 \includegraphics[width=.48\textwidth]{colorbars.pdf}
 \includegraphics[width=.48\textwidth]{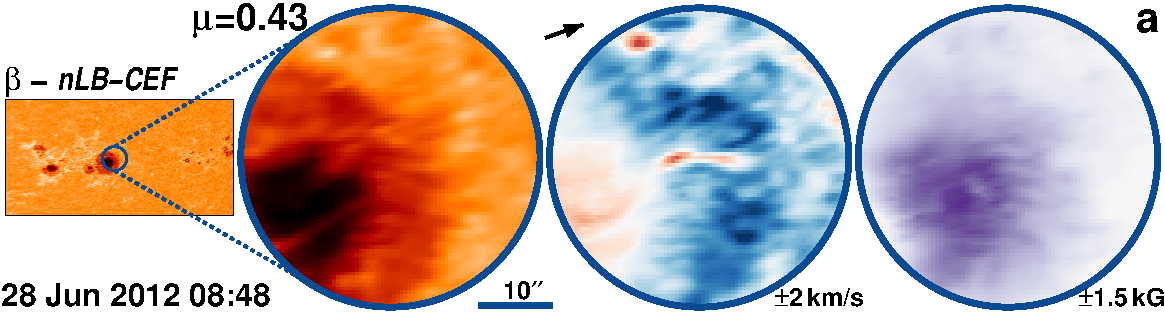}
 \includegraphics[width=.48\textwidth]{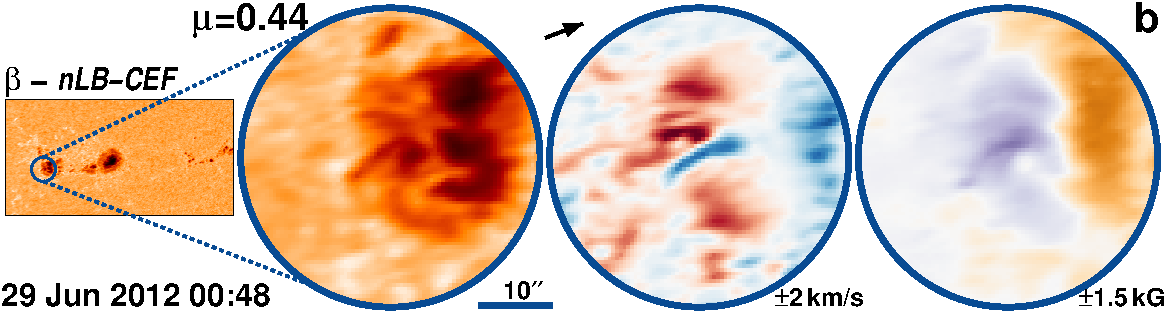}
 \includegraphics[width=.48\textwidth]{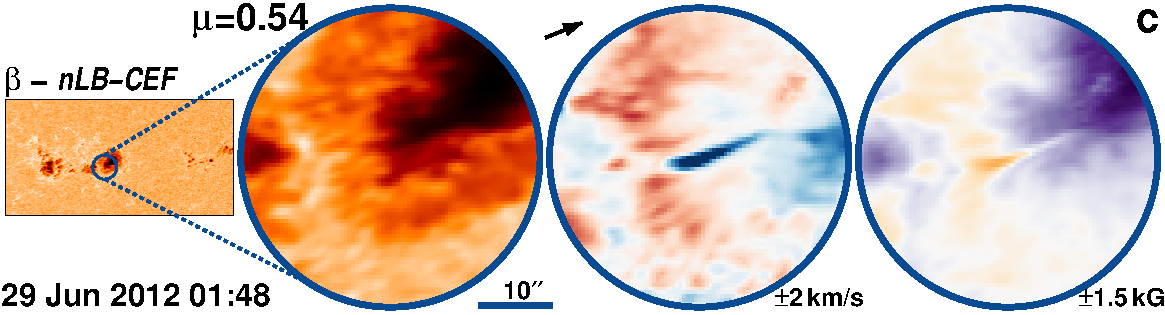}
 \caption{AR\,11515 followed for   10.5 days from 27-Jun-2012 starting at 20:36\,UT.\label{fig:DS98}}
 \end{figure*}

\begin{figure*}[htbp]
 \includegraphics[width=.48\textwidth]{colorbars.pdf}

 \includegraphics[width=.48\textwidth]{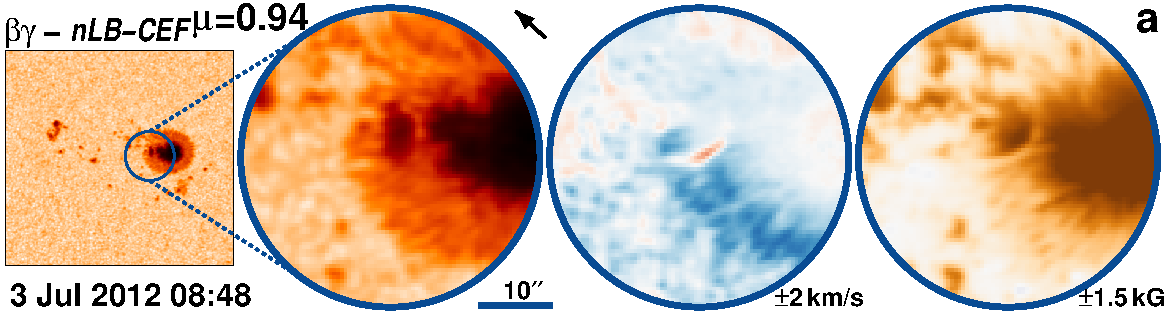}
 \caption{AR\,11513 followed for   11.3 days from 26-Jun-2012 starting at 17:12\,UT.\label{fig:DS99}}
 \end{figure*}

\begin{figure*}[htbp]
 \includegraphics[width=.48\textwidth]{colorbars.pdf}
 \includegraphics[width=.48\textwidth]{colorbars.pdf}
 \includegraphics[width=.48\textwidth]{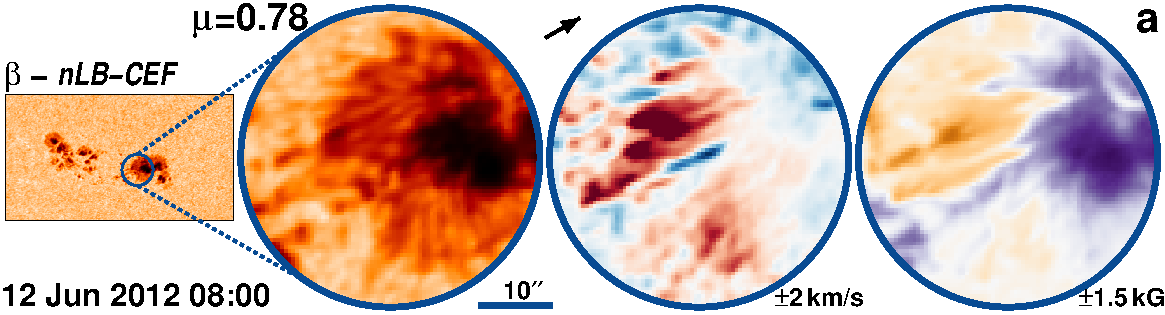}
 \includegraphics[width=.48\textwidth]{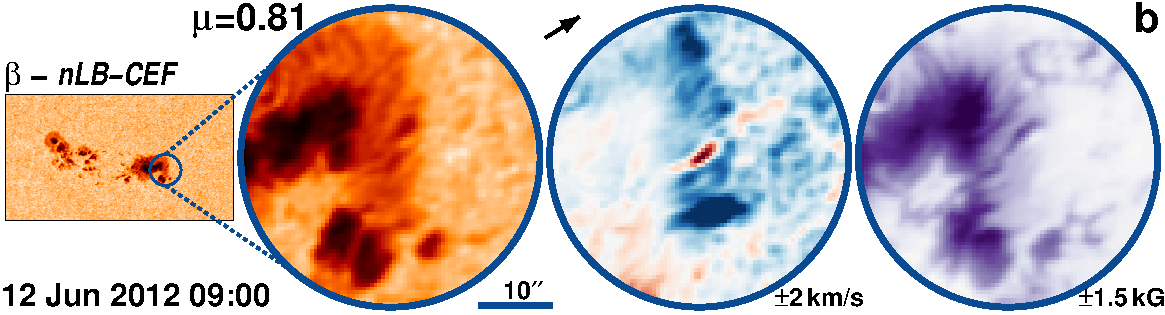}
 \includegraphics[width=.48\textwidth]{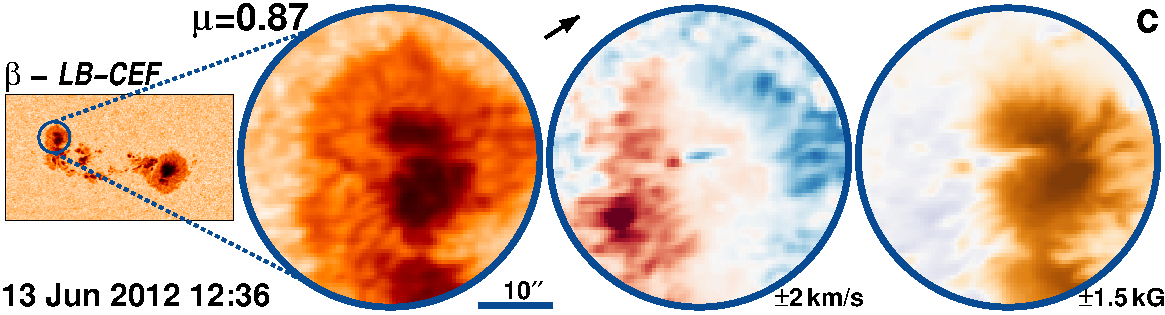}
 \includegraphics[width=.48\textwidth]{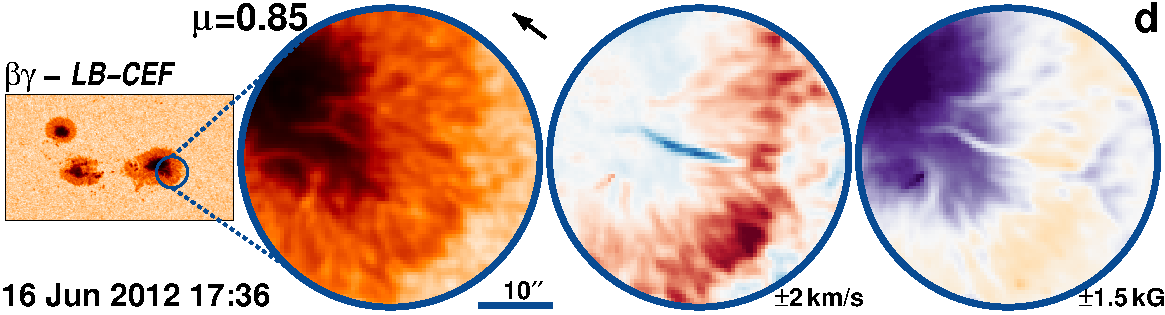}
 \includegraphics[width=.48\textwidth]{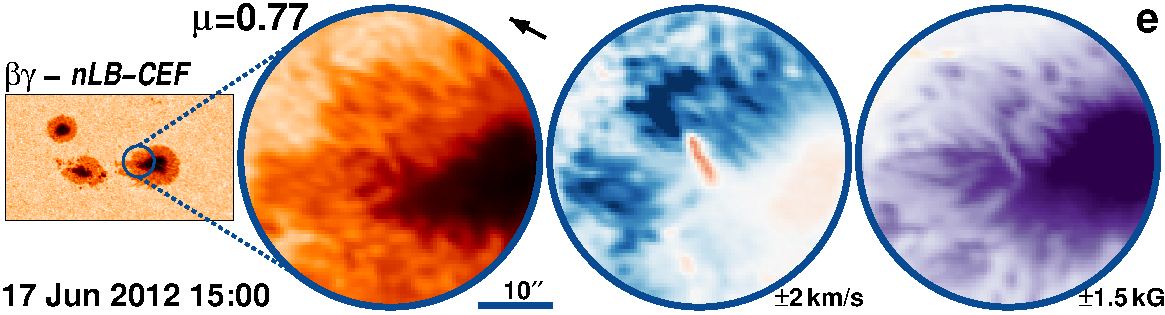}
 \includegraphics[width=.48\textwidth]{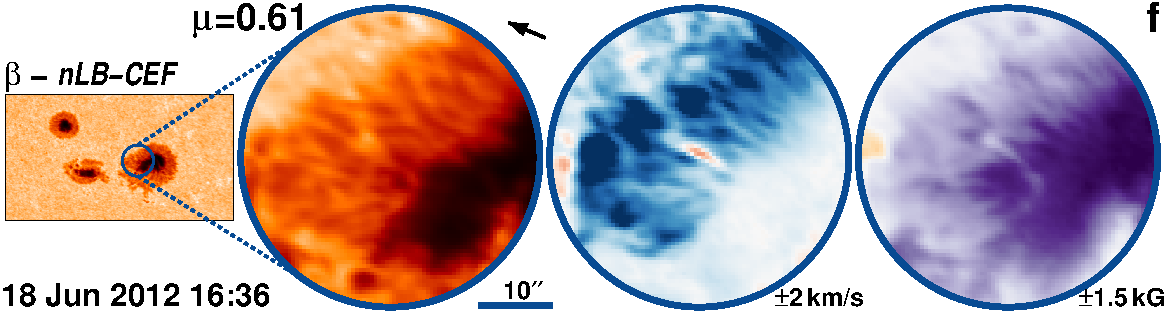}
 \includegraphics[width=.48\textwidth]{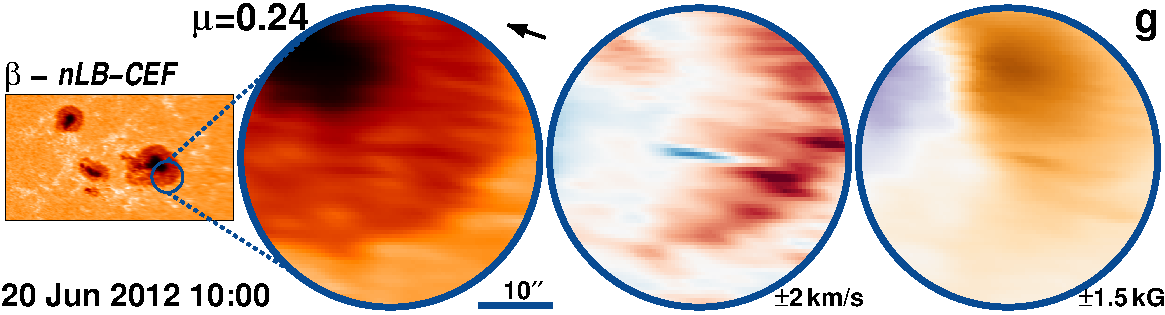}
 \caption{AR\,11504 followed for   11.3 days from  9-Jun-2012 starting at 10:24\,UT.\label{fig:DS100}}
 \end{figure*}

\begin{figure*}[htbp]
 \includegraphics[width=.48\textwidth]{colorbars.pdf}
 \includegraphics[width=.48\textwidth]{colorbars.pdf}
 \includegraphics[width=.48\textwidth]{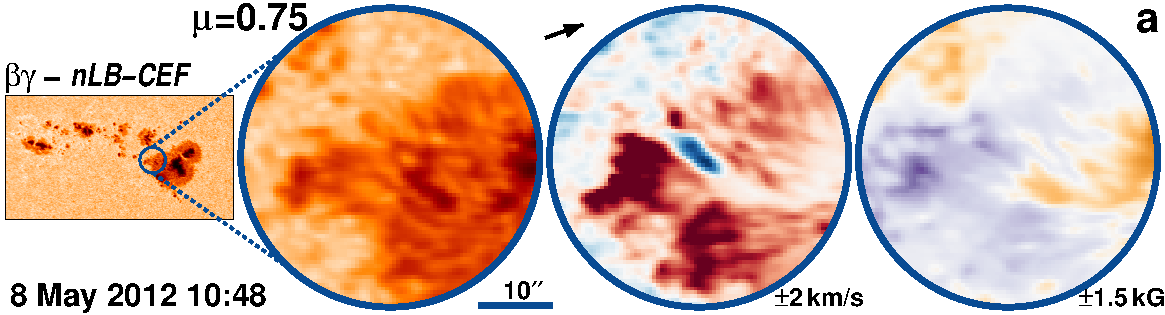}
 \includegraphics[width=.48\textwidth]{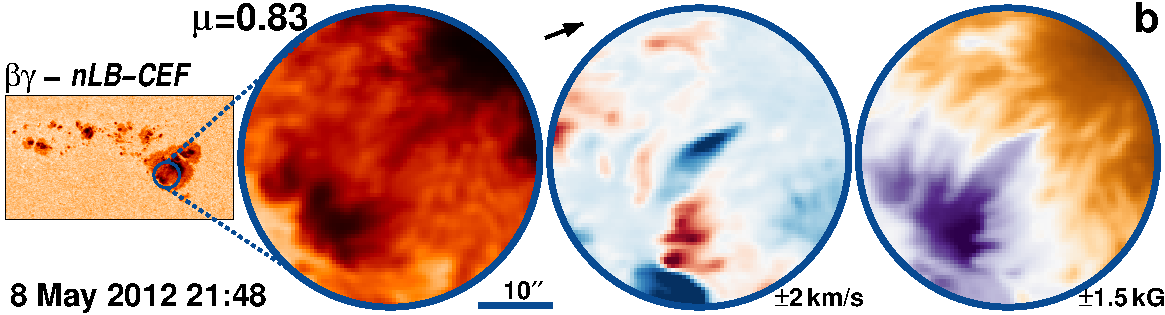}
 \includegraphics[width=.48\textwidth]{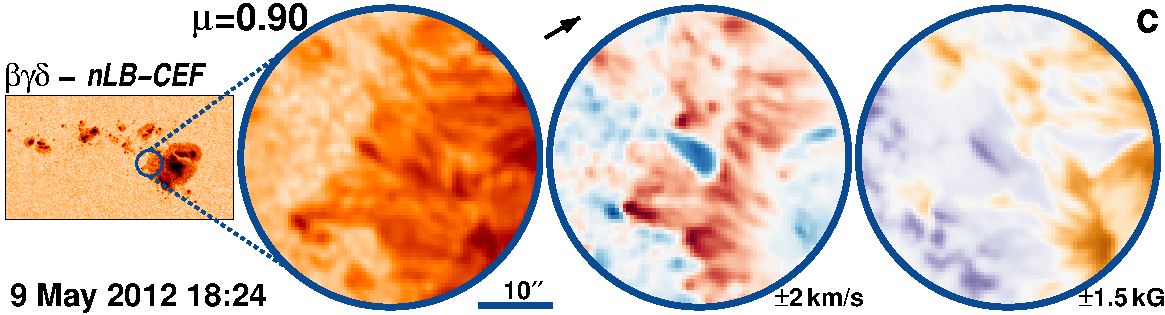}
 \includegraphics[width=.48\textwidth]{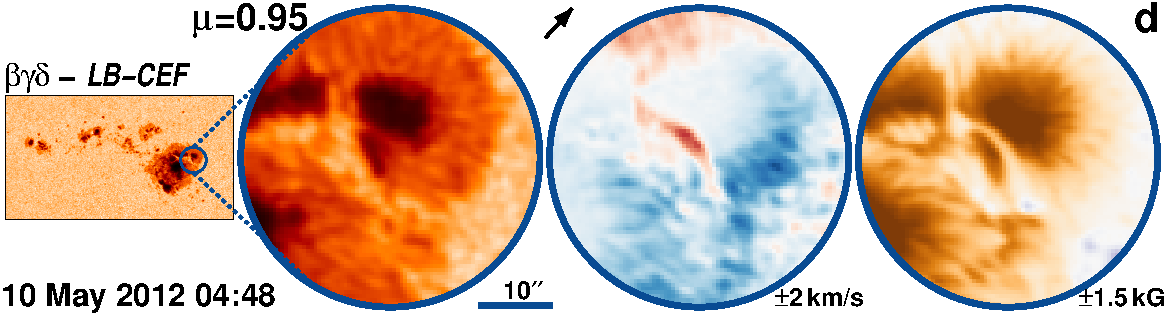}
 \includegraphics[width=.48\textwidth]{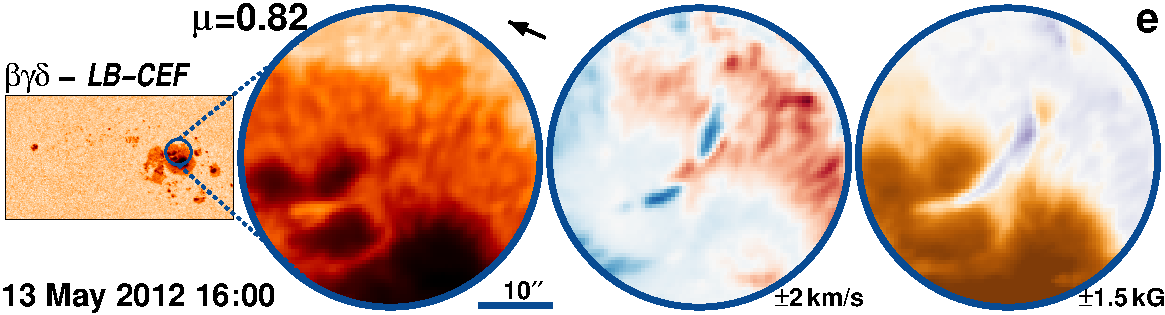}
 \includegraphics[width=.48\textwidth]{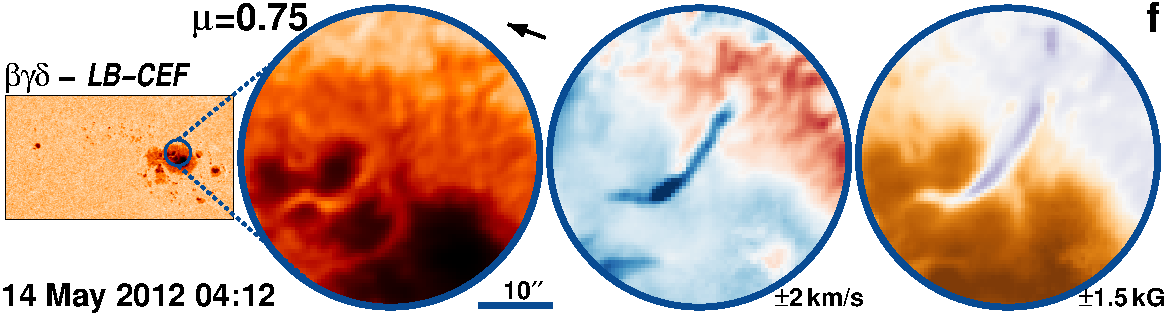}
 \includegraphics[width=.48\textwidth]{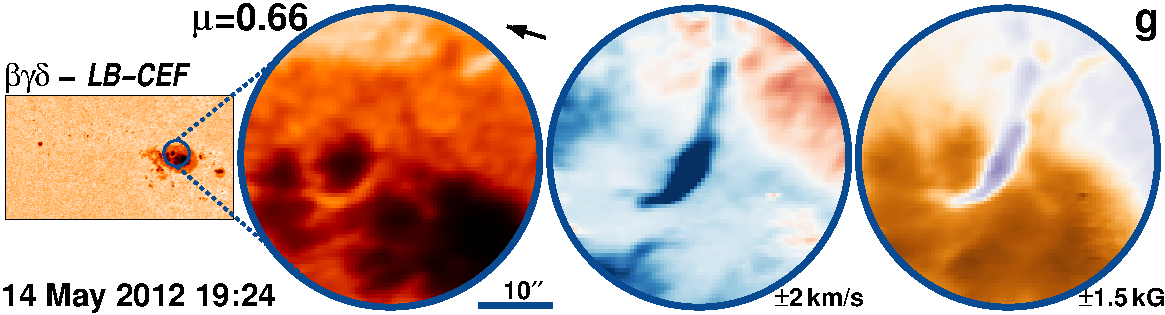}
 \caption{AR\,11476 followed for    8.0 days from  8-May-2012 starting at 00:00\,UT.\label{fig:DS101}}
 \end{figure*}

\begin{figure*}[htbp]
 \includegraphics[width=.48\textwidth]{colorbars.pdf}
 \includegraphics[width=.48\textwidth]{colorbars.pdf}
 \includegraphics[width=.48\textwidth]{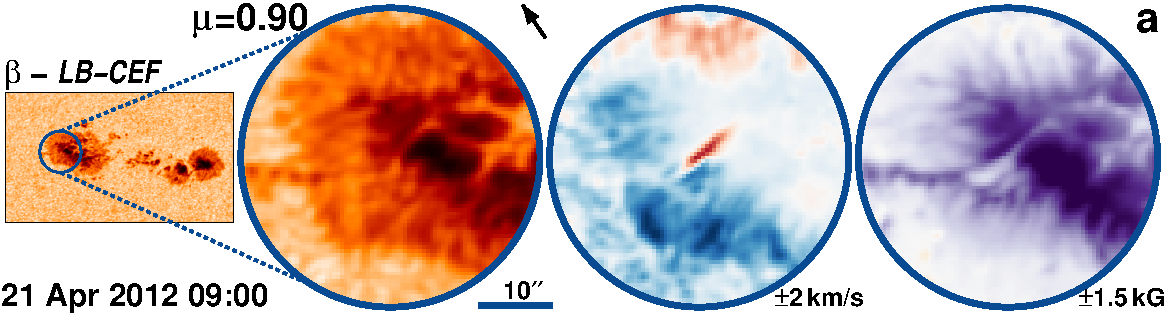}
 \includegraphics[width=.48\textwidth]{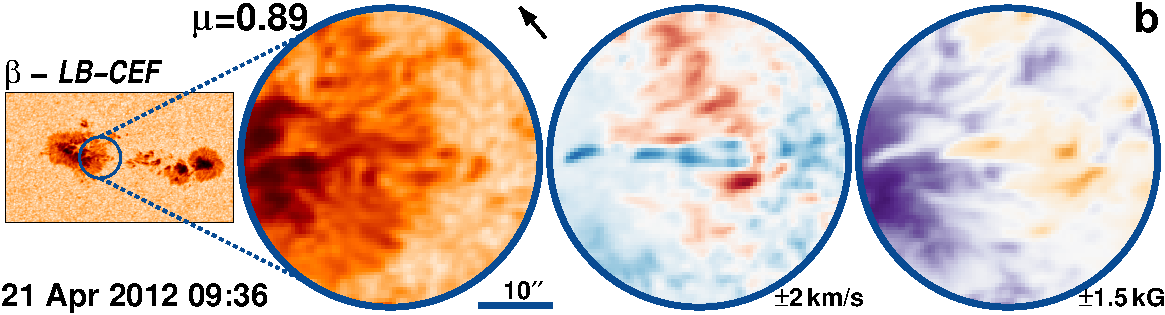}
 \includegraphics[width=.48\textwidth]{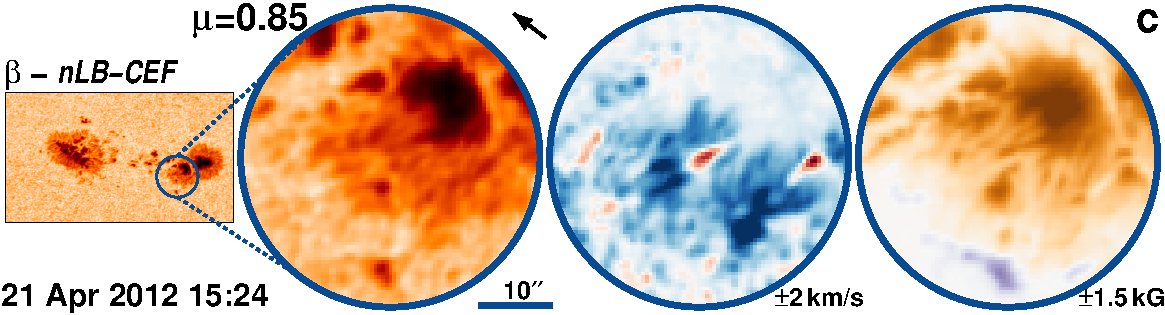}
 \includegraphics[width=.48\textwidth]{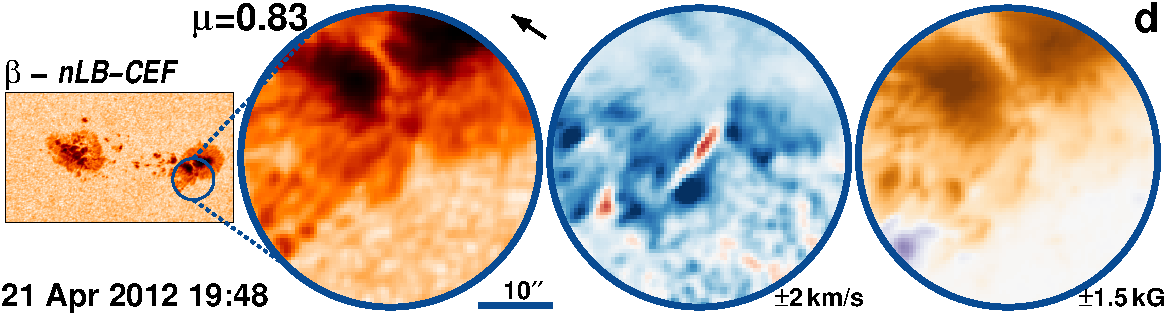}
 \includegraphics[width=.48\textwidth]{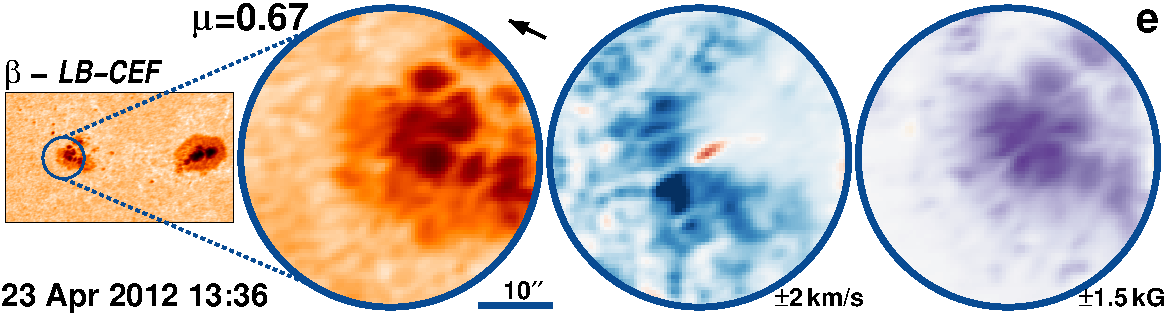}
 \caption{AR\,11460 followed for    8.5 days from 16-Apr-2012 starting at 19:24\,UT.\label{fig:DS103}}
 \end{figure*}

\begin{figure*}[htbp]
 \includegraphics[width=.48\textwidth]{colorbars.pdf}
 \includegraphics[width=.48\textwidth]{colorbars.pdf}
 \includegraphics[width=.48\textwidth]{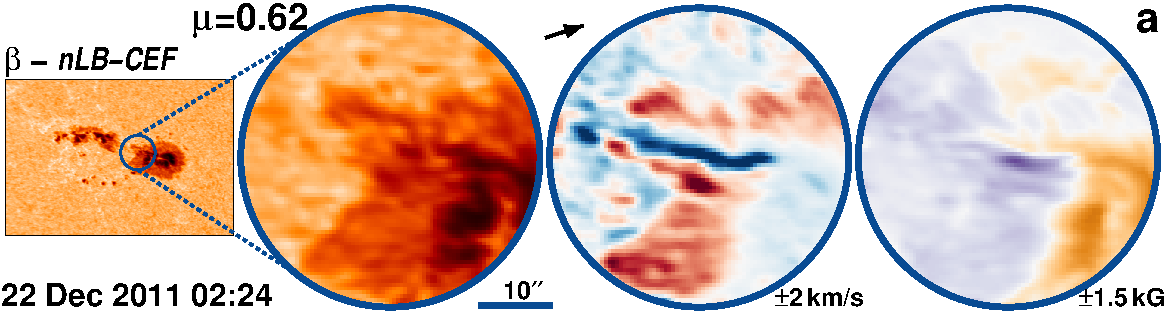}
 \includegraphics[width=.48\textwidth]{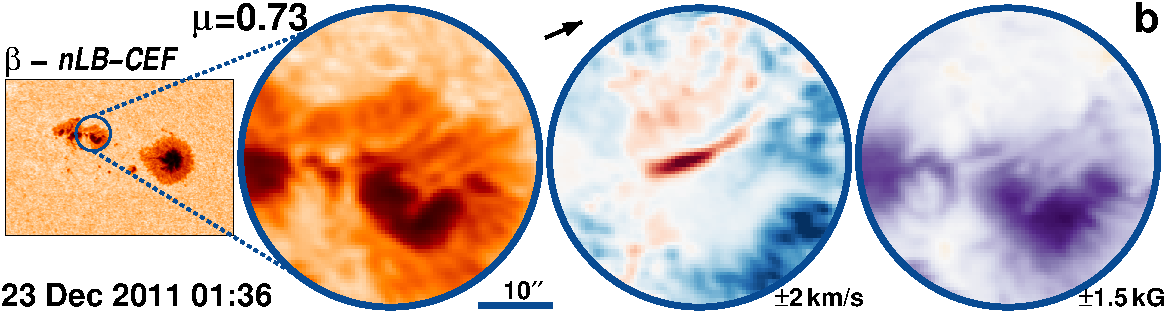}
 \includegraphics[width=.48\textwidth]{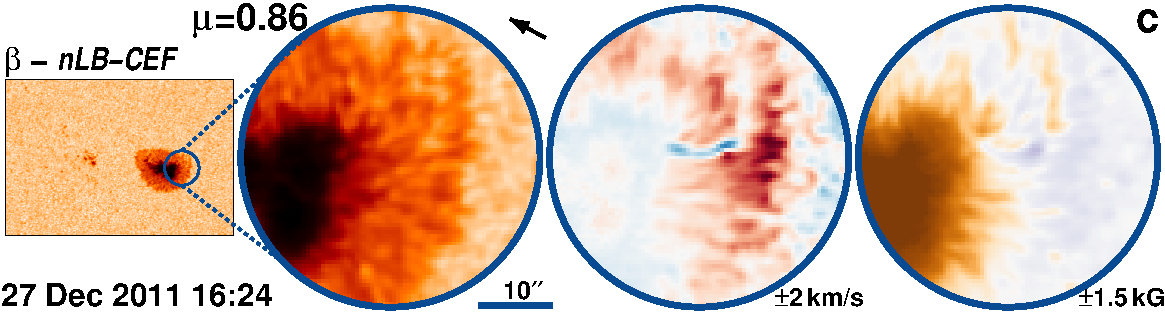}
 \includegraphics[width=.48\textwidth]{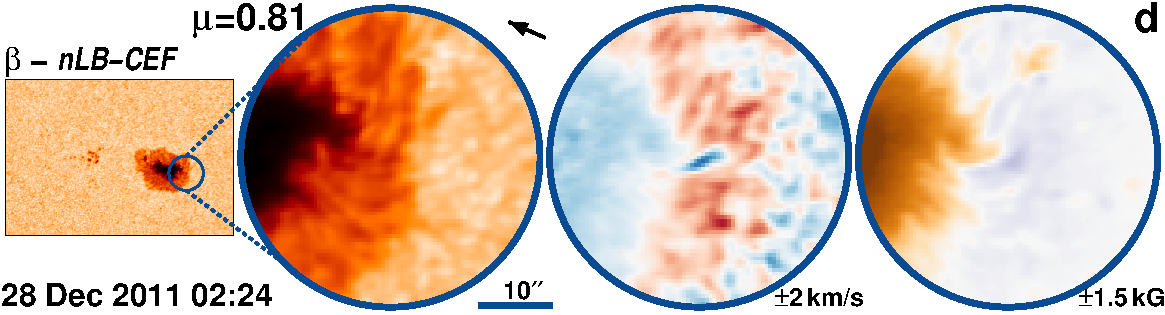}
 \includegraphics[width=.48\textwidth]{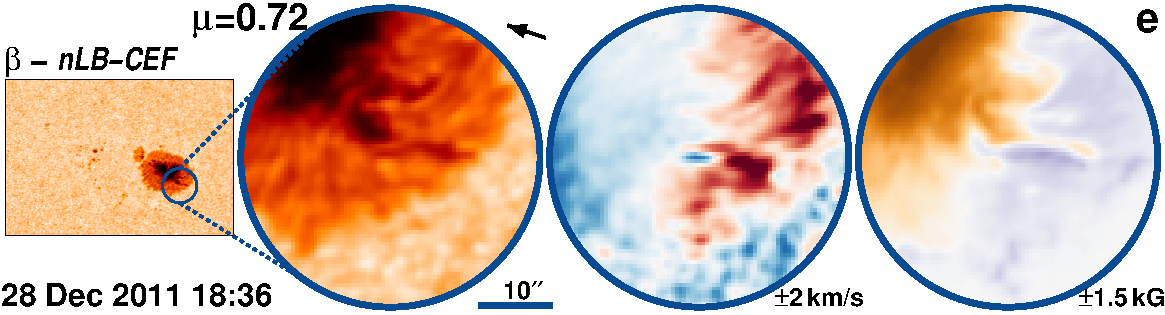}
 \caption{AR\,11384 followed for   11.4 days from 20-Dec-2011 starting at 03:36\,UT.\label{fig:DS105}}
 \end{figure*}

\begin{figure*}[htbp]
 \includegraphics[width=.48\textwidth]{colorbars.pdf}
 \includegraphics[width=.48\textwidth]{colorbars.pdf}
 \includegraphics[width=.48\textwidth]{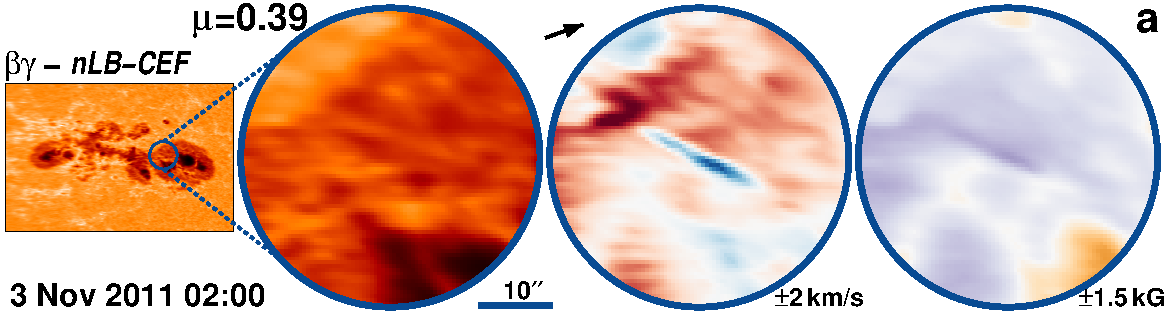}
 \includegraphics[width=.48\textwidth]{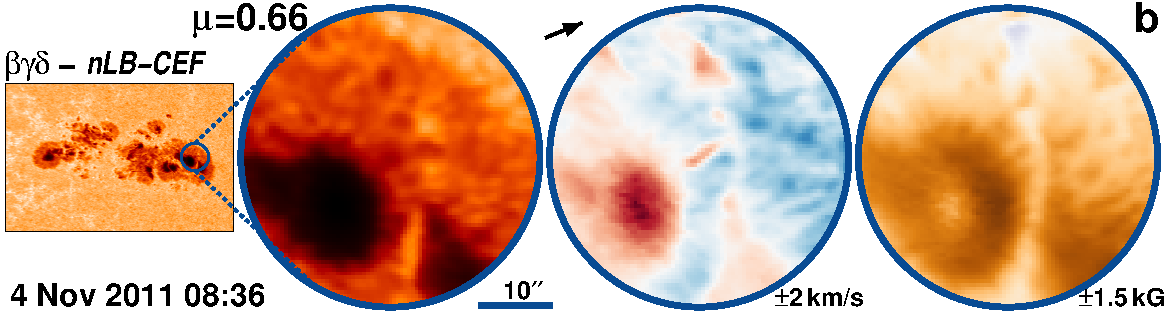}
 \includegraphics[width=.48\textwidth]{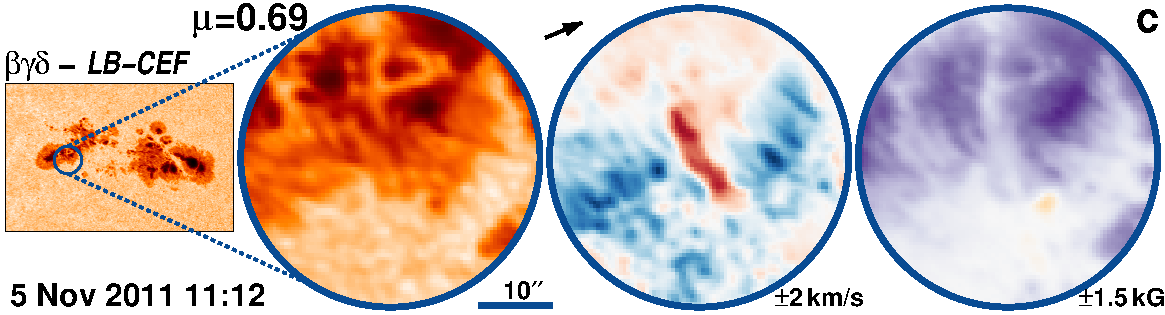}
 \includegraphics[width=.48\textwidth]{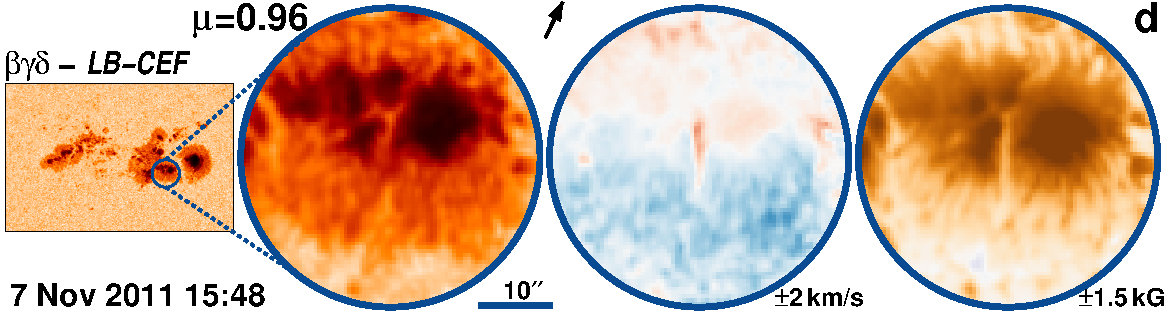}
 \includegraphics[width=.48\textwidth]{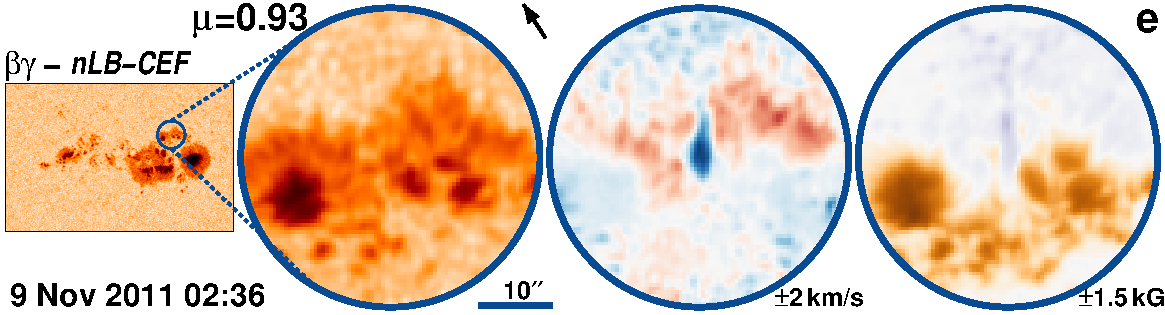}
 \caption{AR\,11339 followed for   11.6 days from  2-Nov-2011 starting at 10:00\,UT.\label{fig:DS106}}
 \end{figure*}

\begin{figure*}[htbp]
 \centering
 \includegraphics[width=.48\textwidth]{colorbars.pdf}
 \includegraphics[width=.48\textwidth]{colorbars.pdf}
 \includegraphics[width=.48\textwidth]{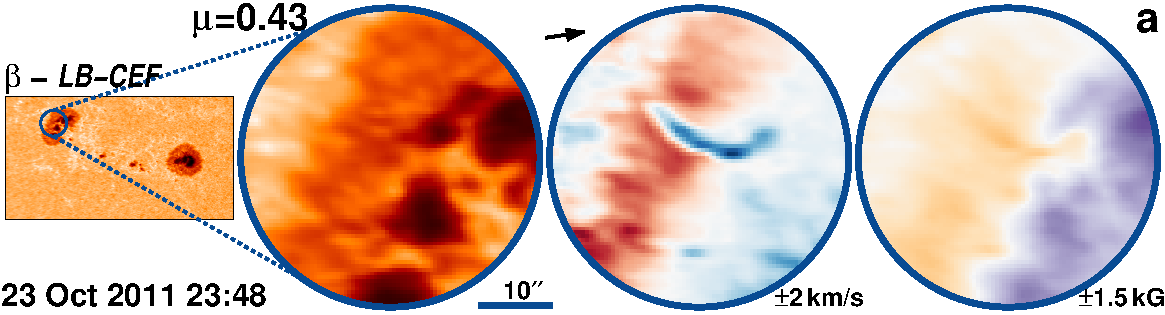}
 \includegraphics[width=.48\textwidth]{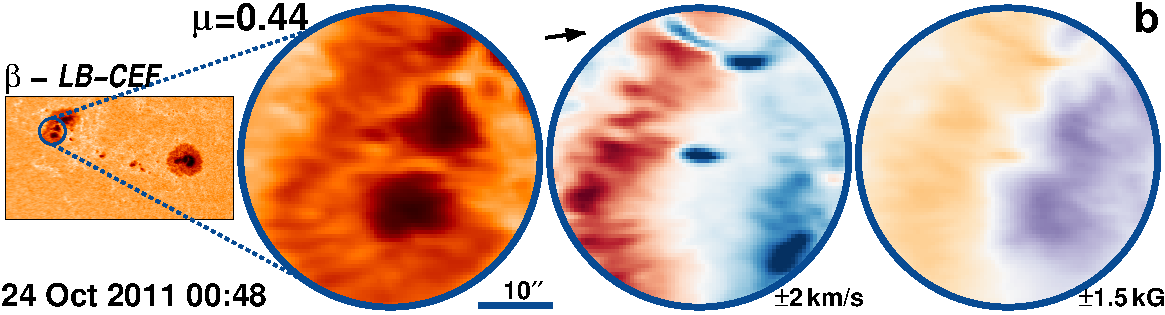}
 \includegraphics[width=.48\textwidth]{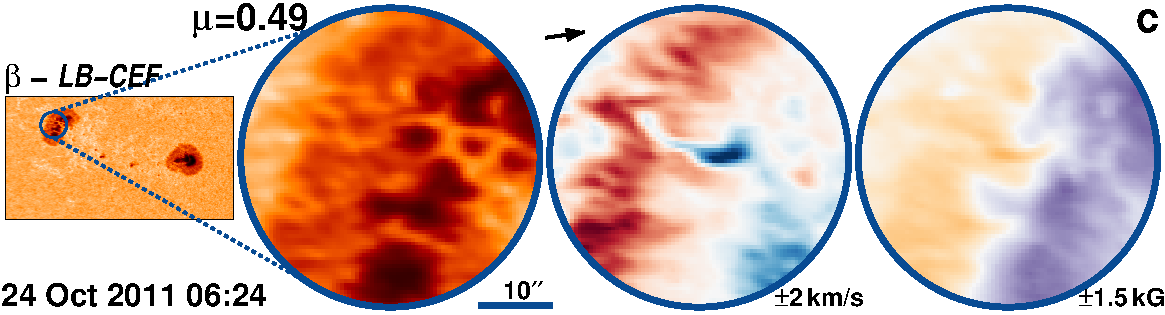}
 \includegraphics[width=.48\textwidth]{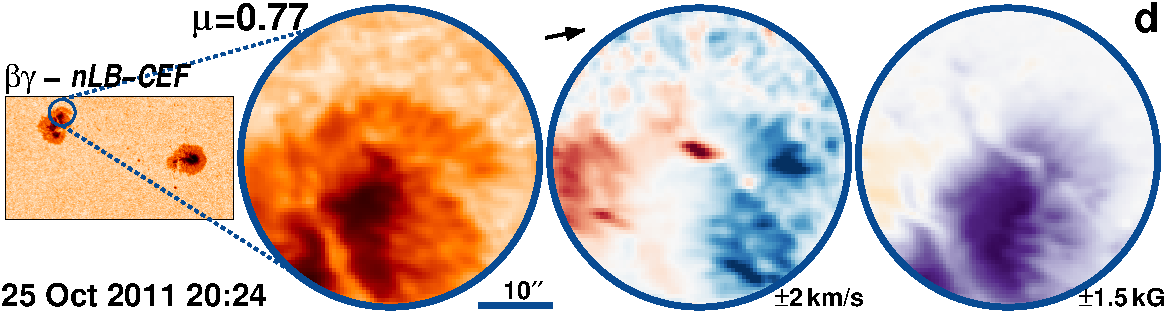}
 \includegraphics[width=.48\textwidth]{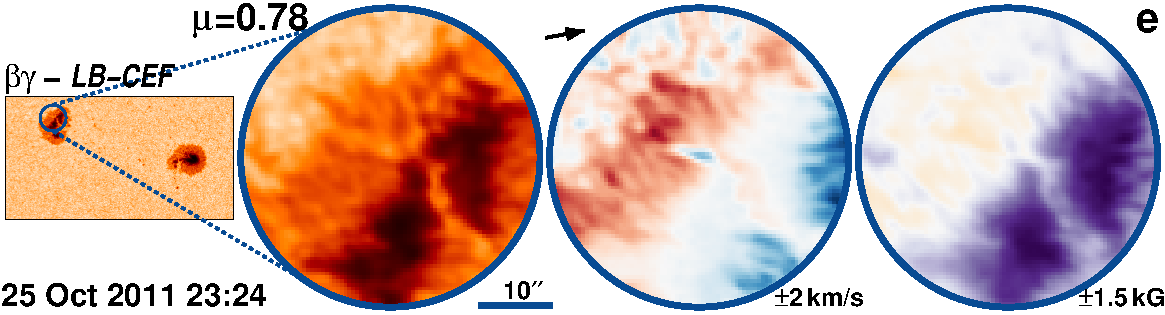}
 \includegraphics[width=.48\textwidth]{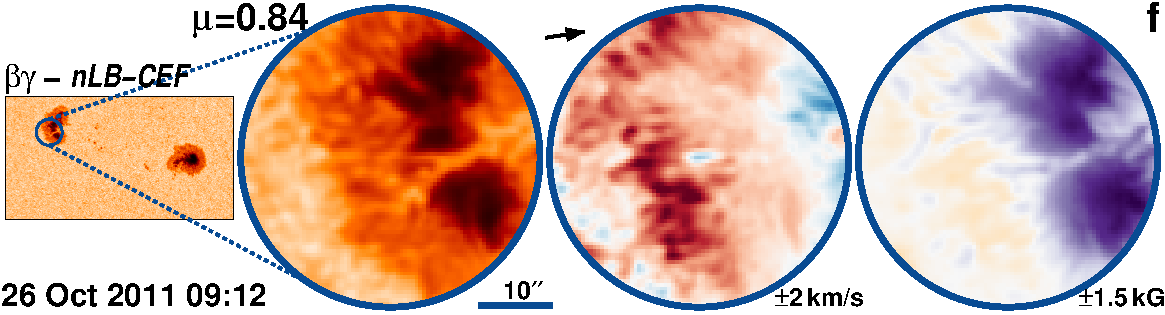}
 \caption{AR\,11330 followed for   10.1 days from 22-Oct-2011 starting at 16:36\,UT.\label{fig:DS107}}
 \end{figure*}

\begin{figure*}[htbp]
 \centering
 \includegraphics[width=.48\textwidth]{colorbars.pdf}
 \includegraphics[width=.48\textwidth]{colorbars.pdf}
 \includegraphics[width=.48\textwidth]{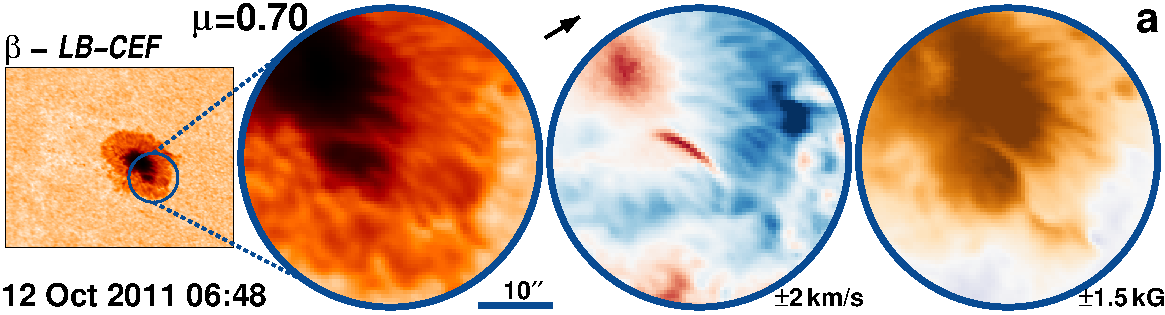}
 \includegraphics[width=.48\textwidth]{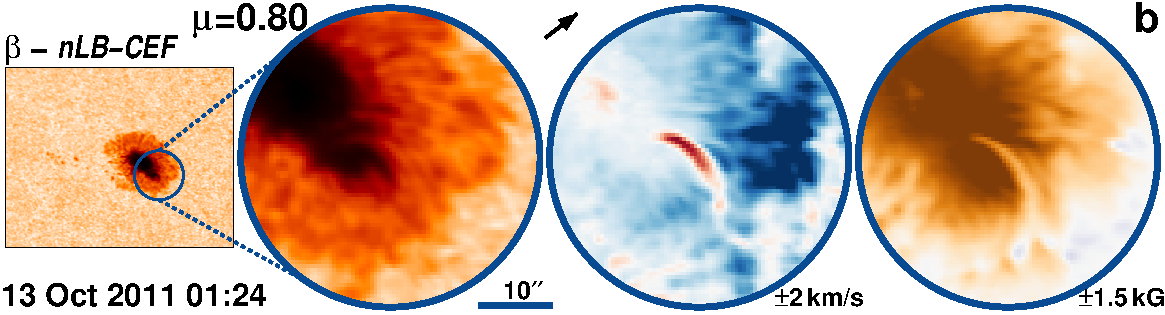}
 \includegraphics[width=.48\textwidth]{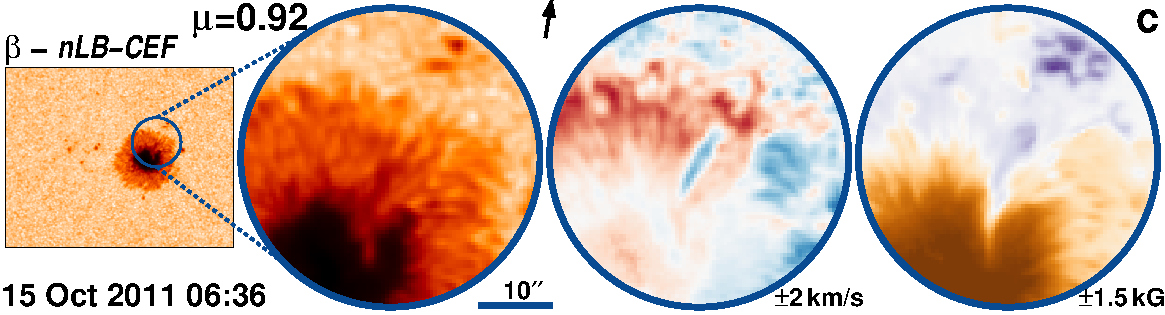}
 \includegraphics[width=.48\textwidth]{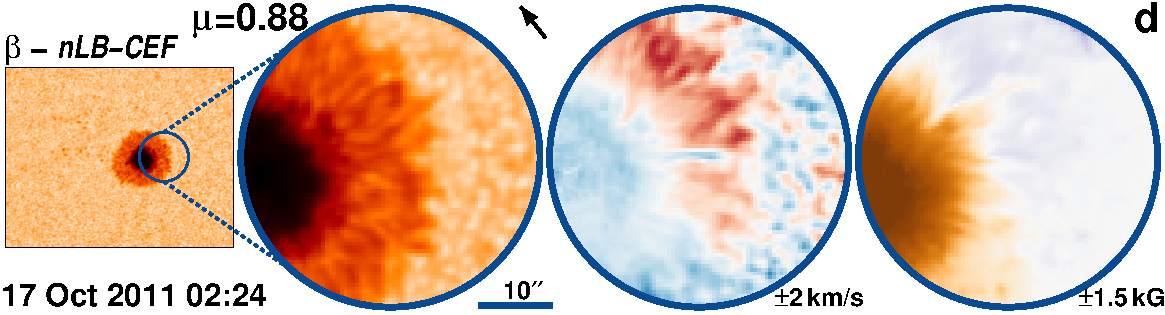}
 \caption{AR\,11314 followed for    8.7 days from 11-Oct-2011 starting at 11:36\,UT.\label{fig:DS108}}
 \end{figure*}

\begin{figure*}[htbp]
 \includegraphics[width=.48\textwidth]{colorbars.pdf}
 \includegraphics[width=.48\textwidth]{colorbars.pdf}
 \includegraphics[width=.48\textwidth]{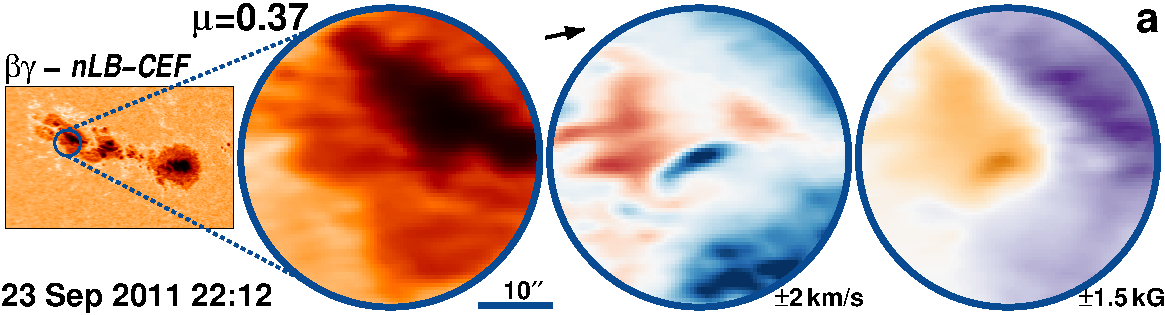}
 \includegraphics[width=.48\textwidth]{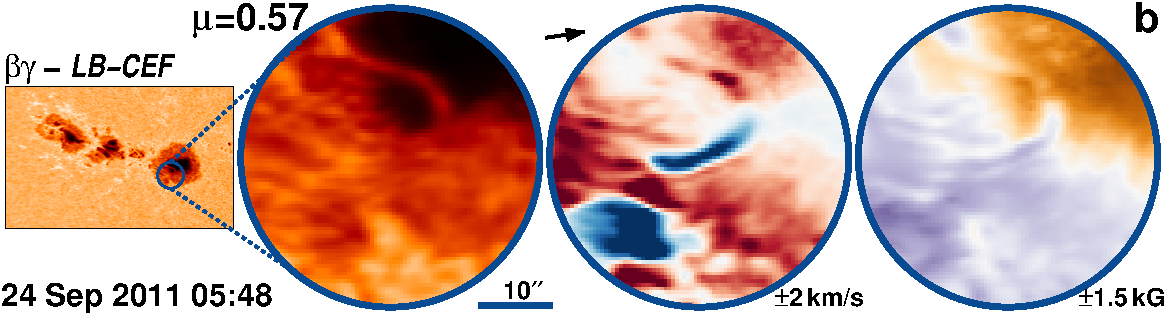}
 \includegraphics[width=.48\textwidth]{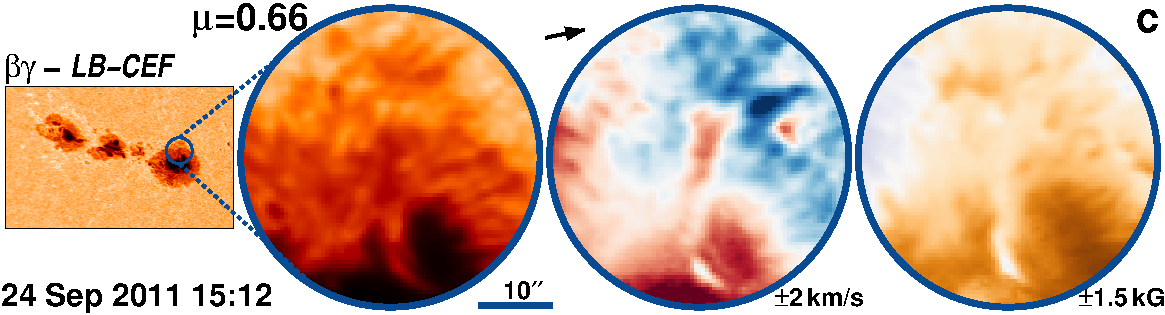}
 \includegraphics[width=.48\textwidth]{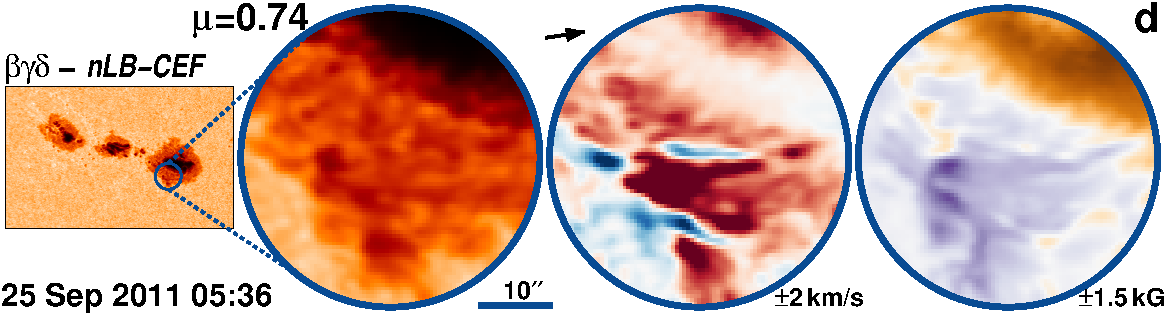}
 \includegraphics[width=.48\textwidth]{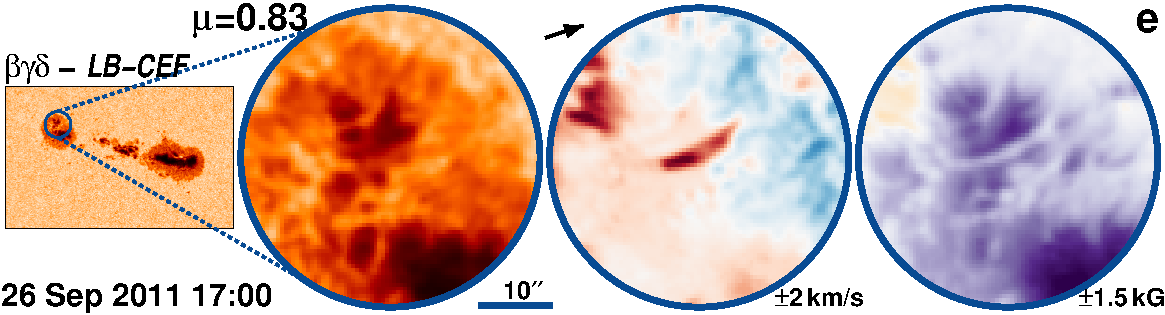}
 \includegraphics[width=.48\textwidth]{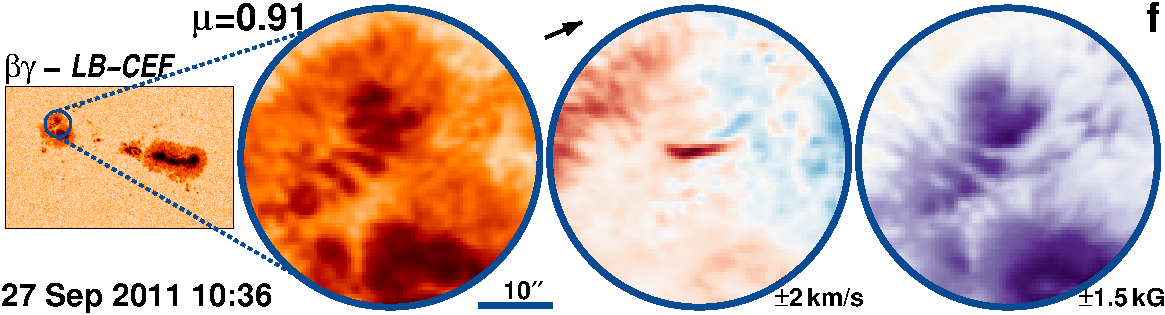}
 \includegraphics[width=.48\textwidth]{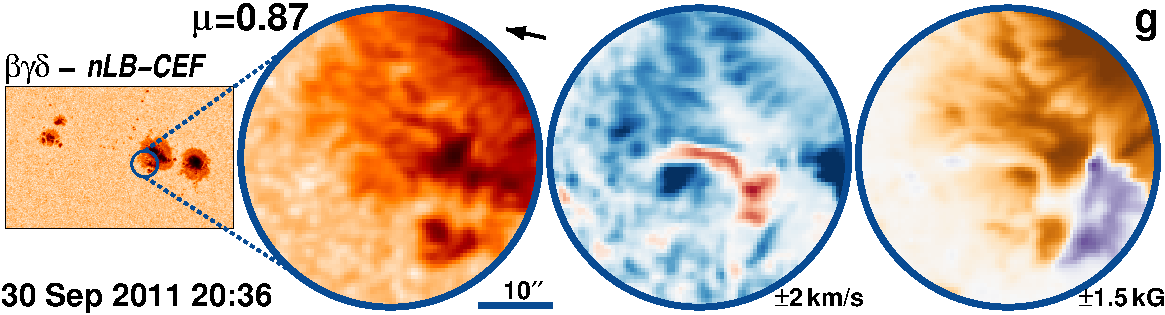}
 \caption{AR\,11302 followed for   11.1 days from 23-Sep-2011 starting at 05:24\,UT.\label{fig:DS110}}
 \end{figure*}

\begin{figure*}[htbp]
 \includegraphics[width=.48\textwidth]{colorbars.pdf}
 \includegraphics[width=.48\textwidth]{colorbars.pdf}
 \includegraphics[width=.48\textwidth]{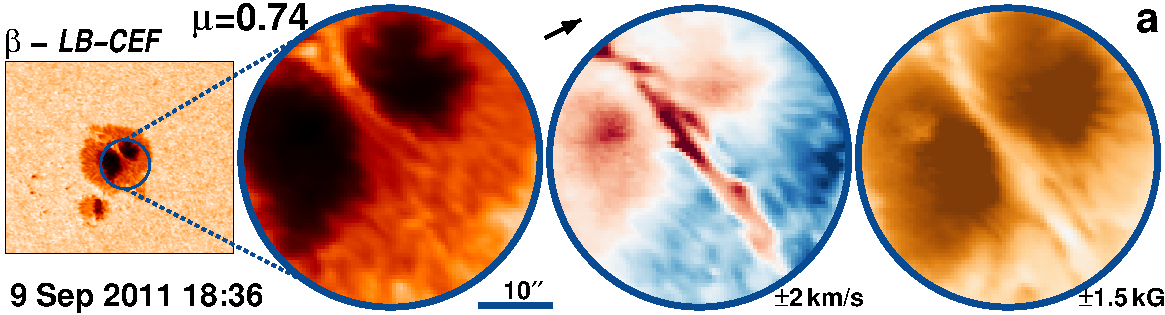}
 \includegraphics[width=.48\textwidth]{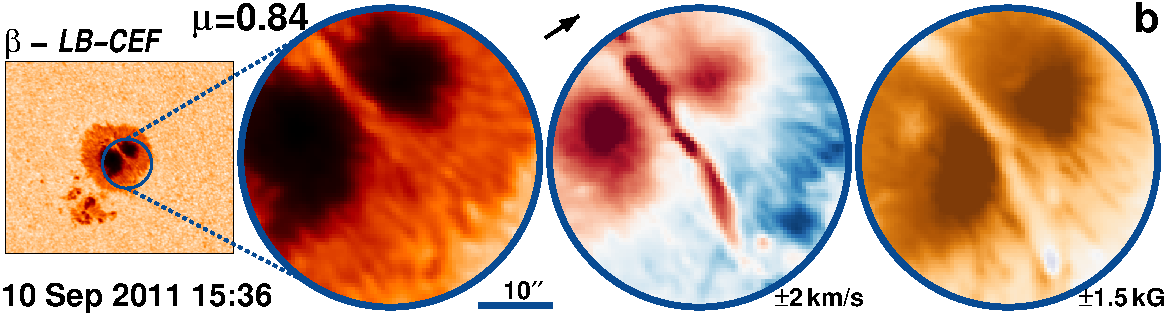}
 \includegraphics[width=.48\textwidth]{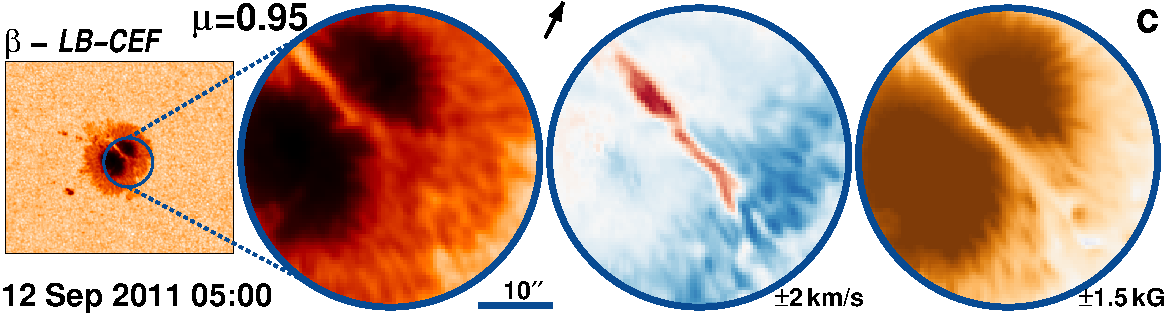}
 \includegraphics[width=.48\textwidth]{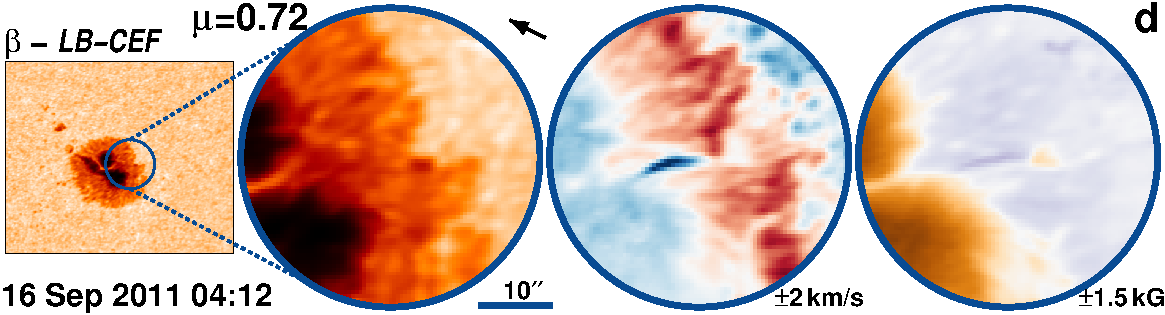}
 \includegraphics[width=.48\textwidth]{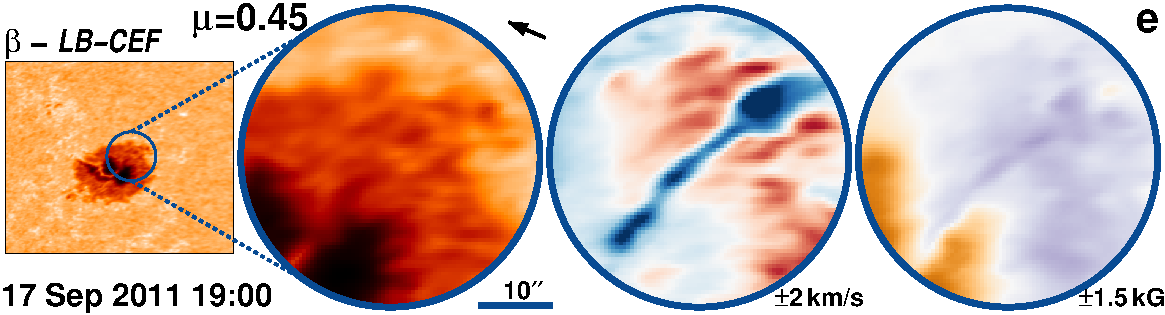}
 \caption{AR\,11289 followed for   10.3 days from  8-Sep-2011 starting at 06:12\,UT.\label{fig:DS111}}
 \end{figure*}

\begin{figure*}[htbp]
 \centering
 \includegraphics[width=.48\textwidth]{colorbars.pdf}
 \includegraphics[width=.48\textwidth]{colorbars.pdf}
 \includegraphics[width=.48\textwidth]{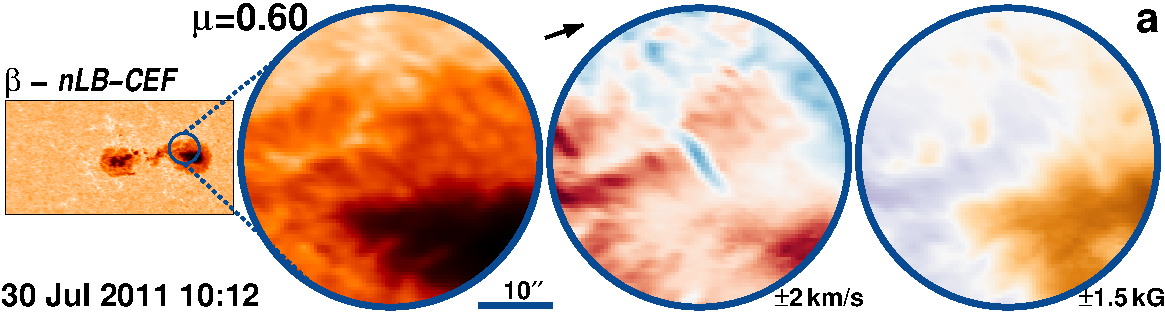}
 \includegraphics[width=.48\textwidth]{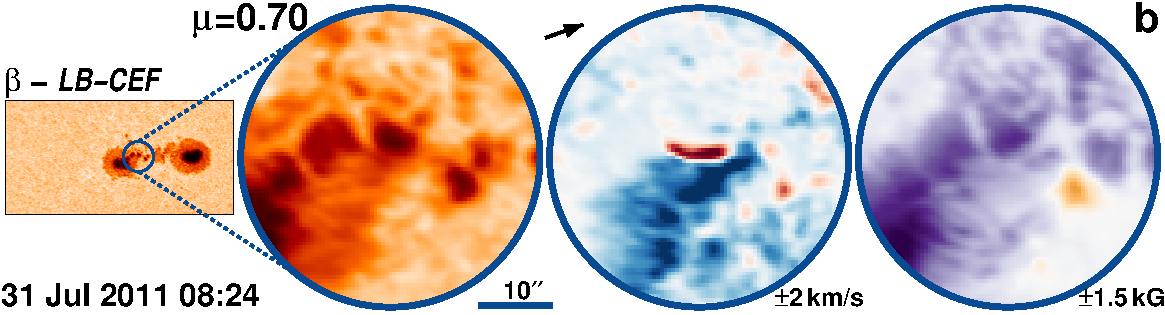}
 \includegraphics[width=.48\textwidth]{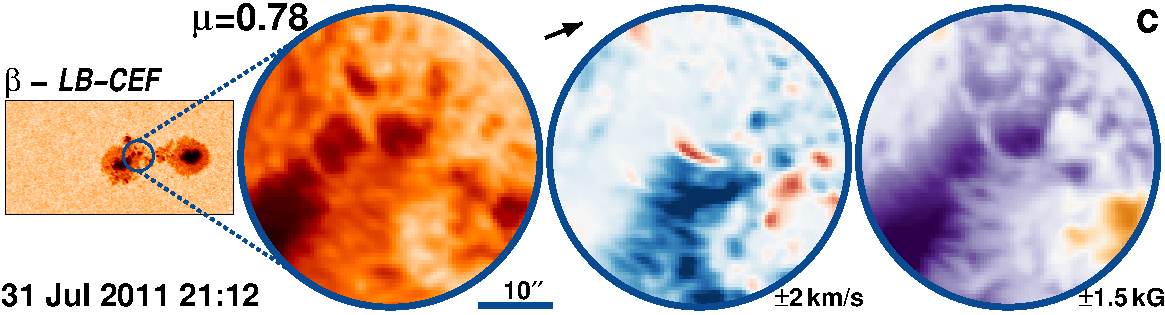}
 \includegraphics[width=.48\textwidth]{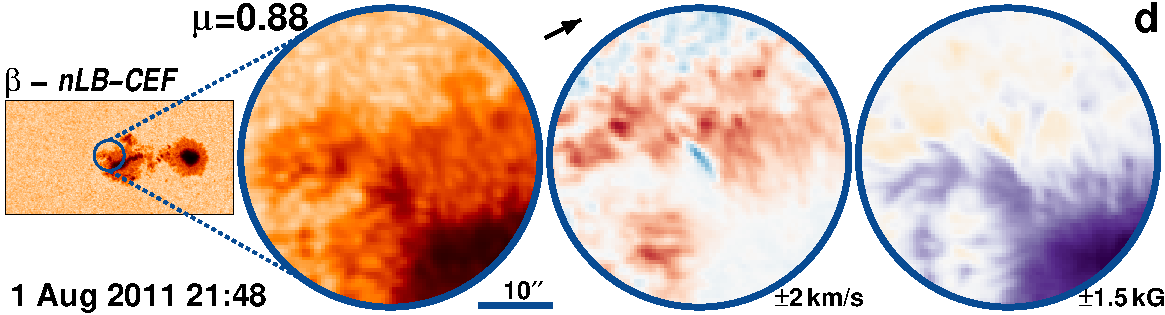}
 \caption{AR\,11263 followed for    9.5 days from 30-Jul-2011 starting at 00:12\,UT.\label{fig:DS112}}
 \end{figure*}

\end{appendix}
\end{document}